\documentclass[prd,twocolumn,showpacs,showkeys,preprintnumbers,floatfix,nofootinbib,superscriptaddress]{revtex4-1}
\usepackage{amsfonts} 
\usepackage{amssymb} 
\usepackage{amsmath} 
\usepackage{graphicx} 
\usepackage{subfigure} 
\usepackage{array} 
\usepackage{dcolumn} 
\usepackage{bm} 
\usepackage{latexsym} 
\usepackage{longtable} 
\usepackage{hyperref} 
\usepackage{bbold}
\usepackage{color}
\usepackage{multirow}
\usepackage{rotating}
\usepackage{comment}

\graphicspath{{./figs/}}

\newcommand{\tsepi}{\mathop{\tau \to \infty}\nolimits}
\newcommand{\tskip}{\mathop{t_{\rm skip}}\nolimits}
\newcommand{\MeV}{\mathop{\rm MeV}\nolimits}
\newcommand{\GeV}{\mathop{\rm GeV}\nolimits}

\newcommand{\fm}{\mathop{\rm fm}\nolimits}
\newcommand{\gsim}{\raisebox{-0.7ex}{$\stackrel{\textstyle >}{\sim}$ }}
\newcommand{\lsim}{\raisebox{-0.7ex}{$\stackrel{\textstyle <}{\sim}$ }}
\newcommand{\FIXME}[1]{}
\newcommand{\expv}[1]{\langle#1\rangle}

\newcommand{\rE}{\mathop{ r_E }\nolimits}
\newcommand{\rM}{\mathop{ r_M }\nolimits}
\newcommand{\rEsq}{\mathop{ \langle r_E^2 \rangle }\nolimits}
\newcommand{\rMsq}{\mathop{ \langle r_M^2 \rangle }\nolimits}
\newcommand{\rDirac}{\mathop{ \langle r_1^2 \rangle }\nolimits}
\newcommand{\rPauli}{\mathop{ \langle r_2^2 \rangle }\nolimits}

\newcommand{\exprE}{\mathop{\langle r_E^2 \rangle}\nolimits}
\newcommand{\exprM}{\mathop{\langle r_M^2 \rangle}\nolimits}

\renewcommand{\Re}{\mathop{\rm Re}}
\renewcommand{\Im}{\mathop{\rm Im}}

\providecommand{\CL}{\nonumber\\}
\providecommand{\abs}[1]{\lvert#1\rvert}
\providecommand{\matrixe}[3]{\langle#1\lvert#2\rvert#3\rangle}

\definecolor{green}{rgb}{0.1, 0.8, 0.1}

\begin{document}


\title{Nucleon Electromagnetic Form Factors in the Continuum Limit from 2+1+1-flavor Lattice QCD}
\author{Yong-Chull~Jang}
\email{ypj@bnl.gov}
\affiliation{Brookhaven National Laboratory, Upton, NY 11973, USA}

\author{Rajan~Gupta}
\email{rajan@lanl.gov}
\affiliation{Los Alamos National Laboratory, Theoretical Division T-2, Los Alamos, NM 87545, USA}

\author{Huey-Wen~Lin}
\email{hwlin@pa.msu.edu}
\affiliation{Department of Physics and Astronomy, Michigan State University, MI, 48824, USA}
\affiliation{Department of Computational Mathematics,  Science and Engineering, Michigan State University, East Lansing, MI 48824}

\author{Boram~Yoon}
\email{boram@lanl.gov}
\affiliation{Los Alamos National Laboratory, Computer Computational and Statistical Sciences, CCS-7, Los Alamos, NM 87545}

\author{Tanmoy~Bhattacharya}
\email{tanmoy@lanl.gov}
\affiliation{Los Alamos National Laboratory, Theoretical Division T-2, Los Alamos, NM 87545, USA}

\collaboration{PNDME Collaboration}
\preprint{LA-UR-19-25275}
\preprint{MSUHEP-19-006}
\date{\today}
\begin{abstract}
We present results for the isovector $(p-n)$ electromagnetic form
factors of the nucleon using eleven ensembles of gauge configurations
generated by the MILC collaboration using the highly improved
staggered quark (HISQ) action with 2+1+1 dynamical flavors. These
ensembles span four lattice spacings $a \approx$ 0.06, 0.09, 0.12 and
0.15~fm and three values of the light-quark masses corresponding to
the pion masses $M_\pi \approx 135, 225$ and $315
\MeV$. High-statistics estimates using the truncated solver method
method allow us to quantify various systematic uncertainties and
perform a simultaneous extrapolation in the lattice spacing, lattice
volume and light-quark masses. We analyze the $Q^2$ dependence of the
form factors calculated over the range $0.05 \lesssim Q^2 \sim
1.4$~GeV${}^2$ using both the model independent $z$-expansion and the
dipole ansatz.  Our final estimates, using the $z$-expansion fit, for
the isovector root-mean-square radius of nucleon are $\rE =
0.769(27)(30)\ {\rm fm}$, $\rM = 0.671(48)(76)\ {\rm fm} $ and
$\mu^{p-n} = 3.939(86)(138)$ Bohr magneton. The first error is the
combined uncertainty from the leading-order analysis, and the second
is an estimate of the additional uncertainty due to using the leading
order chiral-continuum-finite-volume fits. The estimates from the
dipole ansatz, $\rE = 0.765(11)(8)\ {\rm fm}$, $\rM =
0.704(21)(29)\ {\rm fm} $ and $\mu^{p-n} = 3.975(84)(125)$ Bohr
magneton, are consistent with those from the $z$-expansion but with
smaller errors.  Our analysis highlights three points. First, all our
data for form factors from the eleven ensembles and existing lattice
data on, or close to, physical mass ensembles from other
collaborations collapses more clearly onto a single curve when plotted
versus $Q^2/M_N^2$ as compared to $Q^2$ with the scale set by
quantities other than $M_N$. The difference between these two ways of
analyzing the data is indicative of discretization errors, some of
which presumably cancel when the data are plotted versus
$Q^2/M_N^2$. Second, the size of the remaining deviation of this
common curve from the Kelly curve is small and can be accounted for by
statistical and possible systematic uncertainties. Third, to improve
lattice estimates for $\rEsq$, $\rMsq$ and $\mu$, high statistics data
for $Q^2 < 0.1$~GeV${}^2$ are needed.
\end{abstract}
\maketitle
%
%
%
%
\section{Introduction}
\label{sec:into}
\FIXME{sec:into} 

Experiments studying electron scattering off protons and neutrons have
a long history of providing an understanding of the structure of
nucleons~\cite{Perdrisat:2006hj,Punjabi:2005wq}. Quantitative
understanding of the distribution of charge is described by the
electric and magnetic form factors, $G_E$ and
$G_M$~\cite{Bernauer:2013tpr}. Quantities of phenomenological
interest obtained from the slope of the form factors at space-like four momentum transfer squared $Q^2=0$ are
the electric and magnetic charge radii of the nucleons. At present
there is a 6$\sigma$ discrepancy between the electric charge radius of
the proton obtained from electronic energy levels combined with
electron scattering data~\cite{Mohr:2015ccw} versus that from the Lamb
shift in muonic hydrogen $E_{\mu
  p}(2S-2P)$~\cite{Antognini:1900ns,Antognini:2015moa}. A second issue
that needs resolution is the behavior of the ratio $G_E/G_M$ at $Q^2 >
1$~GeV${}^2$~\cite{Ye:2017gyb}, and whether this ratio crosses zero at about
$8$~GeV${}^2$ as indicated by experiments at JLab~\cite{Gayou:2001qd,GEGMratio}. In this work, we focus on
determining the electromagnetic form factors in the range $0.05
\lesssim Q^2 \lesssim 1$~GeV${}^2$ and extracting the charge radii
from them.

The electric and magnetic form factors, $G_E$ and $G_M$, of the
nucleon can be calculated directly from large scale simulations of
lattice QCD. In recent years, advances in algorithms and computing
power have allowed the community to push the calculations towards
physical masses for the light $u$ and $d$ quarks, and on lattice
spacings that are small enough that discretization effects are
expected to be at the few percent
level~\cite{Bhattacharya:2013ehc,Green:2014xba,Capitani:2015sba,Alexandrou:2017msl}.
In this paper we present results from thirteen calculations on eleven
ensembles that cover a range of lattice spacings ($0.06 \, \lsim a \,
\lsim 0.15$~fm), pion masses ($135 \, \lsim M_\pi \, \lsim 320$~MeV)
and lattice volumes ($3.3\, \lsim M_\pi L\, \lsim5.5$). These
ensembles were generated using $2+1+1$-flavors of highly improved
staggered quarks (HISQ)~\cite{Follana:2006rc} by the MILC
collaboration~\cite{Bazavov:2012xda}.  This suite of calculations
allows us to understand and assess various sources of systematic
errors. The analysis is carried out using both the dipole ansatz and
the $z$-expansion, which give consistent estimates for the isovector
mean-square charge radii $\rEsq$ and $\rMsq$ and the magnetic moment $\mu^{p-n}$.

Our final results for the isovector mean-square charge radii $\rEsq$
and $\rMsq$ (also for Dirac, $\rDirac$, and Pauli, $\rPauli$, radii)
defined in Eqs.~\eqref{eq:rdef},~\eqref{eq:rDPdef} and~\eqref{eq:rDP},
and for the magnetic moment $\mu$ are given in
Tab.~\ref{tab:finalresults}. We also present a comparison with other
lattice data obtained close to the physical pion mass and with the
Kelly parameterization of the experimental data~\cite{Kelly:2004hm} in
Fig.~\ref{fig:GEMcomp}.  Our estimates for $\rEsq$, $\rMsq$ and
$\mu^{p-n}$ are about 17\%, 19\%, and 16\% smaller than the
phenomenological values given in Eq.~\eqref{eq:isovectorradii} and the precise experimental value in 
Eq.~\eqref{eq:mu_expt}.  Throughout this paper, we have paid attention
to the size of possible statistical and systematic errors, and find
that a linear combination of these is large enough to explain the deviations.

We analyze the world data for $G_E$ and $G_M$ in
Sec.~\ref{sec:comparison} and find that data from all 13 of our
calculations and those from other collaborations done at or near the
physical pion mass fall roughly onto a single curve when plotted
versus versus $Q^2$ or $Q^2/M_N^2$. However, there is a noticeable
shift between the two curves when compared to the Kelly fit. The
difference between the two ways of analyzing the data is a
discretization artifact: specifically, it is a consequence of the
difference in values of the lattice scale obtained from different
observables. The size of the difference again indicates that the
present underestimate of $\rEsq$, $\rMsq$ and $\mu$ should not be
considered significant. Our overall conclusion is that to
significantly reduce the systematics and improve the precision with
which these observables can be extracted will require high statistics
data at smaller values of the lattice spacing and of $Q^2 < 0.1$~GeV${}^2$.

We stress that the long-term goal of lattice QCD is to directly
predict the form factors and not to reproduce the Kelly curve, a
parameterization of the experimental data. Throughout this paper, we
use the Kelly curve to provide a reference point for comparison, and
for discussing systematics and trends in the lattice data.  We do not 
show an error band on the Kelly curve as it is negligible on the scale of 
the errors in the lattice data. 

This paper is organized as follows. In Sec.~\ref{sec:formfactors}, we
review the theory, computational approach and the status of the
experimental and phenomenological results.  In
Sec.~\ref{sec:Methodology}, we describe the salient features of the
calculation. The fits used to isolate excited-state contamination
(ESC) and extract the form factors are described in
Sec.~\ref{sec:FFanalysis}.  Fits to quantify the $Q^2$ behavior of the
$(p-n)$ form factors are discussed in Sec.~\ref{sec:Q2behavior}, and
the extraction of our final results for the isovector mean-square charge radii,
$\langle r_{E}^2 \rangle$ and $\langle r_M^2 \rangle$, and the
anomalous magnetic moment $\mu^{p-n}$ are presented in
Sec.~\ref{sec:results}.  Comparisons with form factors extracted from
experiments and with previous lattice QCD calculations are made in
Sec.~\ref{sec:comparison}. We end with conclusions in
Sec.~\ref{sec:conclusions}. Some further details of the calculations
are given in four Appendices: lattice parameters in
Appendix~\ref{appendix:parameters}, analysis of nucleon mass in
Appendix~\ref{appendix:NucleonMass}, ESC in
Appendix~\ref{appendix:ESC}, and a review of the experimental data for
the form factors in Appendix~\ref{appendix:expFF}.

\section{Electromagnetic Form Factors of the Nucleon}
\label{sec:formfactors}
\FIXME{sec:formfactors} 

The Dirac, $F_{1}$, and Pauli, $F_{2}$, form factors are extracted
from the matrix elements of the electromagnetic current within the
nucleon state $N$ through the relation
\begin{multline}
\label{eq:VFF}
\left\langle N(\vec{p}_f) | V_\mu^{\rm em} (\vec{q}) | N(\vec{p}_i)\right\rangle  = \\
{\overline u}_N(\vec{p}_f)\left( F_1(Q^2) \gamma_\mu
+\sigma_{\mu \nu} q_\nu
\frac{F_2(Q^2)}{2 M_N}\right)u_N(\vec{p}_i),
\end{multline}
where $\vec{q}=\vec{p}_f-\vec{p}_i$ is the momentum transfer. The
discrete lattice momenta are given by $2 \pi {\bf n} / L a$ with the
entries of the vector ${\bf n} \equiv (n_1, n_2, n_3)$ taking on
integer values, $n_i \in \{0,L\}$. The spacing between the momenta is
controlled by the spatial lattice size, $La$. The normalization used
for the nucleon spinors in Euclidean space is
\begin{equation}
\sum_{s} u_N(\vec p,s) \bar{u}_N(\vec{p},s) =
   \frac{E(\vec{p})\gamma_4-i\vec\gamma\cdot \vec{p} + M}{2 E(\vec{p})} \,.
\label{eq:spinor}
\end{equation}
and in Eq.~\eqref{eq:VFF}, the electromagnetic current is
\begin{equation}
V_\mu^{\rm em} = \frac{2}{3} {\overline u} \gamma_\mu u - \frac{1}{3} {\overline d} \gamma_\mu d \,.
\label{eq:Vem}
\end{equation}
In the isospin symmetric limit, the difference of its matrix elements between a proton and a neutron state 
are related to the isovector form factors of the proton by the relation 
\begin{multline}
\left\langle p(\vec{p}_f) | \overline{u} \gamma_\mu u - \overline{d} \gamma_\mu d | p(\vec{p}_i)\right\rangle  
= \\
\left\langle p(\vec{p}_f) | V_\mu^{\rm em} (\vec{q}) | p(\vec{p}_i)\right\rangle  
- \left\langle n(\vec{p}_f) | V_\mu^{\rm em} (\vec{q}) | n(\vec{p}_i)\right\rangle  \,.
\label{eq:VFF2}
\end{multline}
The quantity we calculate on the lattice is the left hand side of
Eq.~\eqref{eq:VFF2}, i.e., the isovector form factors of the proton.
Throughout this paper, the term isovector form factors of the proton
and the $(p-n)$ form factors refer to the same quantities as defined
in Eq.~\eqref{eq:VFF2}.  These will henceforth be analyzed in terms of
the space-like 4-momentum squared, $Q^2 = {\vec p}^2 - (E-m)^2 = -q^2$.

Another common set of definitions of the electromagnetic form factors,
widely used in the analysis of experimental data, are the Sachs
electric, $G_{E}$, and magnetic, $G_{M}$, form factors that are related to the Dirac and
Pauli form factors as
\begin{align}\label{eq:sachs}
G_E(Q^2) &= F_1(Q^2) - \frac{Q^2}{4M_N^2}F_2(Q^2) \\
G_M(Q^2) &= F_1(Q^2) + F_2(Q^2).
\end{align}
From these, the vector charge is given by 
\begin{equation}
g_V = G_E|_{Q^2=0} = F_1|_{Q^2=0}
\end{equation}
and the difference between the magnetic moment of the proton and the neutron by
\begin{equation}
\mu^{p} - \mu^{n} = G_M|_{Q^2=0} = (F_1 + F_2)|_{Q^2=0} = 1 + \kappa_p - \kappa_n \,.
\label{eq:Pmmdef}
\end{equation}
The anomalous magnetic moments of
the proton and the neutron, in units of the Bohr magneton, 
are known very precisely~\cite{Olive:2016xmw}:
\begin{eqnarray}
\kappa_p  &=& 1.79284735(1)    \qquad\ ({\rm proton})    \,,  \nonumber \\
\kappa_n  &=& -1.91304273(45)   \qquad ({\rm neutron}) \,.
\label{eq:mu_expt}
\end{eqnarray}

The electric and magnetic size of the nucleon are defined as the slope of the form
factors with respect to $Q^2$ at $Q^2=0$~\cite{Miller:2018ybm}:
\begin{equation}
\langle r_{E,M}^2\rangle = -6\frac{d}{dQ^2}\left.\left(\frac{G_{E,M}(Q^2)}{G_{E,M}(0)}\right)\right|_{Q^2=0} \,.
\label{eq:rdef}
\end{equation}
The form factors $G_{E,M}$ are normalized by their values at $Q^2=0$:
$G_E(Q^2=0) \equiv g_V$ and $G_M(Q^2=0)/g_V \equiv \mu$. This definition
makes them independent of the renormalization constant, $Z_V$, of the
lattice vector current, and improves the signal because some of the
systematics cancel in the ratios. Therefore, in this work, we will use
Eq.~\eqref{eq:rdef} when calculating $\exprE$ and $\exprM$. Note that
$Z_V g_V = 1$ as the electric charge is conserved.  A second
independent estimate of $Z_V$, obtained using nonperturbative lattice
calculations in the RI-sMOM scheme, is given in
Ref.~\cite{Gupta:2018qil}, where the difference between the two
estimates was shown to be $\lesssim 3\%$.

One similarly defines the isovector Dirac and Pauli mean-square radii as 
\begin{equation}
\langle r_{1,2}^2\rangle = -6\frac{d}{dQ^2}\left.\left(\frac{F_{1,2}(Q^2)}{F_{1,2}(0)}\right)\right|_{Q^2=0} \,.
\label{eq:rDPdef}
\end{equation}
These are related to $\rEsq$, $\rMsq$ and $\mu \equiv 1 + \kappa $ as 
\begin{align}
\rDirac  &=   \rEsq - \frac{6\kappa}{4M_N^2} \,, \nonumber \\
\kappa \rPauli  &=   \mu \rMsq  - \rEsq + \frac{6\kappa}{4M_N^2} \,. 
\label{eq:rDP}
\end{align}
Our analysis of the lattice data is carried out in terms of $G_E$ and
$G_M$. Results for $\rDirac$ and $\rPauli$ are also given in
Table~\ref{tab:finalresults} in Sec.~\ref{sec:results}, where we
extract $\rEsq$ and $\rMsq$.

The electric root-mean-square charge radius $\rE \equiv \sqrt{\langle
  r_E^2\rangle}$ of the proton has been measured in three ways: (i)
laser spectroscopy of the Lamb shift in muonic
hydrogen~\cite{Antognini:1900ns,Antognini:2015moa,Krauth:2017ijq},
(ii) continuous-wave laser spectroscopy of
hydrogen~\cite{Pohl:2016glp}, and (iii) elastic scattering of
electrons off protons~\cite{Sick:2014sra,Sick:2017aor}. Results using
electrons, i.e., the latter two ways, are included in the CODATA-2014
world average~\cite{Tanabashi:2018oca,Mohr:2015ccw}:
\begin{eqnarray}
r_E^p &=& 0.875(6)~\fm    \qquad\ {\rm CODATA-2014}    ,  \nonumber \\
r_E^p &=& 0.8414(19)~\fm    \quad\ {\rm CODATA-2018}    ,  \nonumber \\
r_E^p &=& 0.8409(4)~\fm   \quad\ \  E_{\mu p}(2S-2P)        \,,
\label{eq:rE_expt}
\end{eqnarray}
and the third result is from muonic hydrogen. 
The large difference between the CODATA-2014 and muonic-hydrogen values
was termed the ``proton radius puzzle''.  The new CODATA-2018 value~\cite{CODATA:2018}
resolves the puzzle in favor of the muonic-hydrogen result.
The magnetic radius of the proton extracted from experiments using
electrons is~\cite{Tanabashi:2018oca,Mohr:2015ccw}
\begin{equation}
r_M^p = 0.776(38)~\fm    \qquad {\rm electrons}       \,.
\label{eq:rM_expt}
\end{equation}
Values for the isovector charge radii, extracted from the experimental
data and used to compare lattice data against, are given in
Eq.~\eqref{eq:isovectorradii} in Appendix~\ref{appendix:expFF}.

To reduce the uncertainty in results from electron scattering
experiments, which have been done down to $Q^2 \approx
0.004$~GeV${}^2$, new experiments to constrain the low $Q^2$ behavior
have been initiated~\cite{Weber:2016plr,Gasparian:2017cgp}.
Similarly, for lattice QCD calculations to help resolve the puzzle, we
need to calculate the form factors to $Q^2 \approx 0.004$~GeV${}^2$ to
extract $\rE$ with better than 1\% accuracy.

A challenge to the direct extraction of $\langle r_i^2\rangle$ from
the lattice data is that the value of the smallest momenta, $2 \pi / L
a$, is large in typical lattice simulations. In our calculations, it
is $\gsim 220$~MeV, and the range of $Q^2$ values, given
in Table~\ref{tab:Q2-ALL}, are between 2--10 $M_\pi^2$. It is,
therefore, traditional to fit the data for the $G_i$ to an ansatz, and
then use the fit to evaluate the derivative given in
Eq.~\eqref{eq:rdef}. Both, using an ansatz and estimating its
parameters from fits to data with $\sqrt{Q^2} \ \gsim 200$~MeV
introduces systematic uncertainties when evaluating the derivative at
$Q^2=0$.  We estimate the dependence of $\langle r_i^2\rangle$ on the
choice of the ansatz by comparing results for each ensemble obtained
using two different fits, the dipole model and the $z$-expansion.

Two alternate approaches are, one, to calculate the form factors at fixed
$Q^2$ and extrapolate these to the continuum limit first and then fit
the $Q^2$ behavior.  Unfortunately, the values of $Q^2$ are different
on each ensemble. Second, combine the dipole or the
$z$-expansion parameterization of the $Q^2$ behavior with the
chiral-continuum-finite volume (CCFV) ansatz for one overall fit. This combined fit is 
discussed in Sec.~\ref{sec:CombinedFit}. The central 
analysis presented here consists of first fitting the data versus $Q^2$ using
the dipole model and the $z$-expansion to extract $\langle
r_i^2\rangle$ and $\mu$ on each ensemble and then get the physical
results from a CCFV fit in $a$, $M_\pi$ and $M_\pi L$ that
addresses the associated systematics.

It is important to note that both the electron scattering
experiments and lattice QCD calculations suffer from paucity of data
close to $Q^2=0$ that impacts the extraction of the charge radii. However, 
there is a large range, $0.004 \lesssim Q^2 \lesssim 1$~GeV${}^2$ over
which accurate experimental data exist. Thus, more than just 
extracting the charge radii, our goal is to directly compare the
lattice and the experimental data over this range of $Q^2$ as
discussed in Sec.~\ref{sec:Q2behavior}. 

An ansatz that is commonly used to fit the experimental data is the
dipole. It arises if one assumes an exponentially falling charge
distribution.  The resulting form factor is characterized by a single
parameter, the mass ${\cal M}$,
\begin{equation}
G_i(Q^2) = \frac{G_i(0)}{(1+Q^2/{\cal M}_i^2)^{2}} \quad \Longrightarrow \quad \langle r_i^2\rangle = \frac{12}{{\cal M}_i^2} \,,
\label{eq:dipole}
\end{equation}
and normalized to $F_1 = G_E = g_V$ at $Q^2=0$. It goes as $Q^{-4}$
in the $Q^2 \to \infty$ limit in accord with perturbation theory~\cite{Lepage:1980fj}.

The second ansatz is a model-independent parameterization called the
$z$-expansion~\cite{Hill:2010yb,Bhattacharya:2011ah}:
\begin{equation}
\frac{G_{E,M}(Q^2)}{G_E(0)} = \sum_{k=0}^{\infty} a_k z(Q^2)^k  \,,
\label{eq:Zexpansion}
\end{equation}
where the $a_k$ are fit parameters and $z$ is defined as
\begin{equation}
  z = \frac{\sqrt{t_\text{cut}+Q^2}-\sqrt{t_\text{cut}+\bar{t_0}}}
           {\sqrt{t_\text{cut}+Q^2}+\sqrt{t_\text{cut}+\bar{t_0}}} \,,
\label{eq:Zdef}
\end{equation}
with $t_\text{cut} = 4 M_\pi^2$ denoting the nearest singularity in
$G_{E,M}(Q^2)$.  In terms of $z$, the domain of analyticity of
$G_E(Q^2)$ is mapped into the unit circle with the branch cut at
$Q^2=-4M_\pi^2$~\cite{Bhattacharya:2011ah}. We analyzed the data with
$\bar{t_0}=0$ and $\bar{t_0}^{\rm
  mid}=\{0.12,\ 0.20,\ 0.40\}$~GeV${}^2$ for the $M_\pi \approx
\{135,\ 220,\ 315\}$~MeV ensembles. By choosing the value of the
constant $\bar{t_0}$ to lie in the middle of the range of $Q^2$ at
which we have data, one reduces $z_{\rm max}$. By reducing the value
of $z_{\rm max}$ we hope to improve the stability of the estimates,
with improvement judged by comparing result from different truncations
of the series.  In practice, for our data set, we find that the
quality of the fits and the results are insensitive to the choice of
$\bar{t_0}$. The final results for the charge radii and magnetic
moment are obtained from fits using $\bar{t_0}^{\rm mid}$.

The values of $Q^2$ for the thirteen calculations are given in
Table~\ref{tab:Q2-ALL}. Note that in four cases the number of nonzero
values are only five.  The data for $G_E(Q^2)$ and $G_M(Q^2)$ versus
$z$ with $\bar{t_0}^{\rm mid}$ are shown in Fig.~\ref{fig:versusZ}. As
discussed in Sec.~\ref{sec:Q2lattice}, we restrict our fits to $Q^2
\le 1$~GeV${}^2$ because the reliability of some of the higher $Q^2$
data is questionable.

To implement the perturbative behavior $G_i(Q^2) \to Q^{-4}$ as $Q^2 \to \infty$~\cite{Lepage:1980fj} in
the $z$-expansion requires $Q^n G_i(Q^2) \to 0$ for $n=0, 1, 2,
3$. These constraints can be incorporated into the z-expansion 
as four sum rules~\cite{Lee:2015jqa}
\begin{equation}
\sum_{k=n}^{k_{\rm max}} k(k-1) \ldots (k-n+1) a_k = 0 \,, \qquad n=0,1,2,3 \,.
\label{eq:sumrule}
\end{equation}
For $n=0$ it reduces to  $\sum_{k=0}^{k_{\rm max}} a_k = 0$.  A priori, using these
sum rules ensures that the $a_k$ are not only bounded but must also
decrease at large $k$~\cite{Lee:2015jqa}.  

A key issue in the $z$-expansion analysis is the value of $k^{\rm
  max}$ required to obtain results with a certain precision. The
analysis of the experimental data carried out in
Appendix~\ref{appendix:expFF} shows that results stabilize for $k_{\rm
  max} \approx 4$ with and without sum rules.  For the lattice data, the
choice has to take into account the number of values of $Q^2$ at which
data have been generated to not over-parameterize the fit.  For
our data and fits without priors, the $a_k$ fluctuate and the higher
order coefficients ($k \ge 4$) are ill determined due to the
over-parameterization of the fits. To avoid the resulting large
fluctuations in $a_k$, we put a bound on them as suggested
in~\cite{Lee:2015jqa}. For $G_E$ and $G_M/5$, we constrain $|a_k|
\lesssim 5.0$ for all $k$ by using Gaussian priors with central value
zero and width five.  With this constraint, results for $\rEsq$,
$\rMsq$ and $\mu$ do not change significantly for $k_{\rm max} \le 3$
and stabilize for $k_{\rm max} \ge 4$ as shown in
Fig.~\ref{fig:stabilityz}.  The convergence of estimates from fits
with sum rules is slower and occurs for $k_{\rm max} \ge 7$ as also
shown in Fig.~\ref{fig:stabilityz}.  We, therefore, use the fits with
$k_{\rm max} =4$ and without sum rules for our final results as they
converge faster. Results with sum rules, which converge for $k_{\rm
  max} \ge 7$, are used only as consistency checks. Since $\rEsq$,
$\rMsq$ and $\mu$ are best extracted from data at small $Q^2$, the sum
rule constraints imposed to guarantee the large $Q^2$ behavior are not
essential for their determination.

Overall, the fits to $G_E(Q^2)$ are more stable than those to
$G_M(Q^2)$.  The main reason is the extra data point at $G_E(Q^2=0)$
which pins down the sign of the slope of $G_E(Q^2)$ at small $Q^2$.
Using a value for $G_M(0)$, derived from the ratio $G_M(Q^2)/G_E(Q^2)$ 
as discussed in Sec.~\ref{sec:extractGM}, greatly improved the stability
of fits to $G_M(Q^2)$.

\section{Lattice Methodology}
\label{sec:Methodology}
\FIXME{sec:Methodology} 

The parameters of the thirteen calculations done on eleven HISQ
ensembles are the same as used in Ref.~\cite{Gupta:2018qil} for the
calculation of isovector charges. To keep the paper self-contained,
the lattice parameters of the calculations and the number of
measurements made are summarized in Table~\ref{tab:ens} in the
appendix~\ref{appendix:parameters}.  The parameters used to generate
the Wilson-clover quark propagators using the multigrid
algorithm~\cite{Babich:2010qb} are also given in
Table.~\ref{tab:cloverparams}.  We remind the reader that two
ensembles, $a06m310$ and $a06m220$, have been analyzed twice with
different smearing parameters giving a total of 13 calculations.
Also, compared to Refs.~\cite{Bhattacharya:2016zcn,Rajan:2017lxk}, six
ensembles ($a12m220S$, $a12m220$, $a12m220L$, $a09m310$, $a09m220$ and
$a09m130W$) have been simulated afresh with randomly chosen source
points on each configuration to increase their statistical
independence, and data at a larger number of momenta have been accumulated.

To increase the statistics cost-effectively, we used the truncated
solver with bias correction method~\cite{Bali:2009hu,Blum:2012uh}.  We
also used the coherent source method to construct sequential
propagators from the sink time slice, at which a zero-momentum nucleon
state is inserted~\cite{Bratt:2010jn,Yoon:2016dij}.

The details of our strategy for the calculations and the analysis have been
published in earlier
works~\cite{Bhattacharya:2016zcn,Rajan:2017lxk,Gupta:2018qil}.  Here
we provide a brief summary of the points relevant to the calculation
of the electric and magnetic form factors: 
\begin{itemize}
\item
All errors are determined using a single elimination jackknife method
over configurations, i.e., we first construct the bias corrected
average for each configuration and then carry out the fits to the two-
and three-point functions within the same jackknife procedure over
these configuration averages.
\item
To control excited-state contamination, we use the same toolkit as in
Ref.~\cite{Gupta:2018qil}. The 2-point functions are fit keeping four
states in the spectral decomposition. The amplitudes and the masses
obtained from these fits are input into the analysis of three-point
functions. The results for the masses are given in Table~\ref{tab:spectrum} in 
Appendix~\ref{appendix:NucleonMass}.
\item
On each ensemble, we calculate the three-point functions at multiple
values of source-sink separation $\tau$. These values of $\tau$, given in
Table~\ref{tab:ens}, are the same as in Ref.~\cite{Gupta:2018qil}.
\item
The insertion of the vector current at definite momenta $\bm{p}$ is
carried out on each time slice $t$ between the source and the sink,
and for each value of $\tau$. These data for the three-point
functions, $C_\Gamma^{(3\text{pt})} (t;\tau;\bm{p}^\prime,\bm{p})$, 
at a large number of values of $t$ and $\tau$ are fit 
using three states in the spectral decomposition:
\begin{align}
  C_\Gamma^{(3\text{pt})}&(t;\tau;\bm{p}^\prime,\bm{p}) = \CL
   &\abs{\mathcal{A}_0^\prime} \abs{\mathcal{A}_0}\matrixe{0^\prime}{\mathcal{O}_\Gamma}{0} e^{-E_0t - M_0(\tau-t)} + \CL
   &\abs{\mathcal{A}_0^\prime} \abs{\mathcal{A}_1}\matrixe{0^\prime}{\mathcal{O}_\Gamma}{1} e^{-E_0t - M_1(\tau-t)} + \CL
   &\abs{\mathcal{A}_1^\prime} \abs{\mathcal{A}_0}\matrixe{1^\prime}{\mathcal{O}_\Gamma}{0} e^{-E_1t -M_0(\tau-t)} + \CL
   &\abs{\mathcal{A}_1^\prime} \abs{\mathcal{A}_1}\matrixe{1^\prime}{\mathcal{O}_\Gamma}{1} e^{-E_1t - M_1(\tau-t)} + \CL 
   &\abs{\mathcal{A}_0^\prime} \abs{\mathcal{A}_2}\matrixe{0^\prime}{\mathcal{O}_\Gamma}{2} e^{-E_0t - M_2(\tau-t)} + \CL
   &\abs{\mathcal{A}_2^\prime} \abs{\mathcal{A}_0}\matrixe{2^\prime}{\mathcal{O}_\Gamma}{0} e^{-E_2t -M_0(\tau-t)} + \CL
   &\abs{\mathcal{A}_1^\prime} \abs{\mathcal{A}_2}\matrixe{1^\prime}{\mathcal{O}_\Gamma}{2} e^{-E_1t - M_2(\tau-t)} + \CL
   &\abs{\mathcal{A}_2^\prime} \abs{\mathcal{A}_1}\matrixe{2^\prime}{\mathcal{O}_\Gamma}{1} e^{-E_2t -M_1(\tau-t)} + \CL
   &\abs{\mathcal{A}_2^\prime} \abs{\mathcal{A}_2}\matrixe{2^\prime}{\mathcal{O}_\Gamma}{2} e^{-E_2t - M_2(\tau-t)} \,, 
   \label{eq:3pt}
\end{align}
where the source point is translated to $t=0$, the operator is
inserted at time $t$, and the nucleon state is annihilated at the sink
time slice $\tau$, which numerically is also the source-sink
separation. In this relation, the numbers refer to the state
$|n\rangle$, a state with superscript ${}^\prime$ denotes that it could have 
nonzero momentum $\bm{p}^\prime$, and the momentum $\bm{p}$ at the
sink is fixed to zero.
\item
With our data, the term $\matrixe{2^\prime}{\mathcal{O}_\Gamma}{2}$
could not be resolved.  So, in all the fits we set the contribution of
the term with $\matrixe{2^\prime}{\mathcal{O}_\Gamma}{2}$ equal to
zero, and call these $3^\ast$-state fits.
\item
In the case of $a12m220S$ data, the $\bm{p}^\prime=0$ data are
analyzed using $3^\ast$-state fits, while the $\bm{p}^\prime\ne 0$
data are fit using two states because the $3^\ast$-state fits for $Q^2 \neq 0$ 
are unstable. Having stated this caveat, we will, for
brevity, use the label $3^\ast$-state to describe the excited-state
fits to all data, even those for this ensemble.
\item
The values of $Q^2$ at which the form factors are calculated are
collected in Table~\ref{tab:Q2-ALL}.  These are obtained using the
nucleon ground-state energy $E_p$ extracted using 4-state fits to the
2-point functions.
\item
To extract the desired matrix element
$\matrixe{0^\prime}{\mathcal{O}_\Gamma}{0}$ using Eq.~\eqref{eq:3pt},
the masses $M_i$, energies $E_i$, and the amplitudes
$\abs{\mathcal{A}_i}$ and $\abs{\mathcal{A}_i^\prime}$ are taken from
the fit to the two-point function within one overall jackknife
procedure. This assumes that the ordering of the coupling to the excited states 
is the same as in two-point functions.  To improve the
signal, the amplitude ${\cal A}_0^\prime$ with which the nucleon
interpolating operator at the source time slice couples to the ground
state $|0^\prime\rangle$ with energy $E_0$ and momentum
$\bm{p}^\prime$ should be large while the coupling to excited states
should be small. We find that for the smearing parameters given in
Table.~\ref{tab:cloverparams}, the signal in all the ten momentum
channels analyzed is good.
\item
Off diagonal terms with nonzero momentum transfer such as
$\abs{\mathcal{A}_i^\prime}
\abs{\mathcal{A}_j}\matrixe{i^\prime}{\mathcal{O}_\Gamma}{j}$ are
related to $\abs{\mathcal{A}_j^\prime}
\abs{\mathcal{A}_i}\matrixe{j^\prime}{\mathcal{O}_\Gamma}{i}$ by a combination of Lorentz boost, 
parity and hermitian transformation provided the tower of states and
the coupling to them are the same on either side of the operator. In
our calculation, the nucleon operator used is 
\begin{equation}
\chi(x) = \epsilon^{abc} \left[ {q_1^a}^T(x) C \gamma_5
            \frac{(1 \pm \gamma_4)}{2} q_2^b(x) \right] q_1^c(x)
\label{eq:nucl_op}
\end{equation}
with color indices $\{a, b, c\}$, charge conjugation matrix
$C=\gamma_0 \gamma_2$, and $q_1$ and $q_2$ denoting the two different
flavors of light Dirac quarks.
The quark propagator is smeared both at the source and the sink using a gauge invariant 
Gaussian smearing procedure~\cite{Gusken:1989ad} described in 
Appendix~\ref{appendix:parameters}. 
The nonrelativistic projection $(1 \pm \gamma_4)/2$, inserted to
improve the
signal~\cite{Bhattacharya:2015wna,Bhattacharya:2016zcn,Yoon:2016jzj},
as well as the smearing of the quark fields, breaks Lorentz
covariance. Also, the sink is explicitly constructed to have
$\vec{p}=0$.  We, therefore, treat all such pair of matrix elements as
independent free parameters in the fits.
\item
The data for 3-point functions at nonzero momentum transfer are not
symmetric about the midpoint, $\tau/2$, between the source and the
sink. Nevertheless, in the simultaneous 3-state fit to the data with
multiple source-sink separations $\tau$ and intermediate times $t$, we
skip the same $\tskip$ points adjacent to the source and the sink for
every $\tau$ to remove points with the largest ESC. Two considerations
motivated this choice: (i) the time slice of the onset of the plateau
in the nucleon effective mass plot is roughly independent of the momentum as
shown in Refs.~\cite{Rajan:2017lxk,Gupta:2018qil}, and (ii) because we
choose the values of $\tskip$ to be as small as possible based on 
the stability of the covariance matrix used in the fits. The values of
$\tskip$ used here are the same as in Ref.~\cite{Gupta:2018qil}.
\item
The vector current in the continuum theory is conserved, however the
local vector current used in our lattice calculations is not.  The
renormalization constant $Z_V$ for this current has been determined in
two ways: (i) nonperturbatively in the RI-sMOM scheme and then
converted to $\overline{MS}$ using perturbation theory and (ii)
measured directly from the matrix element of $V_4$ at $Q^2=0$, i.e.,
$1/g_V$. The two sets of values are compared in
Ref.~\cite{Gupta:2018qil} and differ by up to 3\%. This size of
difference is not unreasonable in our clover-on-HISQ formulation which
has discretization effects starting at $O(\alpha_s a)$. Here, we implement method
(ii) by forming ratios $G_i(Q^2)/G_E(0)$, in which some of the
systematics cancel.  The discretization errors in $\rEsq$, $\rMsq$ and $\mu$ are
addressed by the continuum extrapolation, a part of the CCFV fit.
\end{itemize}
The key input, other than statistical precision of the 3-point data,
that impacts the stability of the n-state fits to control ESC and
obtain the ground state matrix elements is the energy of the first
excited state since the terms with
$\matrixe{1^\prime}{\mathcal{O}_\Gamma}{0}$ and
$\matrixe{0^\prime}{\mathcal{O}_\Gamma}{1}$ give the dominant
contribution.  Once the ground-state matrix elements have been
determined, the procedure for obtaining the form factors from them is
described in the next section.


\begin{table*}[tbp]  
\caption{\FIXME{tab:Q2-ALL} The values of the space-like four-momentum
  squared, $Q^2$, transferred to the ground state nucleon, in units of GeV${}^2$.  
  The data for the thirteen calculations defined in Table~\ref{tab:ens} are labeled 
  by the 3-momentum vector $\vec n$. The ground state energy
  is obtained from a 4-state fit.}
\label{tab:Q2-ALL}
\centering
\renewcommand*{\arraystretch}{1.1}
\begin{ruledtabular}
  \begin{tabular}{cccccccc}
$\vec{n}$ & $a15m310$    & $a12m310$  & $a12m220L$   & $a12m220$    & $a12m220S$  & $a09m310$   & $a09m220$   \\
\hline                                               
$(1,0,0)$ & 0.2519(5)    & 0.1765(5)  & 0.0670(1)    & 0.1047(4)    & 0.1747(15)  & 0.1834(3)   & 0.0861(2)  \\
$(1,1,0)$ & 0.4831(14)   & 0.3415(13) & 0.1318(2)    & 0.2060(15)   & 0.3386(35)  & 0.3558(12)  & 0.1685(4)  \\
$(1,1,1)$ & 0.7034(25)   & 0.4982(24) & 0.1947(4)    & 0.3012(20)   & 0.4905(61)  & 0.5198(43)  & 0.2479(8)  \\
$(2,0,0)$ & 0.9111(60)   & 0.6459(35) & 0.2565(8)    & 0.3909(25)   & 0.6358(87)  & 0.6735(44)  & 0.3244(14) \\
$(2,1,0)$ & 1.1020(67)   & 0.7871(42) & 0.3159(10)   & 0.4824(37)   & 0.774(10)   & 0.8186(79)  & 0.3983(18) \\
$(2,1,1)$ & 1.2971(91)   & 0.9202(52) & 0.3740(13)   & 0.5678(47)   & 0.910(13)   & 0.9610(127) & 0.4703(23) \\
$(2,2,0)$ & 1.6372(215)  & 1.178(9)   & 0.4872(21)   & 0.7321(81)   & 1.178(23)   & 1.1974(92)  & 0.6077(37) \\
$(2,2,1)$ & 1.8026(222)  & 1.293(10)  & 0.5413(25)   & 0.8077(103)  & 1.307(25)   & 1.3229(131) & 0.6743(44) \\
$(3,0,0)$ & 1.7896(289)  & 1.315(19)  & 0.5412(28)   & 0.8064(118)  & 1.238(33)   & 1.3248(168) & 0.6713(46) \\
$(3,1,0)$ & 1.9171(314)  & 1.435(18)  & 0.5950(32)   & 0.8845(124)  & 1.358(36)   & 1.4210(144) & 0.7357(51) \\
\hline
$\vec{n}$ & $a09m130W$ & $a06m310$    & $a06m310W$   & $a06m220$   & $a06m220W$  & $a06m135$  &  \\
\hline                                                                           
$(1,0,0)$ & 0.0492(2)  & 0.1888(13)   & 0.1899(6)    & 0.1101(3)   & 0.1093(3)   & 0.0513(2)  & \\
$(1,1,0)$ & 0.0974(5)  & 0.3648(33)   & 0.3653(15)   & 0.2159(11)  & 0.2132(9)   & 0.1014(6)  &  \\
$(1,1,1)$ & 0.1450(9)  & 0.5322(70)   & 0.5277(29)   & 0.3175(24)  & 0.3130(19)  & 0.1510(12) & \\
$(2,0,0)$ & 0.1913(15) & 0.6828(99)   & 0.6895(48)   & 0.4142(46)  & 0.4120(55)  & 0.1975(15) & \\
$(2,1,0)$ & 0.2373(18) & 0.8457(118)  & 0.8402(65)   & 0.5087(57)  & 0.5045(61)  & 0.2459(22) & \\
$(2,1,1)$ & 0.2824(23) &              &              &             &             & 0.2941(32) & \\
$(2,2,0)$ & 0.3704(33) &              &              &             &             & 0.3866(47) & \\
$(2,2,1)$ & 0.4108(41) &              &              &             &             & 0.4323(51) & \\
$(3,0,0)$ & 0.4067(48) &              &              &             &             & 0.4259(60) & \\
$(3,1,0)$ & 0.4490(50) &              &              &             &             & 0.4703(65) & \\
  \end{tabular}                                                                                             
\end{ruledtabular}                                                                                         
\end{table*}

\begin{table*}[tbp]  
\caption{\FIXME{tab:GEdata-ALL} Results for the bare $G_E(Q^2)$
  extracted from $\Re (V_4)$ are listed for the thirteen calculations
  defined in Table~\protect\ref{tab:ens}.  The results are obtained
  using 4-state fits to the 2-point functions and $3^\ast$-state fits
  to the 3-point functions (2-state fits for the $a12m220S$ ensemble)
  as described in the text. The value, $G_E(0)=1/Z_V$, given in the
  first row provides one estimate of the renormalization constant for
  the vector current.  The momentum transfer $Q^2$, in units of
  GeV${}^2$, associated with each $\vec n$ is given in
  Table~\protect\ref{tab:Q2-ALL}.}
\label{tab:GEdata-ALL}
\centering
\renewcommand*{\arraystretch}{1.1}
\begin{ruledtabular}
  \begin{tabular}{cccccccc}
$\vec{n}$ & $a15m310$  & $a12m310$  & $a12m220L$ & $a12m220$  & $a12m220S$ & $a09m310$  & $a09m220$     \\
\hline                                                                      
$(0,0,0)$ &  1.069(4)  &  1.061(8)  &  1.067(4)  &  1.071(9)  &  1.081(18) &  1.045(3)   &  1.049(4) \\
$(1,0,0)$ &  0.650(4)  &  0.728(8)  &  0.908(12) &  0.840(11) &  0.706(17) &  0.735(4)   &  0.859(5) \\
$(1,1,0)$ &  0.440(4)  &  0.536(9)  &  0.789(12) &  0.666(23) &  0.513(17) &  0.549(6)   &  0.718(7) \\
$(1,1,1)$ &  0.321(4)  &  0.407(10) &  0.694(11) &  0.553(16) &  0.402(20) &  0.423(12)  &  0.614(8) \\
$(2,0,0)$ &  0.261(8)  &  0.332(11) &  0.618(12) &  0.469(15) &  0.324(20) &  0.348(7)   &  0.538(9) \\
$(2,1,0)$ &  0.212(5)  &  0.279(8)  &  0.553(10) &  0.396(17) &  0.279(18) &  0.285(9)   &  0.472(8) \\
$(2,1,1)$ &  0.167(6)  &  0.239(9)  &  0.499(9)  &  0.349(15) &  0.234(17) &  0.240(9)   &  0.417(8) \\
$(2,2,0)$ &  0.140(15) &  0.176(15) &  0.413(8)  &  0.280(19) &  0.156(24) &  0.186(4)   &  0.338(8) \\
$(2,2,1)$ &  0.114(12) &  0.161(12) &  0.380(7)  &  0.260(16) &  0.155(21) &  0.162(4)   &  0.307(7) \\
$(3,0,0)$ &  0.110(30) &  0.157(35) &  0.387(8)  &  0.203(32) &  0.148(42) &  0.177(8)   &  0.315(9) \\
$(3,1,0)$ &  0.088(20) &  0.155(24) &  0.357(7)  &  0.200(23) &  0.154(27) &  0.153(5)   &  0.290(7) \\
\hline                                           
$\vec{n}$ & $a09m130W$ & $a06m310$  & $a06m310W$ & $a06m220$  & $a06m220W$  & $a06m135$     &   \\
\hline                                                                                          
$(0,0,0)$ &  1.052(6)  &  1.043(6)  &  1.035(11) &  1.050(7)  &  1.039(9)   &  1.042(10)    &   \\
$(1,0,0)$ &  0.937(6)  &  0.700(16) &  0.711(9)  &  0.822(8)  &  0.811(9)   &  0.919(10)    &   \\
$(1,1,0)$ &  0.836(6)  &  0.502(21) &  0.521(8)  &  0.670(10) &  0.654(11)  &  0.814(13)    &   \\
$(1,1,1)$ &  0.756(6)  &  0.373(24) &  0.395(9)  &  0.552(14) &  0.536(14)  &  0.716(18)    &   \\
$(2,0,0)$ &  0.680(8)  &  0.306(24) &  0.318(13) &  0.465(17) &  0.440(26)  &  0.664(15)    &   \\
$(2,1,0)$ &  0.624(8)  &  0.232(23) &  0.260(11) &  0.398(17) &  0.384(20)  &  0.588(18)    &   \\
$(2,1,1)$ &  0.571(8)  &            &            &            &             &  0.528(20)    &   \\
$(2,2,0)$ &  0.497(9)  &            &            &            &             &  0.433(21)    &   \\
$(2,2,1)$ &  0.455(9)  &            &            &            &             &  0.399(19)    &   \\
$(3,0,0)$ &  0.439(15) &            &            &            &             &  0.422(21)    &   \\
$(3,1,0)$ &  0.418(12) &            &            &            &             &  0.380(20)    &   \\
\end{tabular}
\end{ruledtabular}
\end{table*}

\begin{table*}[tbp] 
\caption{\FIXME{tab:GEVi-ALL} Results for the bare $G_E(Q^2)$
  extracted from $\Im (V_i)$ are listed for the thirteen  calculations
  defined in Table~\protect\ref{tab:ens}.  The rest is the same as in
  Table~\protect\ref{tab:GEdata-ALL}.}
\label{tab:GEVi-ALL}
\centering
\renewcommand*{\arraystretch}{1.1}
\begin{ruledtabular}
  \begin{tabular}{cccccccc}
$\vec{n}$  & $a15m310$  & $a12m310$   & $a12m220L$ & $a12m220$  & $a12m220S$  & $a09m310$  & $a09m220$   \\
\hline                                                                      
$(1,0,0)$  &  0.610(15) &  0.818(50)  &  0.871(30) &  0.761(61) &  0.774(59) &  0.699(17) &   0.814(30) \\
$(1,1,0)$  &  0.435(10) &  0.592(32)  &  0.786(27) &  0.634(69) &  0.564(38) &  0.536(16) &   0.694(26) \\
$(1,1,1)$  &  0.336(10) &  0.448(27)  &  0.711(28) &  0.575(42) &  0.439(36) &  0.413(29) &   0.619(25) \\
$(2,0,0)$  &  0.262(21) &  0.410(32)  &  0.654(28) &  0.527(42) &  0.432(39) &  0.358(14) &   0.551(27) \\ 
$(2,1,0)$  &  0.205(13) &  0.340(24)  &  0.585(24) &  0.429(36) &  0.316(32) &  0.296(17) &   0.479(22) \\ 
$(2,1,1)$  &  0.158(14) &  0.284(24)  &  0.543(23) &  0.334(37) &  0.249(37) &  0.255(14) &   0.422(21) \\ 
$(2,2,0)$  &  0.135(10) &  0.193(27)  &  0.464(22) &  0.284(37) &  0.181(33) &  0.206(8)  &   0.362(18) \\ 
$(2,2,1)$  &  0.103(25) &  0.166(30)  &  0.429(21) &  0.250(43) &  0.161(45) &  0.182(8)  &   0.308(19) \\ 
$(3,0,0)$  &  0.139(43) &  0.188(66)  &  0.445(25) &  0.249(59) &  0.202(59) &  0.195(15) &   0.342(17) \\
$(3,1,0)$  &  0.083(58) &  0.178(76)  &  0.405(25) &  0.150(74) &  0.220(92) &  0.172(17) &   0.330(25) \\
\hline                                           
$\vec{n}$  & $a09m130W$ & $a06m310$  & $a06m310W$ & $a06m220$   & $a06m220W$ & $a06m135$  &   \\
\hline                                                                                      
$(1,0,0)$  &  0.871(43) &  0.733(59) &  0.641(44) &  0.718(46)  &  0.778(65) &  0.793(65) &   \\
$(1,1,0)$  &  0.791(34) &  0.515(49) &  0.500(29) &  0.600(40)  &  0.660(49) &  0.688(57) &   \\
$(1,1,1)$  &  0.710(31) &  0.367(64) &  0.400(30) &  0.534(43)  &  0.553(51) &  0.626(64) &   \\
$(2,0,0)$  &  0.679(31) &  0.390(48) &  0.270(42) &  0.444(53)  &  0.413(88) &  0.643(49) &   \\
$(2,1,0)$  &  0.625(27) &  0.219(34) &  0.229(37) &  0.420(42)  &  0.366(67) &  0.529(52) &   \\
$(2,1,1)$  &  0.582(27) &            &            &             &            &  0.446(55) &   \\
$(2,2,0)$  &  0.494(26) &            &            &             &            &  0.399(54) &   \\
$(2,2,1)$  &  0.461(26) &            &            &             &            &  0.354(54) &   \\
$(3,0,0)$  &  0.467(38) &            &            &             &            &  0.367(60) &   \\
$(3,1,0)$  &  0.435(33) &            &            &             &            &  0.311(58) &   \\
\end{tabular}
\end{ruledtabular}
\end{table*}

\begin{table*}[tbp] 
\caption{\FIXME{tab:GMdata-ALL} Results for the bare magnetic form
  factor ${G}_M(Q^2)$ for the thirteen calculations defined in
  Table~\protect\ref{tab:ens}.  Values of $G_M(0)$ are obtained by a
  linear extrapolation of the data for $G_M(Q^2)/(G_E(Q^2) \times
  Z_V)$ to $Q^2 = 0$ as discussed in the text. The rest is the same as in
  Table~\protect\ref{tab:GEdata-ALL}. }
\label{tab:GMdata-ALL}
\centering
\renewcommand*{\arraystretch}{1.1}
\begin{ruledtabular}
  \begin{tabular}{cccccccc}
$\vec{n}$ & $a12m310$   & $a12m310$  & $a12m220L$ & $a12m220$  & $a12m220S$ & $a09m310$ & $a09m220$ \\
\hline                            
$(0,0,0)$ &  4.596(61)  &  4.553(107)&  4.538(107)&  4.465(144)& 4.597(217) &  4.324(32) &   4.505(76) \\
$(1,0,0)$ &  2.968(29)  &  3.318(51) &  4.018(61) &  3.657(90) & 3.139(85)  &  3.207(19) &   3.749(51) \\
$(1,1,0)$ &  2.160(32)  &  2.597(43) &  3.557(41) &  3.082(76) & 2.352(75)  &  2.513(24) &   3.249(40) \\
$(1,1,1)$ &  1.665(26)  &  2.092(46) &  3.172(32) &  2.660(68) & 1.940(82)  &  2.041(42) &   2.863(36) \\
$(2,0,0)$ &  1.255(45)  &  1.728(59) &  2.874(31) &  2.251(75) & 1.566(98)  &  1.684(32) &   2.468(46) \\
$(2,1,0)$ &  1.155(28)  &  1.532(41) &  2.615(30) &  2.009(66) & 1.389(74)  &  1.471(35) &   2.225(38) \\
$(2,1,1)$ &  0.959(31)  &  1.363(31) &  2.407(29) &  1.818(67) & 1.257(68)  &  1.283(41) &   2.034(36) \\
$(2,2,0)$ &  0.817(50)  &  1.106(54) &  2.032(37) &  1.643(76) & 1.164(92)  &  1.074(18) &   1.698(35) \\
$(2,2,1)$ &  0.760(47)  &  0.969(55) &  1.913(32) &  1.448(75) & 0.963(82)  &  0.982(19) &   1.554(36) \\
$(3,0,0)$ &  0.717(102) &  1.241(128)&  1.883(41) &  1.422(88) & 0.851(136) &  0.947(35) &   1.585(44) \\
$(3,1,0)$ &  0.734(33)  &  0.911(104)&  1.771(41) &  1.424(89) & 0.917(111) &  0.913(49) &   1.467(36) \\
\hline
$\vec{n}$ & $a09m130W$ & $a06m310$  & $a06m310W$ & $a06m220$  & $a06m220W$  & $a06m135$    & \\
\hline                                                                                       
$(0,0,0)$ & 4.297(82)  &  4.163(168)&  4.303(134)&  4.138(102)&  4.293(142) &  4.229(123)  & \\
$(1,0,0)$ & 3.956(67)  &  3.083(73) &  3.181(70) &  3.405(65) &  3.505(99)  &  3.824(105)  & \\
$(1,1,0)$ & 3.547(50)  &  2.440(59) &  2.491(57) &  2.865(55) &  2.873(83)  &  3.413(90)   & \\
$(1,1,1)$ & 3.281(47)  &  1.984(73) &  2.020(60) &  2.493(59) &  2.411(89)  &  3.051(97)   & \\
$(2,0,0)$ & 2.992(48)  &  1.591(83) &  1.655(69) &  2.214(68) &  2.196(117) &  2.838(89)   & \\
$(2,1,0)$ & 2.820(39)  &  1.341(83) &  1.459(58) &  1.862(66) &  1.837(104) &  2.612(84)   & \\
$(2,1,1)$ & 2.616(37)  &            &            &            &             &  2.340(97)   & \\
$(2,2,0)$ & 2.286(39)  &            &            &            &             &  2.064(96)   & \\
$(2,2,1)$ & 2.156(39)  &            &            &            &             &  1.850(97)   & \\
$(3,0,0)$ & 2.158(59)  &            &            &            &             &  1.849(110)  & \\
$(3,1,0)$ & 2.041(46)  &            &            &            &             &  1.764(96)   & \\
\end{tabular}
\end{ruledtabular}
\end{table*}

\section{Extracting form factors from matrix elements}
\label{sec:FFanalysis}
\FIXME{sec:FFanalysis} 

The following ratios, ${\cal R}_{\mu}$, of the
three-point to the two-point correlation functions,
\begin{align}
{\cal R}_{\mu}&(t, \tau, \bm{p}^\prime, \bm{p}) =   \frac{C^{(3\text{pt})}_{\mu}(t,\tau;\bm{p}^\prime,\bm{p})}{C^{(2\text{pt})}(\tau,\bm{p}^\prime)} \, \times \, \nonumber \\
&
  \left[ \frac{C^{(2\text{pt})}(t,\bm{p}^\prime) C^{(2\text{pt})}(\tau,\bm{p}^\prime) C^{(2\text{pt})}(\tau-t,\bm{p})}{C^{(2\text{pt})}(t,\bm{p}) C^{(2\text{pt})}(\tau,\bm{p}) C^{(2\text{pt})}(\tau-t,\bm{p}^\prime)}
  \right]^{1/2} \,,
\label{eq:ratio}
\end{align}
give the desired ground state matrix elements (ME) 
$\matrixe{0^\prime}{\mathcal{O}_\Gamma}{0}$, introduced in
Eq.~\ref{eq:3pt}, in the limits $t \to \infty$ and $(\tau-t) \to
\infty$.  In the calculation of the nucleon three-point functions, we
use the spin projection operator ${\cal P}_3 = (1 + \gamma_4)(1 +
i\gamma_5 \gamma_3)/2$.  With this ${\cal P}_3 $, and the vector
current defined in Eqs.~\eqref{eq:Vem} and~\eqref{eq:VFF2} with Euclidean
$\gamma_\mu$, the following quantities have a signal and give
either the electric or the magnetic form factors:
\begin{align}
  \sqrt{2E_p(M_N+E_p)} \Re ({\cal R}_{i}) =&\; - \epsilon_{ij3} q_j {G}_M  \,,
  \label{eq:GM1} \\
  \sqrt{2E_p(M_N+E_p)} \Im ({\cal R}_{i}) =&\;   q_i {G}_E  \,,
  \label{eq:GE1} \\
  \sqrt{2E_p(M_N+E_p)} \Re ({\cal R}_{4}) =&\;   (M_N+E_p) {G}_E  \,. 
  \label{eq:GE4} 
\end{align}
Note that, in practice, these ratios are used only to plot the data. Our results are obtained by 
making n-state fits to the correlation functions. 

Exploiting the cubic symmetry under spatial rotations, we construct
two averages over equivalent 3-point correlators before doing fits to
get the ground state matrix elements: over $ \Re ({ C}_{1})$ and $ \Re
({ C}_{2})$ for $G_M(Q^2)$ and over $ \Im ({ C}_{1})$, $ \Im ({
  C}_{2})$ and $ \Im ({ C}_{3})$ for $G_E(Q^2)$. We label these form
factors as $ G_M^{V_i}$ and $ G_E^{V_i}$. Together with $ G_E^{V_4}$
extracted from Eq.~\eqref{eq:GE4}, they constitute the three form
factors analyzed. Their extraction is straightforward as each of the
three is given by a distinct three-point function. It is important to
note that the discretization artifacts and the excited-state
contaminations in each can be very different.\looseness-1

The data for the ratio defined in Eq.~\eqref{eq:ratio} and the results
of $3^\ast$ fits to the three 3-point correlators are illustrated in
Figs.~\ref{fig:GE-ESC-a09m130W}--\ref{fig:GM-ESC-vol} and
Figs.~\ref{fig:GM-ESC-smear310}--\ref{fig:GM-ESC-smear220}.  The ideal
expected behavior of all 3-point functions with large $t$ and
$\tau-t$, is a flat region near $\tau/2$ that becomes independent of
$\tau$. Our data show that this is not manifest even at $\tau \approx
1.4$~fm. We, therefore, use $3^\ast$-state fits to data at the various
values of $t$ and $\tau$ to obtain estimates of the
ground state matrix elements.  Results for the three
sets of form factors, $ G_E^{V_4}$, $ G_E^{V_i}$ and $ G_M^{V_i}$, extracted from these 
matrix elements using Eqs.~\eqref{eq:GM1},~\eqref{eq:GE1} and~\eqref{eq:GE4} 
are given in Tables~\ref{tab:GEdata-ALL},~\ref{tab:GEVi-ALL}
and~\ref{tab:GMdata-ALL} for the thirteen calculations.

\subsection{Extraction of $G_E(Q^2)$}
\label{sec:extractGE}
\FIXME{sec:extractGE} 

The pattern of the ESC in the extraction of $G_E^{V_i}$ versus $G_E^{V_4}$ can be, and
is found to be, very different as shown in Figs.~\ref{fig:GE-ESC-a09m130W}
and~\ref{fig:GE-ESC-a06m135} for the two physical mass ensembles.  The
data for $G_E^{V_4}$ show a clear monotonic but slow convergence from above,
and a flattish region near the middle. The estimates of the $\tau \to
\infty$ values given by the $3^\ast$ fits are found to be stable under
variations in $\tskip$ and the values of $\tau$ included in the fits.

The data for $G_E^{V_i}$ show much larger ESC and the ME 
$\matrixe{0^\prime}{\mathcal{O}_\Gamma}{1}$ and
$\matrixe{1^\prime}{\mathcal{O}_\Gamma}{0}$ are an order of magnitude
larger for $n^2 = 1$ as compared to those from $G_E^{V_4}$. The resulting pattern versus
$t$ is essentially linear for each $\tau$. As $\tau$ is increased,
this ``line'' rotates towards becoming flat, but the rotation is
slow. The pivot point is approximately the point of intersection of
the various $\tau$ lines and converges to the ground state estimate as
$t$ and $(\tau -t) \to \infty$.

The difference in the shape of the ESC between $G_E^{V_i}$ and
$G_E^{V_4}$ can be explained by the behavior of the transition matrix
elements under parity transformation and hermitian conjugation. The
imaginary parts of the matrix elements of $V_i$ at nonzero momentum
pick up a negative sign under the combined transformations. As a
result, for example, the term $\abs{\mathcal{A}_0^\prime}
\abs{\mathcal{A}_1}\matrixe{0^\prime}{\mathcal{O}_\Gamma}{1} e^{-E_0t
  - M_1(\tau-t)}$ has opposite sign to that of its partner
$\abs{\mathcal{A}_1^\prime}
\abs{\mathcal{A}_0}\matrixe{1^\prime}{\mathcal{O}_\Gamma}{0} e^{-E_1t
  -M_0(\tau-t)}$.  Thus, each such pair of terms give a ``sinh''-like
correction, that makes the data looks like a straight line at an angle
to the extracted ground-state result. On the other hand, the matrix
elements in the related pairs of terms from the real parts of $V_i$
and $V_4$ have the same sign, and therefore exhibit a ``cosh''-like
correction.  Even in this case, the magnitudes of the two ME
in such pairs of terms are not the same. Therefore, in fits
to the three-point data using Eq.~\eqref{eq:3pt}, we leave all the
matrix elements as free parameters. In fact, in practice, it is the
product of the amplitudes and the ME, such as
$\abs{\mathcal{A}_0^\prime}
\abs{\mathcal{A}_1}\matrixe{0^\prime}{\mathcal{O}_\Gamma}{1}$, that
are free parameters in the fits. In these cases, only the energies are
free parameters and these are taken from the two-point functions.

It is also evident from Figs.~\ref{fig:GE-ESC-a09m130W}
and~\ref{fig:GE-ESC-a06m135} that the ESC in $ G_E^{V_i}$ is the
largest at the smallest nonzero momentum, i.e., the ``angle'' the
data make with the horizontal line is the largest. On the other hand,
the ESC in $ G_E^{V_4}$ increases with momentum. By comparing the data
in the two figures, we also conclude that the ESC increases with
decreasing $a$ for both $ G_E^{V_i}$ and $ G_E^{V_4}$.  

A consequence of this difference in the ESC behavior is that the
errors in $G_E^{V_i}$ are 3--10 times larger than in $G_E^{V_4}$ (see
data in Tables~\ref{tab:GEdata-ALL} and~\ref{tab:GEVi-ALL}). Also,
since one cannot extract a value for $G_E(Q^2=0)$ using the operators
$V_{i}$ due to kinematic constraint, the fits to $G_E^{V_i}$ versus $Q^2$, discussed in
Sec.~\ref{sec:Q2lattice}, are less stable because they are not
anchored at $Q^2=0$. As a result, the extraction of the electric
charge radius from the $G_E^{V_i}$ data has much larger errors.
Because of these two reasons, it has been common to analyze only 
$G_E^{V_4}(Q^2)$. With our high-statistics data, we are 
able to compare the ESC, the efficacy of the $3^\ast$ fits, and 
the discretization errors between $G_E^{V_4}$ and $G_E^{V_i}$.

\begin{figure*}[tpb] 
\centering
\subfigure{
\includegraphics[width=0.24\linewidth]{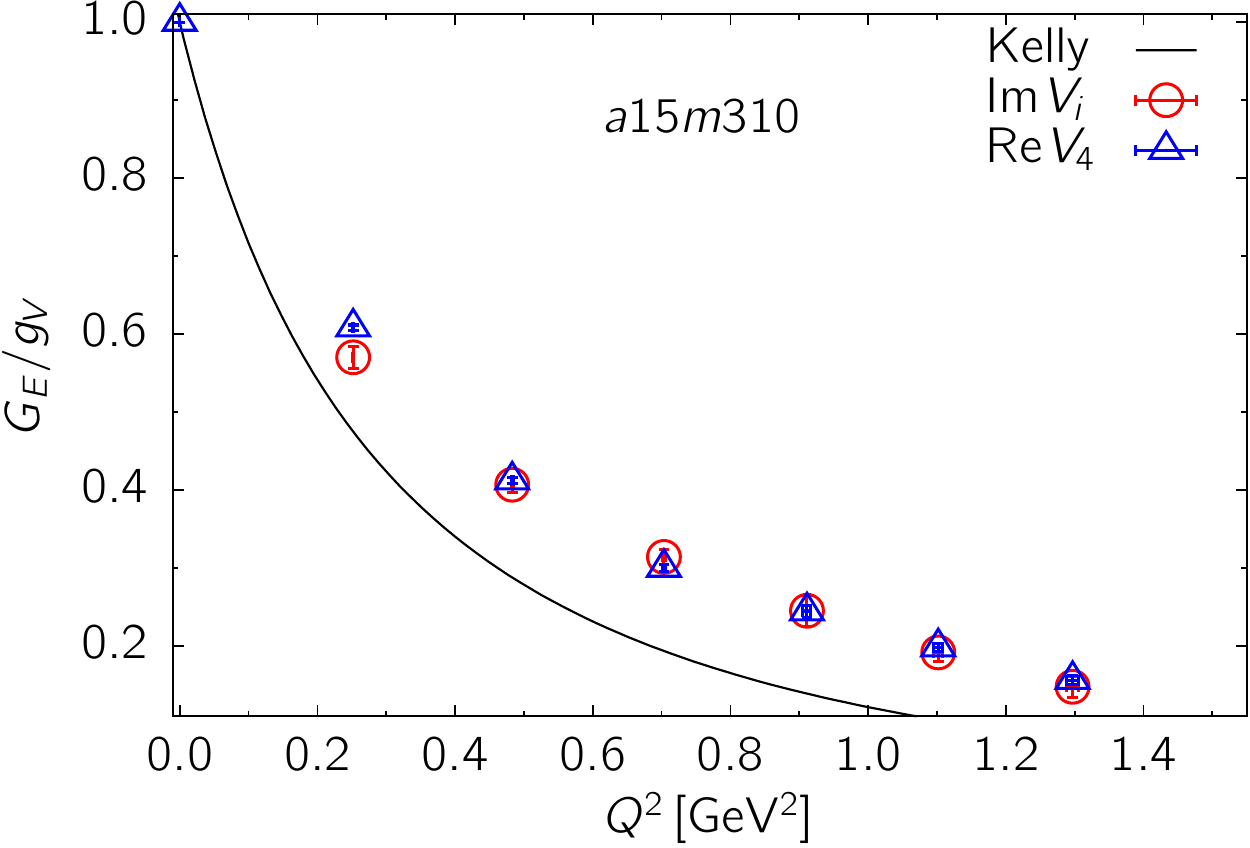}
\includegraphics[width=0.24\linewidth]{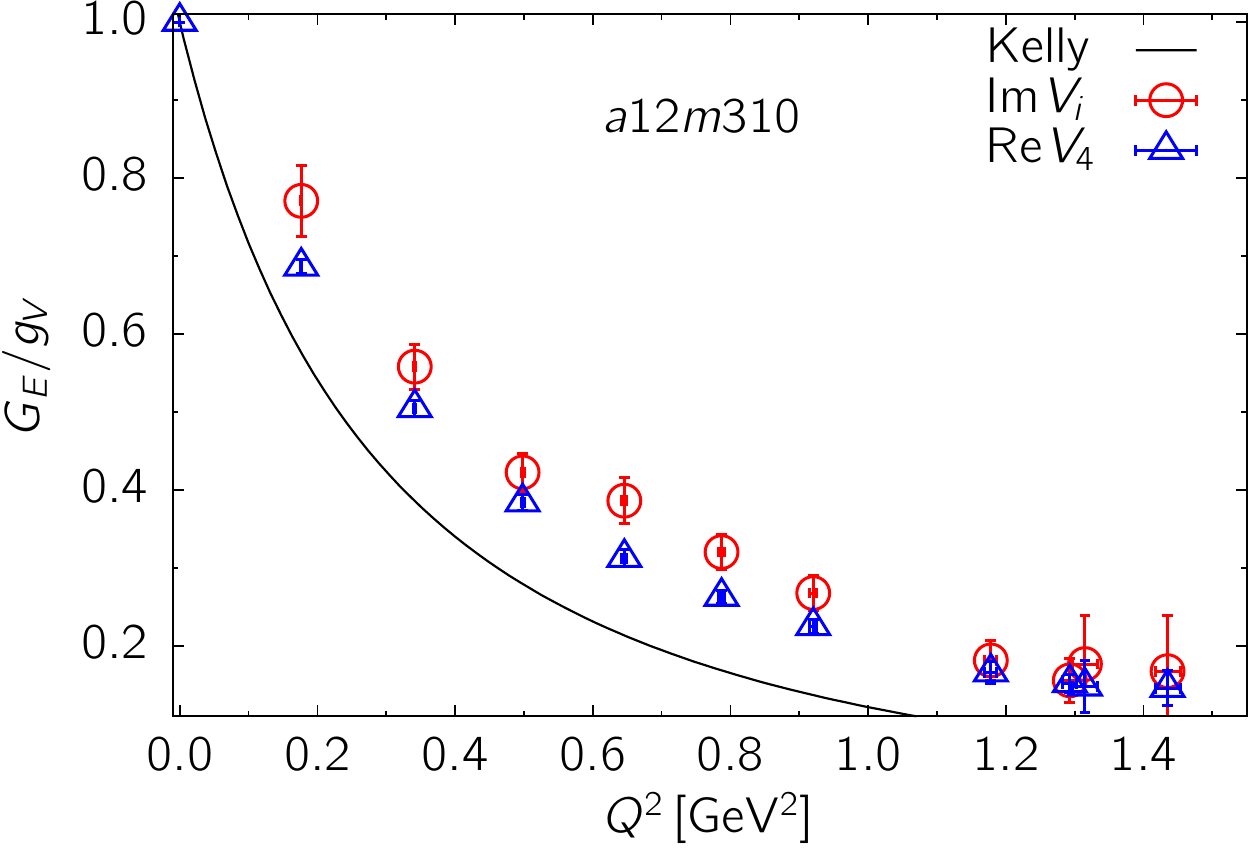}
\includegraphics[width=0.24\linewidth]{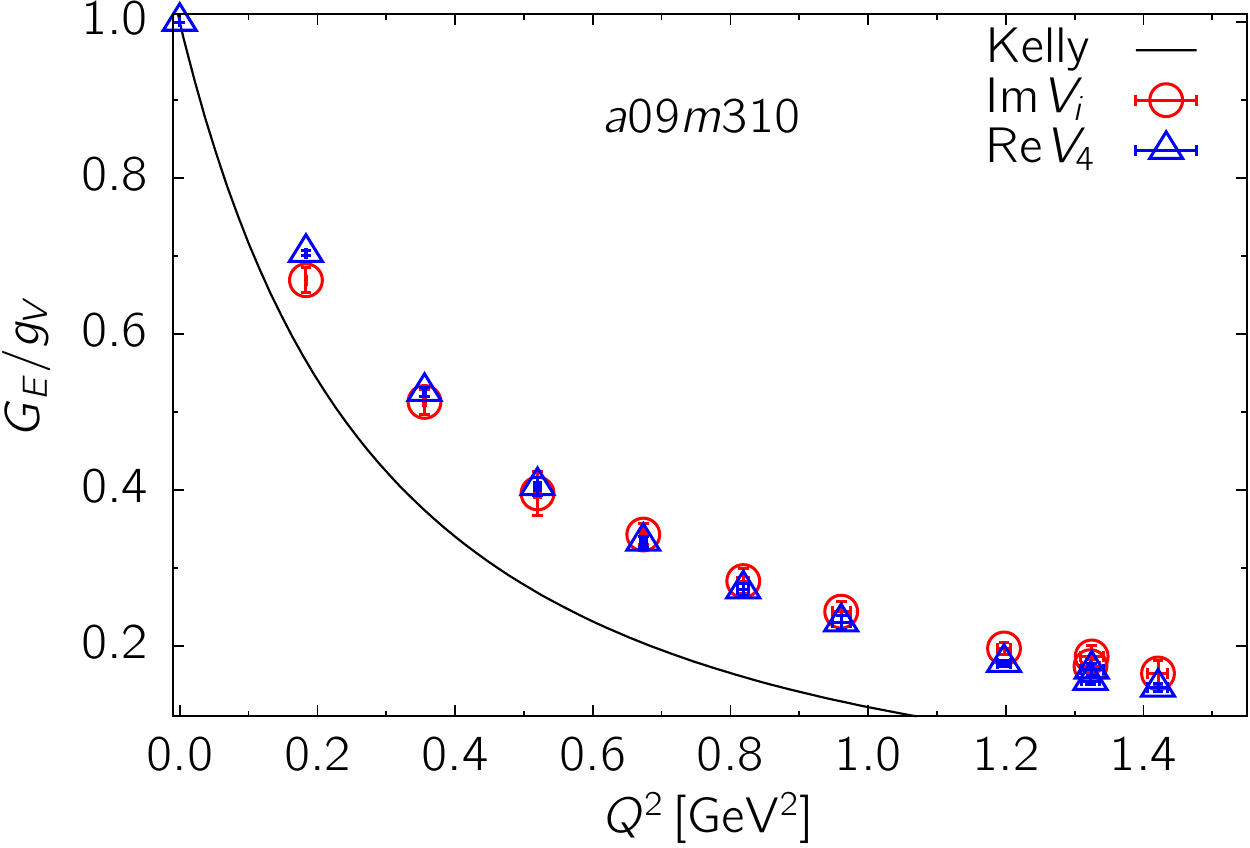}
\includegraphics[width=0.24\linewidth]{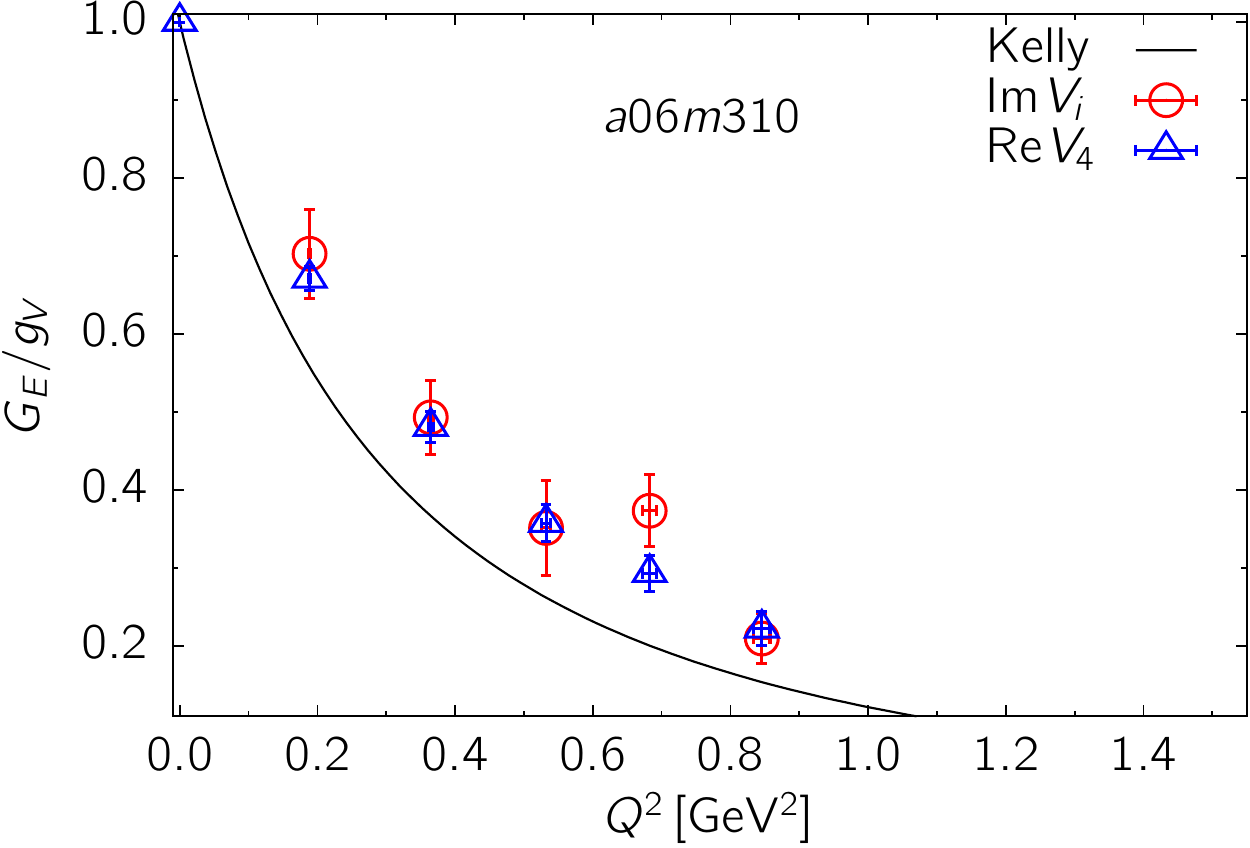}
}
\subfigure{
\includegraphics[width=0.24\linewidth]{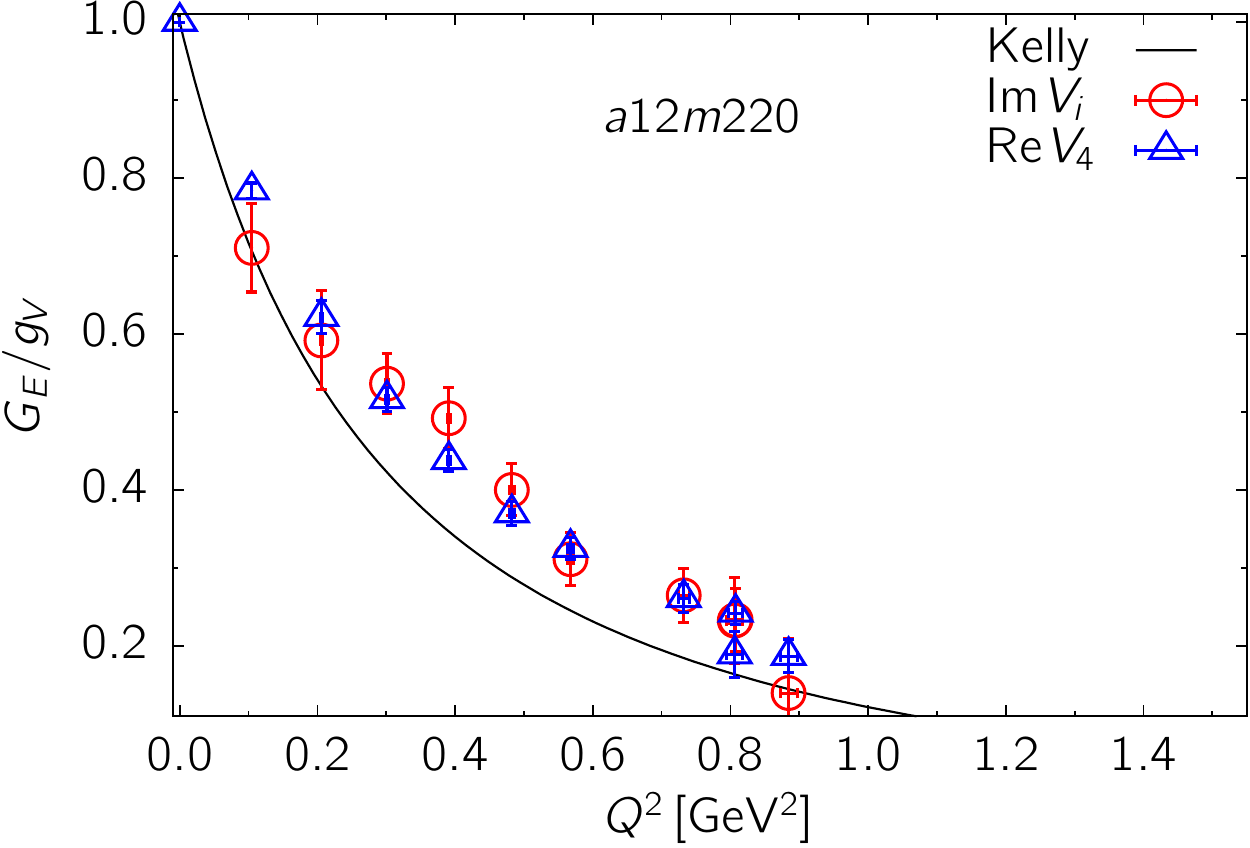}
\includegraphics[width=0.24\linewidth]{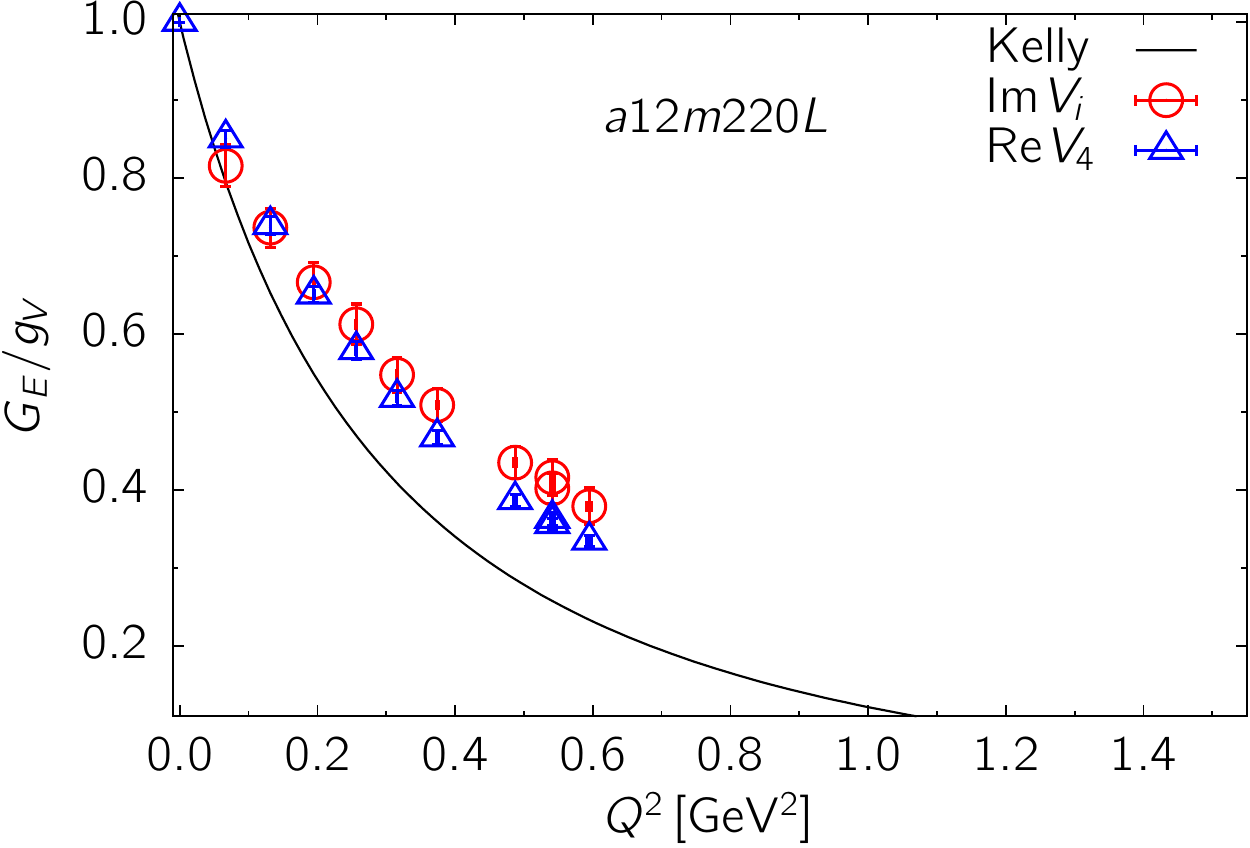}
\includegraphics[width=0.24\linewidth]{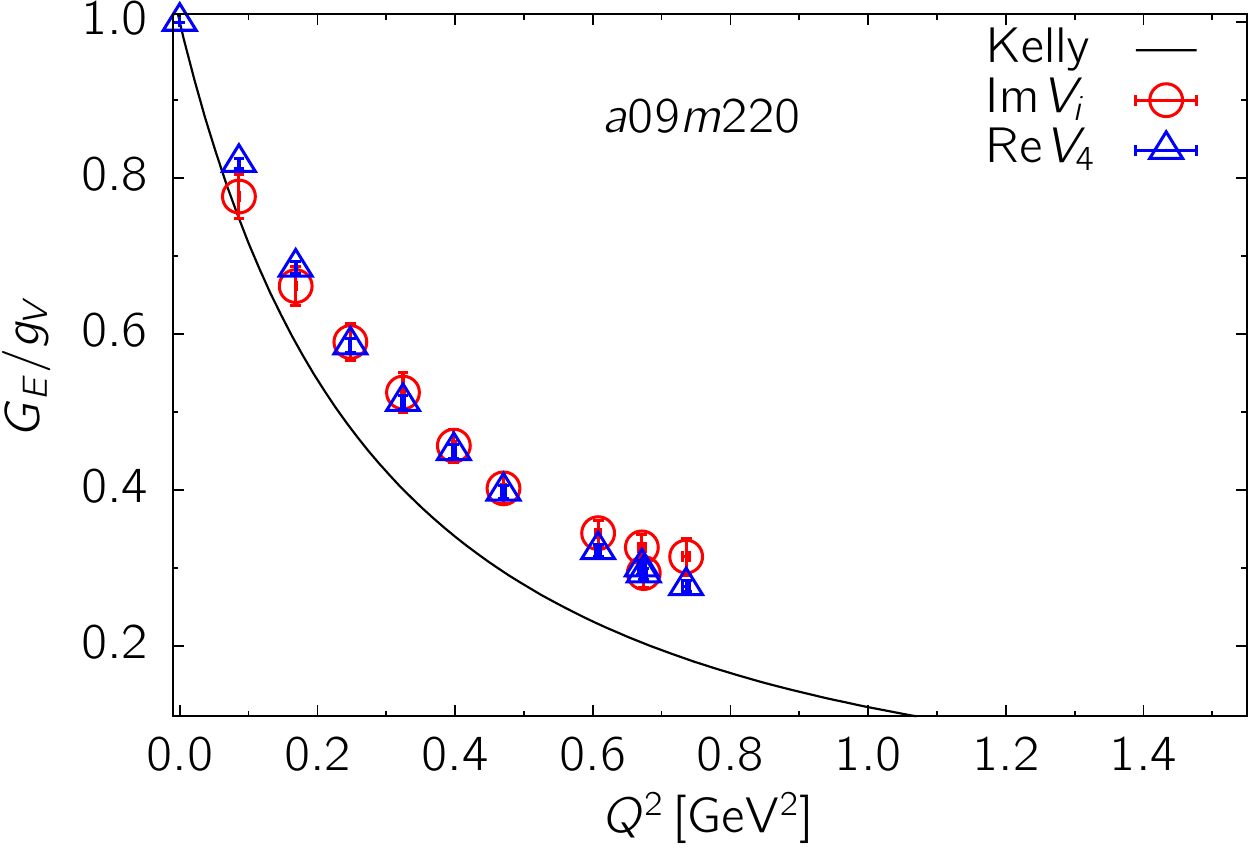}
\includegraphics[width=0.24\linewidth]{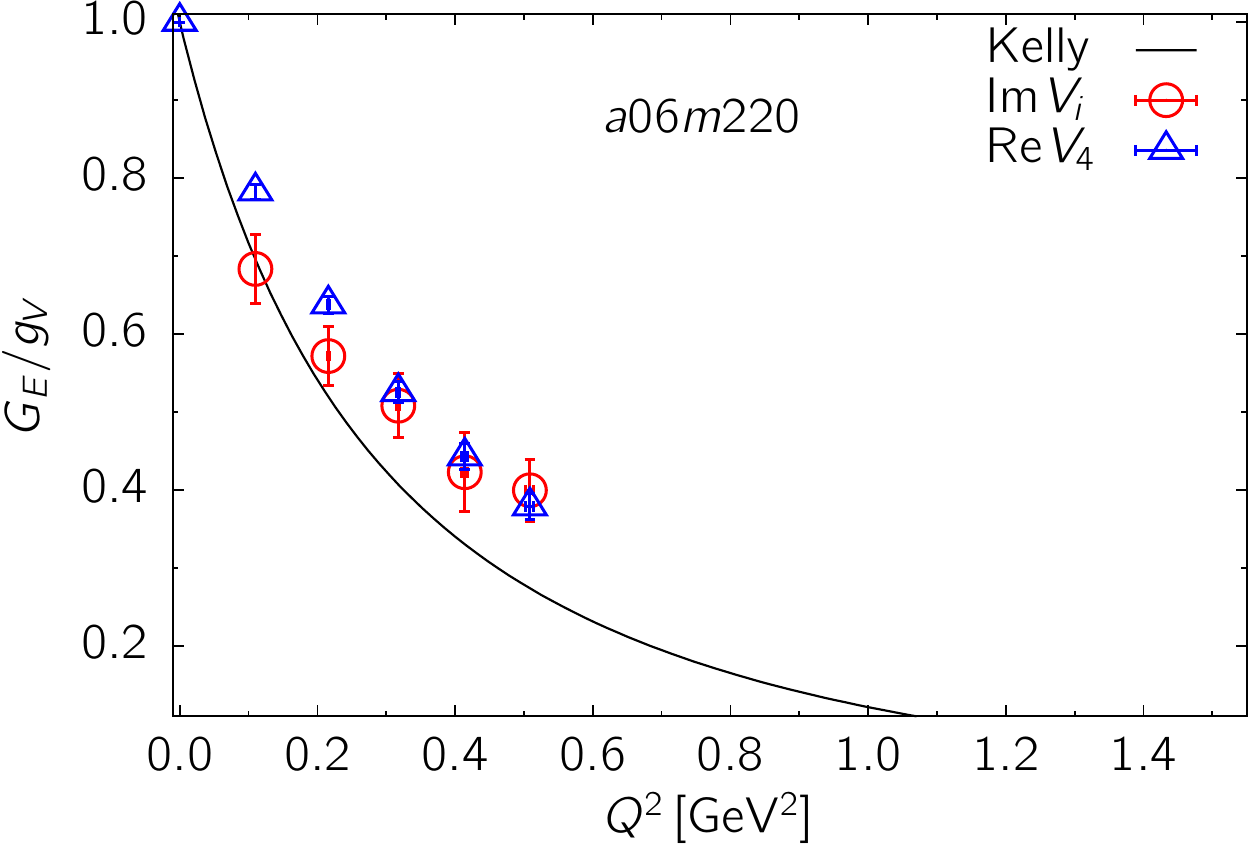}
}
\subfigure{
\includegraphics[width=0.24\linewidth]{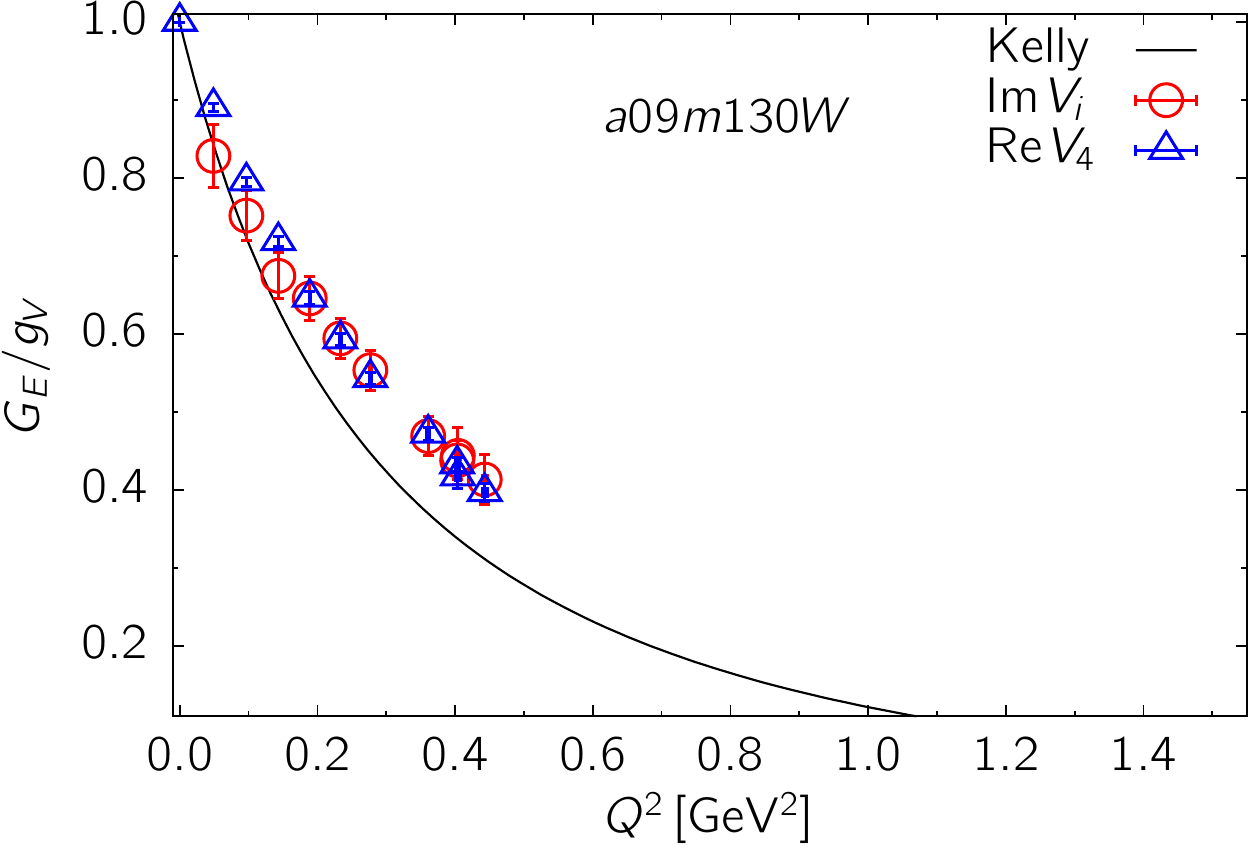} 
\includegraphics[width=0.24\linewidth]{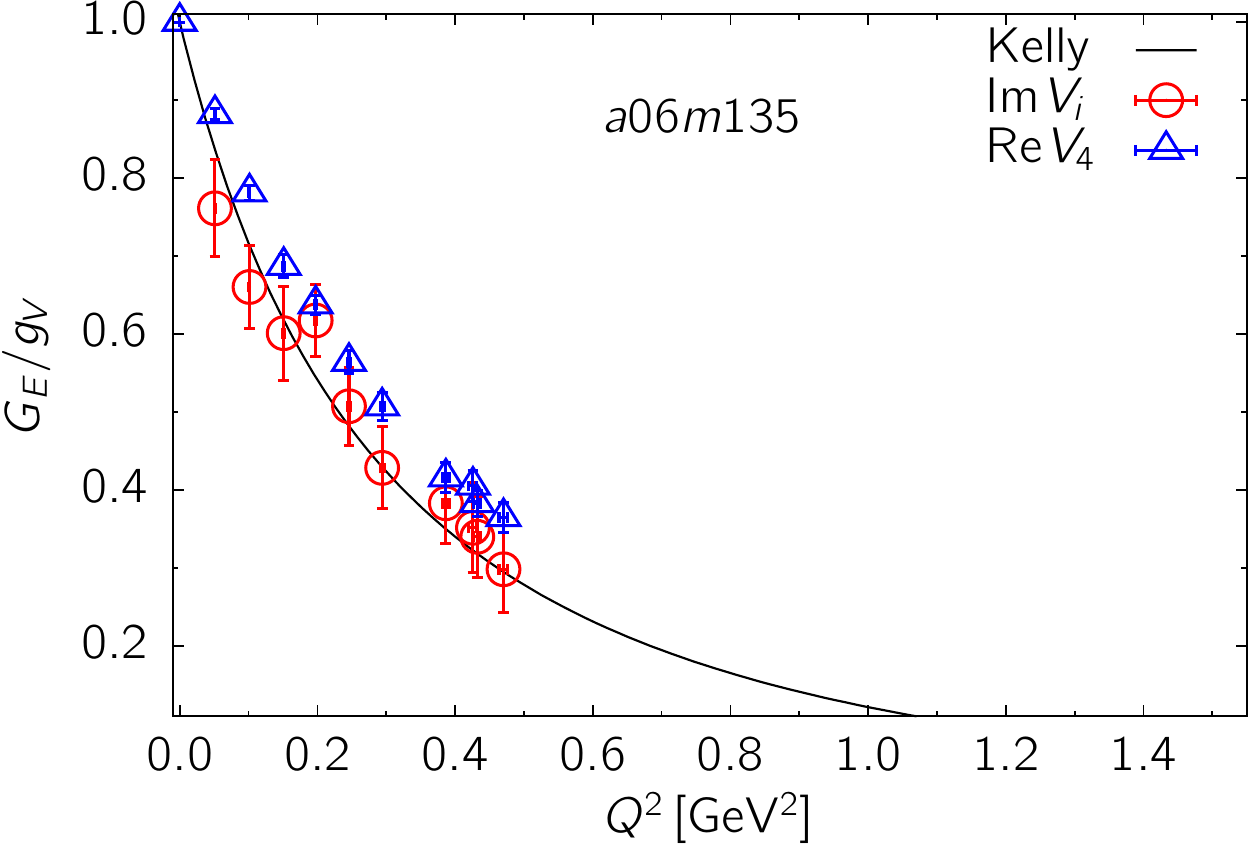} 
}
\subfigure{
\includegraphics[width=0.24\linewidth]{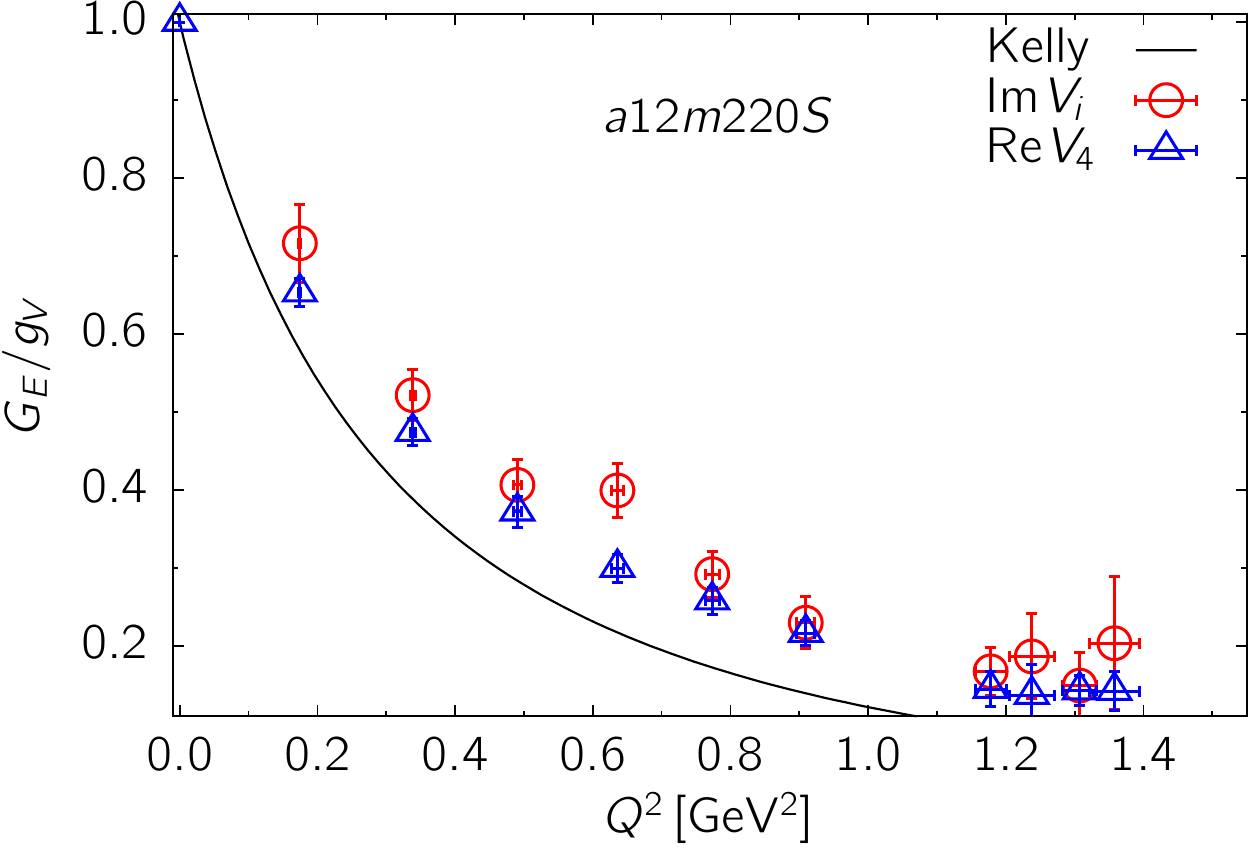}
\includegraphics[width=0.24\linewidth]{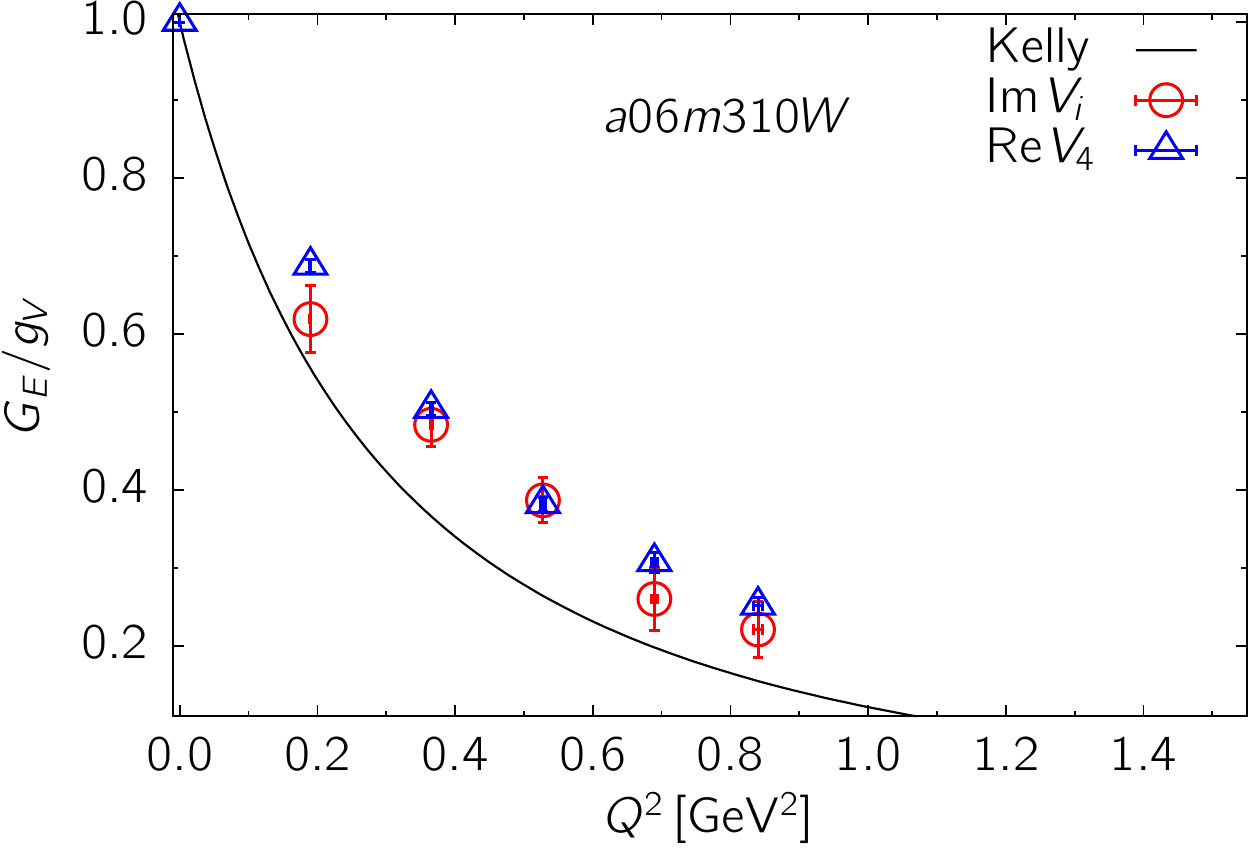}
\includegraphics[width=0.24\linewidth]{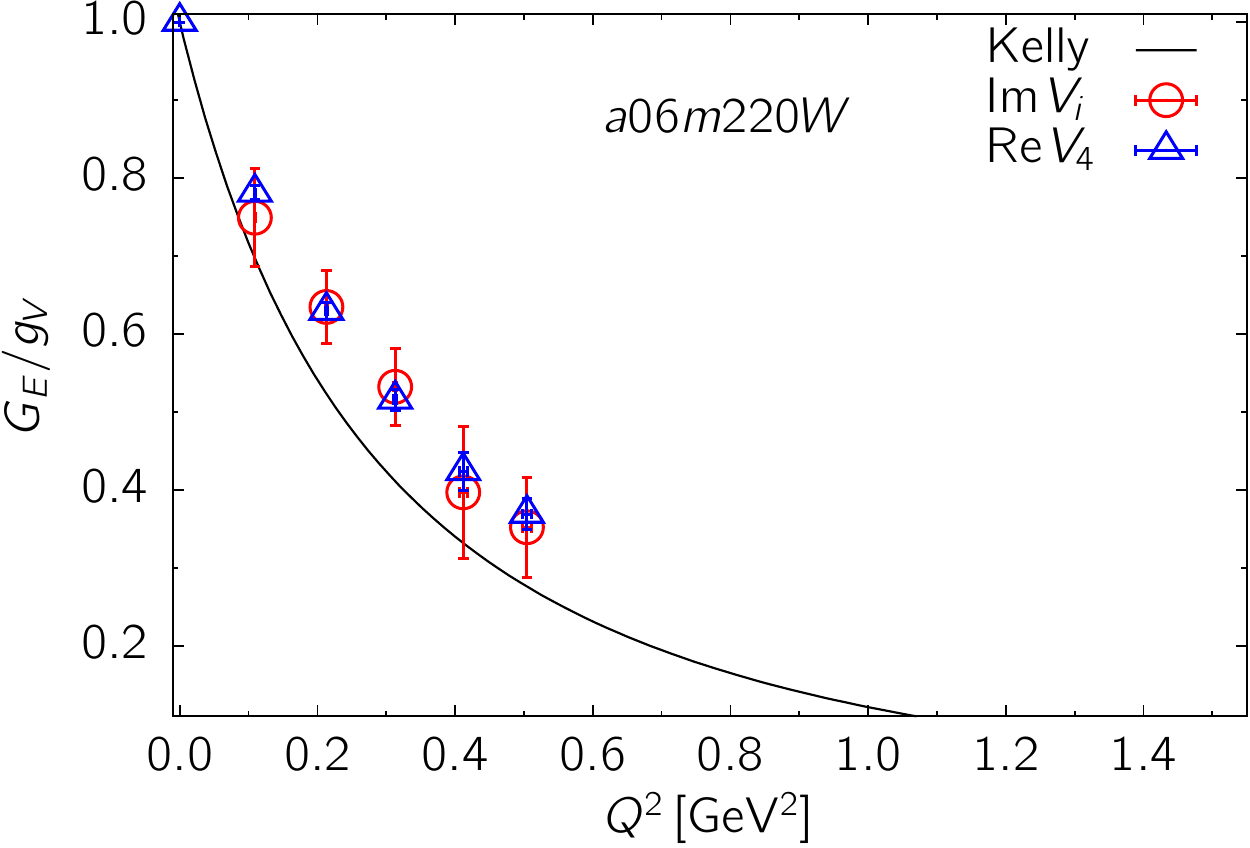}
}
\caption{\FIXME{fig:ViV4comp} Comparison of the renormalized form
  factor $G_E^{V_i}(Q^2)/g_V$ (red circles) versus
  $G_E^{V_4}(Q^2)/g_V$ (blue triangles). The first row gives data for
  the $M_\pi \approx 310$~MeV ensembles; the second row for the $M_\pi
  \approx 220$~MeV ensembles; the third for the two physical mass
  ensembles $a09m130W$ and $a06m135$; and the data for the remaining
  three calculations are shown in the fourth row. The solid black line
  shows the Kelly fit to the experimental isovector, $G_E^{p-n}$, data.}
\label{fig:ViV4comp}
\end{figure*}

A comparison of results for $ G_E^{V_i}$ and $ G_E^{V_4}$ is presented
in Fig.~\ref{fig:ViV4comp} for the thirteen
calculations. As stated above, the errors in $ G_E^{V_i}$ are much 
larger than those in $ G_E^{V_4}$, however, there are two
additional noteworthy patterns.  First, the data for $ G_E^{V_i}$ for
$Q^2 \lesssim 0.2$~GeV${}^2$ on the $a12m220$, $a09m220$, $a06m220$ and
the two physical mass ensembles $a09m130W$ and $a06m135$, have the
largest errors and mostly lie below those from $G_E^{V_4}$. On the
other hand, the data for $Q^2 \gtrsim 0.2$~GeV${}^2$ overlap in most
cases. Our conclusion, based on these data, is that for $Q^2 \gtrsim
0.2$~GeV${}^2$ the two measurements can be considered to have the same
mean but with different variance.

The pattern of data at $Q^2 \lesssim 0.2$~GeV${}^2$ is puzzling and
we do not have an explanation for the larger errors or the systematic
differences.  In particular, we cannot discern whether they are due
to residual ESC, statistical fluctuations and/or different
discretization errors.  In summary, while our high-statistics data
have allowed us to quantify the larger errors and fluctuations in
$G_E^{V_i}$, we do not have a resolution for the
difference. Operationally, using a weighted average of the nonzero
$Q^2$ data from $G_E^{V_i}$ and $G_E^{V_4}$, i.e., assuming that the
differences are statistical fluctuations, gives results that are
essentially identical to those from $G_E^{V_4}$.  We, therefore, analyze only the
data from $G_E^{V_4}$ in the rest of this paper.  To establish full
control over all systematics, future calculations should demonstrate
consistency between $G_E^{V_i}$ and $G_E^{V_4}$.\looseness-1


%

\subsection{Extraction of $G_M(Q^2)$}
\label{sec:extractGM}
\FIXME{sec:extractGM} 

Examples of the size and shape of the ESC in the extraction of
$G_M^{V_i}$ are shown in Figs.~\ref{fig:GM-ESC-p1}
and~\ref{fig:GM-ESC-p5}. For small momentum transfer, the convergence
is monotonic from below as shown in Fig.~\ref{fig:GM-ESC-p1} for
$n^2 =1$.  The ESC is observed to grow with decreasing $a$ and
$M_\pi$.

The pattern of convergence changes with $Q^2$: for small $n^2$ it is from below but by about $n^2 =6$,
it has changed to from above in most cases as illustrated in
Fig.~\ref{fig:GM-ESC-p5}.  As a result, removing ESC increases the value of
$G_M(Q^2)$ at small momentum transfers and decreases it at larger
momenta. Consequently, if ESC is not removed, both the magnetic
charge radius and the magnetic moment extracted are underestimated.

The results of the $3^\ast$-fits to the data for the bare form factor
$G_M(Q^2)$ are summarized in Table~\ref{tab:GMdata-ALL}.  A key
shortcoming of the analysis of the lattice $G_M(Q^2)$ is the lack of
data at $Q^2 = 0$. To overcome this, we note that the ratio $G_M/G_E$,
shown in Fig.~\ref{fig:GEGMratio}, is, within errors, linear in $Q^2$
for $Q^2 \lesssim 0.6$~GeV${}^2$.  We, therefore make a linear fit to
the ratio of the form factor data, $G_M(Q^2)/G_E(Q^2)$, with momenta
up to $\vec n = (2,1,1) $ to obtain an estimate for the renormalized
$G_M(Q^2=0)$. The corresponding unrenormalized values, which we call
derived $G_M(Q^2=0)$, are also given in
Table~\ref{tab:GMdata-ALL}. These values are indistinguishable from
those obtained from taking a ratio of the two correlators and then
making a linear fit versus $Q^2$ to these data. Including these values
of $G_M(Q^2=0)$ improved the stability of the $z$-expansion fits. Note
that, the extrapolation of $G_M/G_E$, inclusion of the extrapolated
value of $G_M(0)$, and the fit to $G_M$ are done within a single
jackknife loop, therefore, the statistical errors are accounted for
correctly.

To estimate the importance of using the derived point $G_M(Q^2=0)$,
which anchors the fits to data, especially on ensembles with largish
values of the minimum $Q^2$, we performed the following test.  We fit
the nonzero $Q^2$ data for $G_E^{V_4}$ to extract the value and the
slope at $Q^2=0$ for each ensemble. Comparing the value for $g_V$ from
this fit with the data given in Table~\ref{tab:GEdata-ALL}, we find
the magnitude of the difference for the dipole and $z^{4}$ fit is
between 0.01--0.04 for the 13 calculations. The difference in the
slope, $\rEsq$, compared to the data in Table~\ref{tab:rE-results} is
up to 9\% for the dipole fit and up to 20\% for the $z^{4}$ fit. Based
on this test, it is not unreasonable that an uncertainty of similar
size can be present in the extraction of $\mu$ and $\rMsq$.  Thus, to get
high precision results without resorting to a derived value for
$G_M(0)$ or without using priors, requires having data at smaller
values of $Q^2$.

\subsection{Dependence of $G_E(Q^2)$ and $G_M(Q^2)$ on the lattice parameters}
\label{sec:dependence}
\FIXME{sec:dependence} 

In Figs.~\ref{fig:GE-vsM}--\ref{fig:GE-GM-vsS}, we explore the
dependence of the renormalized form factors $G_E^{V_4}(Q^2)/g_V$ and
$G_M^{V_i}(Q^2)/g_V$, which we henceforth label $G_E(Q^2)/g_V$ and
$G_M(Q^2)/g_V$ for brevity, as a function of the pion mass, lattice
spacing, lattice volume and the smearing size. The significant
features are:
\begin{itemize}
\item
The dependence of $G_E(Q^2)/g_V$ on the pion mass, keeping the lattice
spacing roughly constant, is shown in Fig.~\ref{fig:GE-vsM}.  The data show a
steeper fall off as the quark mass is lowered.  The behavior of
$G_M(Q^2)/g_V$ is similar as shown in Fig.~\ref{fig:GM-vsM}.
\item
The data for $G_E(Q^2)/g_V$ do not show any significant dependence on
the lattice spacing $a$ for fixed pion mass as shown in
Fig.~\ref{fig:GE-vsa}. A similar insensitivity to change in $a$ is
exhibited by $G_M(Q^2)/g_V$ as shown in Fig.~\ref{fig:GM-vsa}.
\end{itemize}

Estimates for $\rMsq$ from $z$-expansion fits without including our
derived value for $G_M(0)$ are, in many cases unstable even for the
$z^3$ or $z^{3+4}$ fits, i.e., estimates for $\rM$ become negative.
We conclude that the fits in these cases are over-parameterized. Including the derived
value of $G_M(0)$ and imposing the constraint on $a_k$ discussed in
Sec.~\ref{sec:formfactors} greatly improved the $z$-expansion fits. On the other
hand, the dipole fits give consistent estimates with or without using
a value for $G_M(0)$.  Our final results for both types of $Q^2$ fits
are obtained including the $G_M(0)$ points.

Lastly, the comparison of the lattice data with the Kelly fit to the
experimental data is shown in Figs.~\ref{fig:GE-vsa}
and~\ref{fig:GM-vsa}. Both $G_E(Q^2)/g_V$ and $G_M(Q^2)/g_V$ move
towards the Kelly curve as $M_\pi$ and $a$ are reduced. However,
$G_E(Q^2)/g_V$ from the two physical mass ensembles still shows
significant deviations from the Kelly fit. The data for $G_M(Q^2)/g_V$
show a different curvature from the Kelly curve and points with $Q^2
\lesssim 0.2$~GeV from the physical mass ensembles move below the
Kelly curve. This change in behavior in $G_M(Q^2)/g_V$ results in an
underestimate of both $\rMsq$ and the magnetic moment $\mu^{p-n}$ as
discussed in Sec.~\ref{sec:results}.

\subsubsection{Dependence on lattice size}
\label{sec:volume}
\FIXME{sec:volume} 

Simulations on large lattices are not only important for reducing
finite volume effects but also provide the simplest solution to obtaining data
at smaller $Q^2$ for fixed $a$ and $M_\pi$. To demonstrate the
improvement possible, we compare data from the $a12m220S$, $a12m220$
and $a12m220L$ ensembles in Fig.~\ref{fig:GM-ESC-vol} in
Appendix~\ref{appendix:ESC}. As the data move to smaller $Q^2$ with increasing 
$L$, the statistical quality of the signal also improves for
a fixed number of measurements.

In Fig.~\ref{fig:GE-GM-vsV}, we show $G_E$ and $G_M$ versus $Q^2$ for
these three ensembles. The data on the two larger volumes, $a12m220$
($M_\pi L = 4.38$) and $a12m220L$ ($M_\pi L = 5.49$), overlap for both
$G_E(Q^2)/g_V$ and $G_M(Q^2)/g_V$, indicating that finite volume
effects are small for $M_\pi L \gtrsim 4.4$. On the smaller volume
$a12m220S$ ($M_\pi L = 3.29$), $G_M(Q^2)/g_V$ falls off faster with $Q^2$. 

In Fig.~\ref{fig:Vol-rErM}, we compare the results of three fits to
$G_E(Q^2)$ and $G_M(Q^2)$ given in Tables~\ref{tab:GEdata-ALL}
and~\ref{tab:GMdata-ALL} versus $Q^2$ for these three ensembles. For
the $z$-expansion fits, the results for $\rEsq$ and $\rMsq$ from the two
larger volumes are consistent within $1\sigma$, while those on
$a12m220S$ differ.  We find no significant difference in the dipole
fits. These comparisons indicate that finite volume corrections are
smaller than the statistical errors on the two larger volumes
corresponding to $M_\pi L \gtrsim 4.4$. For this reason, we carry out
CCFV fits including (11-point fit) and discarding the $a12m220S$ point
($10^\ast$-point fit). Operationally, the fits are insensitive to the
$12m220S$ point due to the larger errors in it. Nevertheless, our final results,
presented in Sec.~\ref{sec:results}, are from the 11-point fit.

The bottom line is that increasing $L$ for fixed $M_\pi $ and $a$
improves the analysis in a number of ways because the values of $Q^2$
for a given ${\vec n}$ decrease. First, the statistical errors for a
fixed number of measurements decrease. The reduction in errors roughly
compensates for the increase in cost of each measurement due to a
larger volume. Second, with the decrease in $Q^2$, the ESC in
$G_E^{V_4}$ becomes smaller, while that in $G_M$ becomes easier to
control using n-state fits. Lastly, the extraction of $\rEsq$, $\rMsq$
and $\mu$ improves since the fit parameters are determined from data with values
of $Q^2$ closer to zero.

\subsubsection{Dependence on smearing size}
\label{sec:smearing}
\FIXME{sec:smearing} 

In Figs.~\ref{fig:GM-ESC-smear310} and~\ref{fig:GM-ESC-smear220} in
Appendix~\ref{appendix:ESC}, we compare the ESC in $G_E^{V_4}$ and
$G_M^{V_i}$ for two different smearing sizes using data from the
$a06m310$ and $a06m220$ ensembles. The data show that the ESC is
smaller with the larger smearing size.

\begin{figure*}[tpb] 
\centering
\subfigure{
\includegraphics[width=0.47\linewidth]{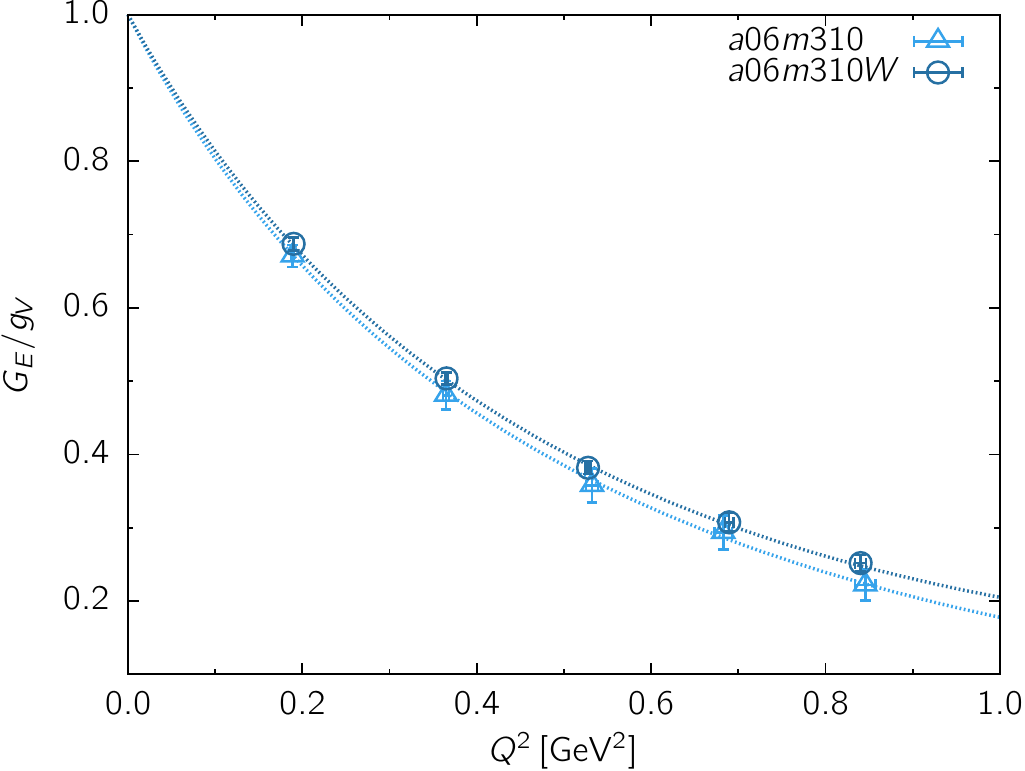}
\includegraphics[width=0.47\linewidth]{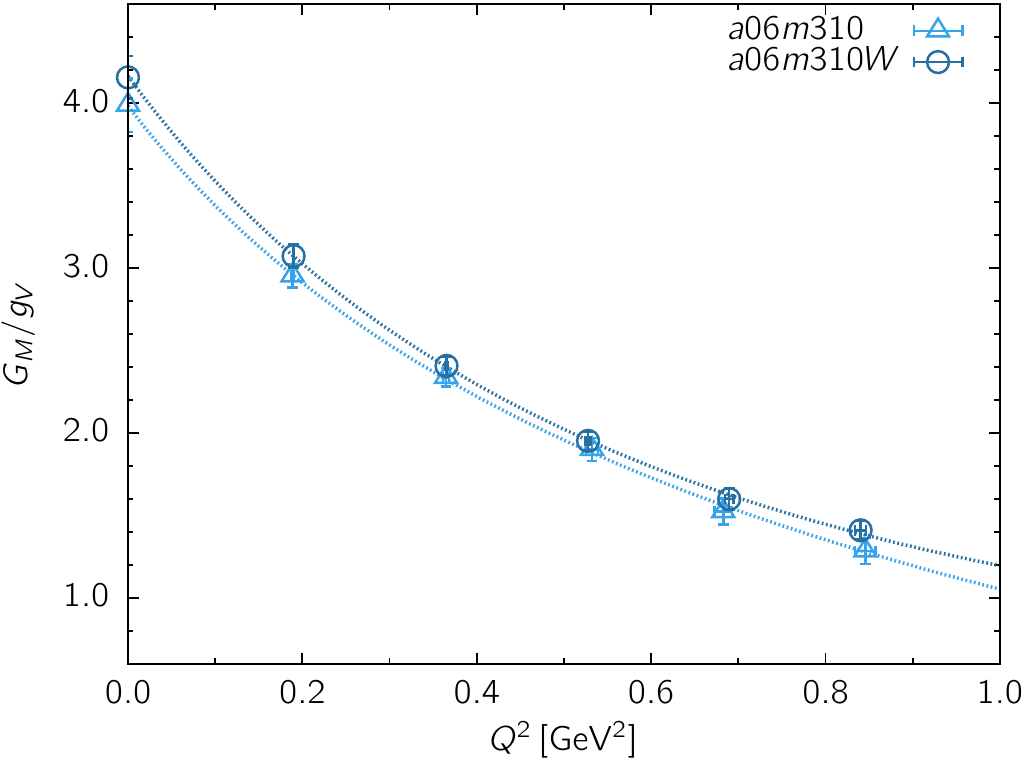}
}
\subfigure{
\includegraphics[width=0.47\linewidth]{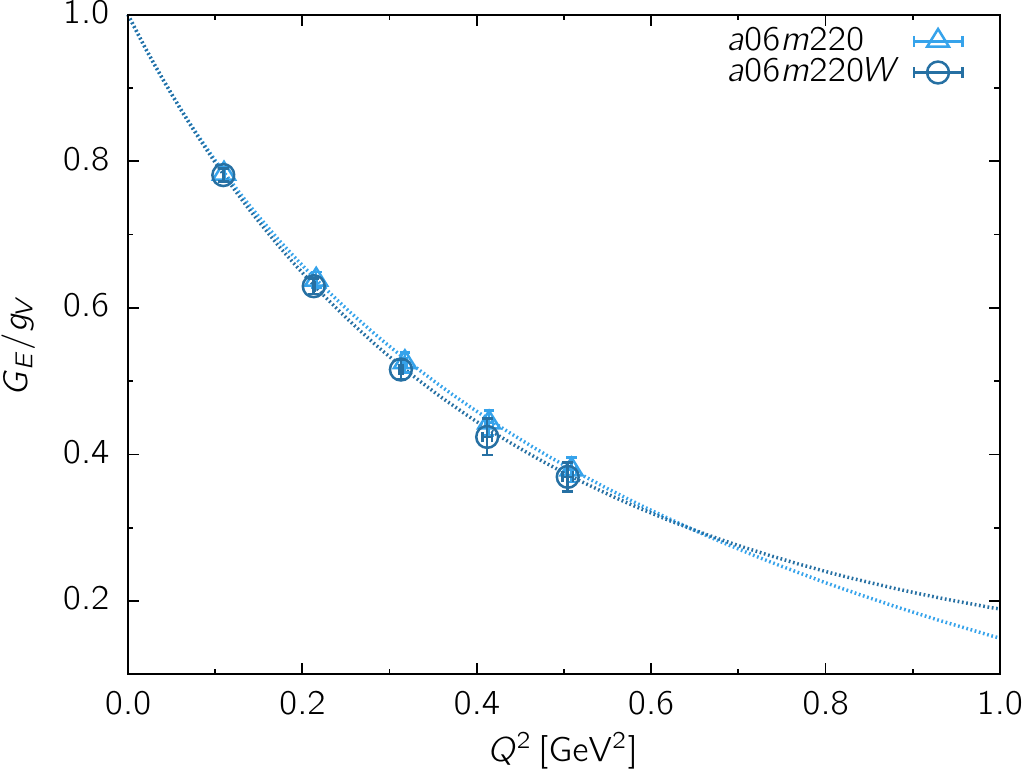} 
\includegraphics[width=0.47\linewidth]{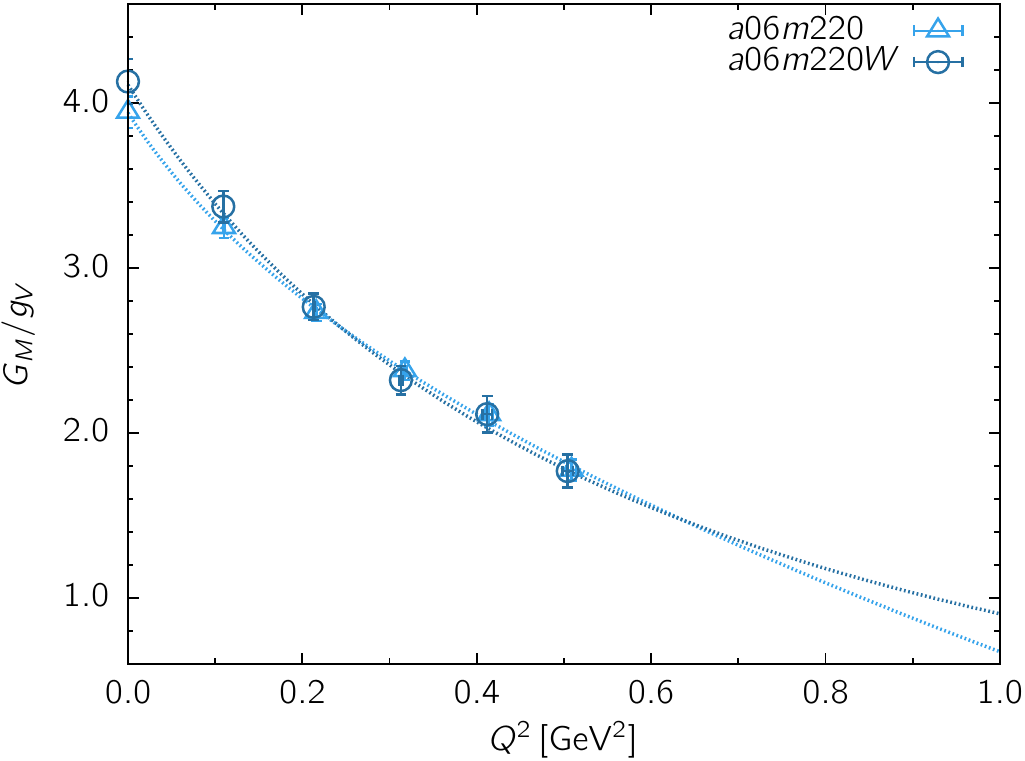}
}
\caption{\FIXME{fig:GE-GM-vsS} The data for the renormalized
  $G_E(Q^2)/g_V$ (left) and $G_M(Q^2)/g_V$ (right) form factors
  plotted versus $Q^2$ for the two ensembles $a06m310$ (top) and
  $a06m220$ (bottom) analyzed with two different source/sink smearing
  parameters given in Table~\protect\ref{tab:cloverparams}.  The dotted lines show the $z^{4}$
  fit. }
\label{fig:GE-GM-vsS}
\end{figure*}

The results of the dipole, $z^4$ and $z^{5+4}$ fits to $G_E(Q^2)$ and $G_M(Q^2)$ versus $Q^2$
for these two ensembles are shown in Fig.~\ref{fig:Smear-rErM}.
Results for $\rEsq$ and $\rMsq$ are consistent within $1\sigma$ for
the two smearings. The data in Fig.~\ref{fig:GE-GM-vsS}, however, show
that estimates of $\mu$ can differ by about 5\% between the two
calculations with different smearing size.  This level of difference
can be explained by a combination of statistical and possible systematic
uncertainties.

\subsubsection{Dependence on lattice scale setting}
\label{sec:scale}
\FIXME{sec:scale} 

The two places the lattice scale enters our calculation is in
converting $Q^2 a^2$ to physical units and in the CCFV fits. In
Table~\ref{tab:ens}, we give the values of $a$ for the HISQ ensembles
obtained by the MILC collaboration using the Sommer scale
$r_1$~\cite{Bazavov:2012xda,Sommer:2014mea}.  In Table~\ref{tab:spectrum} in
Appendix~\ref{appendix:NucleonMass}, we give the value of $M_N$
obtained on each ensemble using these values of $a$ and fit them using
the leading order CCFV fit defined in Eq.~\eqref{eq:MNfitB}. The result
in the continuum limit is $M_N = 976(20)$~MeV.  The deviation of about 4\% from the
experimental value indicates a systematic uncertainty of 2--6\% in
the scale obtained from $r_1$ versus $M_N$, the latter analyzed using the leading 
order CCFV fit. The question then is, how does this difference impact
the analysis of the form factors and the extraction of $\rEsq$,
$\rMsq$ and $\mu$?  

\begin{figure}[tpb] 
\centering
\subfigure{
\includegraphics[width=0.97\linewidth]{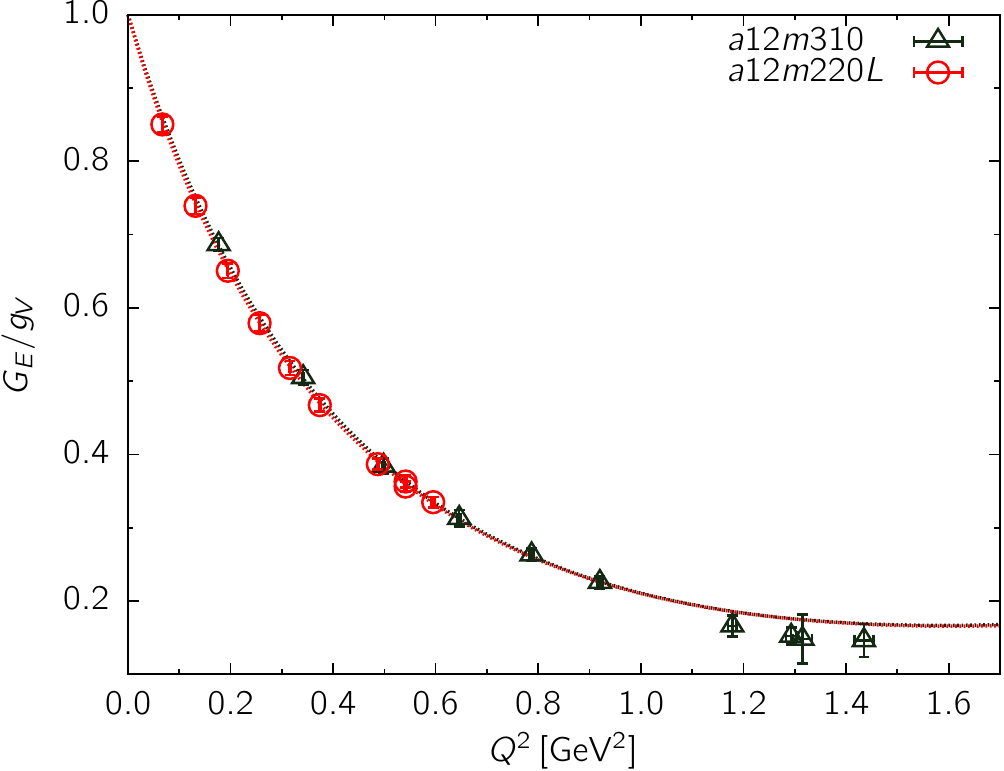} 
}
\subfigure{
\includegraphics[width=0.97\linewidth]{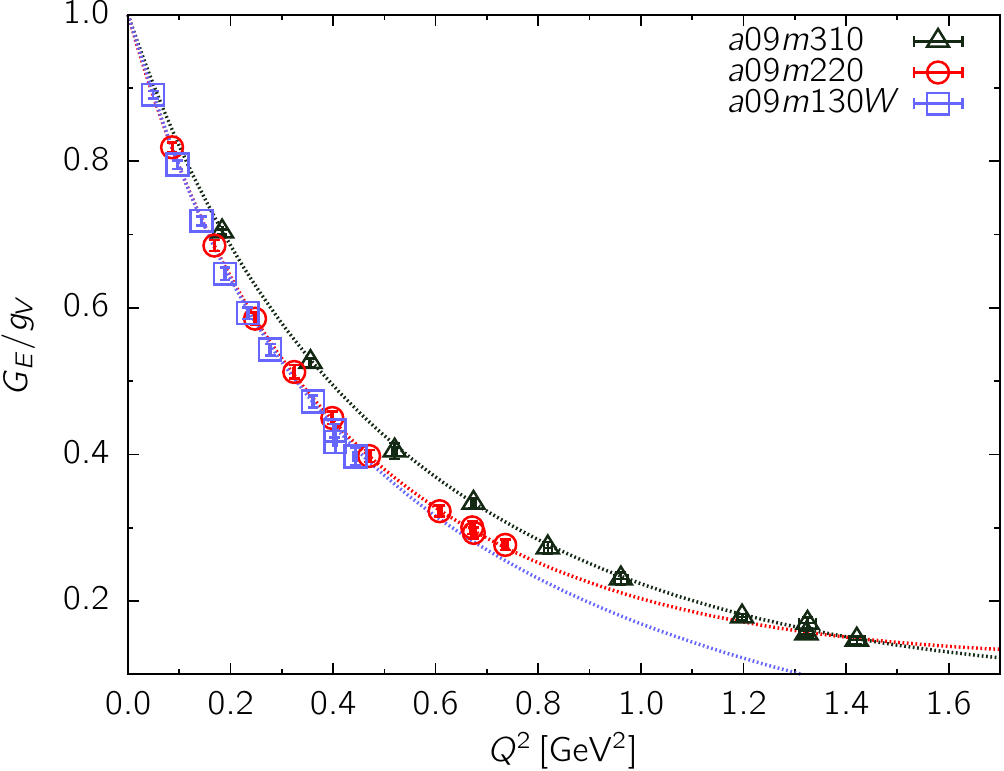} 
}
\subfigure{
\includegraphics[width=0.97\linewidth]{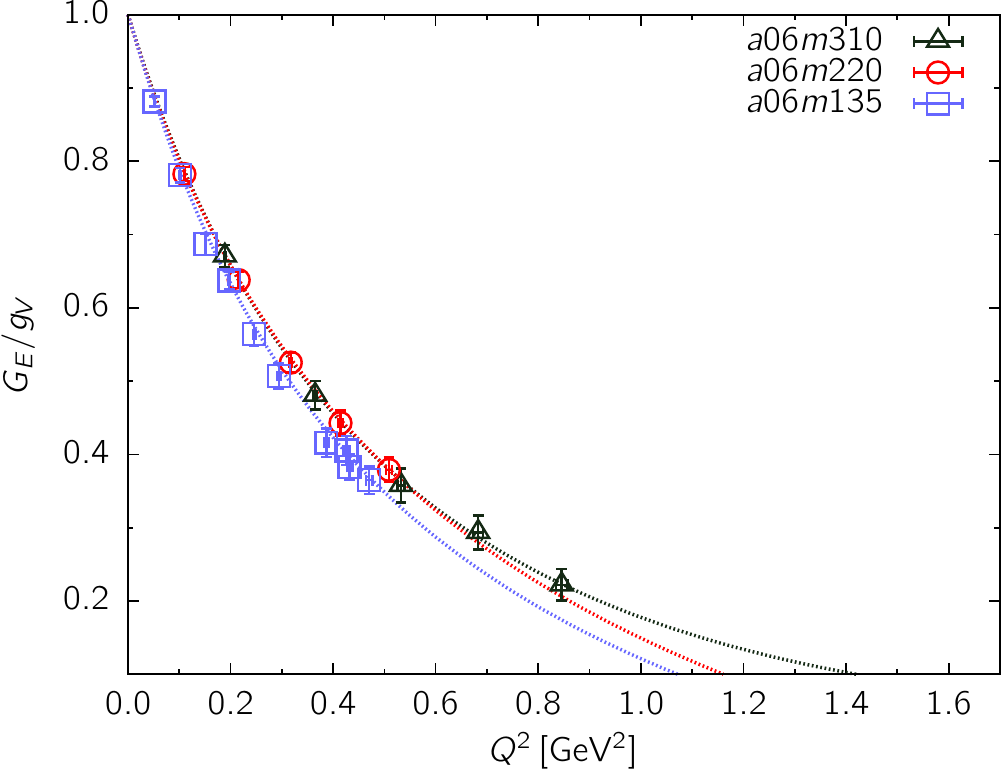} 
}
  \caption{\FIXME{fig:GE-vsM} The data for the renormalized electric
    form factor $G_E(Q^2)/g_V$ versus $Q^2$ plotted to highlight the
    dependence on $M_\pi^2$ for fixed $a$. The dotted lines show the
    $z^{4}$ fit.  The top figure is for the $a\approx 0.12$~fm
    ensembles, the middle for the $a\approx 0.09$~fm ensembles, and
    the bottom for the $a\approx 0.06$~fm ensembles.  The color scheme
    used is black triangles for the $M_\pi \approx 310$~MeV, red circles
    for $M_\pi \approx 220$~MeV, and blue squares for the $M_\pi \approx
    135$~MeV ensembles data.}
\label{fig:GE-vsM}
\end{figure}

\begin{figure}[tpb]    
\centering
\subfigure{
\includegraphics[width=0.97\linewidth]{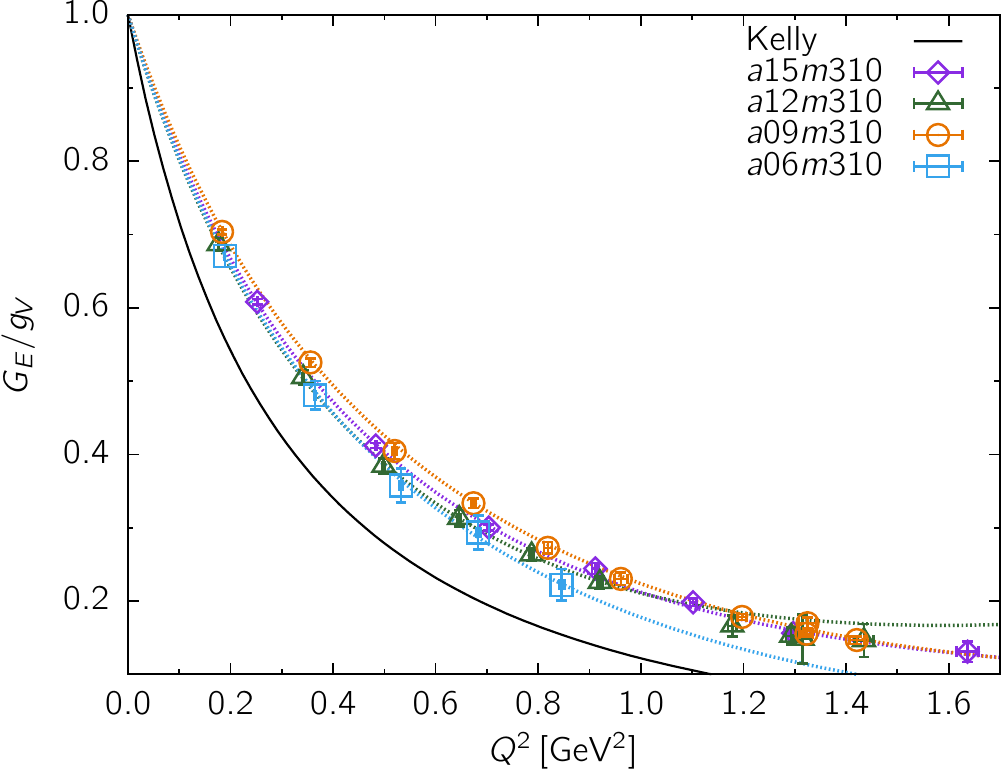}
}
\subfigure{
\includegraphics[width=0.97\linewidth]{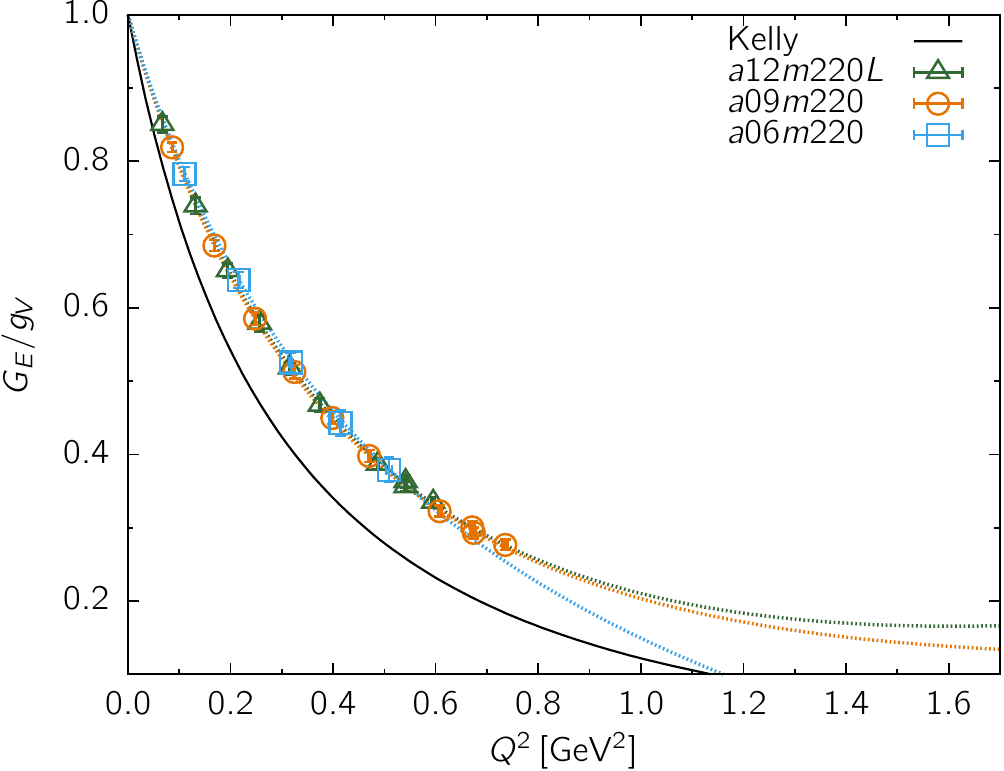}
}
\subfigure{
\includegraphics[width=0.97\linewidth]{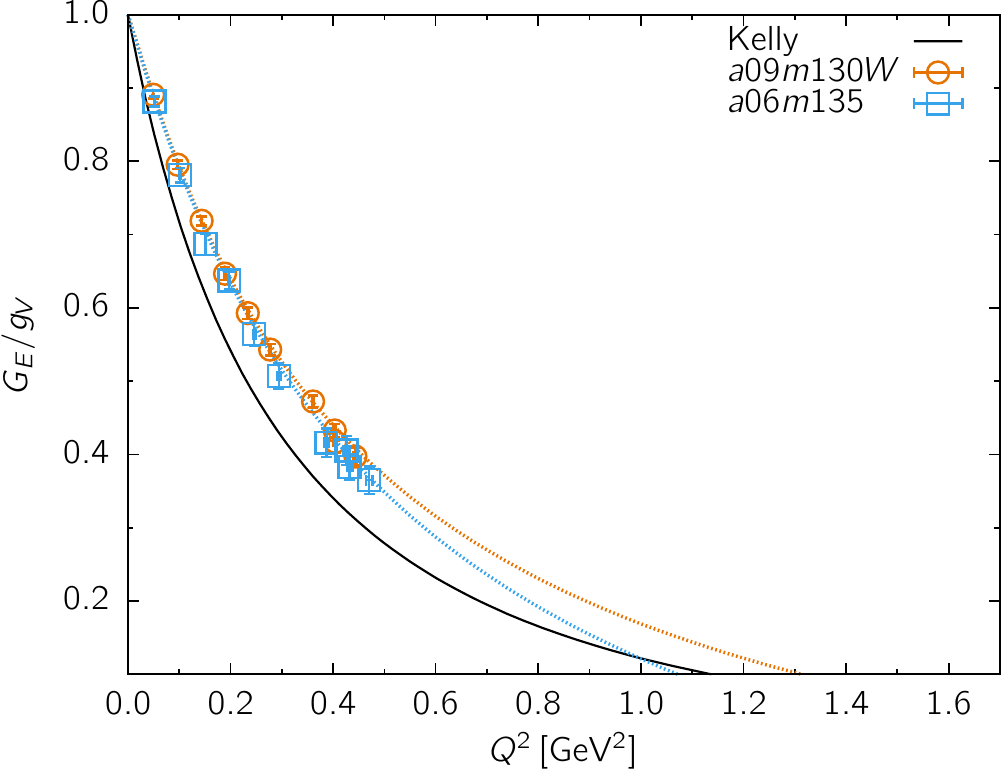}
}
  \caption{\FIXME{fig:GE-vsa} The data and fits for the renormalized
    electric form factor $G_E(Q^2)/g_V$ versus $Q^2$ plotted to
    highlight the dependence on $a$ for fixed $M_\pi$. The dotted
    lines show the $z^{4}$ fit and the solid line is
    the Kelly fit to the experimental $G_E^{p-n}$ data. The top figure is for the
    $M_\pi \approx 310$~MeV ensembles, the middle for the $M_\pi
    \approx 220$~MeV ensembles, and the bottom for the $M_\pi \approx
    135$~MeV ensembles. The symbols used are: purple diamond for the
    $a \approx 0.15$~fm, green triangles for the $a \approx 0.12$~fm,
    orange circles for $a \approx 0.09$~fm and blue squares for the $a
    \approx 0.06$~fm ensembles data.}
\label{fig:GE-vsa}
\end{figure}

\begin{figure}[tpb] 
\centering
\subfigure{
\includegraphics[width=0.97\linewidth]{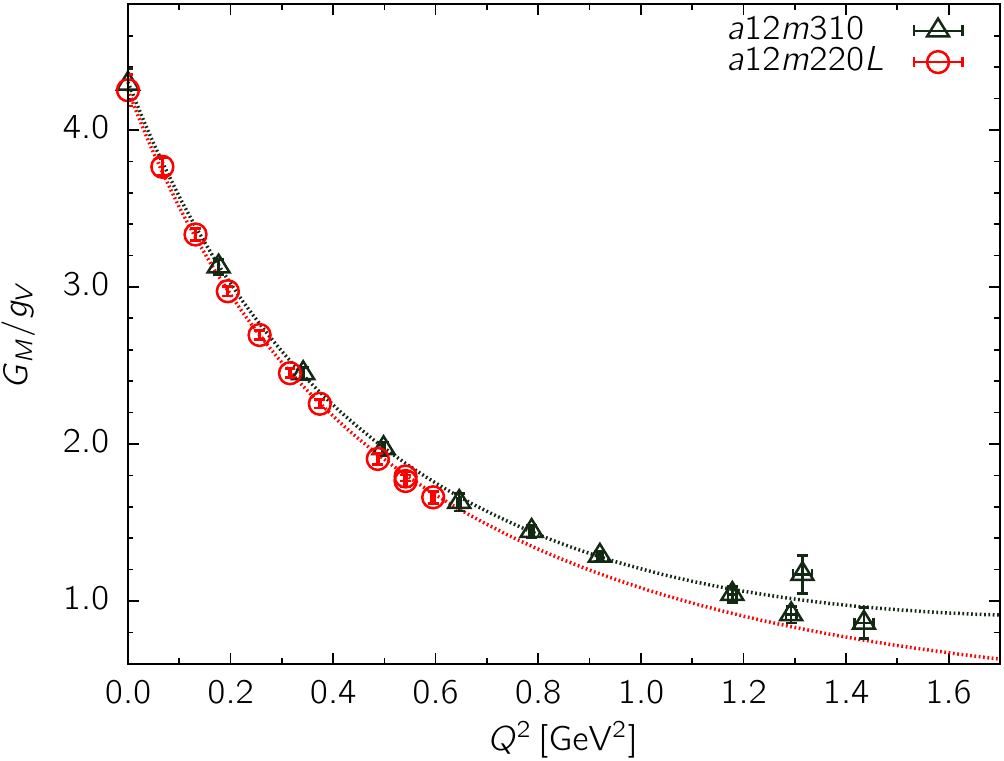} 
}
\subfigure{
\includegraphics[width=0.97\linewidth]{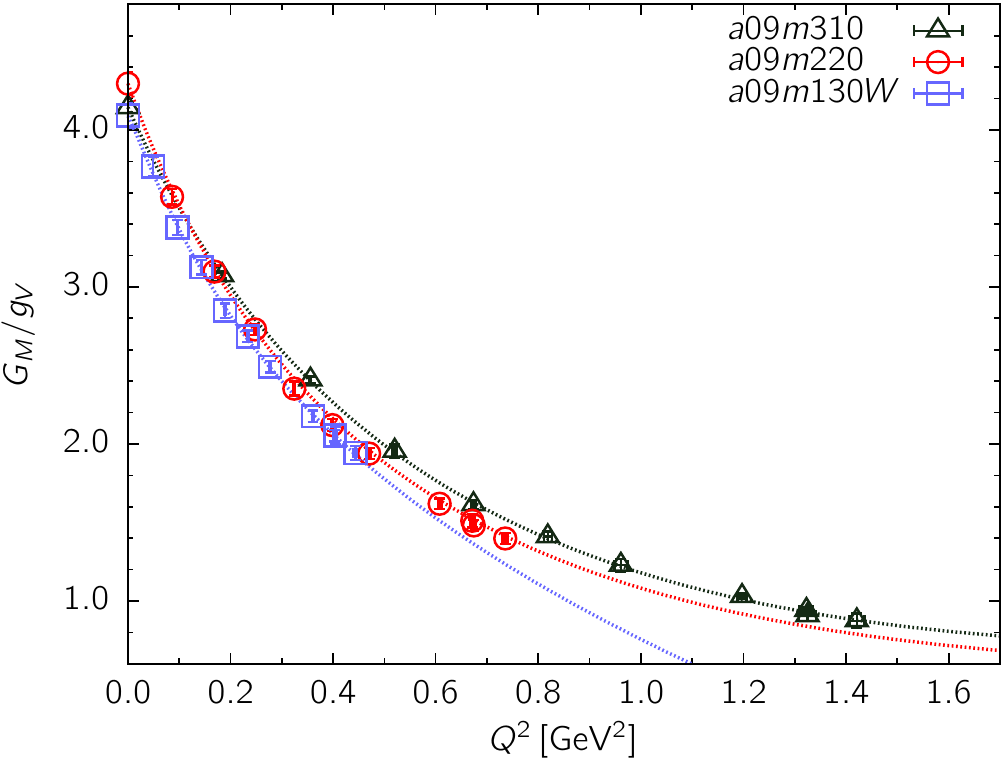} 
}
\subfigure{
\includegraphics[width=0.97\linewidth]{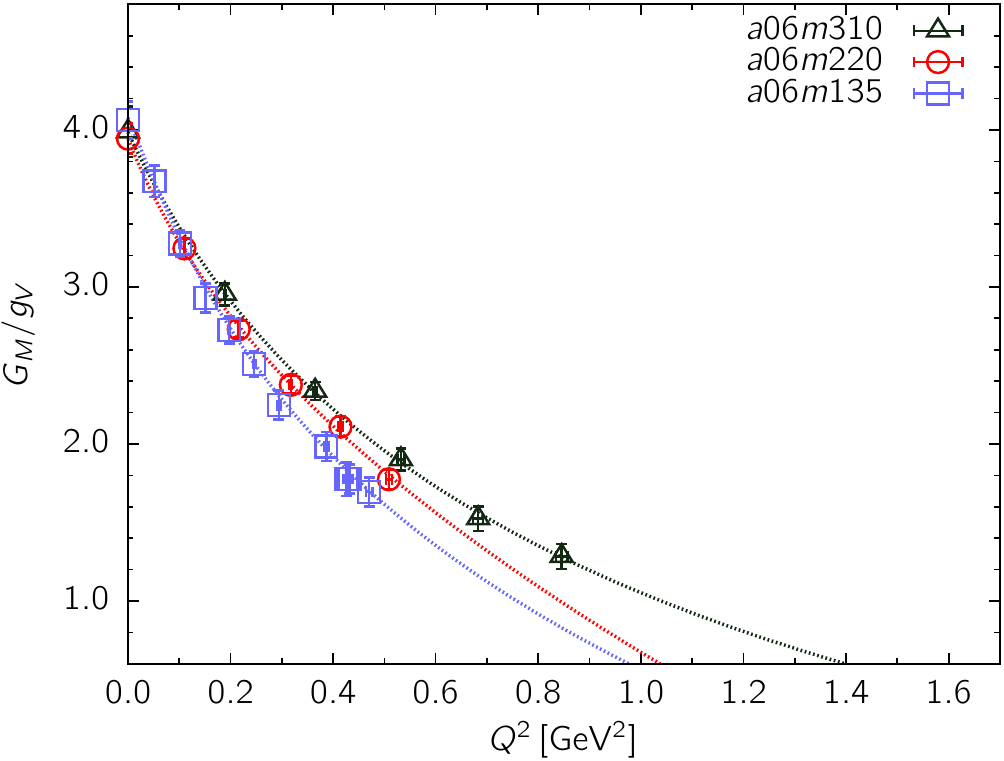} 
}
\caption{\FIXME{fig:GM-vsM} The data and $z^4$ fits to the
  renormalized magnetic form factor $G_M(Q^2)/g_V$ plotted versus
  $Q^2$ to highlight the dependence on $M_\pi^2$ for fixed $a$. The
  top figure is for the $a\approx 0.12$~fm ensembles, the middle for
  the $a\approx 0.09$~fm ensembles, and the bottom for the $a\approx
  0.06$~fm ensembles.  The symbols used are: black triangles for the
  $M_\pi \approx 310$~MeV, red circles for $M_\pi \approx 220$~MeV, and purple
  squares for the $M_\pi \approx 130$~MeV ensembles.  }
\label{fig:GM-vsM}
\end{figure}

\begin{figure}[tpb] 
\centering
\subfigure{
\includegraphics[width=0.97\linewidth]{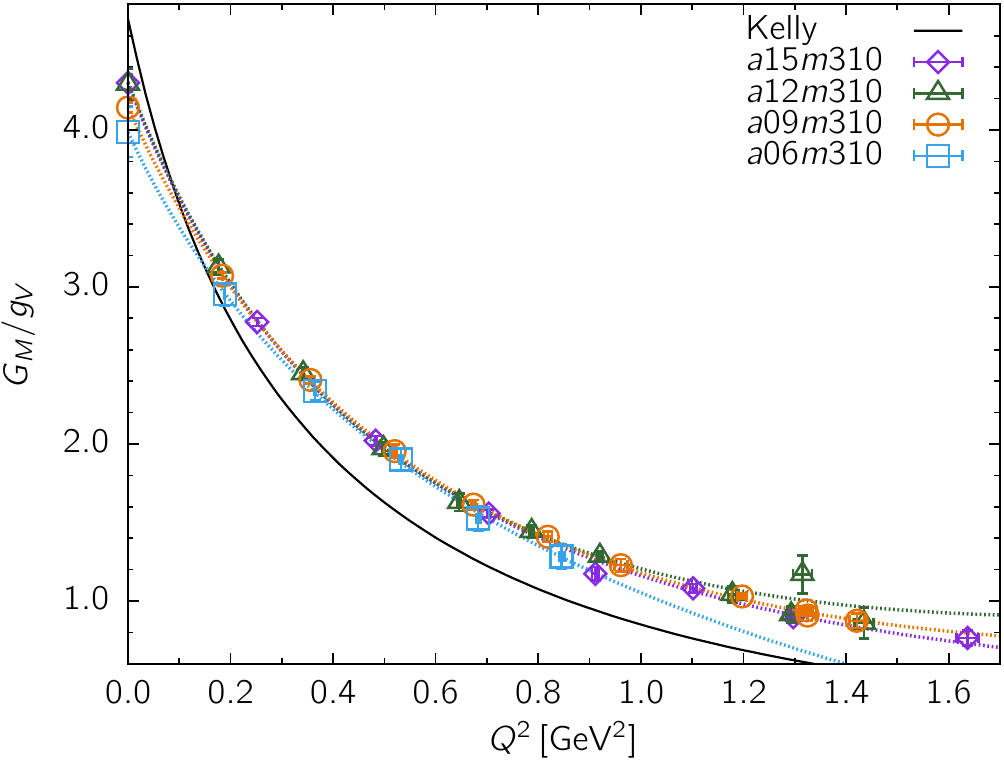}
}
\subfigure{
\includegraphics[width=0.97\linewidth]{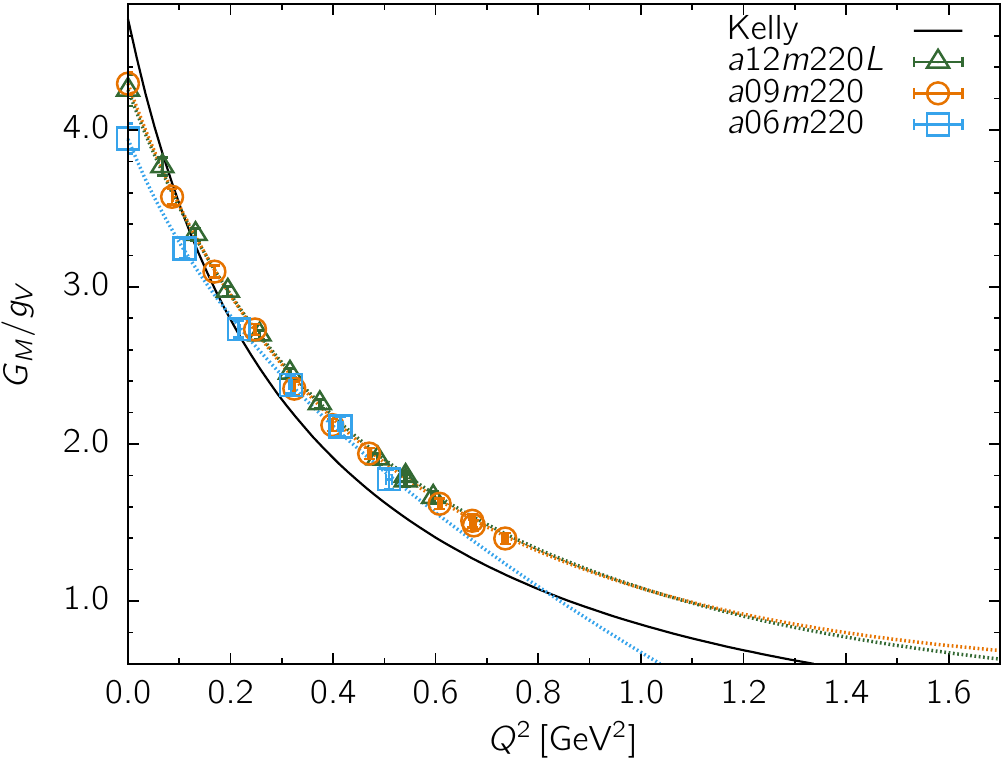}
}
\subfigure{
\includegraphics[width=0.97\linewidth]{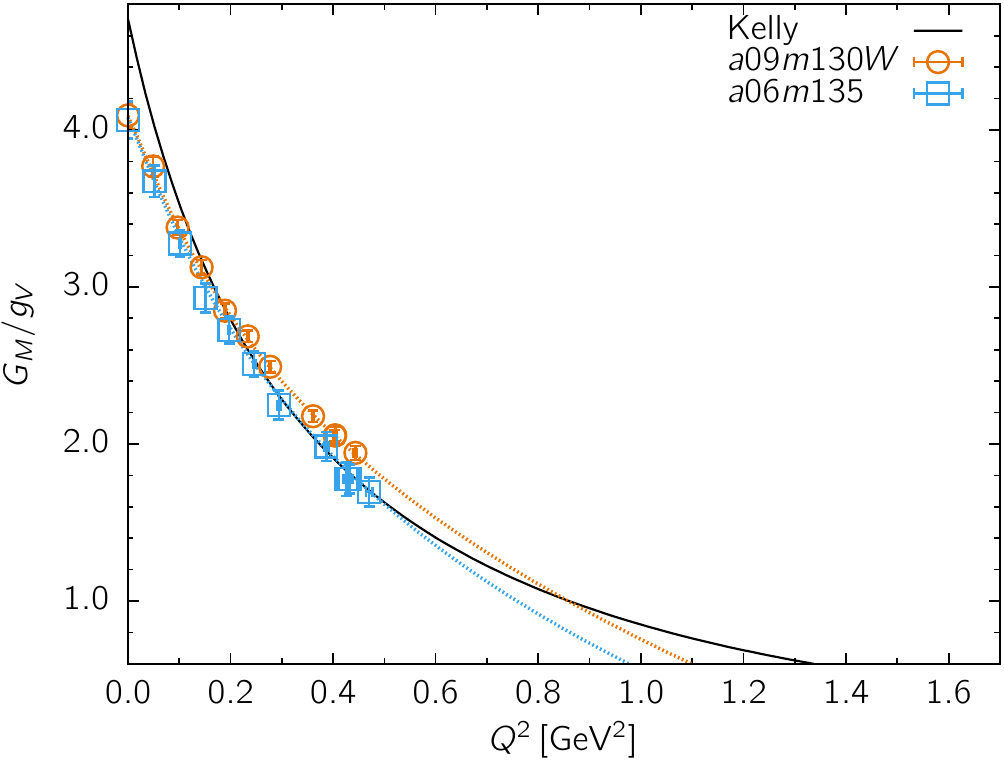}
}
\caption{\FIXME{fig:GM-vsa} The data and the $z^4$ fits (dotted lines)
  for the renormalized magnetic form factor $G_M(Q^2)/g_V$ plotted
  versus $Q^2$ to highlight the dependence on $a$ for fixed
  $M_\pi$. The solid line is the Kelly fit to the isovector
  combination, $(p-n)$, of the experimental data. The top figure is
  for the $M_\pi \approx 310$~MeV ensembles, the middle for the $M_\pi
  \approx 220$~MeV ensembles, and the bottom for the $M_\pi \approx
  130$~MeV ensembles. The symbols used are: magenta diamonds for the
  $a \approx 0.15$ fm, green triangles for the $a \approx 0.12$ fm,
  orange circles for $a \approx 0.09$ fm and blue squares for the $a
  \approx 0.06$~fm ensembles.}
\label{fig:GM-vsa}
\end{figure}

The lattice data plotted in Figs.~\ref{fig:GE-vsM}--\ref{fig:GM-vsa}
show that the dependence of the form factors on $M_\pi^2$ and $a$ is
small.  To explore the dependence further, we remove the use of $a$
taken from the analysis of the Sommer scale $r_1$ on the HISQ ensembles by plotting the data
versus $Q^2/M_N^2$ in Fig.~\ref{fig:Kelly} (bottom) where the lattice
values of $M_N$ are used to construct the dimensionless ratio
$Q^2/M_N^2$ for the lattice data and $M_N=939$~MeV for the Kelly
curve. The relative movement between the data and the Kelly curve,
when plotted versus $Q^2/M_N^2$ as compared to $Q^2$, brings the data
closer together onto a single curve as can be seen by comparing the
top and bottom set of panels. For the physical mass ensembles, the
size of the relative movement of data depends only on the
discretization errors, i.e., the value of $M_N$ at that value of $a$,
assuming finite volume corrections are negligible. Presuming a
cancellation of some of the systematics when the data are plotted
versus $Q^2/M_N^2$, this comparison indicates that the observed larger
deviation from the Kelly curve, when the data are plotted versus
$Q^2$, can be explained partly as a systematic effect due to
discretization errors, i.e., variations in the lattice scale set using
different observables. This systematic is avoided if data at a given
$Q^2$ are first extrapolated to $M_\pi = 135$~MeV and $a=0$ and then
compared with the Kelly curve. An attempt at doing this is described
in Sec.~\ref{sec:CombinedFit}.

It is important to note that $\rEsq$, $\rMsq$ and $\mu$ extracted for
each ensemble are unchanged whether one calculates them using $Q^2$ or
$Q^2/M_N^2$ as the independent variable in Eq.~\eqref{eq:rdef}.  The
result would be different if the product $M_N^2 \rEsq$ is calculated on each
ensemble, and extrapolated to the continuum limit 
first, and the result divided by the experimental value for $M_N$.  We discuss this 
analysis in Sec.~\ref{sec:Dimless}.

Having made clear that part of the noticeable spread in the behavior of
the form factors shown in Fig.~\ref{fig:Kelly} can be accounted for,
in a large part, as due to discretization errors, the question
is--what is the more robust way of analyzing the data?  Should we use
the scale set using $r_1$ or work with dimensionless variables in
units of $M_N$?  While our analysis has exposed this systematic, our
conclusion is that a larger data set, or the use of a lattice action
with much smaller discretization errors or a better determined
extrapolation ansatz are needed to significantly reduce such
systematics. Having highlighted the size of this systematic
uncertainty, most of the analysis presented below is carried out
versus $Q^2$. We provide comparison with results plotted versus
$Q^2/M_N^2$ at appropriate places, and analyze data for $M_N^2 \rEsq$
and $M_N^2 \rMsq$ in Sec.~\ref{sec:Dimless}.

%

\section{Characterizing the $Q^2$ behavior of the form factors}
\label{sec:Q2behavior}

\FIXME{sec:Q2behavior}

In order to extract the charge radii defined in Eq.~\eqref{eq:rdef}
and the magnetic moment in Eq.~\eqref{eq:Pmmdef}, we need to
parameterize the form factors versus $Q^2$. The two fits we explore
are the dipole and the $z$-expansion truncated at some power $k$ as
discussed in Sec.~\ref{sec:formfactors}.  Since the dipole ansatz is
the solution to an exponentially falling charge distribution (thus a
model) and the $z$-expansion involves a truncation plus a constraint on
the size of the coefficients $a_k$, it behooves us to first test these
ans\"atze on the high-precision experimental data as discussed next.

\subsection{Experimental data for the form factors and their $Q^2$ behavior}
\label{sec:Q2expt}

\FIXME{sec:Q2expt} 

Electromagnetic form factors of nucleons are extracted from
differential cross-sections measured in the scattering of electrons
off nuclei. The process of going from measurements of the differential
cross-sections to nucleon form factors is nontrivial and involves
modeling~\cite{Perdrisat:2006hj,Lee:2015jqa,Yan:2018bez}. As already
stated, we have two reasons to analyze the experimental data: to
compare them against the lattice data over the range $0 < Q^2 \lesssim
0.8$~GeV${}^2$, and to test the efficacy of the dipole and
$z$-expansion fit ans\"atze. For these purposes, we have collected together compiled
experimental data for the proton and the neutron in
Appendix~\ref{appendix:expFF} (see Figs.~\ref{fig:FFproton} and~\ref{fig:FFneutron}). 
From these, we have
determined the Kelly parameterization for the isovector combinations,
$G_E^p-G_E^n$ and $G_M^p - G_M^n$.  Henceforth, for brevity, we will
continue to use $G_E(Q^2)/g_V$ and $G_M(Q^2)/g_V$ to represent the $(p-n)$
combinations when comparing the lattice and the experimental data.

Next, we test the fit ans\"atze on the experimental data.  The results
for $\exprE$ and $\exprM$ for the proton are shown in
Fig.~\ref{fig:FFproton}. Based on the $\chi^2/{\rm DOF}$, the dipole
fit works surprisingly well for $G_E(Q^2)$, and the deviation from the
data is less than a percent over the range $0 < Q^2 \le
1$~GeV${}^2$. This difference is far less than the precision of our
lattice data. For $G_M(Q^2)$, the deviation is larger (up to 6\%) and
the $\chi^2/DOF$ of the fit is poor.  In the the $z$-expansion fits
with constraints, results for $\rEsq$, $\rMsq$ and $\mu$ stabilize for
$k \ge 5$ as shown in Fig.~\ref{fig:stabilityEXP}.

Based on this analysis, and as noted in Appendix~\ref{appendix:expFF},
one should not expect a match between our lattice and the experimental
data to better than about 5\% or be able to resolve differences
between the dipole and the $z$-expansion fits at or below this
level. These comparisons provide a framework for our lattice analyses
using the $z$-expansion: extract $\rEsq$ and $\rMsq$ from $k=4$ to
avoid over-parameterization for some of the ensembles. 

In the final estimates, we have assigned an additional systematic uncertainty to
account for the fact that the CCFV fits have been made using just the
leading order corrections. This is discussed further in Sec.~\ref{sec:results}.

\begin{figure*}    
\centering
\subfigure{
\includegraphics[width=0.47\linewidth]{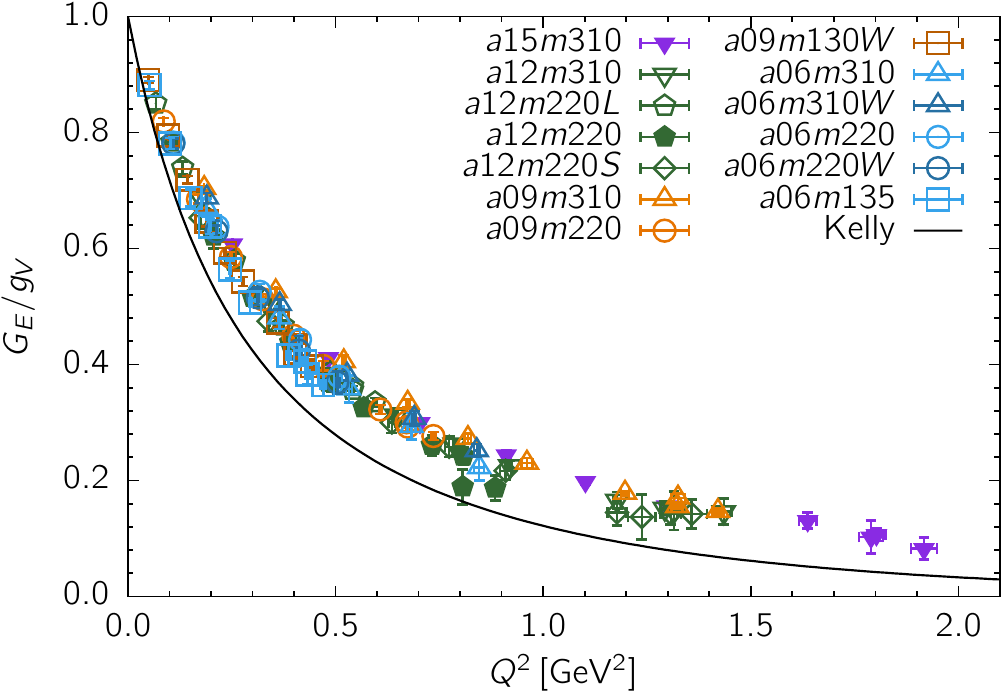}
\includegraphics[width=0.47\linewidth]{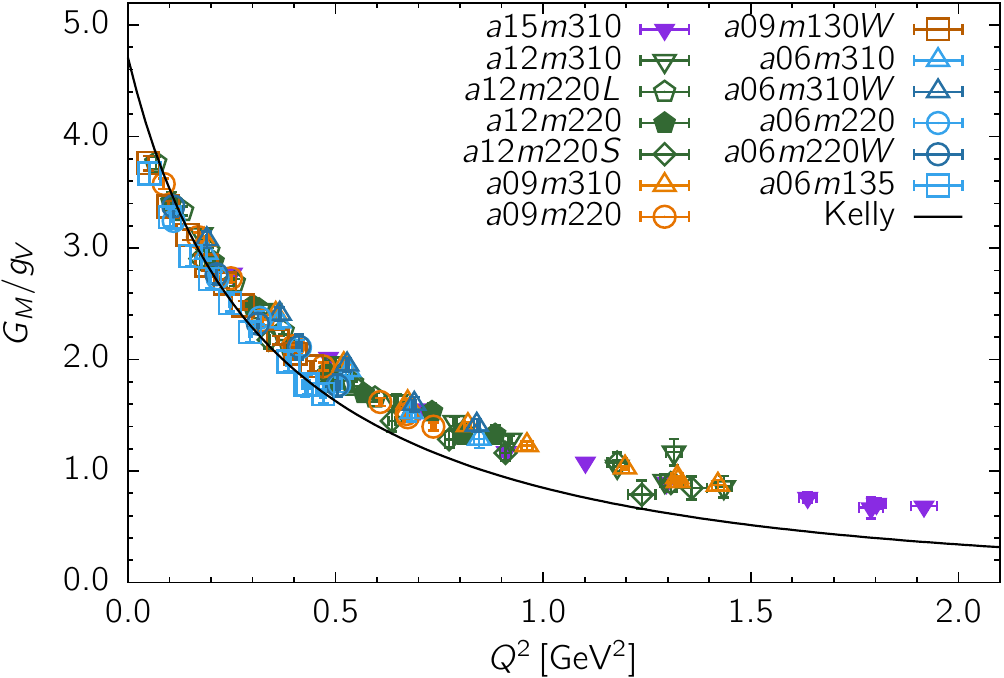}
}
\subfigure{
\includegraphics[width=0.47\linewidth]{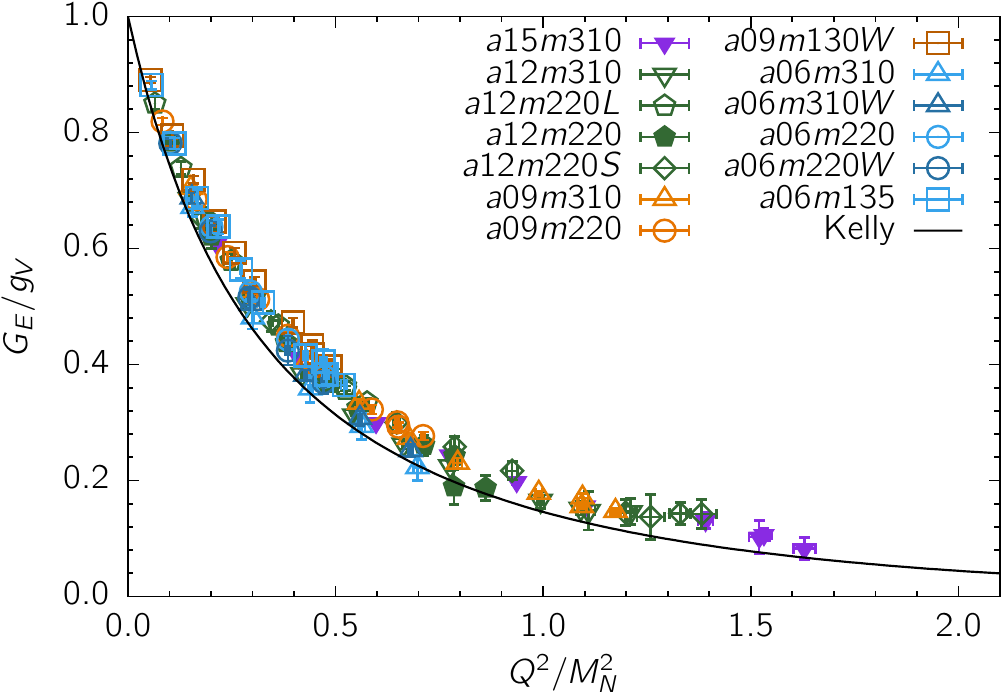}
\includegraphics[width=0.47\linewidth]{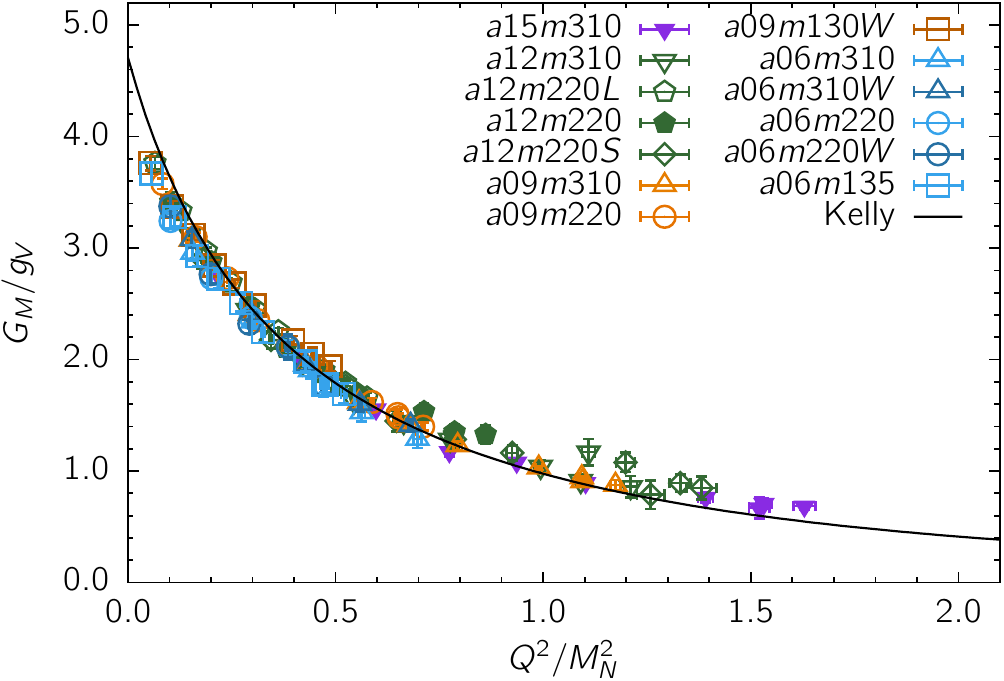}
}
\caption{\FIXME{fig:Kelly} The lattice data for the renormalized
  isovector form factors $G_E^{p-n}(Q^2)/g_V$ (left) and
  $G_M^{p-n}(Q^2)/g_V$ (right) from all thirteen calculations plotted
  versus $Q^2$ expressed in units of GeV${}^2$ (top) and versus $Q^2/M_N^2$
  with $M_N$ taken from the lattice (bottom). The solid black line is
  the Kelly fit to the $(p-n)$ experimental data with $M_N=939$~MeV.
}
\label{fig:Kelly}
\end{figure*}
%

\subsection{Analysis of the lattice QCD data for the form factors}
\label{sec:Q2lattice}
\FIXME{sec:Q2lattice} 

A comparison of the form factors $G_E(Q^2)/g_V$ and $G_M(Q^2)/g_V$
from all thirteen simulations with the Kelly parameterization of the
experimental $(p-n)$ data is shown in Fig.~\ref{fig:Kelly}.  The data
for $G_E(Q^2)/g_V$ lie above the Kelly curve with those from the two
physical mass ensembles being the closest as shown in
Fig.~\ref{fig:GE-vsa}, whereas the data for $G_M(Q^2)/g_V$ lies about
the Kelly curve for $Q^2 \gtrsim 0.2$~GeV${}^2$ and then falls below
it for smaller $Q^2$ as highlighted in Fig.~\ref{fig:GM-vsa}.  In both
cases, these deviations from the Kelly curve impact the slope at $Q^2
= 0$, i.e., both $\rE$ and $\rM$ come out smaller than the
phenomenological estimates. More importantly, the very precisely measured magnetic moment,
$G_M(0) = \mu_p - \mu_n$, is underestimated by about 16\%. As
remarked above in Sec.~\ref{sec:extractGE} and Sec.~\ref{sec:extractGM},
removing the ESC using the $3^\ast$-fits increases the value of all
three, nevertheless, the final results presented in
Sec.~\ref{sec:results} are smaller than the experimental values.
Furthermore, deviations of the lattice form factors from the Kelly
curve are apparent over a range of $Q^2$. 

As discussed in Sec.~\ref{sec:scale}, part of the difference between
the Kelly curve and the data is due to the mismatch in the scale set
by $r_1$ and $M_N$.  This is highlighted in Fig.~\ref{fig:Kelly} where
data are plotted versus $Q^2$ (top panels), evaluated using the
lattice scale set by $r_1$, and versus the dimensionless variable
$Q^2/M_N^2$ (bottom panels). Note that this change of variable does
not impact the results for $\rEsq$, $\rMsq$ and $\mu^{p-n}$ on each
individual ensemble and thus their extrapolated values.

\begin{figure*}    
  \centering
\subfigure{
\includegraphics[height=1.65in,trim={0.0cm 0.40cm 0 0},clip]{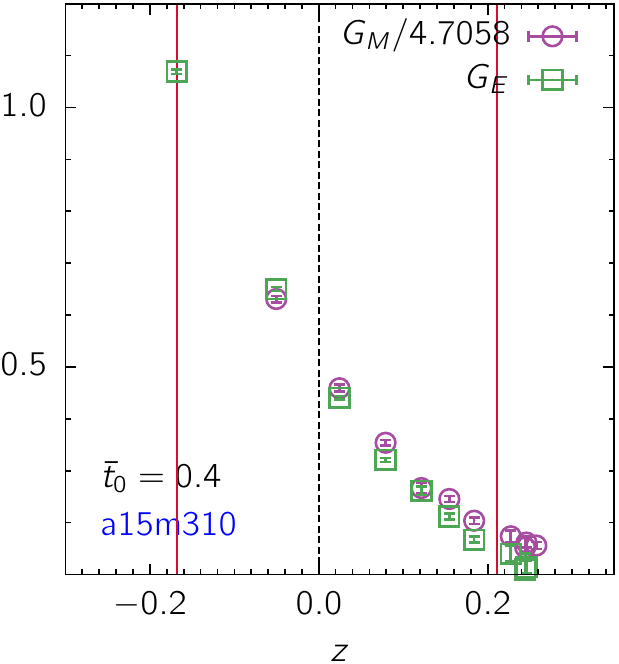} 
\includegraphics[height=1.65in,trim={0.0cm 0.40cm 0 0},clip]{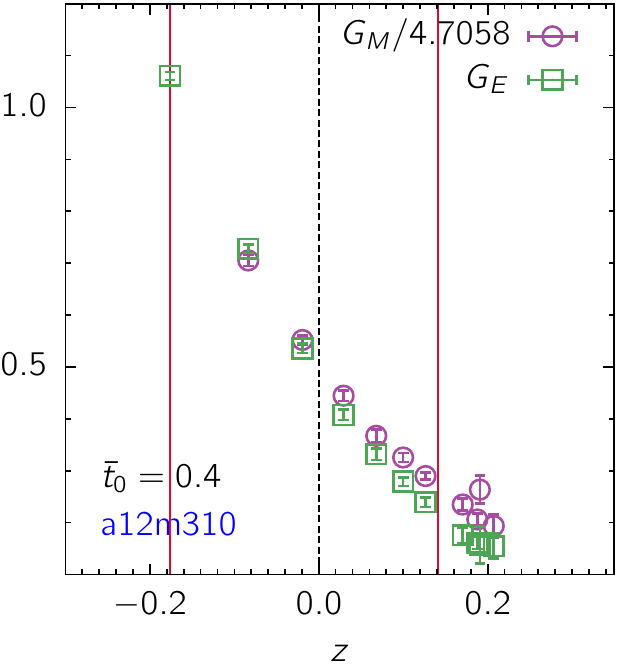} 
\includegraphics[height=1.65in,trim={0.0cm 0.40cm 0 0},clip]{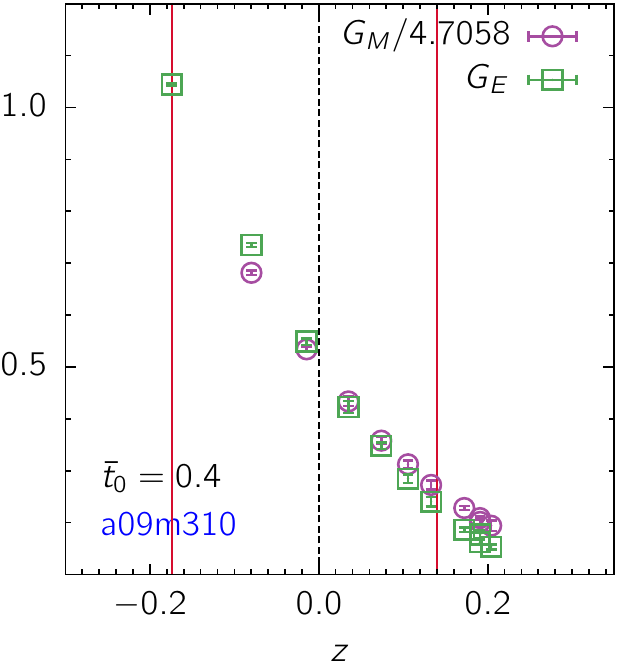} 
\includegraphics[height=1.65in,trim={0.0cm 0.40cm 0 0},clip]{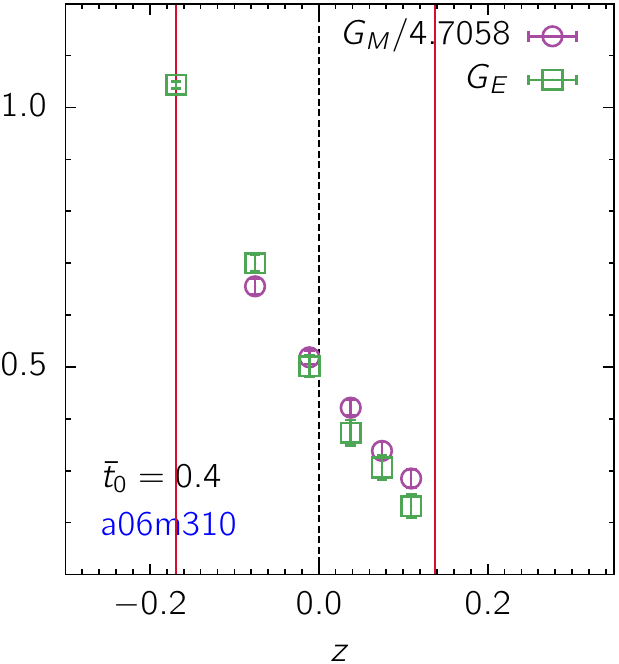} 
}
\subfigure{
\includegraphics[height=1.65in,trim={0.0cm 0.40cm 0 0},clip]{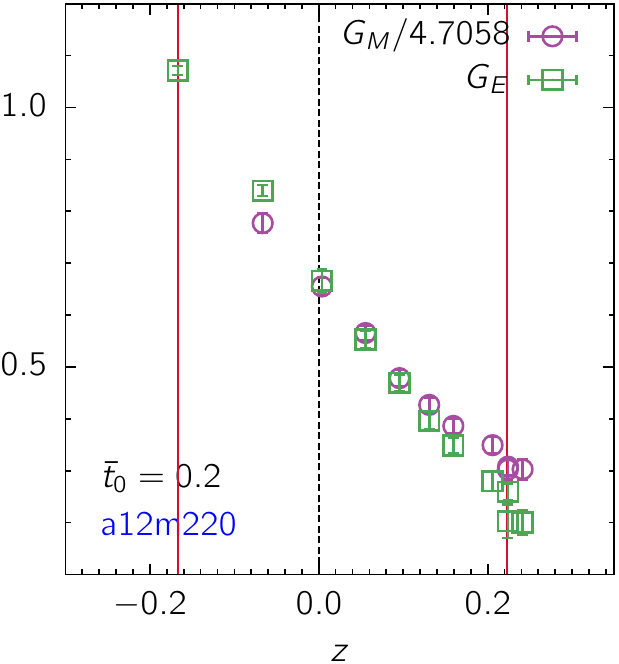} 
\includegraphics[height=1.65in,trim={0.0cm 0.40cm 0 0},clip]{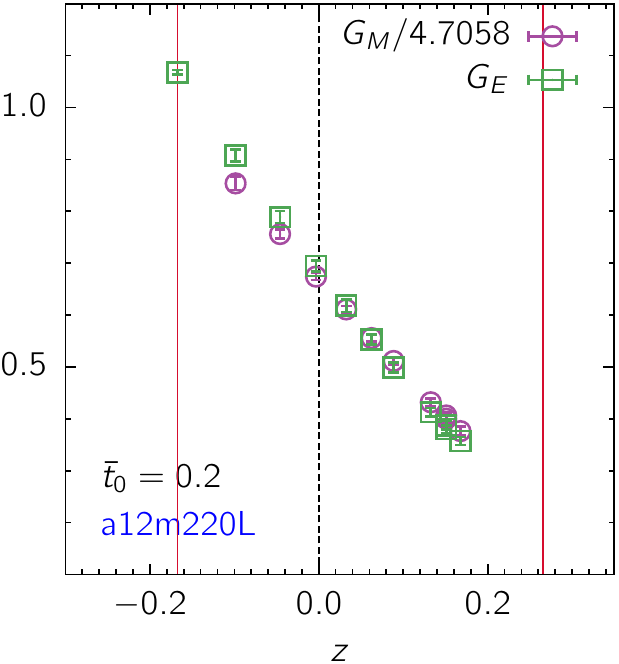} 
\includegraphics[height=1.65in,trim={0.0cm 0.40cm 0 0},clip]{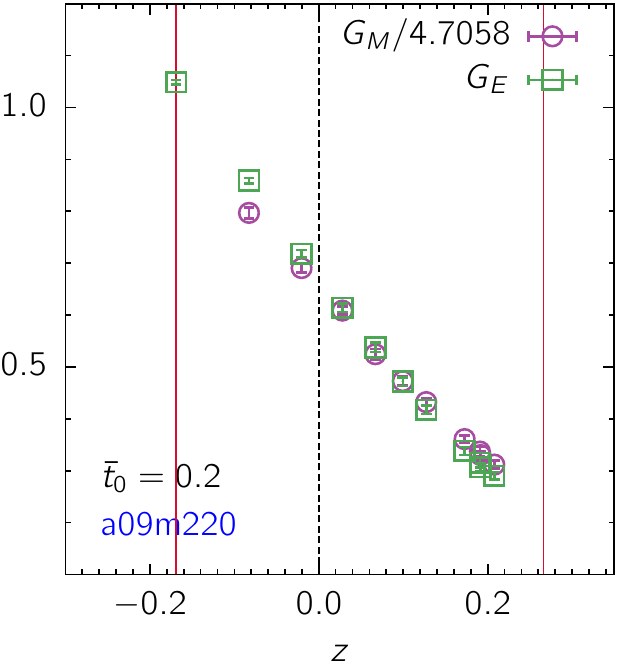} 
\includegraphics[height=1.65in,trim={0.0cm 0.40cm 0 0},clip]{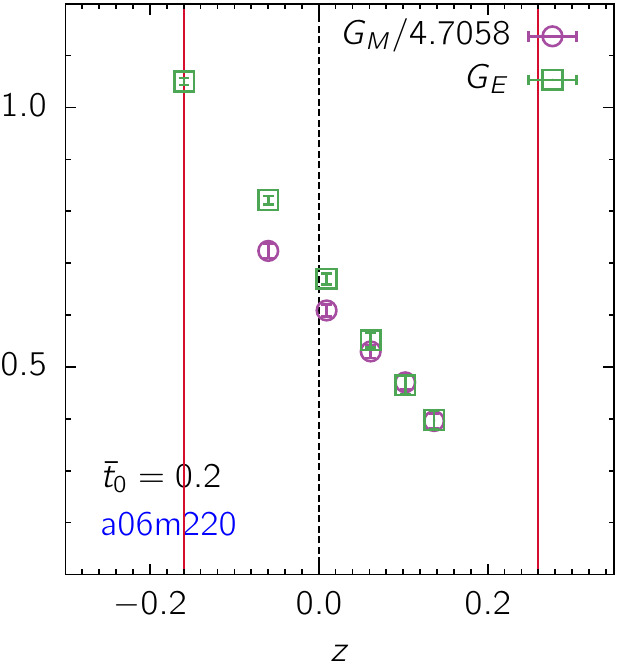} 
}
\subfigure{
\includegraphics[height=1.65in,trim={0.0cm 0.20cm 0 0},clip]{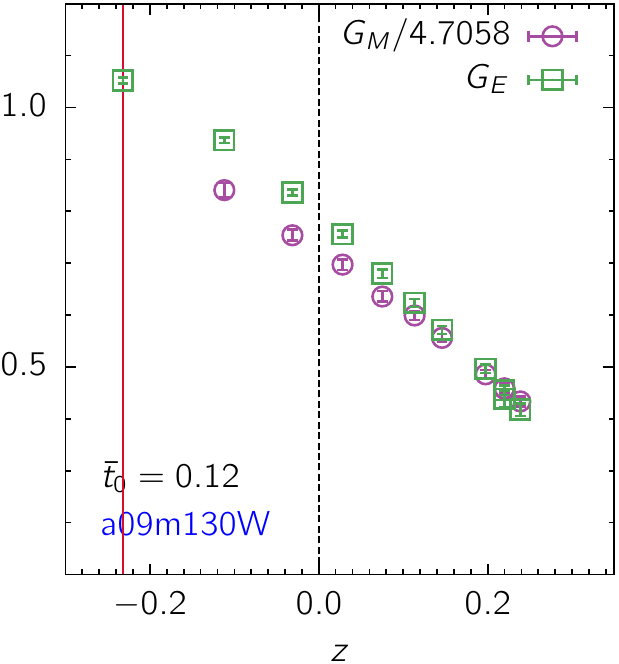} 
\includegraphics[height=1.65in,trim={0.0cm 0.20cm 0 0},clip]{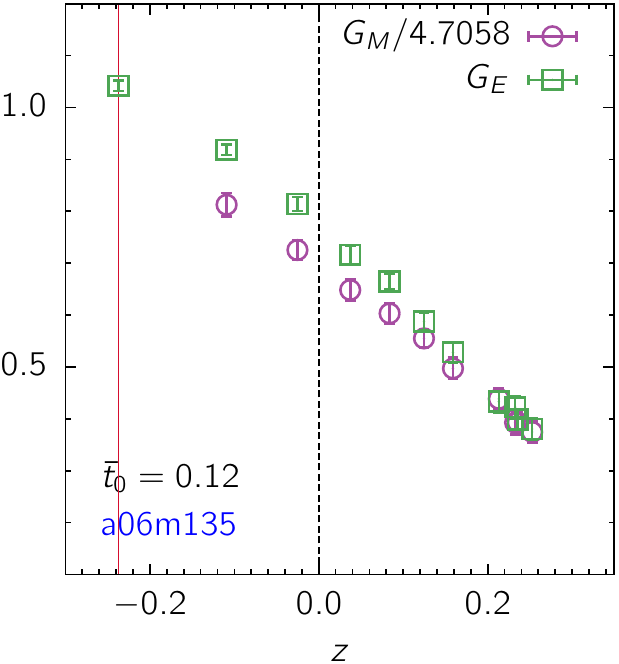} 
}
\subfigure{
\includegraphics[width=0.24\linewidth]{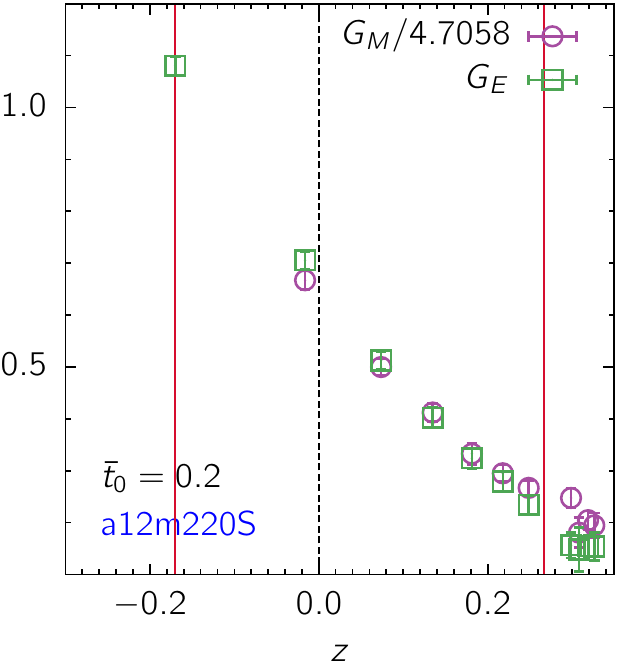}
\includegraphics[width=0.24\linewidth]{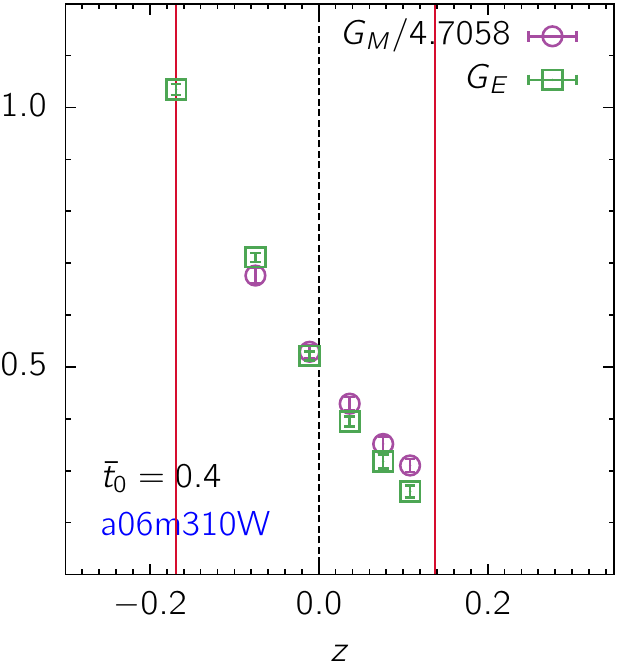}
\includegraphics[width=0.24\linewidth]{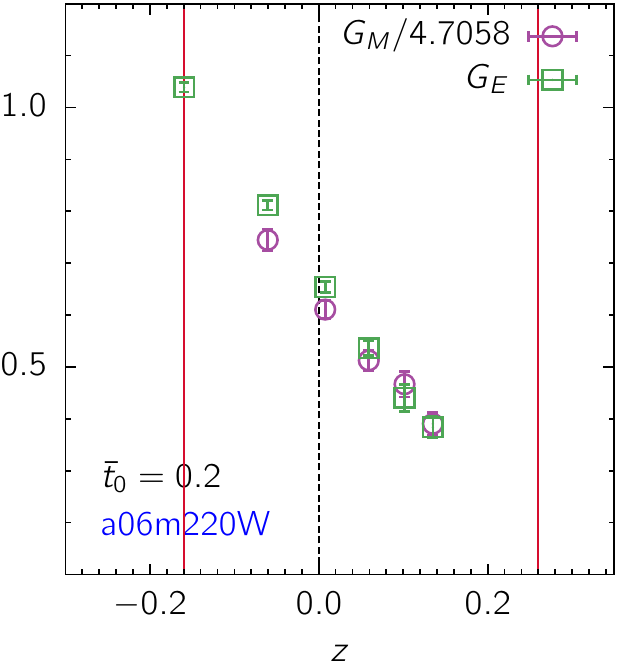}
}
\caption{\FIXME{fig:versusZ} The data for $G_E$ and $G_M/(\mu =4.7058)$ 
  plotted versus $z$ for the 13 calculations. The vertical red line on the left corresponds to
  $Q^2=0$, while on the right to $Q^2=1$~GeV${}^2$ except for $a15m310$
  ($Q^2 = 1.4$~GeV${}^2$) and $a12m220$ ($Q^2 = 0.8$~GeV${}^2$). In
  the two physical mass cases, the right vertical red line lies
  outside the panel.}
\label{fig:versusZ}
\end{figure*}

The data for $G_E$ and $G_M/(\mu =4.7058)$ versus $z$ are shown in
Fig.~\ref{fig:versusZ}. Our overall strategy for extracting $\rEsq$,
$\rMsq$ and $\mu^{p-n}$ is the following: We first determine by eye the largest value of $Q^2$ up to
which the data are smooth in $z$.  Next, since we are interested in
the value and slope of the fits at $Q^2=0$, we restricted the data to
$Q^2\le 1$~GeV${}^2$, except for the $a15m310$ ($Q^2\le
1.4$~GeV${}^2$) and $a12m220$ ($Q^2\le 0.8$~GeV${}^2$) ensembles.  The
allowed range $0$--$Q^2|_{\rm max}$, where $Q^2|_{\rm max}$ is the
largest value allowed by the cuts defined above, is marked by the two
vertical red lines in Fig.~\ref{fig:versusZ}.  With these cuts, the
points at all $Q^2$ are retained for most of the ensembles. Only the
high $Q^2$ data for $a15m310$, $a12m310$, $a12m220S$ and $a12m220$
ensembles are removed. These show a break in the smooth behavior in $z$
as is clear from Fig.~\ref{fig:versusZ}. Going back to the ESC analysis, the reliability of these points
is questionable since the data have large errors and the ESC fits were poor.  

\begin{figure*}         
\centering
\includegraphics[width=0.46\linewidth]{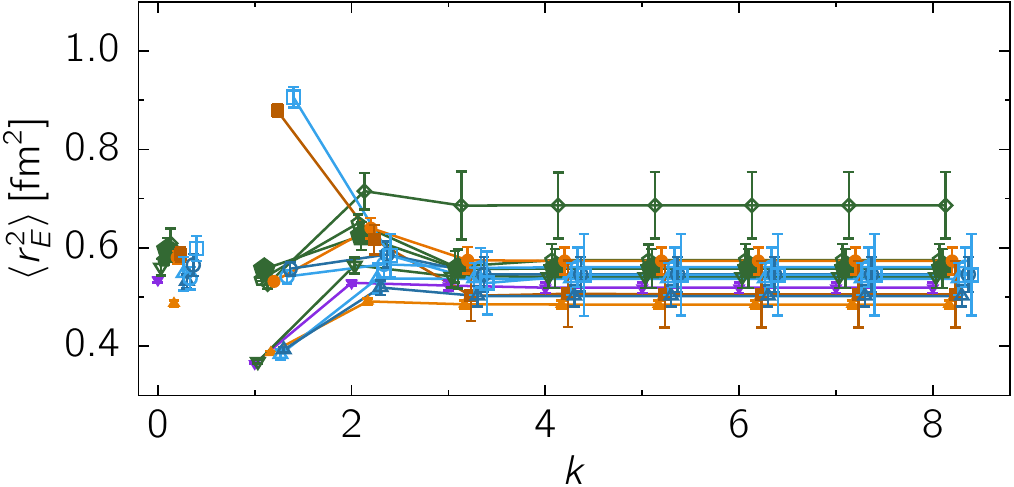}
\includegraphics[width=0.46\linewidth]{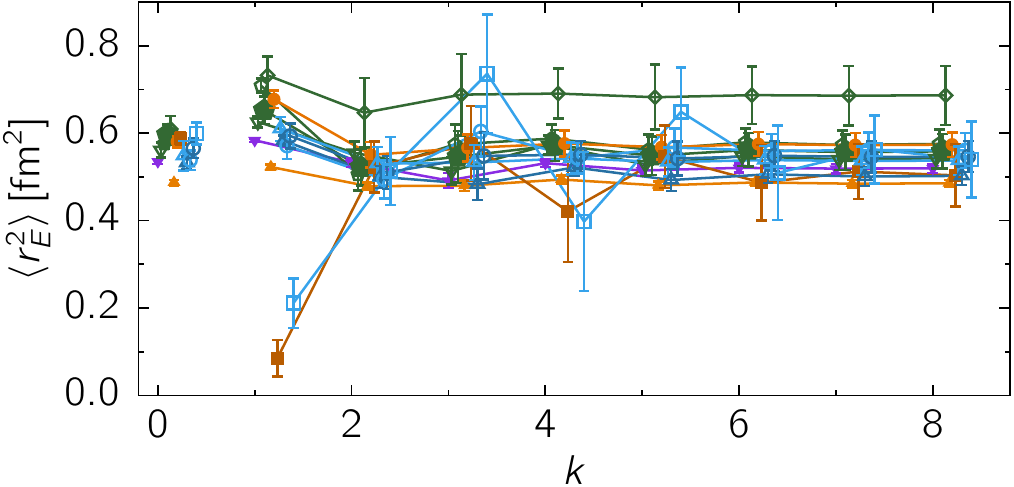} \\
\includegraphics[width=0.46\linewidth]{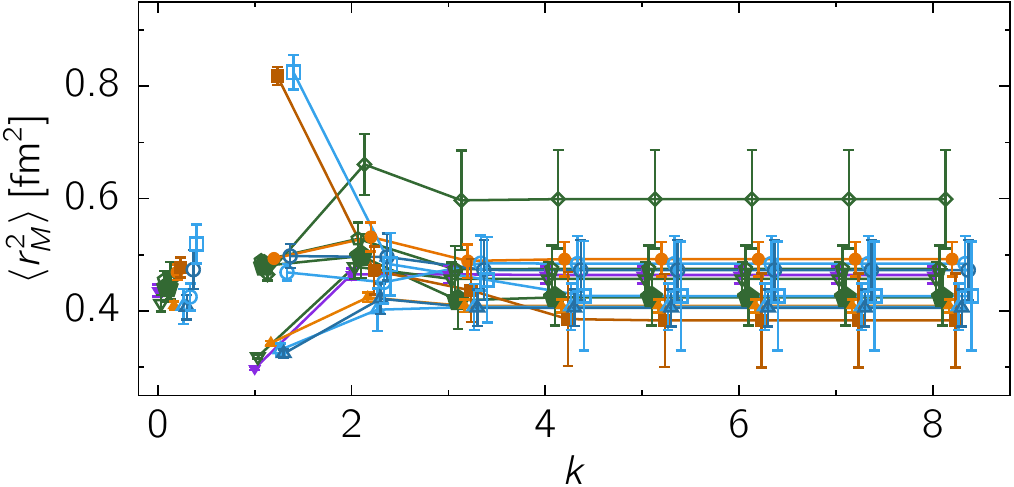}
\includegraphics[width=0.46\linewidth]{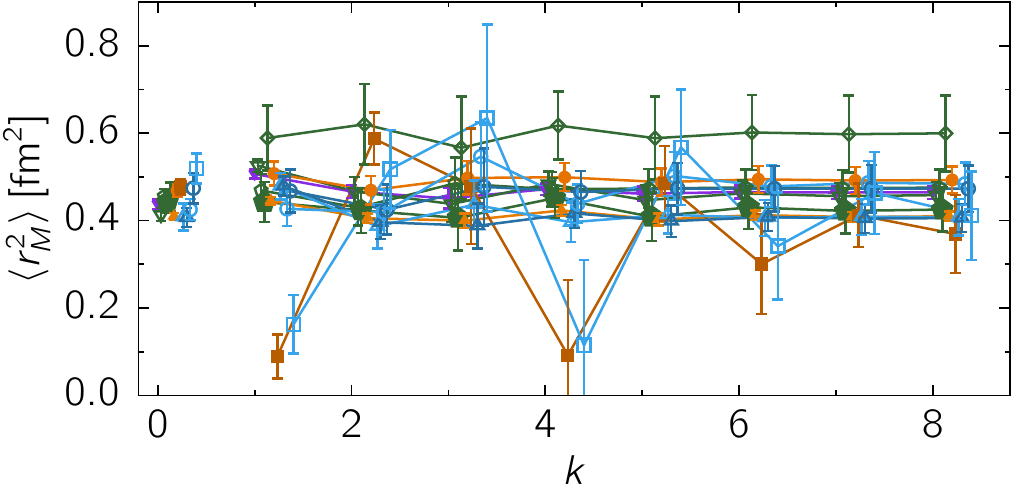}\\
\includegraphics[width=0.46\linewidth]{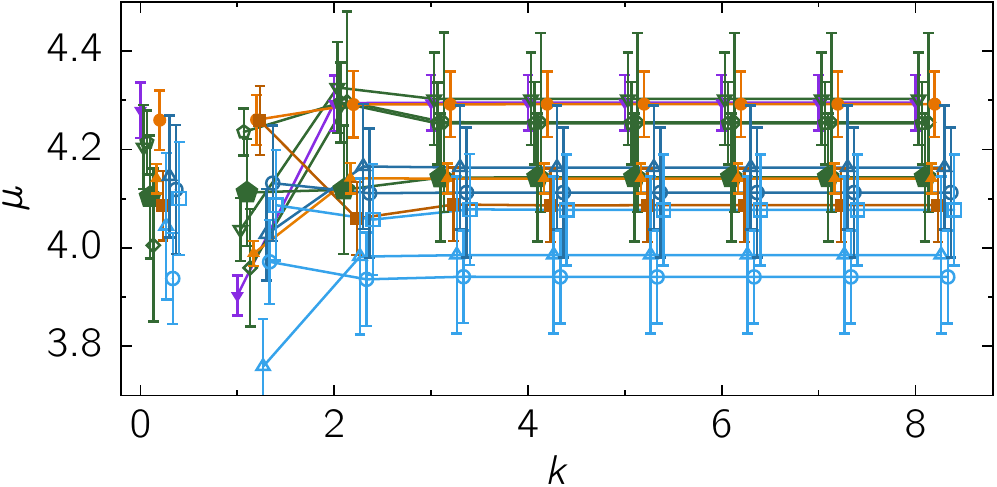}
\includegraphics[width=0.46\linewidth]{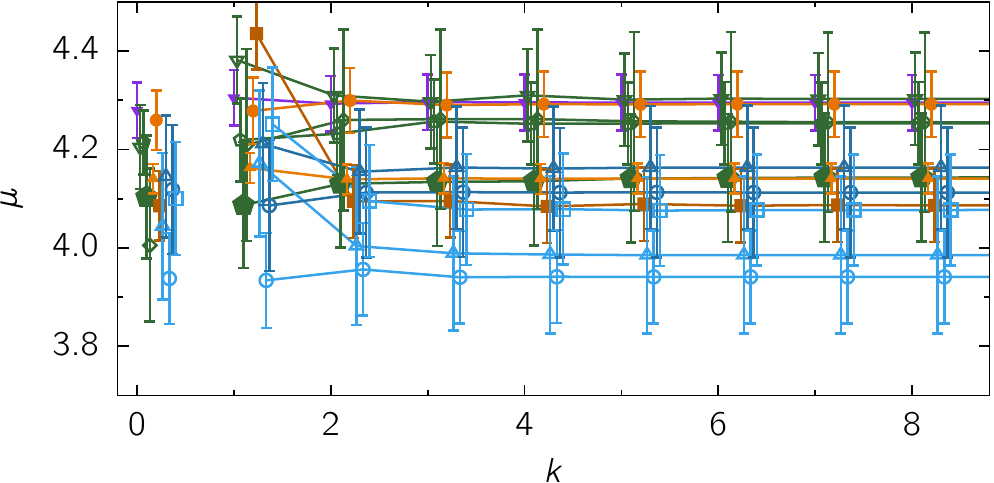}
\caption{\FIXME{fig:stabilityz} Estimates of $\rEsq$, $\rEsq$ and
  $\mu$ from each of the thirteen calculations are shown as a function
  of order $z^k$ (left) and $z^{k+4}$ (right) of the truncation of the $z$-expansion.  The dipole
  results is shown at $k=0$. For clarity, the data from the 13
  calculations (using same symbol and color code as in
  Fig.~\protect\ref{fig:Kelly}) are shifted slightly along the
  x-axis for clarity. }
\label{fig:stabilityz}
\end{figure*}

The results from the $z$-expansion fits are stable for $k \ge 4$ as
shown in Fig.~\ref{fig:stabilityz}. Results from fits including the
sum rules are similar, except that stability is reached only for $k
\ge 7$.  Estimates from fits with and without sum rules are consistent,
however the errors are larger with the sum rules. The values and
$\chi^2/{\rm DOF}$ of the dipole fits have been stable under increase
in statistics for all thirteen calculations. On the other hand, the
results from the $z$-expansion fits required high statistics to
exhibit convergence with the order of the truncation.

\begin{figure*}    
  \centering
\subfigure{
\includegraphics[height=1.65in,trim={0.0cm 0.40cm 0 0},clip]{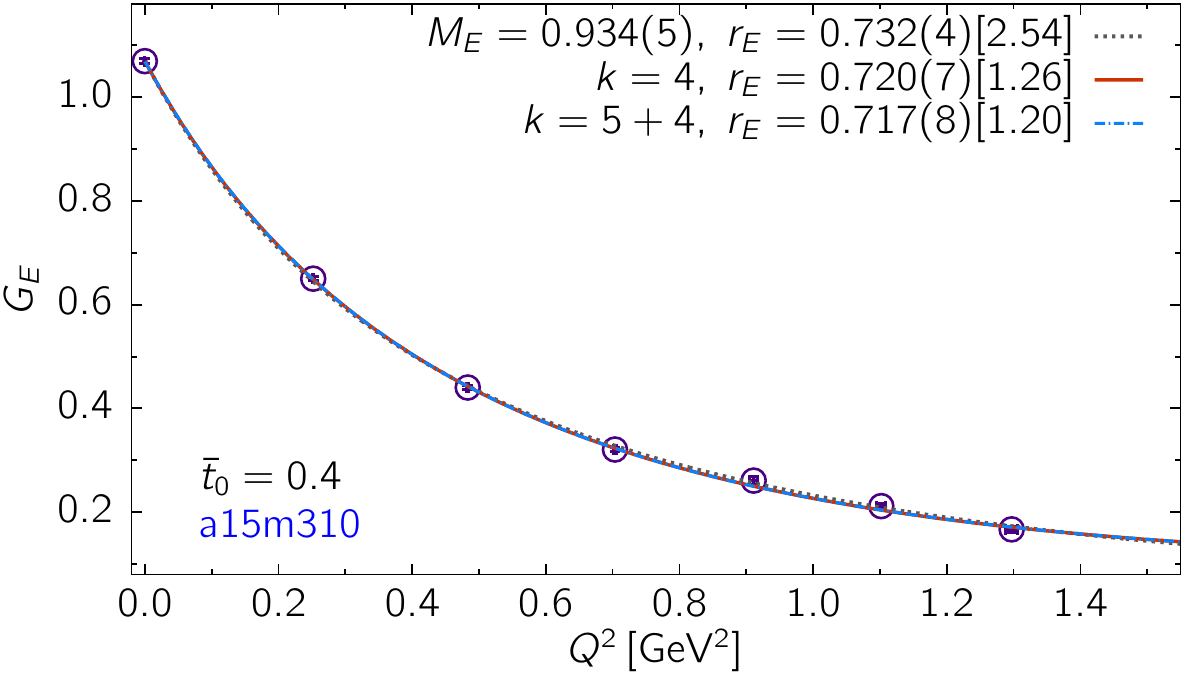} 
\includegraphics[height=1.65in,trim={0.0cm 0.40cm 0 0},clip]{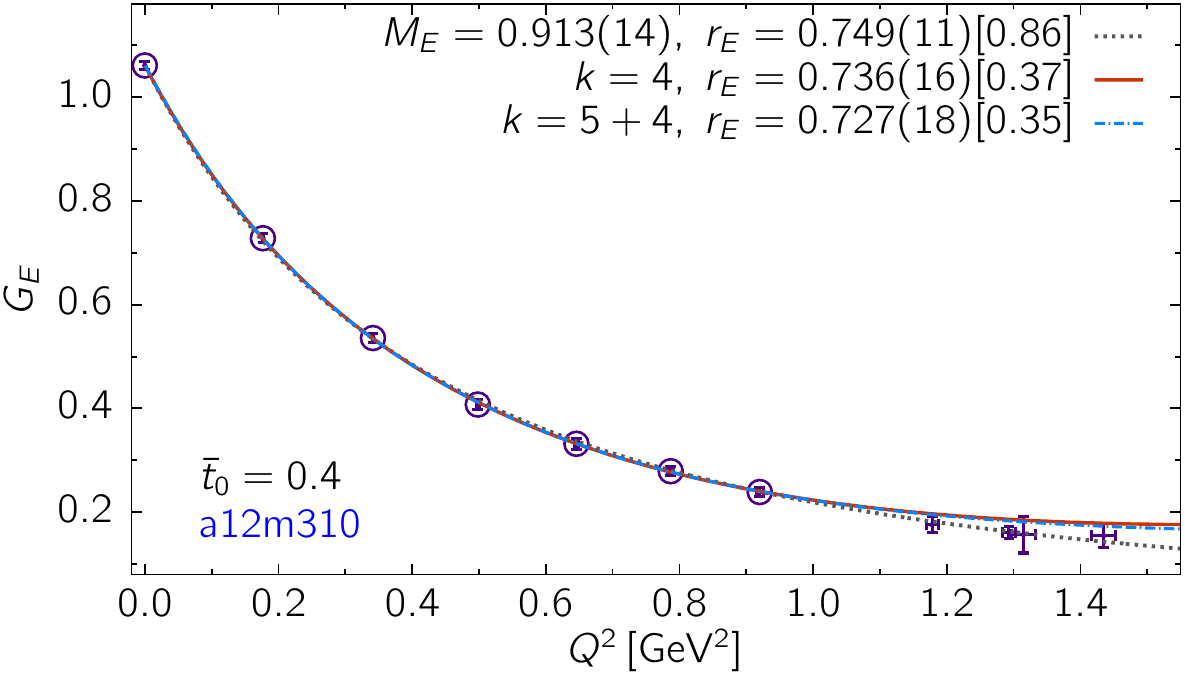}
}
\subfigure{
\includegraphics[height=1.65in,trim={0.0cm 0.40cm 0 0cm},clip]{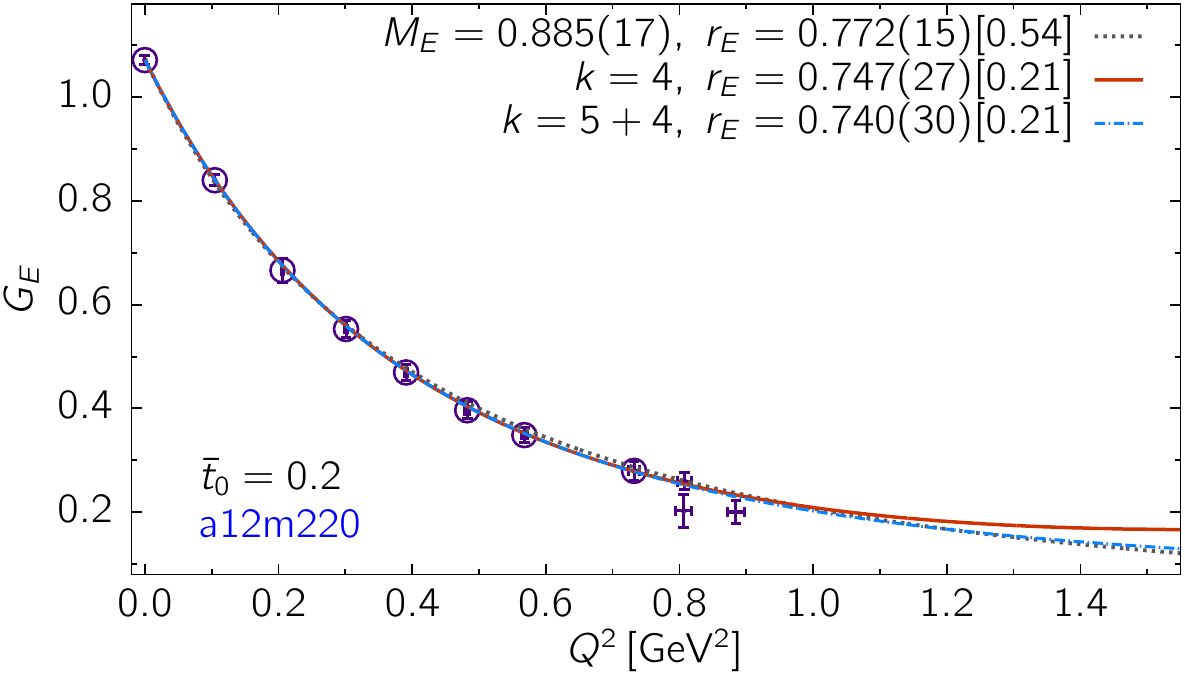}
\includegraphics[height=1.65in,trim={0.0cm 0.40cm 0 0cm},clip]{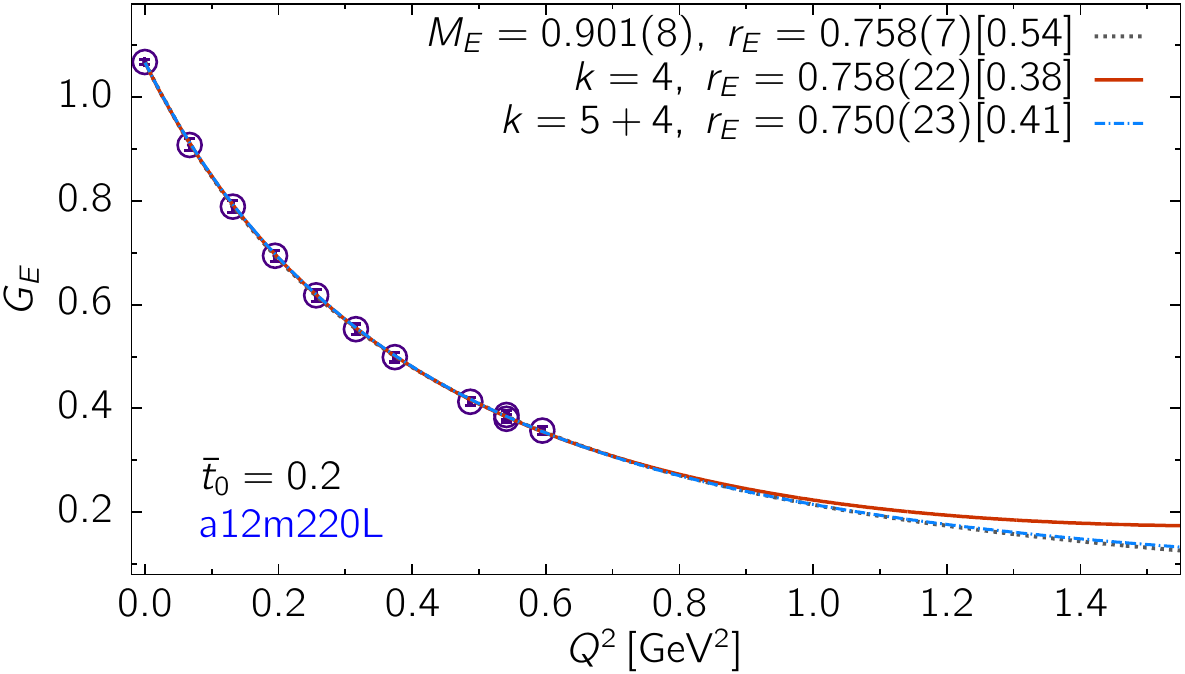}
}
\subfigure{
\includegraphics[height=1.65in,trim={0.0cm 0.40cm 0 0},clip]{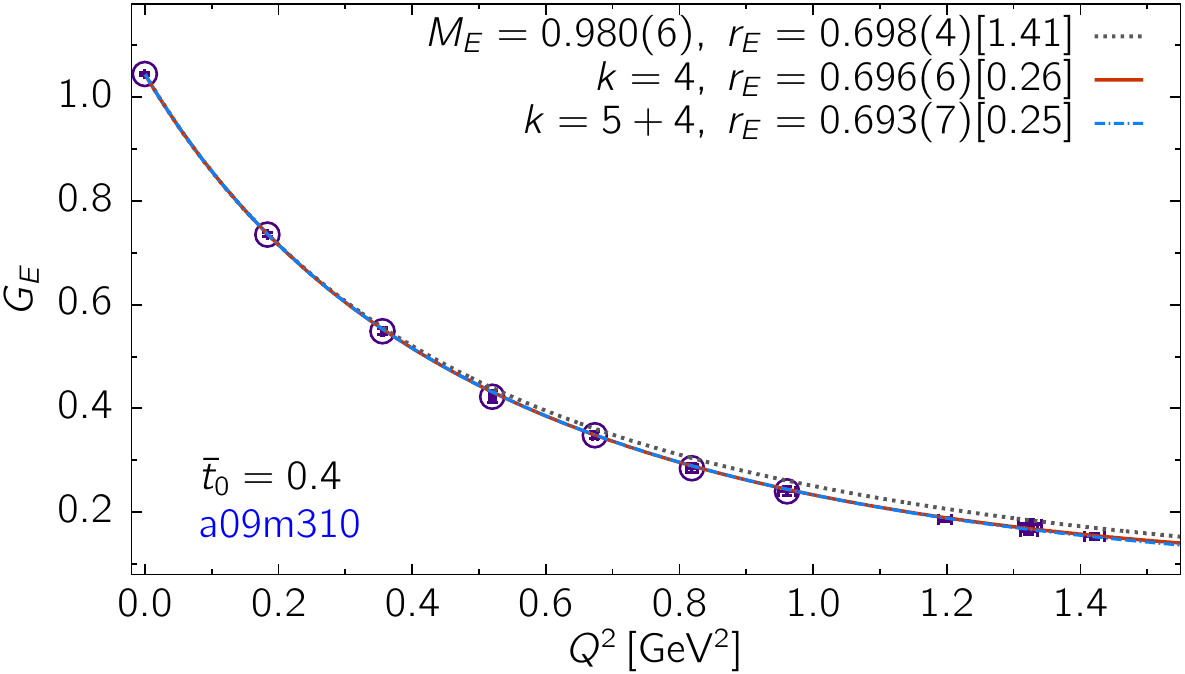}
\includegraphics[height=1.65in,trim={0.0cm 0.40cm 0 0},clip]{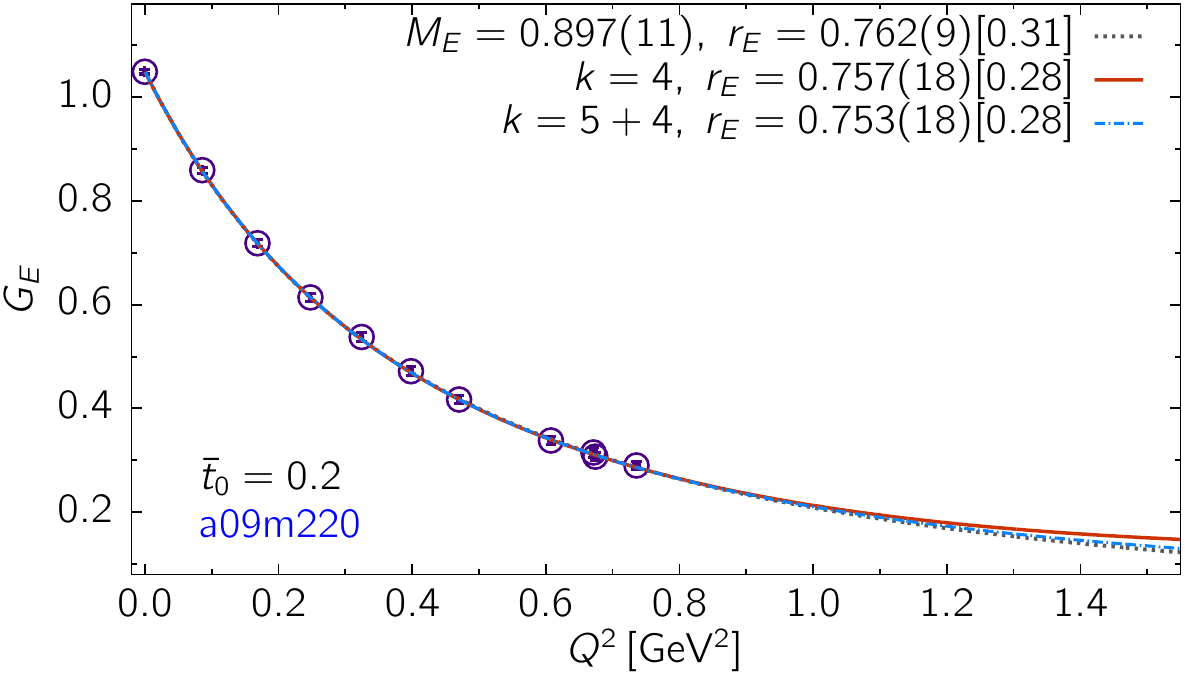}
}
\subfigure{
\includegraphics[height=1.65in,trim={0.0cm 0.40cm 0 0},clip]{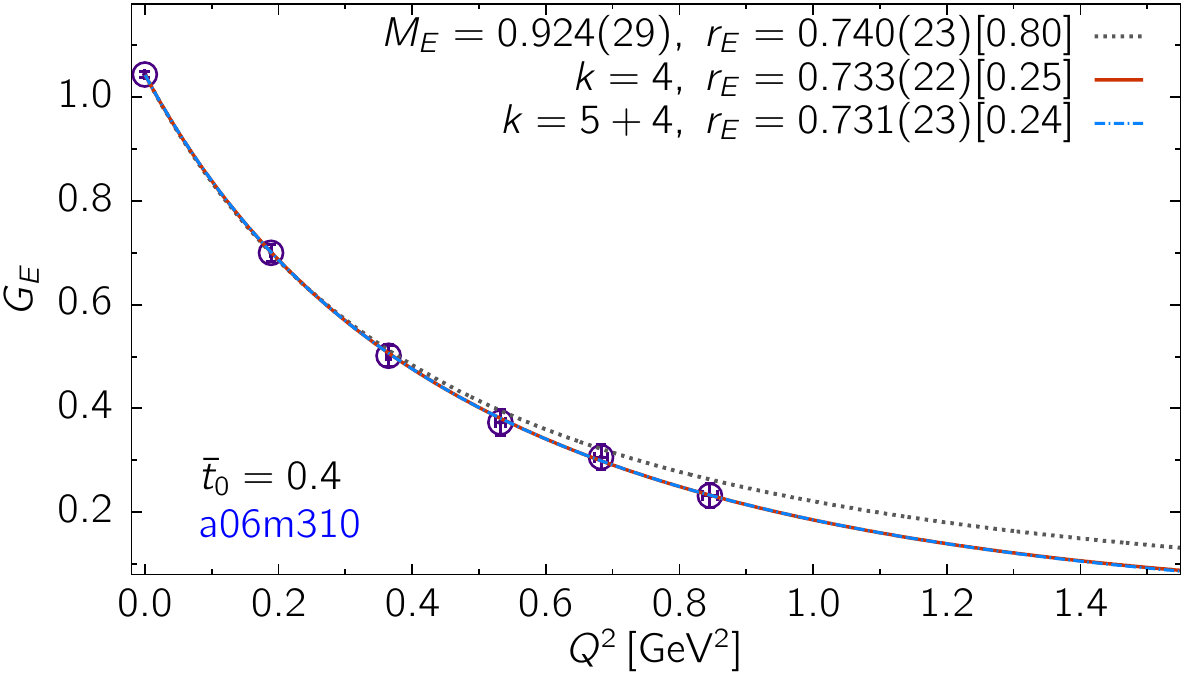}
\includegraphics[height=1.65in,trim={0.0cm 0.40cm 0 0},clip]{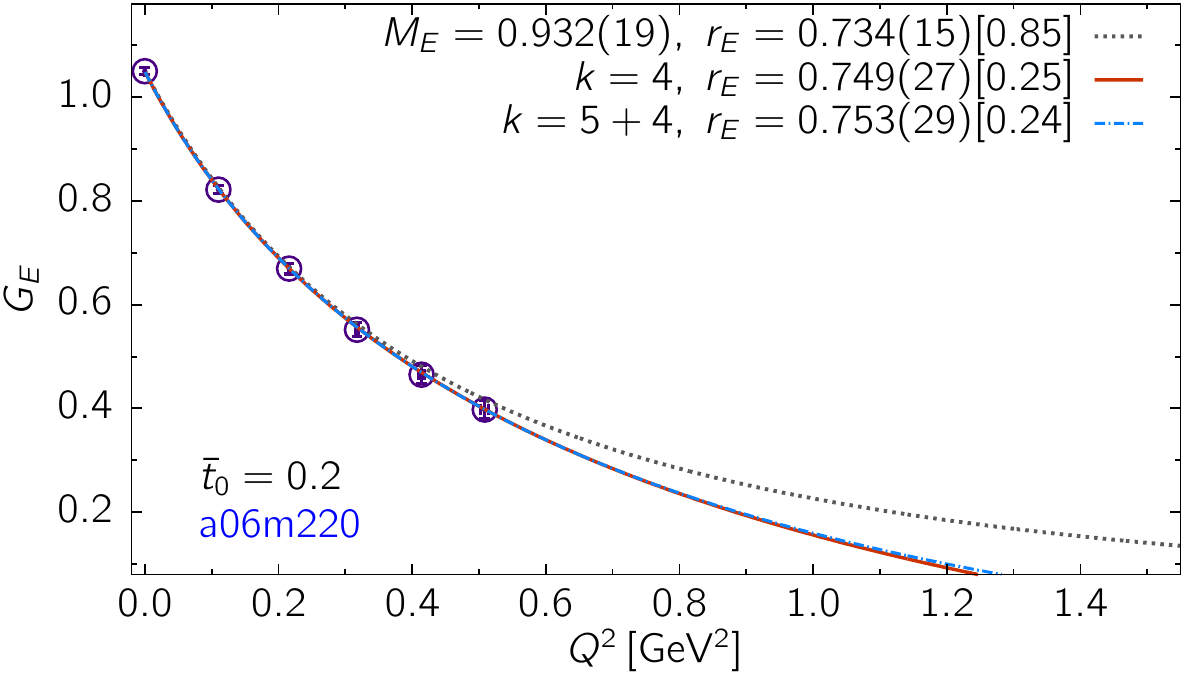}
}
\subfigure{
\includegraphics[width=0.44\linewidth]{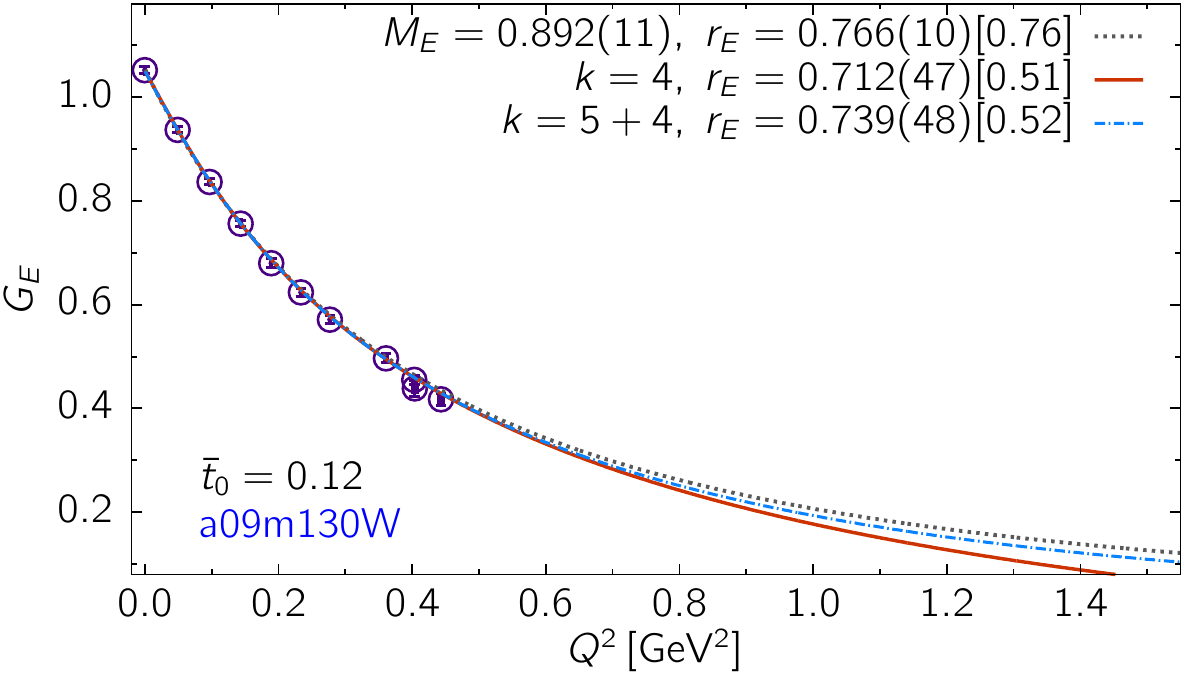}
\includegraphics[width=0.44\linewidth]{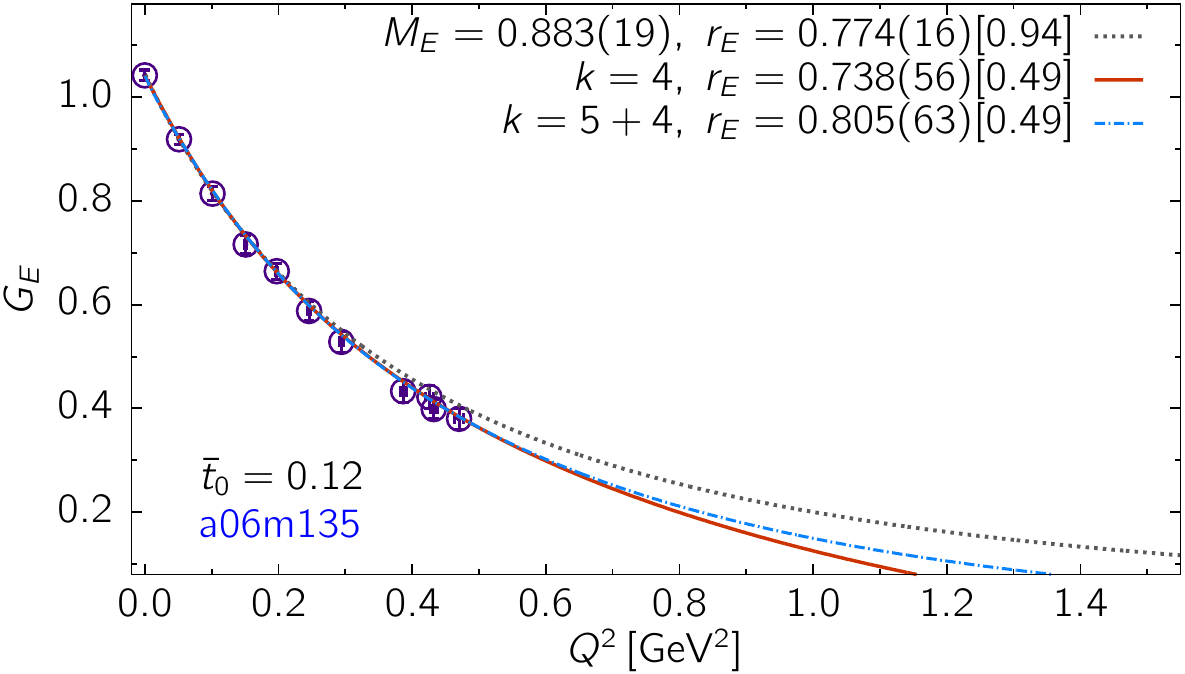}
}
\caption{\FIXME{fig:10fits-rE} Results of the dipole, $z^4$ and $z^{5+4}$ fits 
to the unrenormalized isovector $G_E(Q^2)$
  versus $Q^2$ (GeV${}^2$) for ten ensembles. The top two panels
  show data from the $a15m310$ and $a12m310$ ensembles; 
  the second row from $a12m220$ and $a12m220L$ ensembles; the
  third row from $a09m310$ and $a09m220$; the fourth row from
  $a06m310$ and $a06m220$; and the fifth row from the two physical
  mass ensembles $a09m130$ and $a06m135$. Estimates of the dipole mass
  ${\cal M}_E$ (GeV) and the charge radius $\rE$ (fm) from the three fits are
  given in the labels. The numbers within the square parentheses are the
  $\chi^2/{\rm DOF}$ of the fit. Data points without circles around them 
  are not included in the fits as explained in the text.} 
\label{fig:10fits-rE}
\end{figure*}

\begin{figure*}   
\centering
\subfigure{
\includegraphics[height=1.65in,trim={0.0cm 0.40cm 0 0},clip]{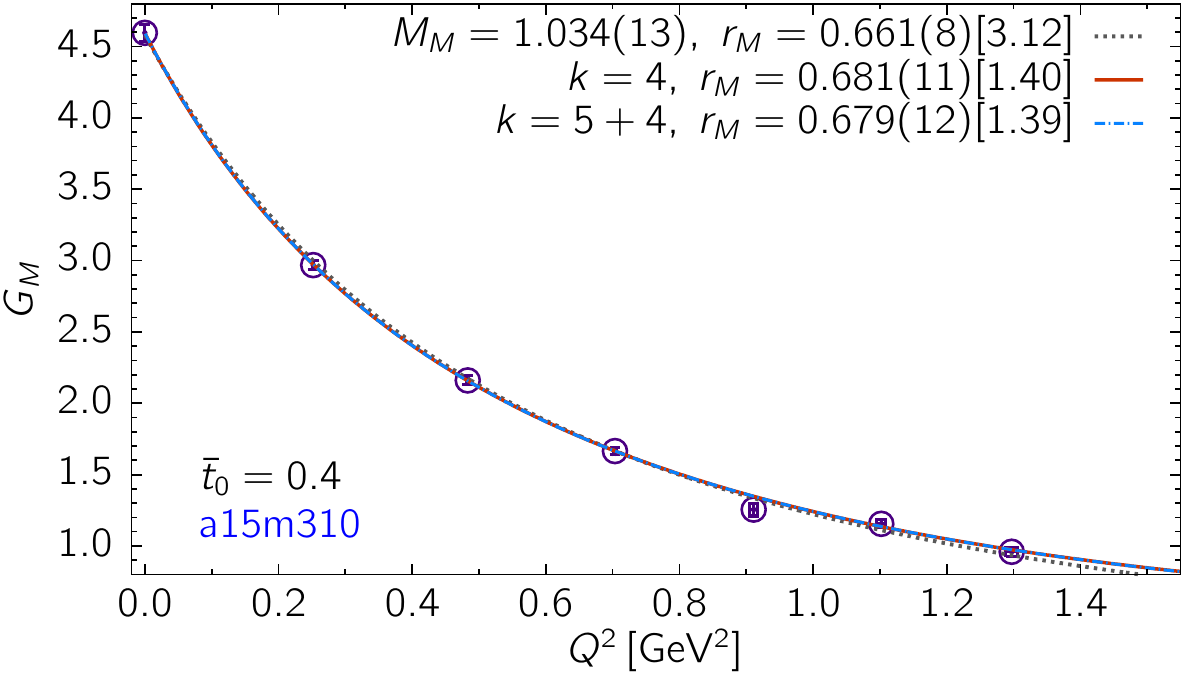}
\includegraphics[height=1.65in,trim={0.0cm 0.40cm 0 0},clip]{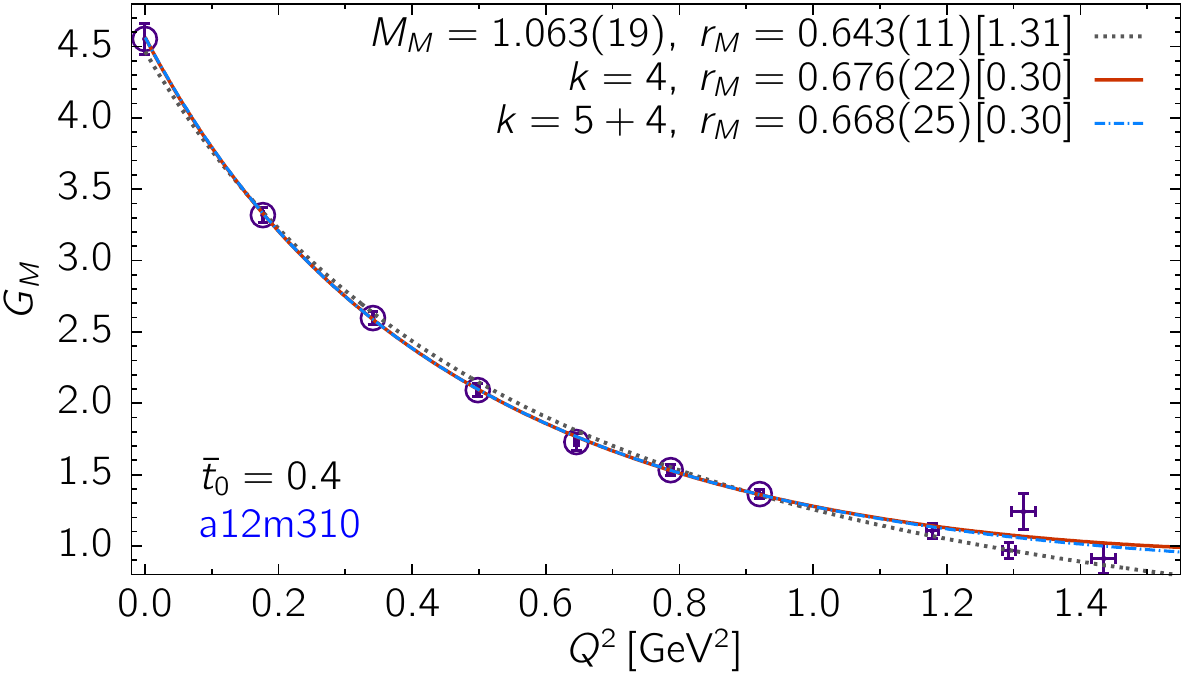}
}
\subfigure{
\includegraphics[height=1.65in,trim={0.0cm 0.40cm 0 0},clip]{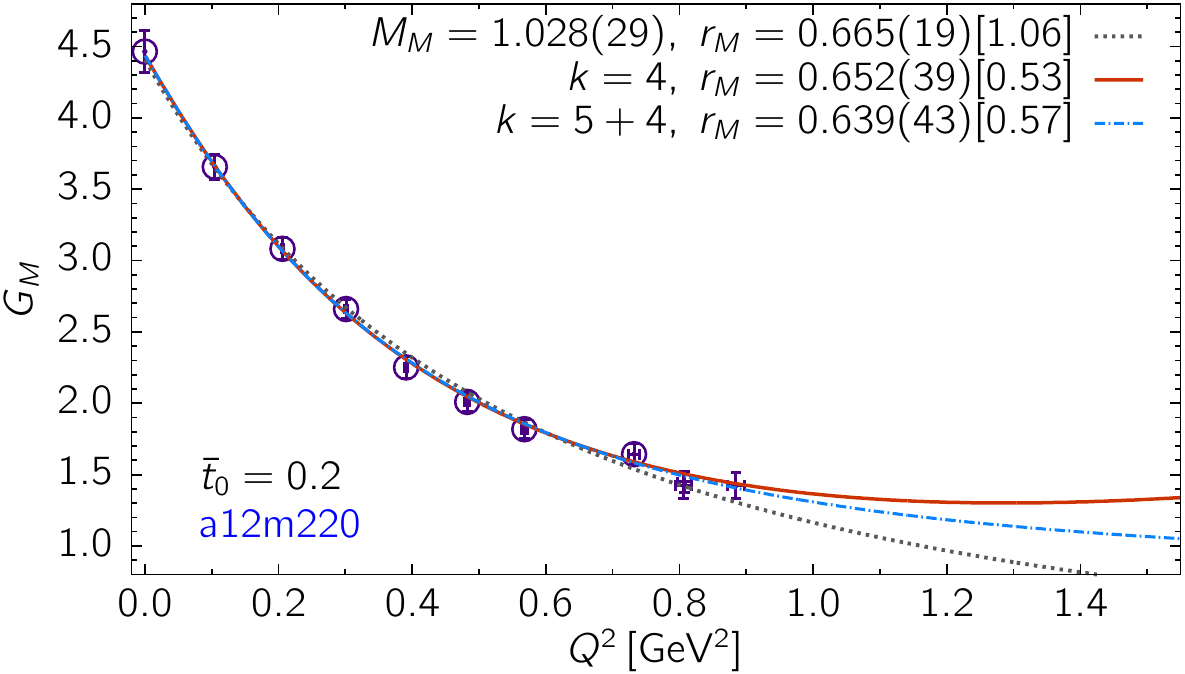}
\includegraphics[height=1.65in,trim={0.0cm 0.40cm 0 0},clip]{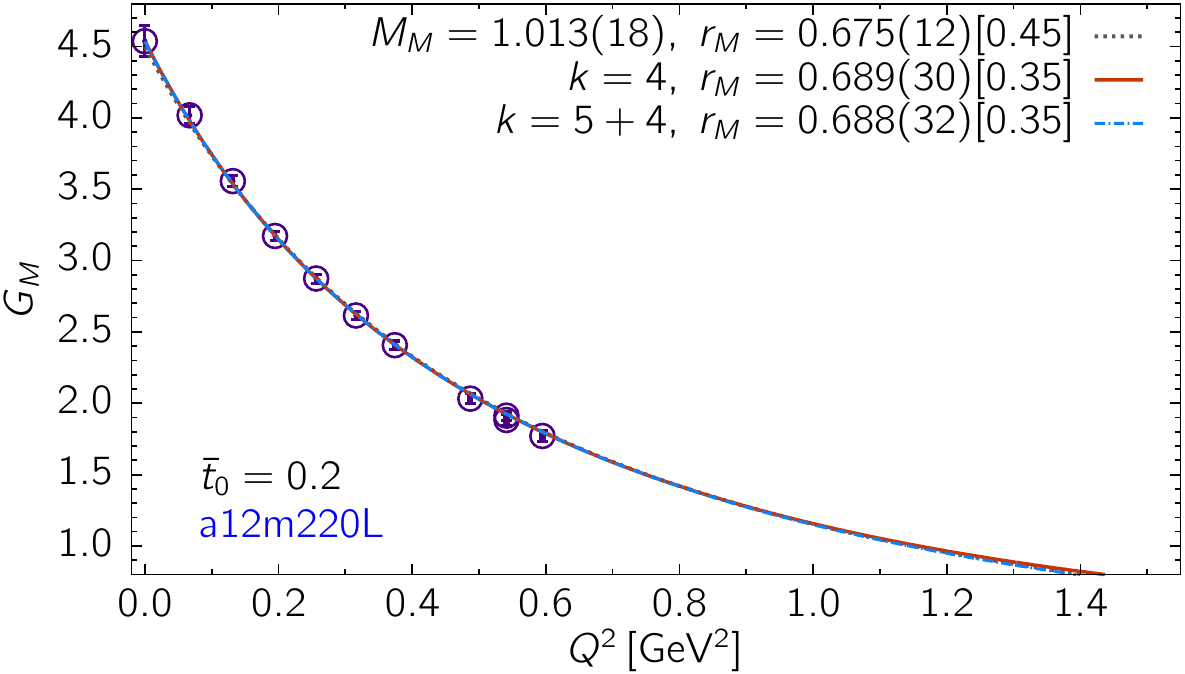}
}
\subfigure{
\includegraphics[height=1.65in,trim={0.0cm 0.40cm 0 0},clip]{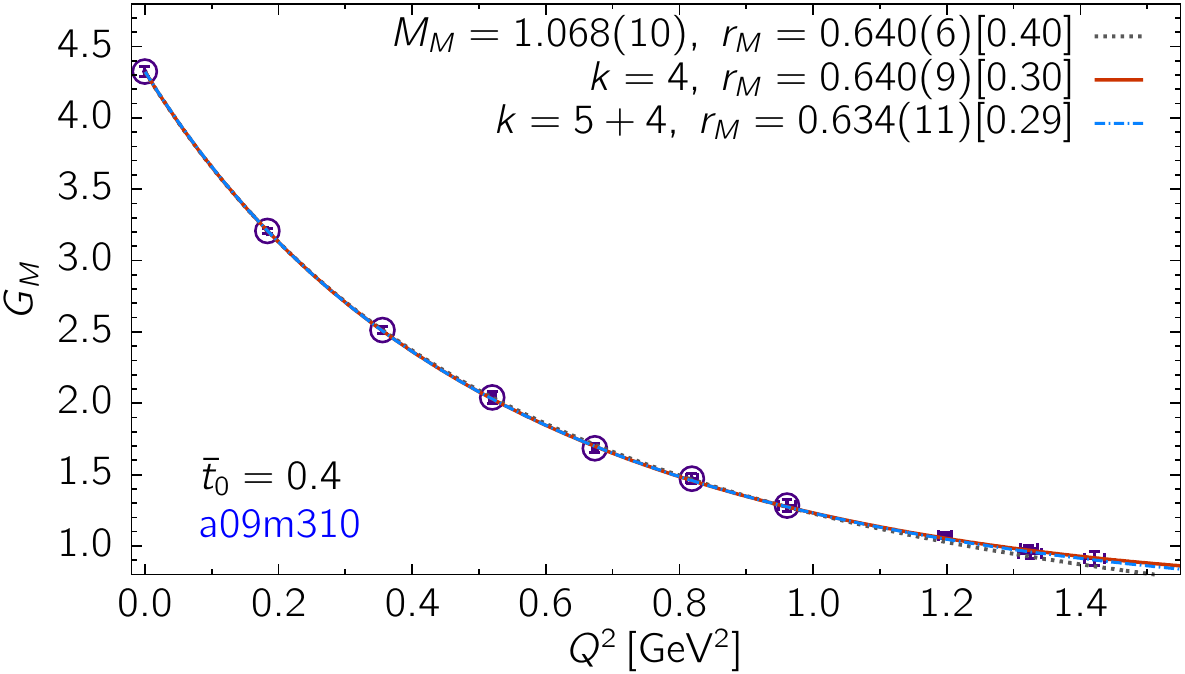}
\includegraphics[height=1.65in,trim={0.0cm 0.40cm 0 0},clip]{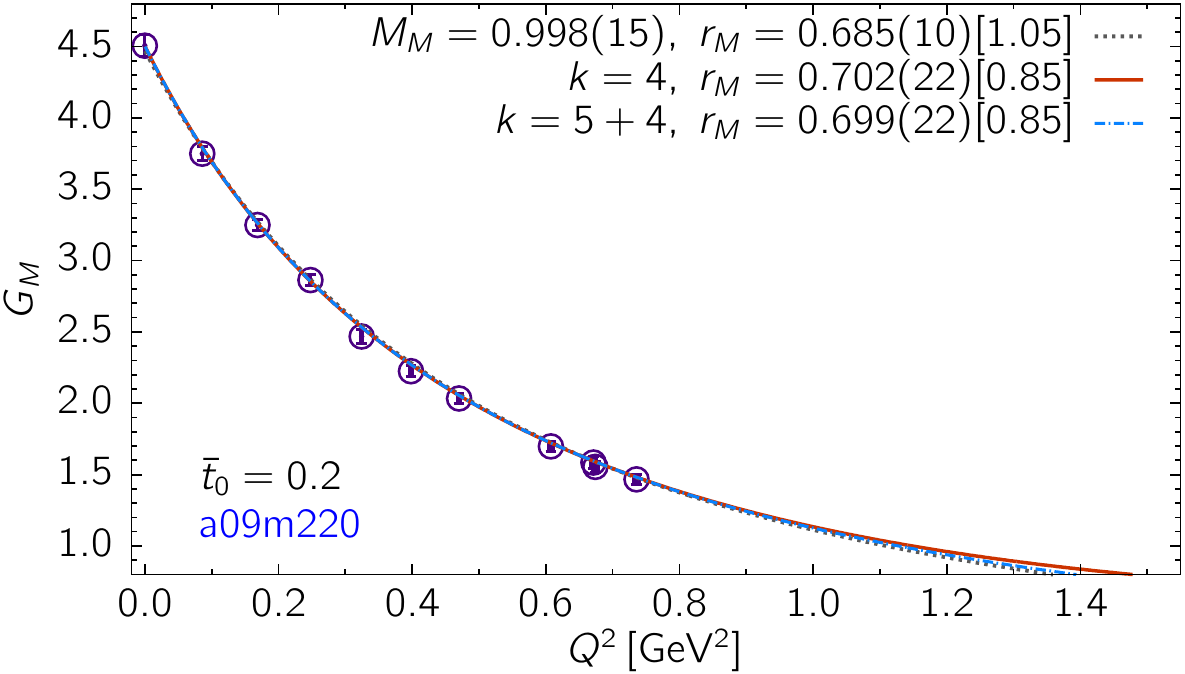}
}
\subfigure{
\includegraphics[height=1.65in,trim={0.0cm 0.40cm 0 0},clip]{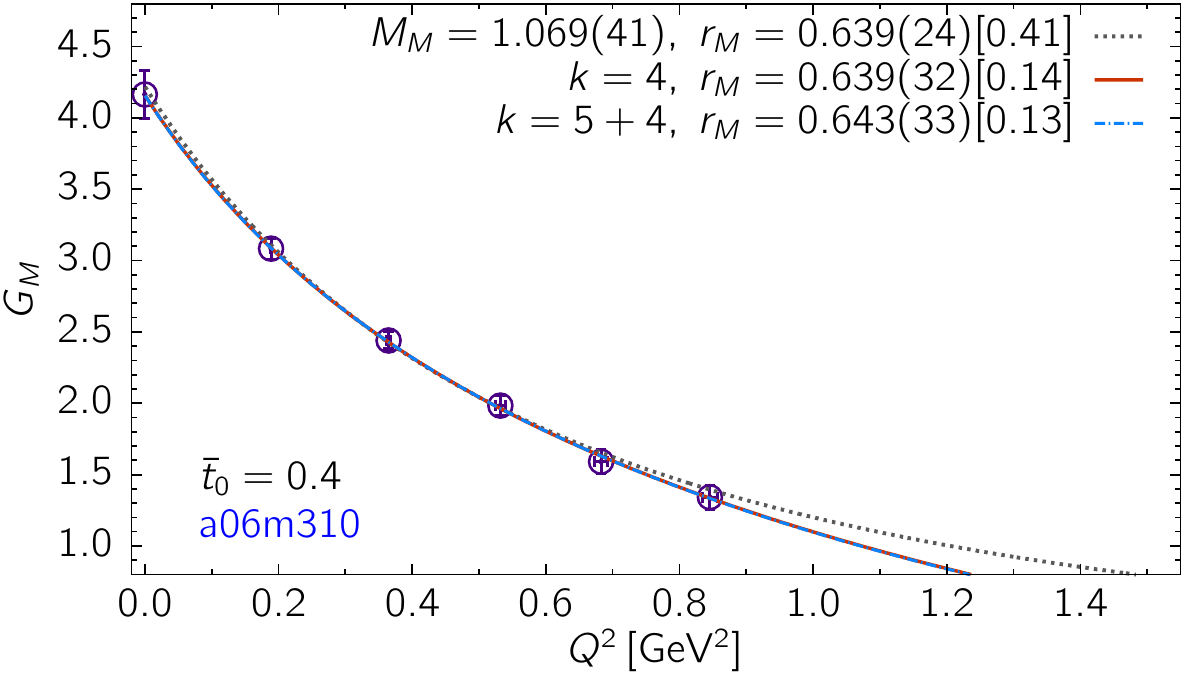}
\includegraphics[height=1.65in,trim={0.0cm 0.40cm 0 0},clip]{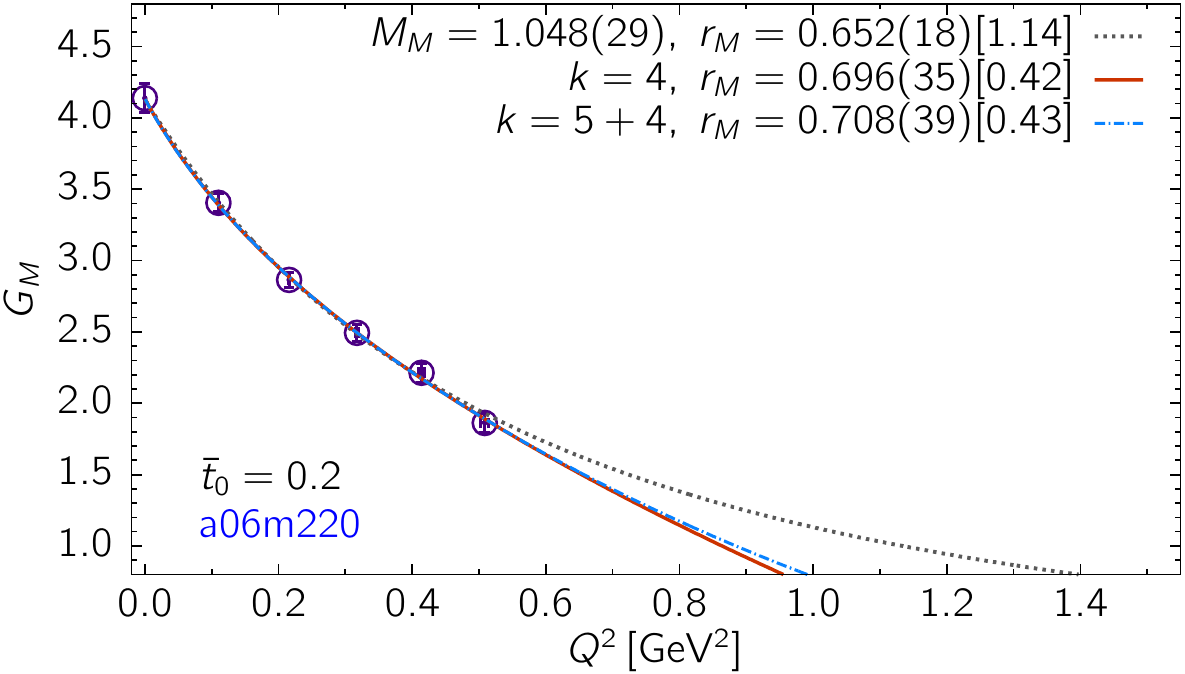}
}
\subfigure{
\includegraphics[width=0.44\linewidth]{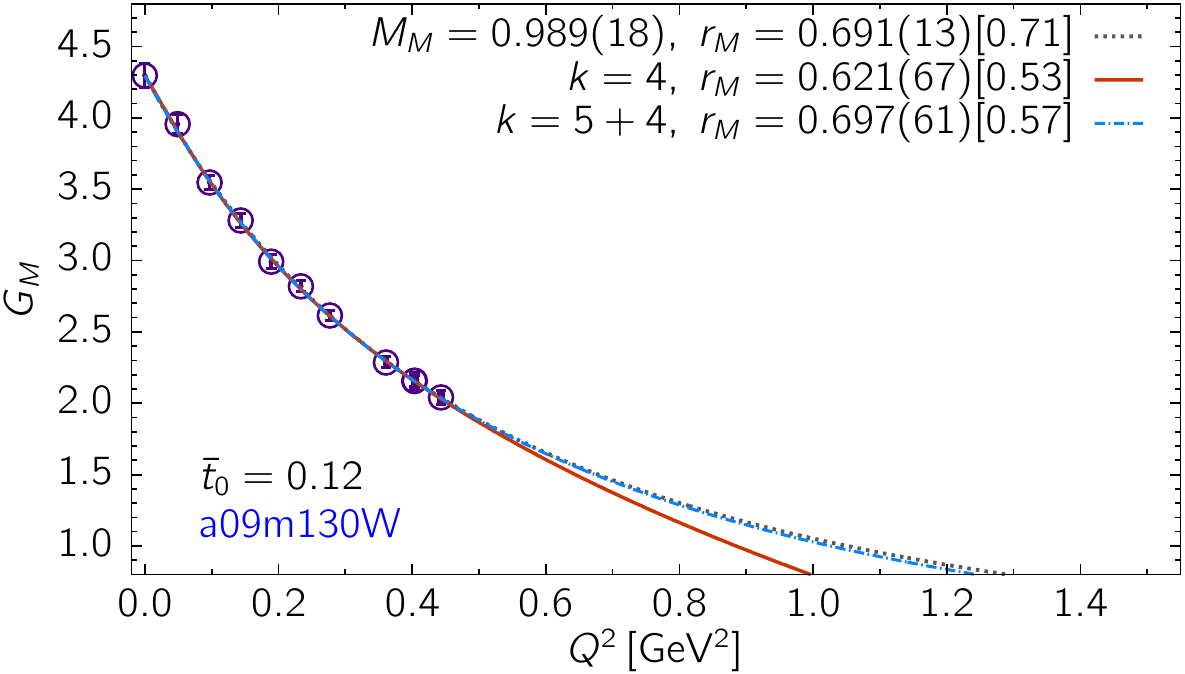}
\includegraphics[width=0.44\linewidth]{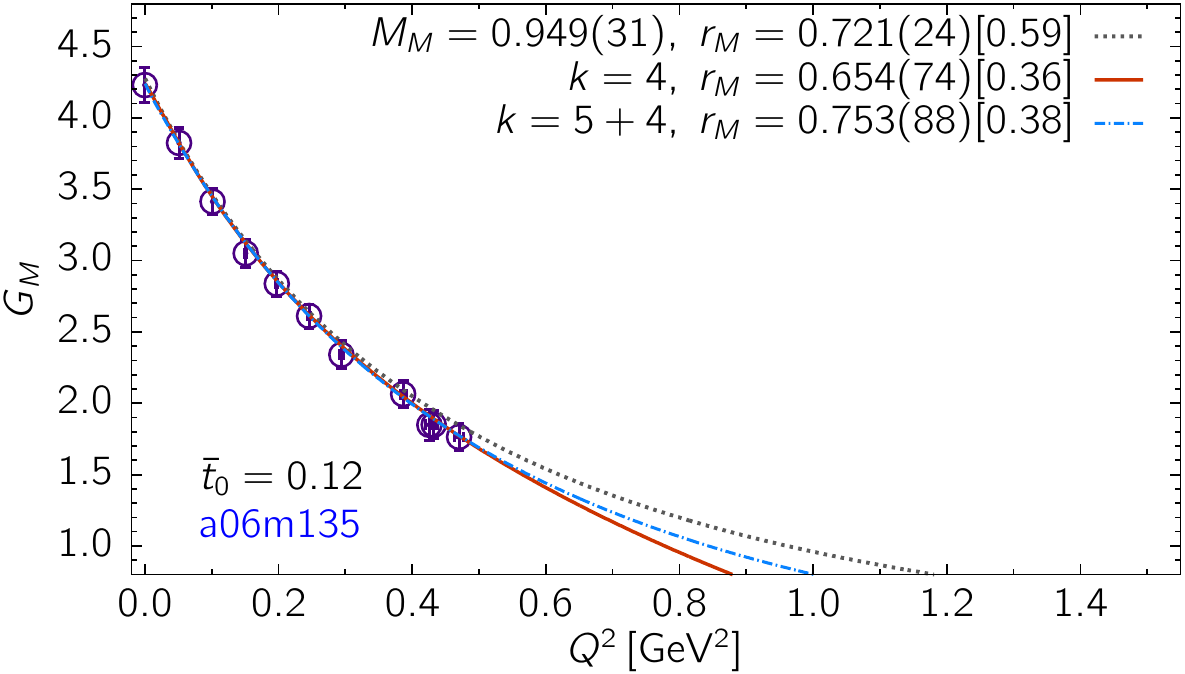}
}
\caption{\FIXME{fig:10fits-rM} Results of the dipole, $z^4$ and  $z^{5+4}$
  fits to the unrenormalized isovector $G_M(Q^2)$ versus $Q^2$
  (GeV${}^2$) for ten ensembles. The rest is the same as in
  Fig.~\protect\ref{fig:10fits-rE}. }
\label{fig:10fits-rM}
\end{figure*}

The results from seven fit ansatz are collected together in
Tables~\ref{tab:rE-results},~\ref{tab:rM-results}
and~\ref{tab:mu-results}. Overall, the seven estimates are consistent
within errors.  Since $\rEsq$, $\rMsq$ and $\mu^{p-n}$ should be
extracted from the small $Q^2$ behavior, our final results are from
the $z^4$ fits. Estimates with sum rules are used only as consistency
checks. The dipole, $z^4$ and $z^{5+4}$ fits and results are shown in
Figs.~\ref{fig:10fits-rE} and~\ref{fig:10fits-rM} for ten ensembles.

\begin{figure}   
\centering
\subfigure{
\includegraphics[width=0.97\linewidth]{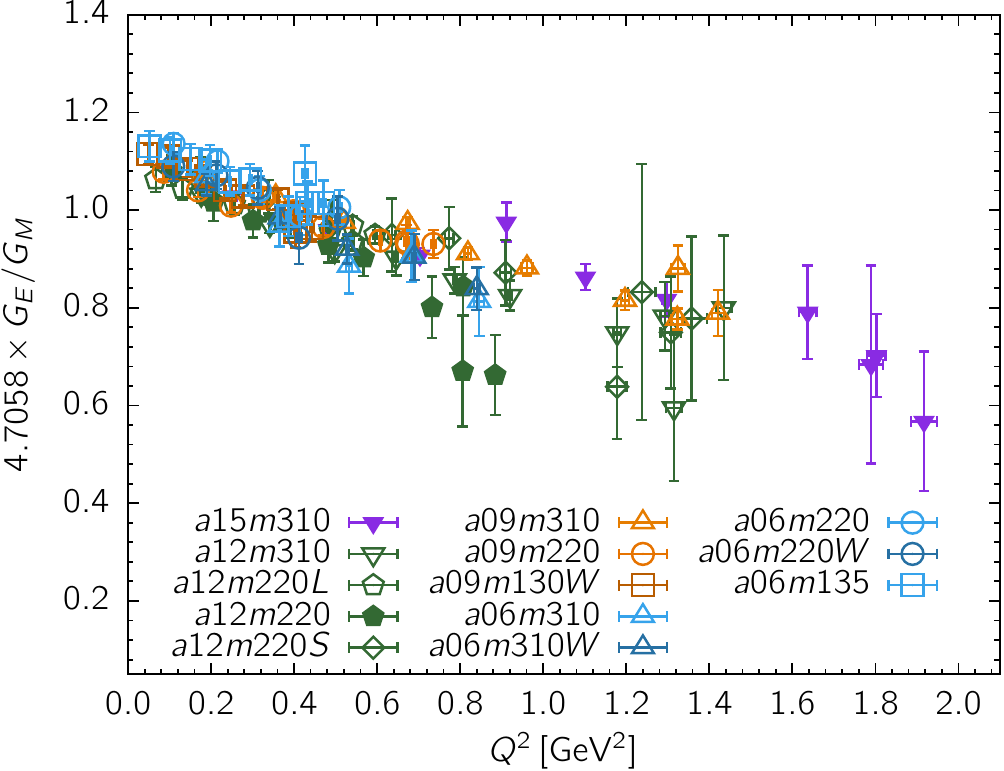}
}
\caption{\FIXME{fig:GEGMratio} The ratio of isovector form factors
  $G_E(Q^2)/G_M(Q^2)$ multiplied by the experimental value of 
  the magnetic moment $\mu^{p-n} = 4.7058$. The deviation from unity
  at $Q^2=0$ is the amount by which the lattice data underestimates
  $\mu^{p-n}$. }
\label{fig:GEGMratio}
\end{figure}

The data for the ratio $\mu^{p-n} \times G_E(Q^2)/G_M(Q^2)$ are shown
in Fig.~\ref{fig:GEGMratio}.  Experimental data indicate that this
ratio for the proton is estimated to cross zero around
$Q^2=8$~GeV${}^2$~\cite{Punjabi:2014tna}. Our data for the isovector
combination do show a negative slope over the region $Q^2 \lesssim
0.6$~GeV${}^2$, nevertheless, data at larger $Q^2$ are needed to
determine if and where the ratio crosses zero.  As discussed in
Sec.~\ref{sec:extractGM}, we have used this ``linear'' behavior at
small $Q^2$ to estimate $G_M(0)$ from the ratio, including which
helped stabilize the fits to $G_M(Q^2)$.  In the next section, we
discuss the continuum-chiral-finite-volume (CCFV) fits used to get the
physical estimates.

\begin{table*}   
\caption{\FIXME{tab:rE-results} The isovector electric mean-square charge radius
  $\rEsq$ in units of fm${}^2$ from the seven fits (dipole, $z^3$, $z^{3+4}$,
  $z^4$, $z^{4+4}$, $z^5$ and $z^{5+4}$) to the isovector form factor
  $G_E(Q^2)$. The bottom half of the table gives the results of the
  four CCFV fits discussed in the text, with and without the leading
  finite volume term $c_4^E$ and for each $Q^2$ fit.  }
\label{tab:rE-results}
  \centering
  \renewcommand{\arraystretch}{1.2}
  \begin{ruledtabular}
    \begin{tabular}{l|ccccccccc}
Ensemble              & Dipole        & $z^3$         & $z^4$         & $z^5$         & $z^{3+4}$     & $z^{4+4}$     & $z^{5+4}$     \\ \hline
$a15m310$             &  0.535(6)     &  0.523(10)    &  0.519(10)    &  0.519(10)    &  0.492(17)    &  0.531(8)     &  0.514(12)   \\
$a12m310$             &  0.561(17)    &  0.542(23)    &  0.542(23)    &  0.542(23)    &  0.513(31)    &  0.573(22)    &  0.528(26)   \\
$a12m220L$            &  0.575(10)    &  0.562(32)    &  0.575(33)    &  0.575(33)    &  0.574(45)    &  0.588(32)    &  0.562(35)   \\
$a12m220$             &  0.596(23)    &  0.557(40)    &  0.558(40)    &  0.558(40)    &  0.546(59)    &  0.572(35)    &  0.548(44)   \\
$a12m220S$            &  0.609(30)    &  0.686(69)    &  0.686(67)    &  0.686(67)    &  0.688(93)    &  0.690(57)    &  0.682(74)   \\
$a09m310$             &  0.487(6)     &  0.485(8)     &  0.485(8)     &  0.485(8)     &  0.480(12)    &  0.494(10)    &  0.480(10)   \\
$a09m220$             &  0.580(14)    &  0.575(26)    &  0.574(28)    &  0.573(28)    &  0.566(31)    &  0.576(29)    &  0.568(28)   \\
$a09m130W$            &  0.587(15)    &  0.503(51)    &  0.507(67)    &  0.506(67)    &  0.577(85)    &  0.421(115)   &  0.546(72)   \\
$a06m310$             &  0.548(34)    &  0.537(32)    &  0.537(32)    &  0.537(32)    &  0.533(41)    &  0.542(36)    &  0.534(33)   \\
$a06m310W$            &  0.532(14)    &  0.502(21)    &  0.502(21)    &  0.502(21)    &  0.483(36)    &  0.522(20)    &  0.493(25)   \\
$a06m220$             &  0.538(22)    &  0.560(40)    &  0.561(40)    &  0.561(40)    &  0.604(57)    &  0.543(35)    &  0.567(44)   \\
$a06m220W$            &  0.565(22)    &  0.546(35)    &  0.546(35)    &  0.546(35)    &  0.548(54)    &  0.551(32)    &  0.542(39)   \\
$a06m135$             &  0.599(25)    &  0.529(64)    &  0.545(83)    &  0.545(82)    &  0.735(135)   &  0.398(159)   &  0.649(101)  \\
\hline
13-pt $c_4^E\neq 0$   &  0.592(17)    &  0.570(39)    &  0.596(40)    &  0.595(40)    &  0.658(55)    &  0.604(39)    &  0.601(43)   \\
13-pt $c_4^E=0$       &  0.581(13)    &  0.565(30)    &  0.597(34)    &  0.597(34)    &  0.674(47)    &  0.609(36)    &  0.618(37)   \\
\hline
11-pt $c_4^E\neq 0$   &  0.586(17)    &  0.564(39)    &  0.591(41)    &  0.590(41)    &  0.653(56)    &  0.604(41)    &  0.597(44)   \\
11-pt $c_4^E=0$       &  0.572(14)    &  0.554(32)    &  0.588(36)    &  0.587(36)    &  0.665(49)    &  0.606(39)    &  0.609(39)   \\
\hline
10-pt $c_4^E\neq 0$   &  0.587(36)    &  0.552(69)    &  0.567(70)    &  0.567(70)    &  0.632(93)    &  0.530(68)    &  0.592(76)   \\
10-pt $c_4^E=0$       &  0.558(24)    &  0.540(48)    &  0.570(53)    &  0.570(53)    &  0.662(74)    &  0.554(58)    &  0.618(59)   \\
\hline
10$^\ast$-pt $c_4^E\neq 0$  &  0.595(18)    &  0.587(40)    &  0.609(42)    &  0.608(42)    &  0.668(57)    &  0.607(41)    &  0.609(45)   \\
10$^\ast$-pt $c_4^E=0$      &  0.571(14)    &  0.546(32)    &  0.577(36)    &  0.577(36)    &  0.657(50)    &  0.587(40)    &  0.600(40)   \\
    \end{tabular}
  \end{ruledtabular}
\end{table*}

\begin{table*}
\caption{\FIXME{tab:rM-results} Isovector magnetic charge radius
  $\rMsq$ in units of fm${}^2$ from the seven fits to the isovector form factor
  $G_M(Q^2)$.  The derived value for $G_M(0)$ is included in the fits
  as discussed in the text. The rest is the same as in
  Table~\protect\ref{tab:rE-results}. }
\label{tab:rM-results}
  \centering
  \renewcommand{\arraystretch}{1.2}
  \begin{ruledtabular}
    \begin{tabular}{cccccccc}
Ensemble              & Dipole        & $z^3$         & $z^4$         & $z^{5}$       & $z^{3+4}$     & $z^{4+4}$     &  $z^{5+4}$     \\ \hline
$a15m310$             &  0.437(11)    &  0.466(15)    &  0.464(15)    &  0.464(15)    &  0.451(23)    &  0.472(13)    &  0.461(17)   \\
$a12m310$             &  0.414(15)    &  0.457(30)    &  0.457(30)    &  0.457(30)    &  0.438(41)    &  0.484(28)    &  0.447(34)   \\
$a12m220L$            &  0.456(16)    &  0.475(41)    &  0.475(41)    &  0.475(41)    &  0.485(58)    &  0.472(39)    &  0.473(44)   \\
$a12m220$             &  0.442(25)    &  0.419(51)    &  0.425(50)    &  0.424(50)    &  0.406(74)    &  0.451(46)    &  0.408(56)   \\
$a12m220S$            &  0.454(33)    &  0.597(89)    &  0.599(87)    &  0.599(87)    &  0.566(118)   &  0.617(77)    &  0.588(95)   \\
$a09m310$             &  0.410(8)     &  0.409(11)    &  0.409(11)    &  0.409(11)    &  0.400(17)    &  0.423(12)    &  0.402(14)   \\
$a09m220$             &  0.469(14)    &  0.489(30)    &  0.492(31)    &  0.492(31)    &  0.498(38)    &  0.499(32)    &  0.488(31)   \\
$a09m130W$            &  0.478(18)    &  0.437(56)    &  0.386(84)    &  0.384(83)    &  0.478(132)   &  0.092(173)   &  0.485(85)   \\
$a06m310$             &  0.409(31)    &  0.408(41)    &  0.408(41)    &  0.408(41)    &  0.436(55)    &  0.396(46)    &  0.413(43)   \\
$a06m310W$            &  0.407(21)    &  0.406(32)    &  0.406(32)    &  0.405(32)    &  0.388(52)    &  0.416(33)    &  0.398(36)   \\
$a06m220$             &  0.425(24)    &  0.485(49)    &  0.484(49)    &  0.484(49)    &  0.546(78)    &  0.438(44)    &  0.502(55)   \\
$a06m220W$            &  0.474(35)    &  0.473(53)    &  0.473(53)    &  0.473(53)    &  0.477(88)    &  0.465(48)    &  0.473(59)   \\
$a06m135$             &  0.519(34)    &  0.456(76)    &  0.427(97)    &  0.427(97)    &  0.634(214)   &  0.115(195)   &  0.567(133)  \\
\hline
13-pt $c_4^M\neq 0$   &  0.497(29)    &  0.454(61)    &  0.462(64)    &  0.461(64)    &  0.560(90)    &  0.443(65)    &  0.496(68)   \\
13-pt $c_4^M=0$       &  0.482(19)    &  0.449(41)    &  0.458(49)    &  0.457(49)    &  0.575(77)    &  0.441(59)    &  0.520(54)   \\
\hline
11-pt $c_4^M\neq 0$   &  0.495(29)    &  0.445(62)    &  0.450(65)    &  0.449(65)    &  0.548(92)    &  0.434(67)    &  0.486(69)   \\
11-pt $c_4^M=0$       &  0.480(20)    &  0.436(43)    &  0.443(51)    &  0.441(51)    &  0.558(80)    &  0.432(63)    &  0.505(56)   \\
\hline
10-pt $c_4^M\neq 0$   &  0.537(44)    &  0.471(100)   &  0.467(102)   &  0.466(102)   &  0.617(141)   &  0.363(102)   &  0.539(110)  \\
10-pt $c_4^M=0$       &  0.497(28)    &  0.444(62)    &  0.447(71)    &  0.446(71)    &  0.610(112)   &  0.378(87)    &  0.549(80)   \\
\hline
10$^\ast$-pt $c_4^M\neq 0$  &  0.503(32)    &  0.481(65)    &  0.484(67)    &  0.484(67)    &  0.563(94)    &  0.449(67)    &  0.505(71)   \\
10$^\ast$-pt $c_4^M=0$      &  0.480(20)    &  0.431(43)    &  0.433(52)    &  0.432(51)    &  0.550(80)    &  0.407(64)    &  0.498(57)   \\
    \end{tabular}
  \end{ruledtabular}
\end{table*}

\begin{table*}
\caption{\FIXME{tab:mu-results} Isovector magnetic moment of the
  nucleon, $\mu^{p-n} \equiv \mu_p - \mu_n$, in units of the Bohr
  magneton from the seven fits to the isovector form factor $G_M(Q^2)$
  and including the derived value for $G_M(0)$.  The rest is the same
  as in Table~\protect\ref{tab:rE-results}. }
\label{tab:mu-results}
  \centering
  \renewcommand{\arraystretch}{1.2}
  \begin{ruledtabular}
    \begin{tabular}{cccccccc}
Ensemble              & Dipole        & $z^3$         & $z^4$         & $z^{5}$       & $z^{3+4}$     &  $z^{4+4}$    &  $z^{5+4}$   \\ \hline
$a15m310$             &  4.280(57)    &  4.295(57)    &  4.295(57)    &  4.295(57)    &  4.296(57)    &  4.296(57)    &  4.295(57)   \\
$a12m310$             &  4.205(85)    &  4.303(94)    &  4.303(94)    &  4.303(94)    &  4.297(94)    &  4.310(94)    &  4.301(94)   \\
$a12m220L$            &  4.215(65)    &  4.253(84)    &  4.253(84)    &  4.253(84)    &  4.257(85)    &  4.252(83)    &  4.253(84)   \\
$a12m220$             &  4.103(125)   &  4.143(130)   &  4.143(130)   &  4.143(130)   &  4.134(130)   &  4.135(130)   &  4.141(130)  \\
$a12m220S$            &  4.005(155)   &  4.256(182)   &  4.255(182)   &  4.255(182)   &  4.262(182)   &  4.262(182)   &  4.257(182)  \\
$a09m310$             &  4.141(30)    &  4.141(31)    &  4.141(31)    &  4.141(31)    &  4.141(31)    &  4.140(31)    &  4.141(31)   \\
$a09m220$             &  4.260(60)    &  4.292(66)    &  4.292(66)    &  4.292(66)    &  4.291(66)    &  4.292(67)    &  4.292(66)   \\
$a09m130W$            &  4.086(71)    &  4.088(75)    &  4.087(75)    &  4.087(75)    &  4.095(74)    &  4.084(75)    &  4.089(75)   \\
$a06m310$             &  4.044(149)   &  3.985(159)   &  3.985(159)   &  3.985(159)   &  3.989(157)   &  3.986(160)   &  3.985(159)  \\
$a06m310W$            &  4.145(124)   &  4.163(126)   &  4.163(126)   &  4.163(126)   &  4.163(125)   &  4.161(126)   &  4.163(125)  \\
$a06m220$             &  3.938(93)    &  3.941(94)    &  3.941(94)    &  3.941(94)    &  3.940(94)    &  3.941(94)    &  3.941(94)   \\
$a06m220W$            &  4.119(131)   &  4.113(132)   &  4.113(132)   &  4.113(132)   &  4.113(132)   &  4.112(132)   &  4.113(132)  \\
$a06m135$             &  4.100(115)   &  4.078(113)   &  4.077(113)   &  4.077(113)   &  4.078(112)   &  4.079(113)   &  4.076(113)  \\
\hline
13-pt $c_4^\mu\neq 0$ &  3.962(79)    &  3.930(81)    &  3.929(81)    &  3.929(81)    &  3.932(81)    &  3.927(81)    &  3.930(81)   \\
13-pt $c_4^\mu=0$     &  3.950(79)    &  3.918(80)    &  3.917(80)    &  3.917(80)    &  3.920(80)    &  3.915(80)    &  3.917(80)   \\
\hline
11-pt $c_4^\mu\neq 0$ &  3.975(84)    &  3.940(86)    &  3.939(86)    &  3.939(86)    &  3.942(86)    &  3.937(86)    &  3.939(86)   \\
11-pt $c_4^\mu=0$     &  3.968(84)    &  3.933(86)    &  3.932(86)    &  3.932(86)    &  3.935(85)    &  3.930(86)    &  3.933(86)   \\
\hline
10-pt $c_4^\mu\neq 0$ &  4.167(151)   &  3.982(164)   &  3.982(164)   &  3.982(164)   &  3.987(164)   &  3.976(164)   &  3.982(164)  \\
10-pt $c_4^\mu=0$     &  3.999(122)   &  3.879(129)   &  3.877(129)   &  3.877(129)   &  3.884(129)   &  3.874(129)   &  3.879(129)  \\
\hline
10$^\ast$-pt $c_4^\mu\neq 0$  &  3.970(85)    &  3.942(86)    &  3.941(86)    &  3.941(86)    &  3.945(86)    &  3.940(86)    &  3.942(86)   \\
10$^\ast$-pt $c_4^\mu=0$    &  3.964(84)    &  3.933(86)    &  3.932(86)    &  3.932(86)    &  3.936(85)    &  3.930(86)    &  3.933(86)   \\
    \end{tabular}
  \end{ruledtabular}
\end{table*}

\section{Results for  $\langle r_E^2 \rangle$,  $\langle r_M^2 \rangle$ and $\mu$}
\label{sec:results}
\FIXME{sec:results} 

\begin{figure*}[tb]   
{
    \includegraphics[width=0.32\linewidth]{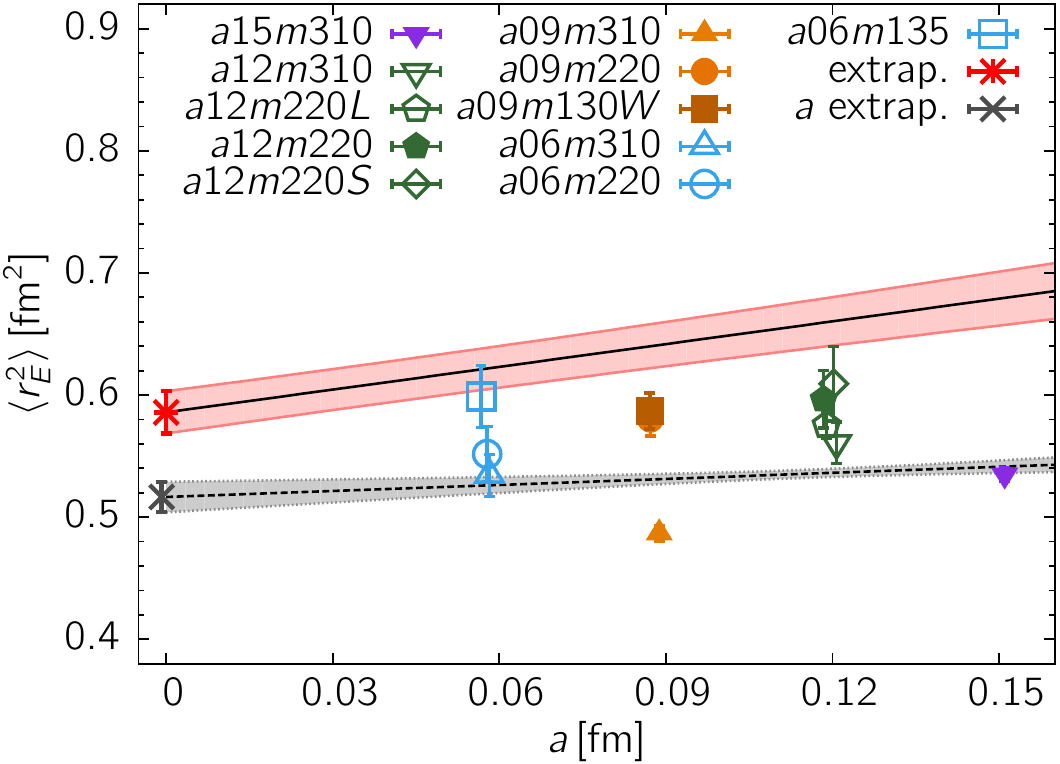}
    \includegraphics[width=0.32\linewidth]{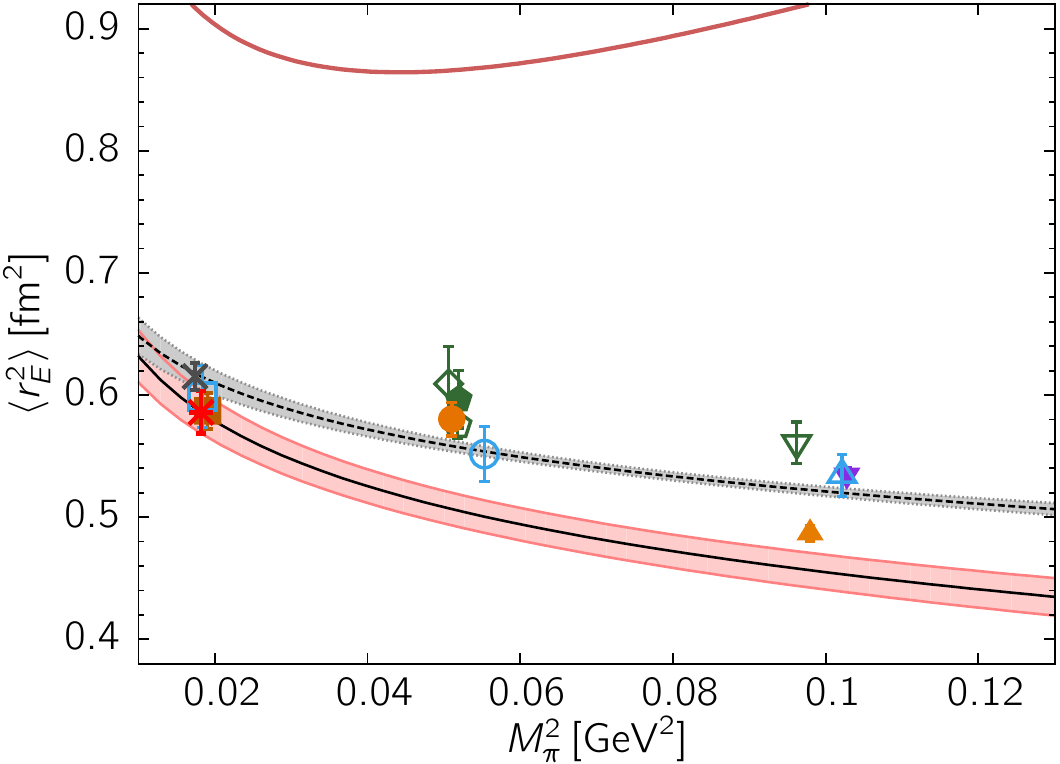}
    \includegraphics[width=0.32\linewidth]{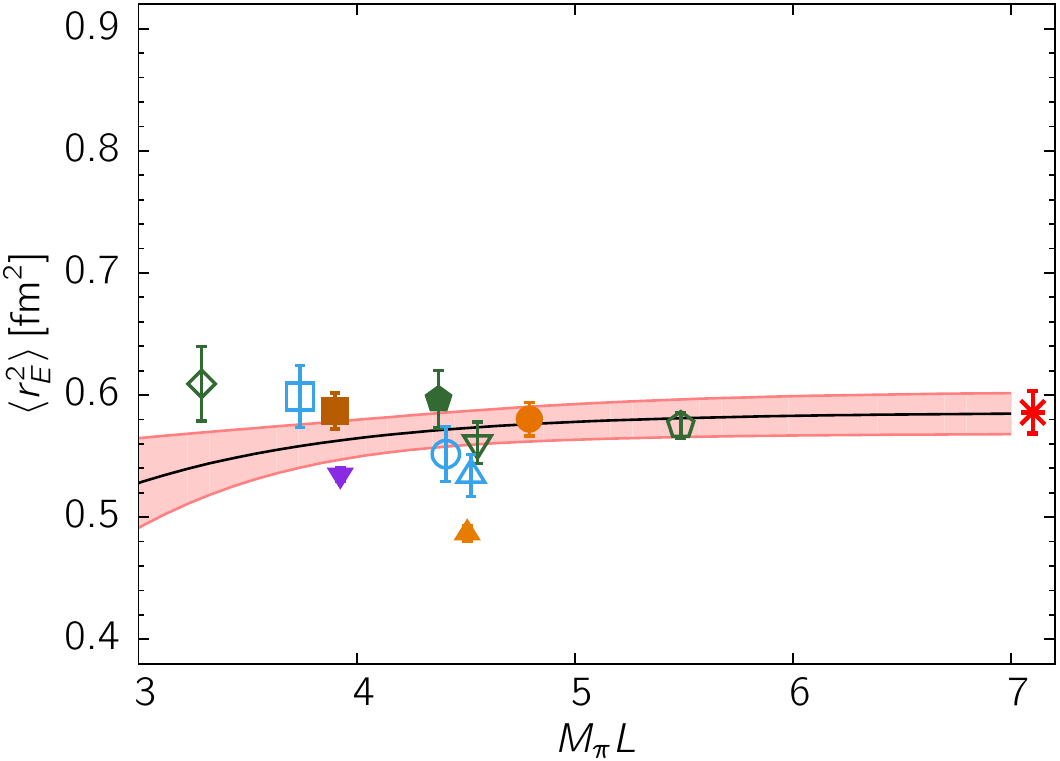}
}
{
    \includegraphics[width=0.32\linewidth]{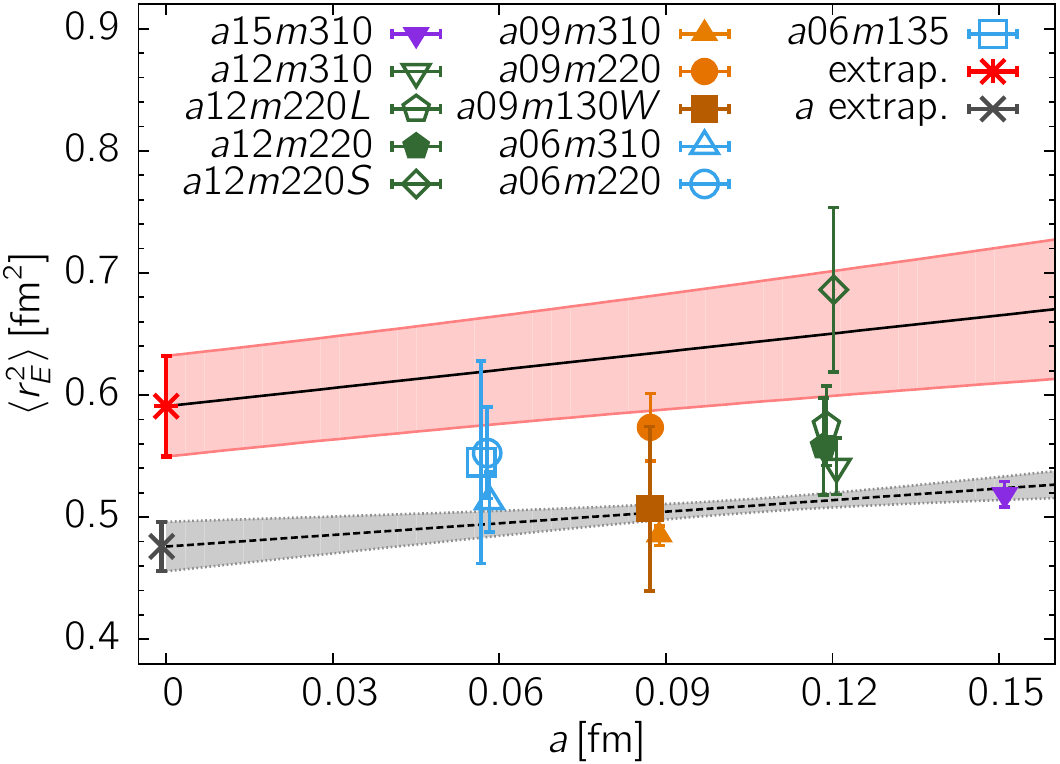}
    \includegraphics[width=0.32\linewidth]{Figs-CCFV/rE_mpisq_dip_a_log3_fv4_nolabel_incl_ChPT}
    \includegraphics[width=0.32\linewidth]{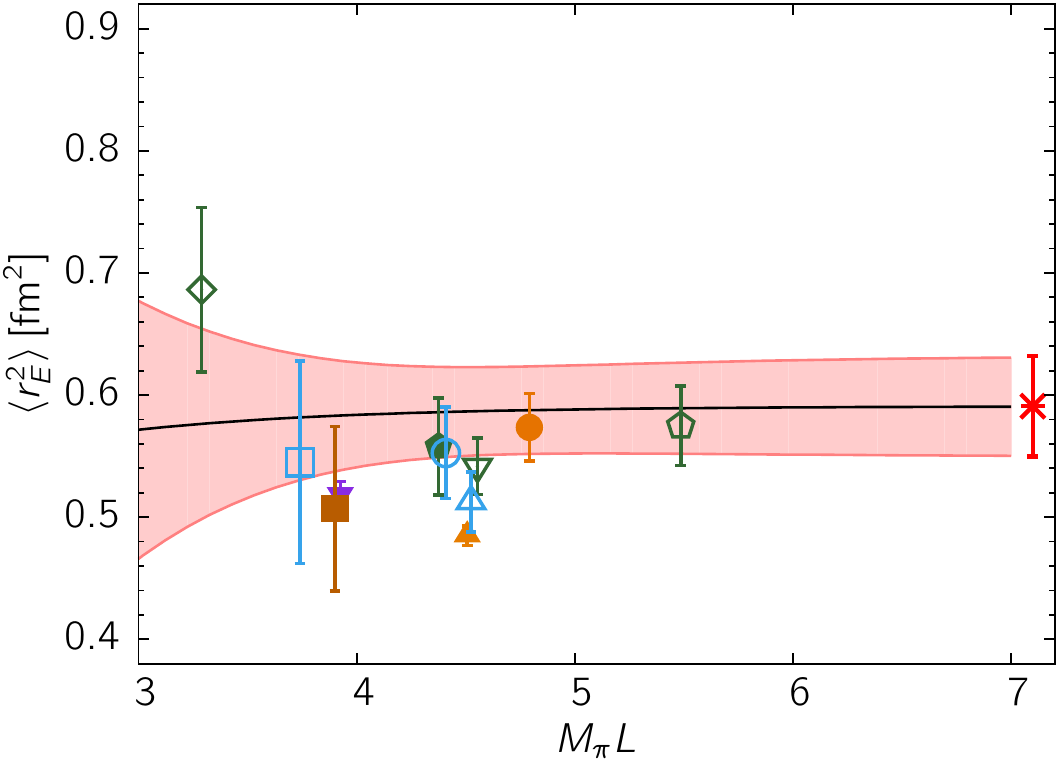}
}
{
    \includegraphics[width=0.32\linewidth]{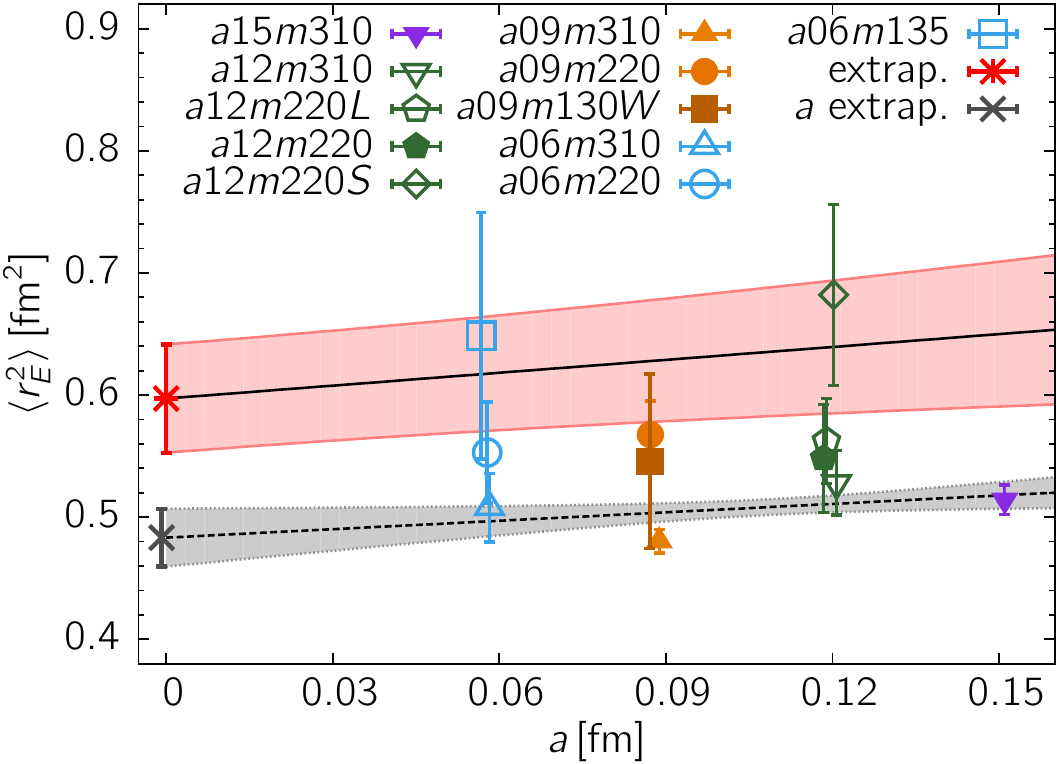}
    \includegraphics[width=0.32\linewidth]{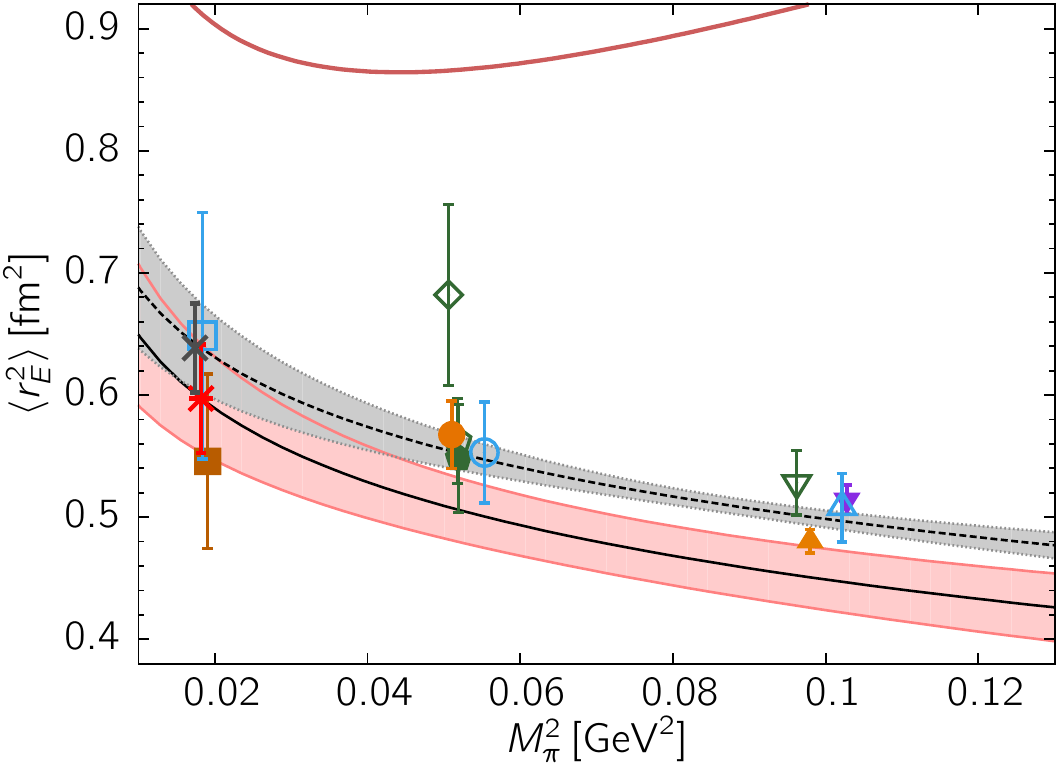}
    \includegraphics[width=0.32\linewidth]{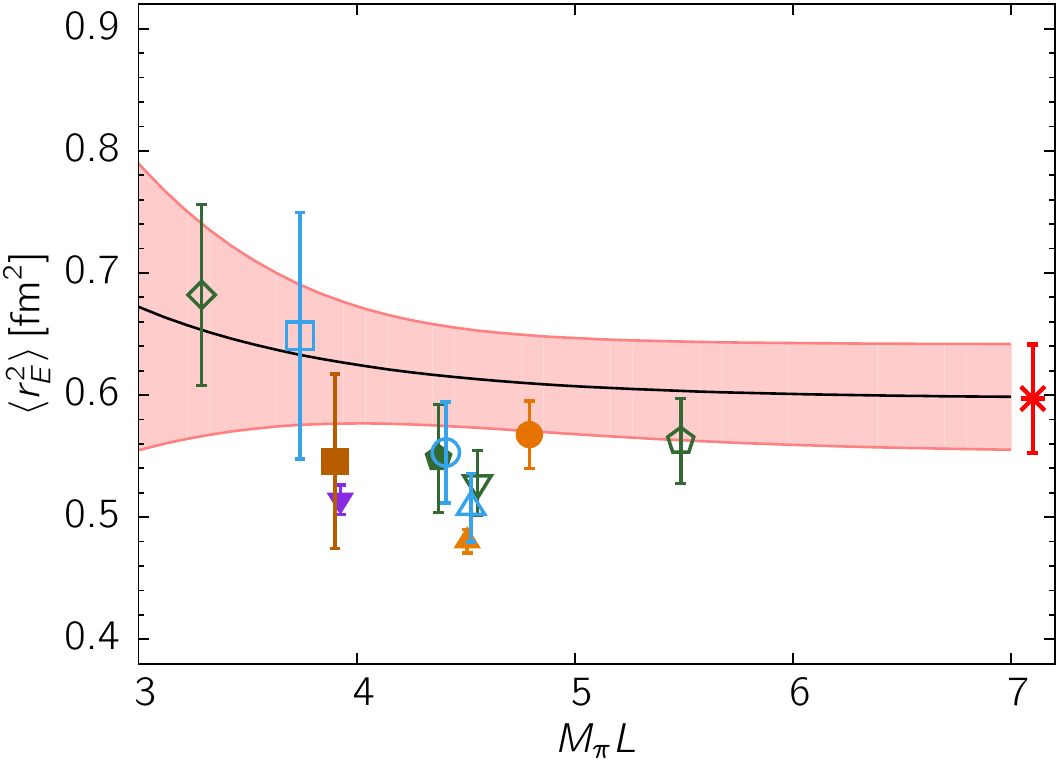}
}
  \caption{\FIXME{fig:rE-extrap11} The 11-point CCFV fits for $\langle
    r_E^2 \rangle$ to the dipole (top), $z^4$ (middle) and $z^{5+4}$
    (bottom) data given in Table~\ref{tab:rE-results}.  In each panel,
    the CCFV fit (pink band) is shown versus a single variable with
    the other two variables set to their values at the physical
    point. The extrapolated values of $\langle r_E^2 \rangle$ are
    shown using the symbol red star. Fits in a single variable ($a$ or
    $M_\pi$) are shown as gray bands and the corresponding
    extrapolated value by a black star. The solid red line is the prediction of $\chi$PT using the expressions 
    given in Ref.~\protect\cite{Kubis:2000zd}.}
\label{fig:rE-extrap11}
\end{figure*}

\begin{figure*}[tb] 
{
    \includegraphics[width=0.32\linewidth]{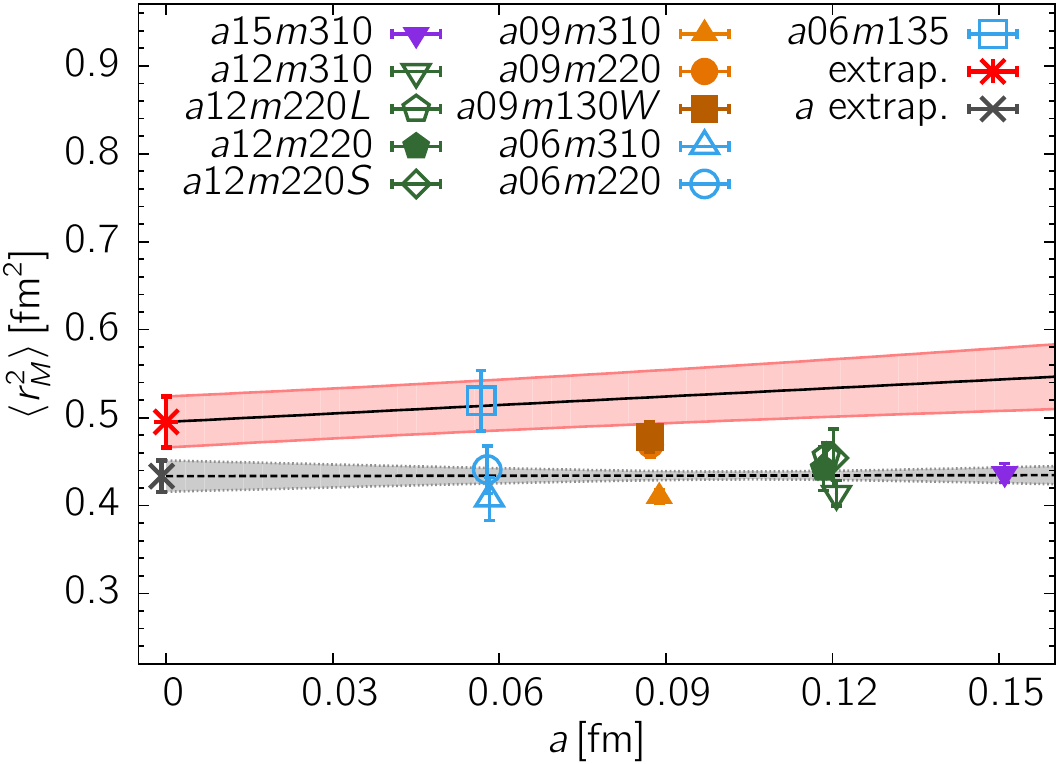}
    \includegraphics[width=0.32\linewidth]{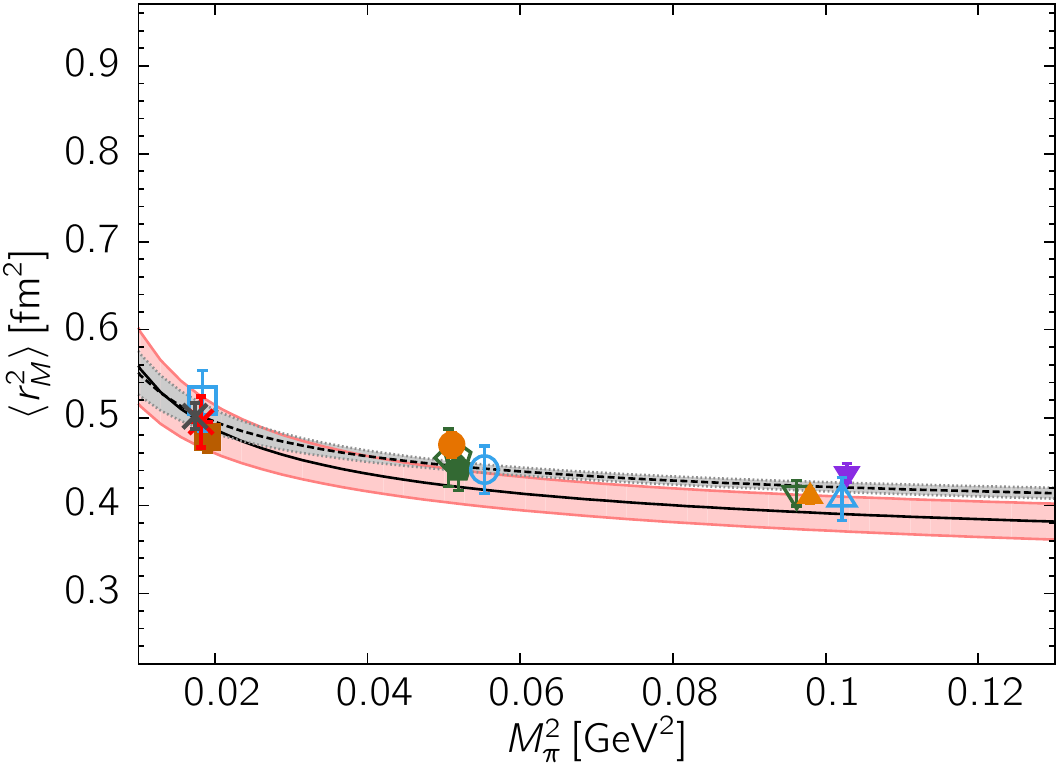}
    \includegraphics[width=0.32\linewidth]{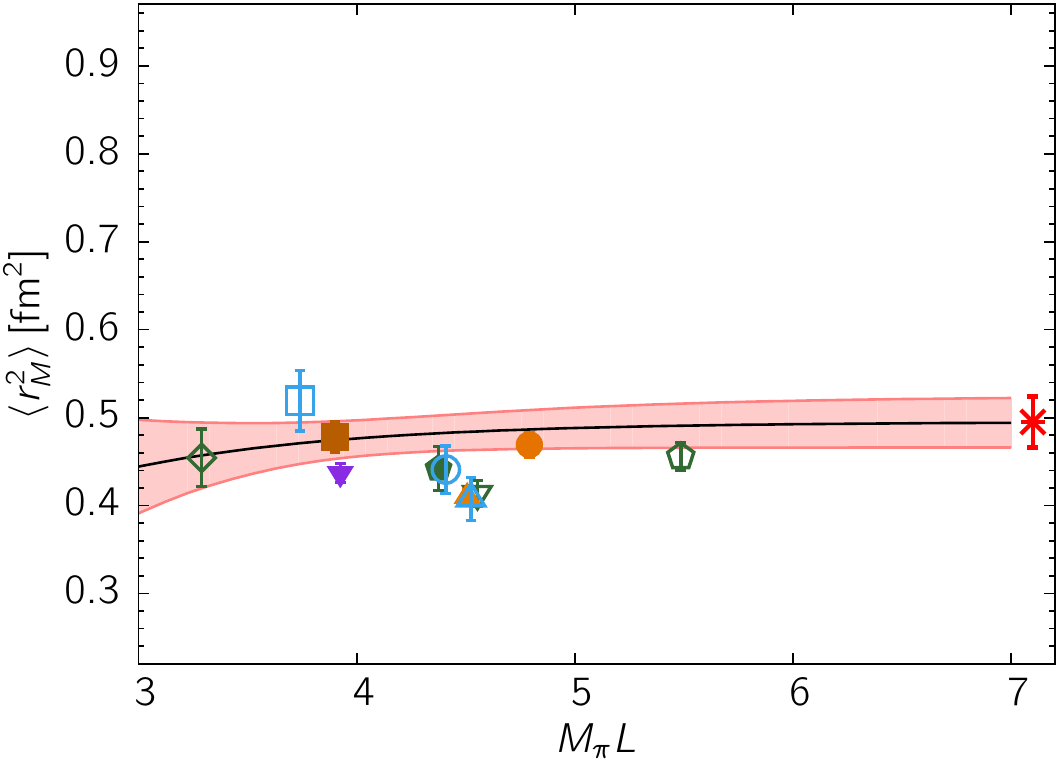}
}
{
    \includegraphics[width=0.32\linewidth]{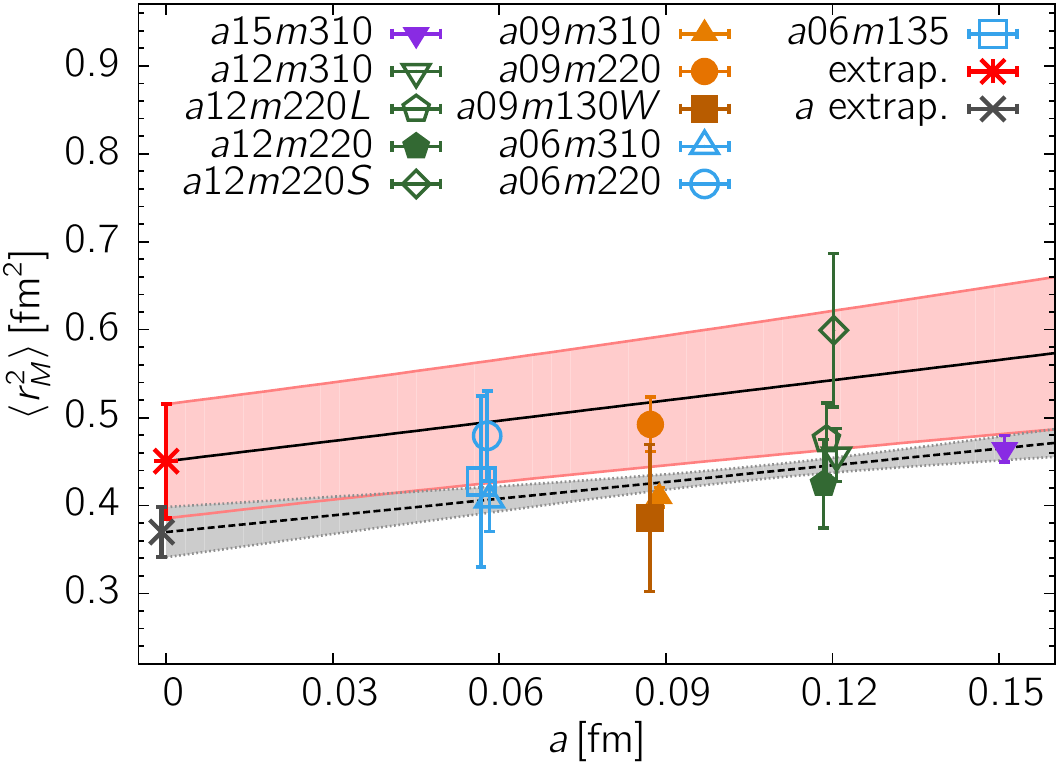}
    \includegraphics[width=0.32\linewidth]{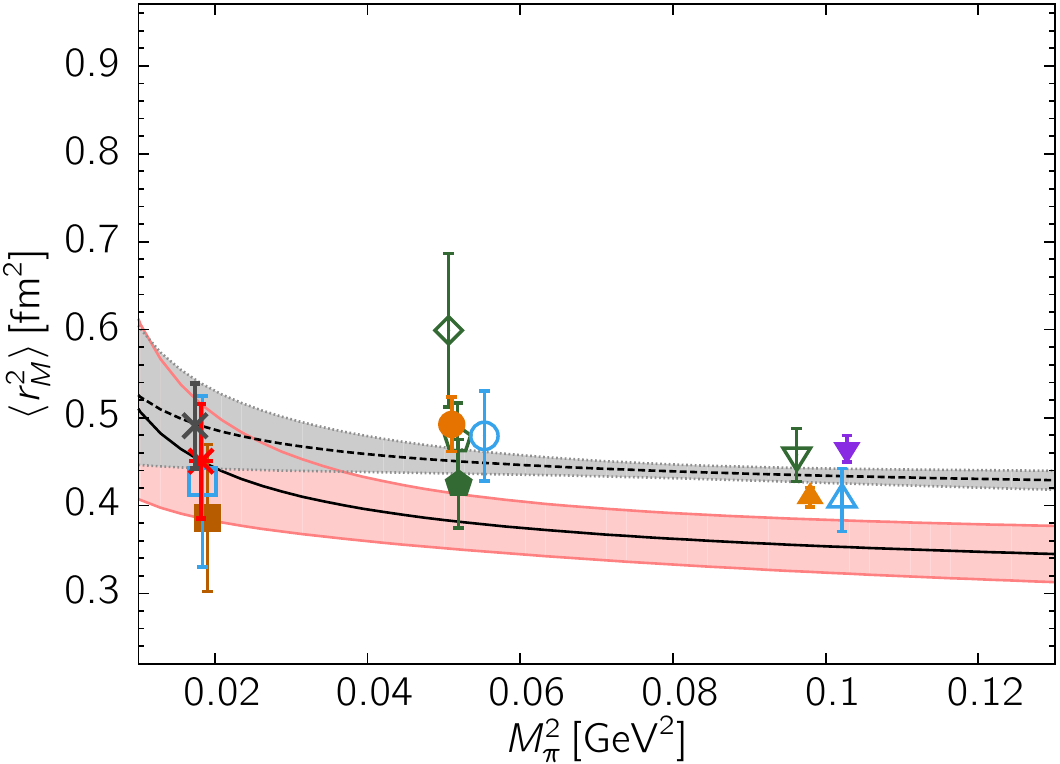}
    \includegraphics[width=0.32\linewidth]{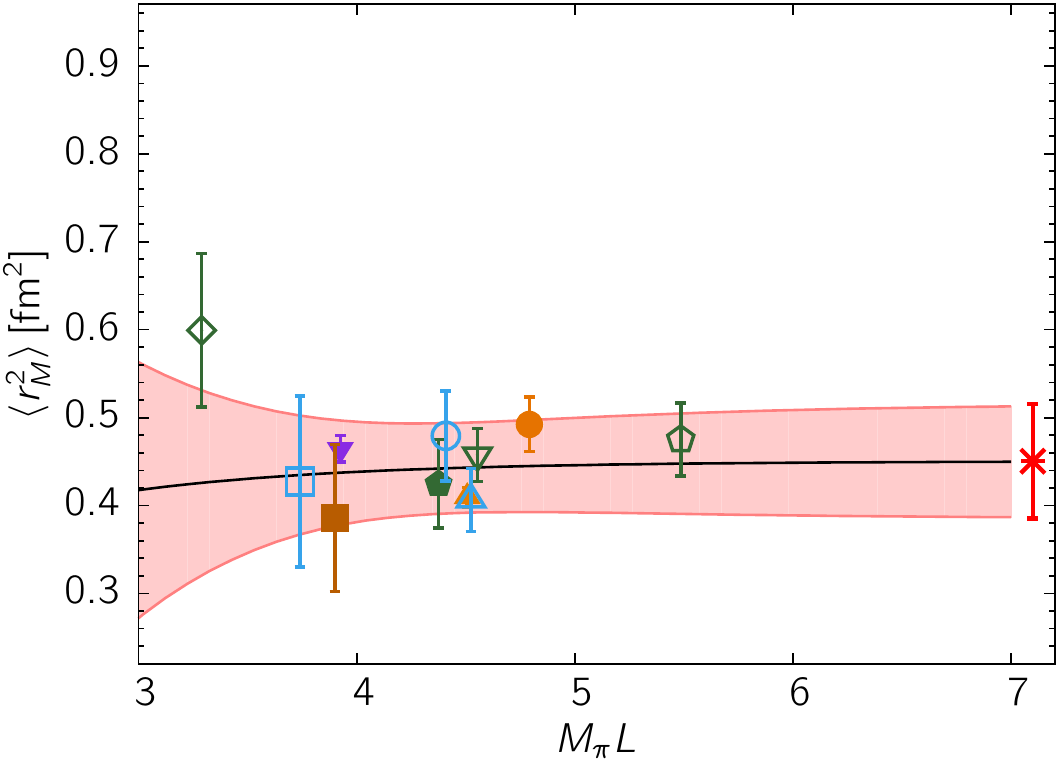}
}
{
    \includegraphics[width=0.32\linewidth]{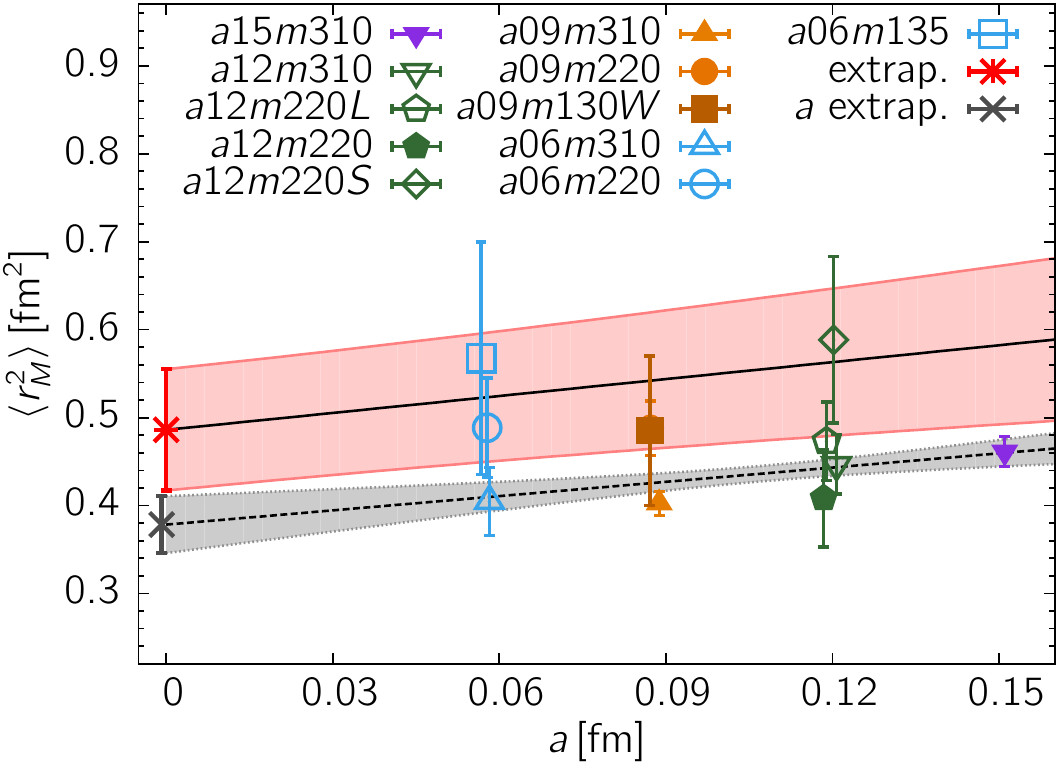}
    \includegraphics[width=0.32\linewidth]{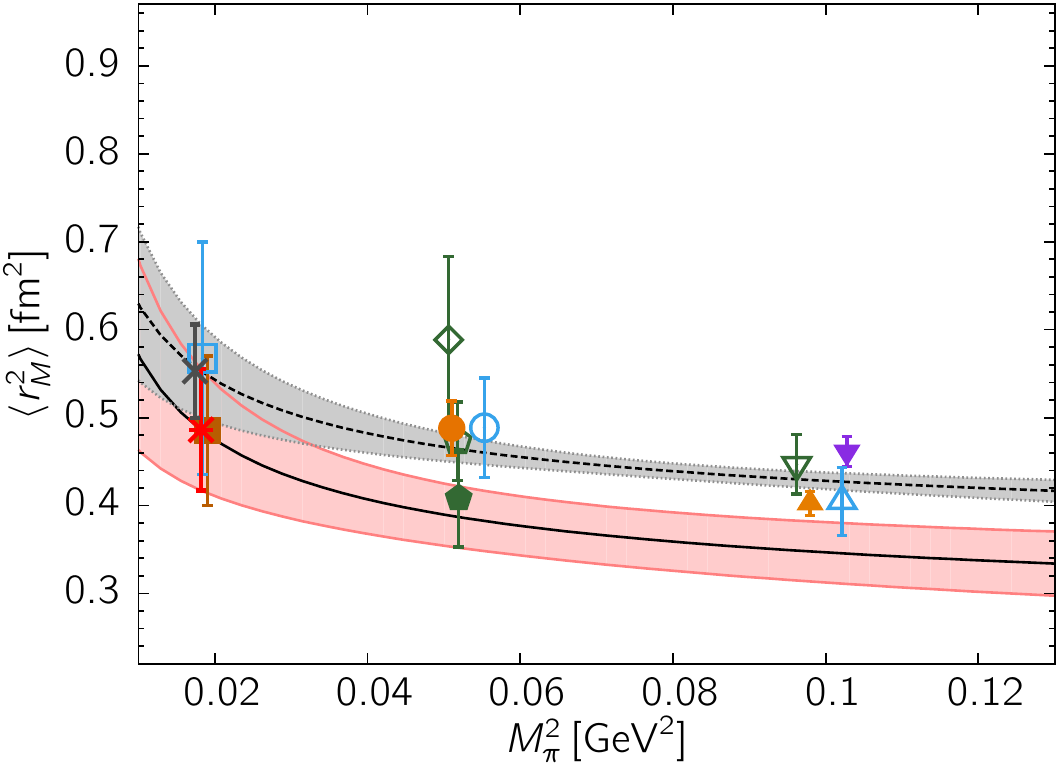}
    \includegraphics[width=0.32\linewidth]{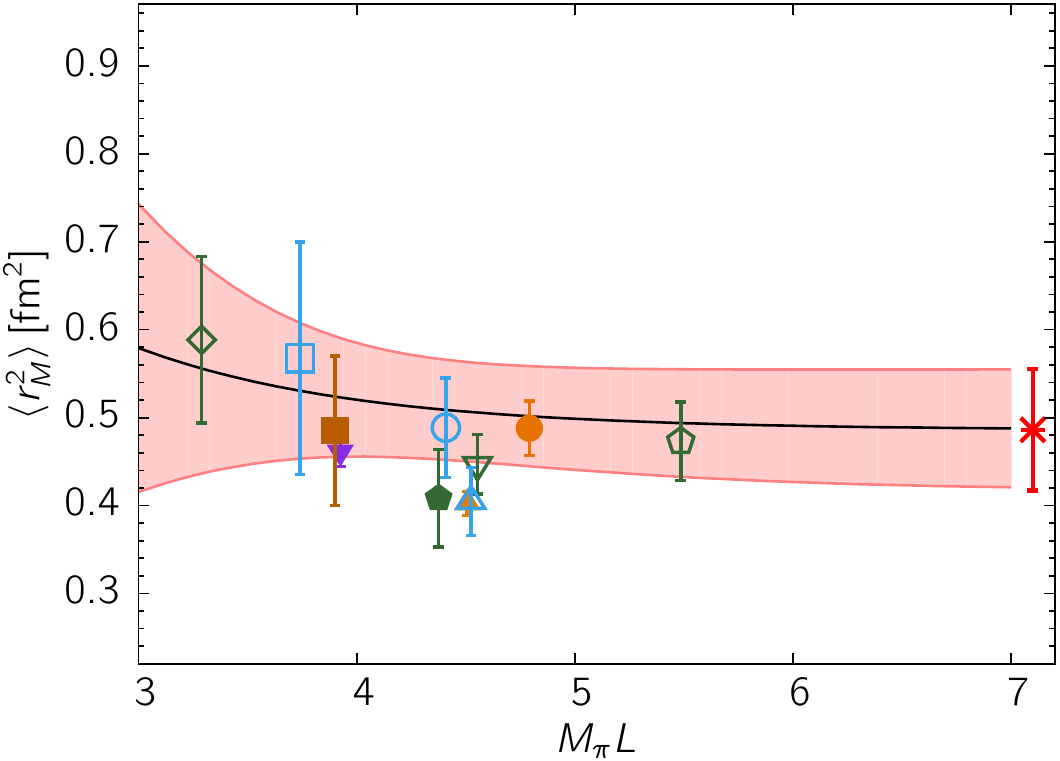}
}
  \caption{\FIXME{fig:rM-extrap11} The 11-point CCFV fits for $\langle
    r_M^2 \rangle$ to the dipole (top), $z^4$ (middle) and $z^{5+4}$(bottom) data
    given in Table~\ref{tab:rM-results}.  Rest is the same as in
    Fig.~\protect\ref{fig:rE-extrap11}.  }
\label{fig:rM-extrap11}
\end{figure*}

\begin{figure*}[tb]   
{
    \includegraphics[width=0.32\linewidth]{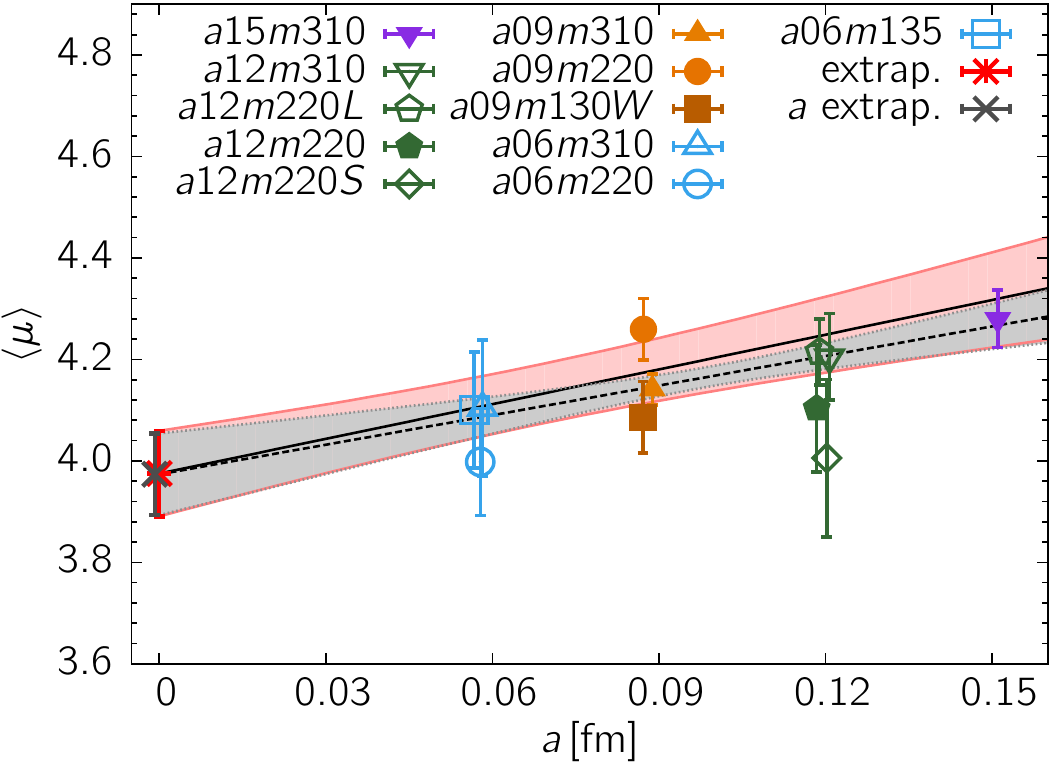}
    \includegraphics[width=0.32\linewidth]{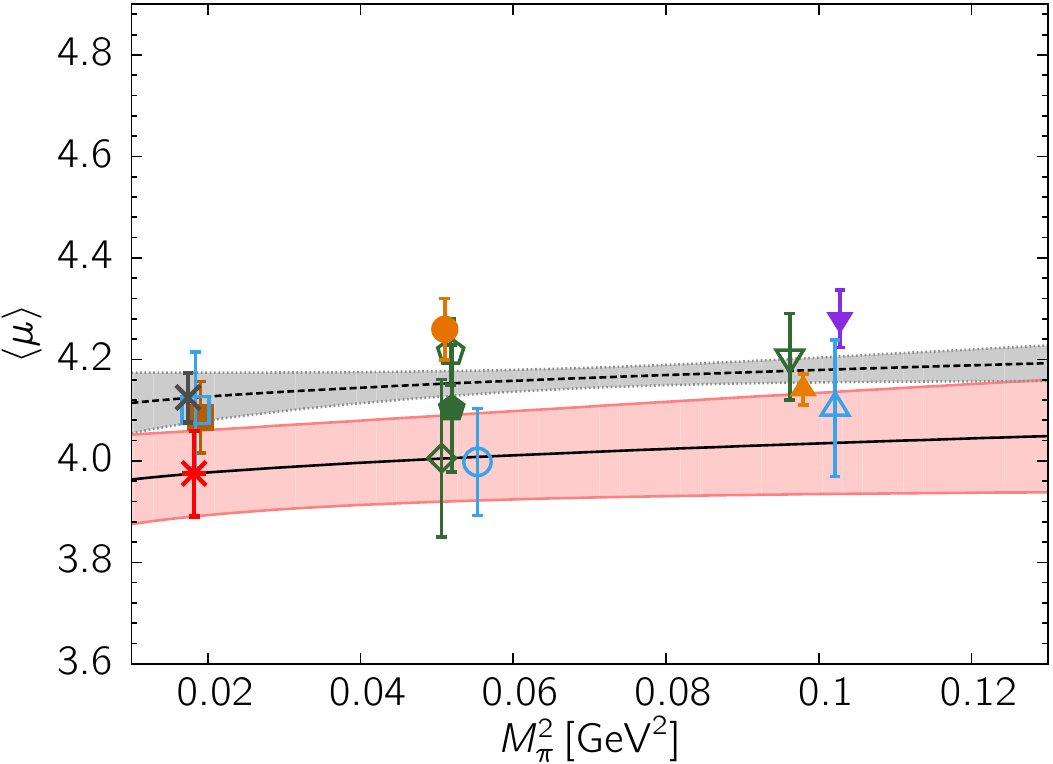}
    \includegraphics[width=0.32\linewidth]{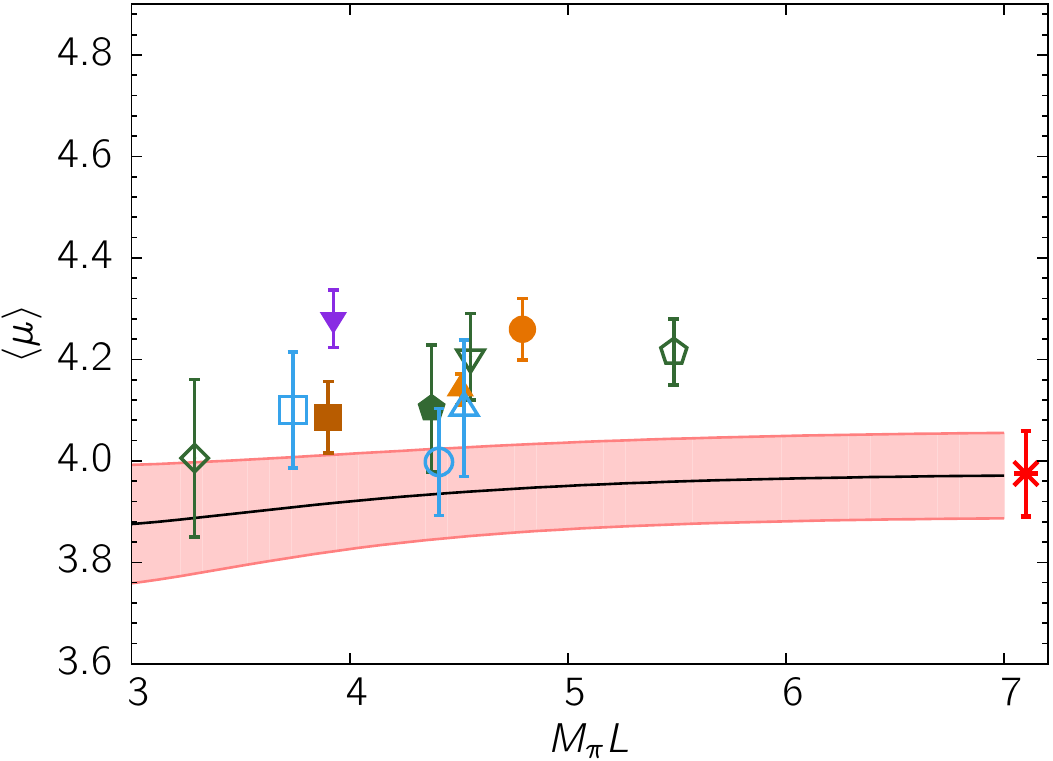}
}
{
    \includegraphics[width=0.32\linewidth]{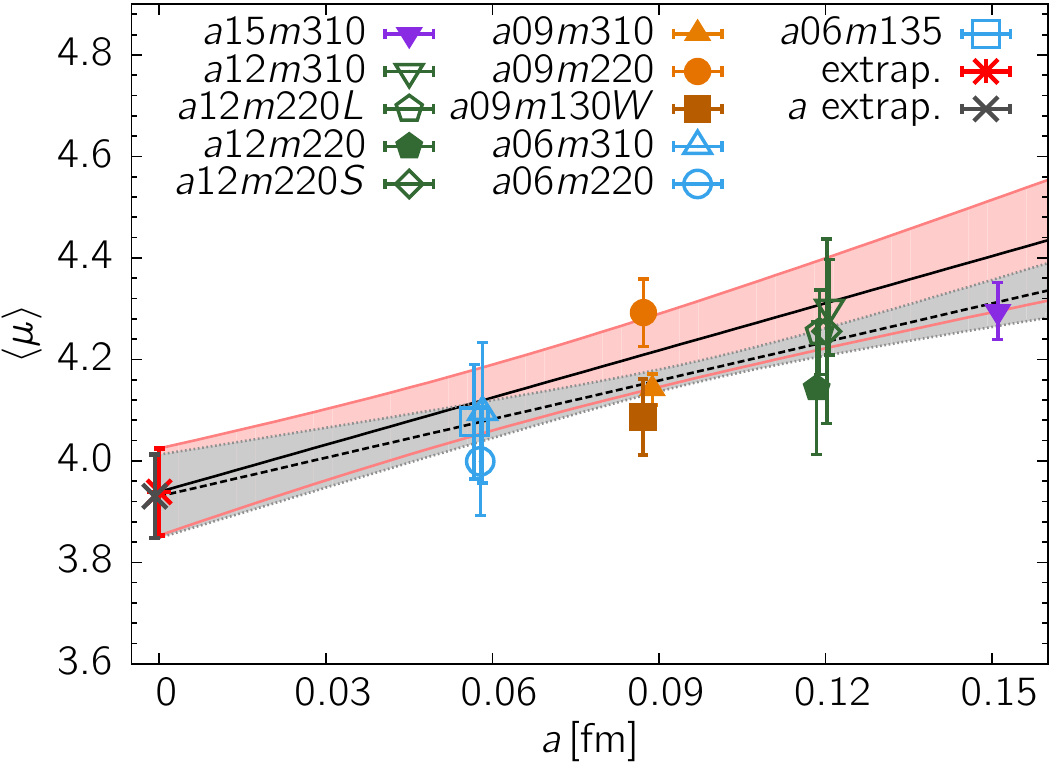}
    \includegraphics[width=0.32\linewidth]{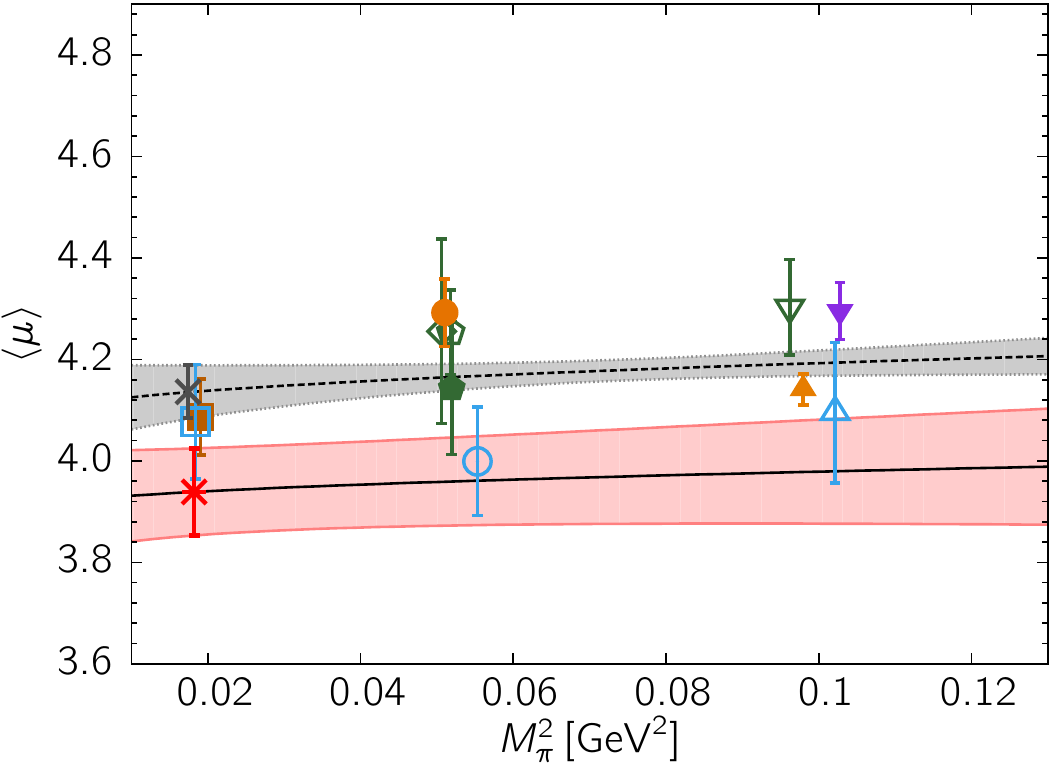}
    \includegraphics[width=0.32\linewidth]{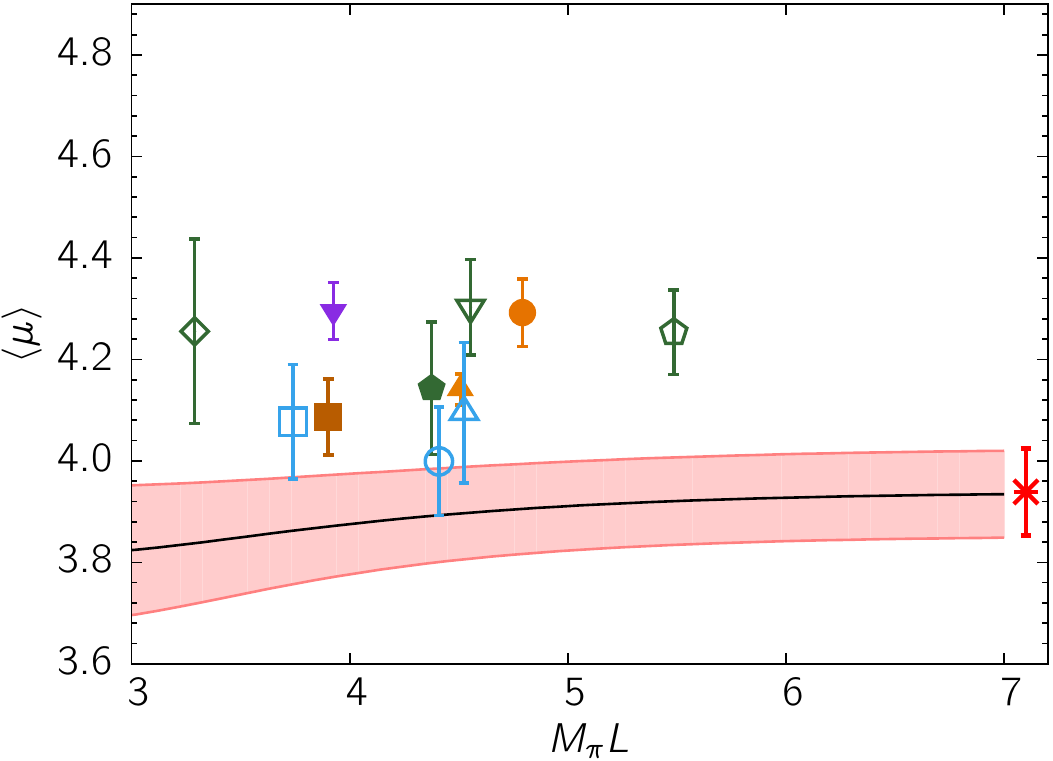}
}
{
    \includegraphics[width=0.32\linewidth]{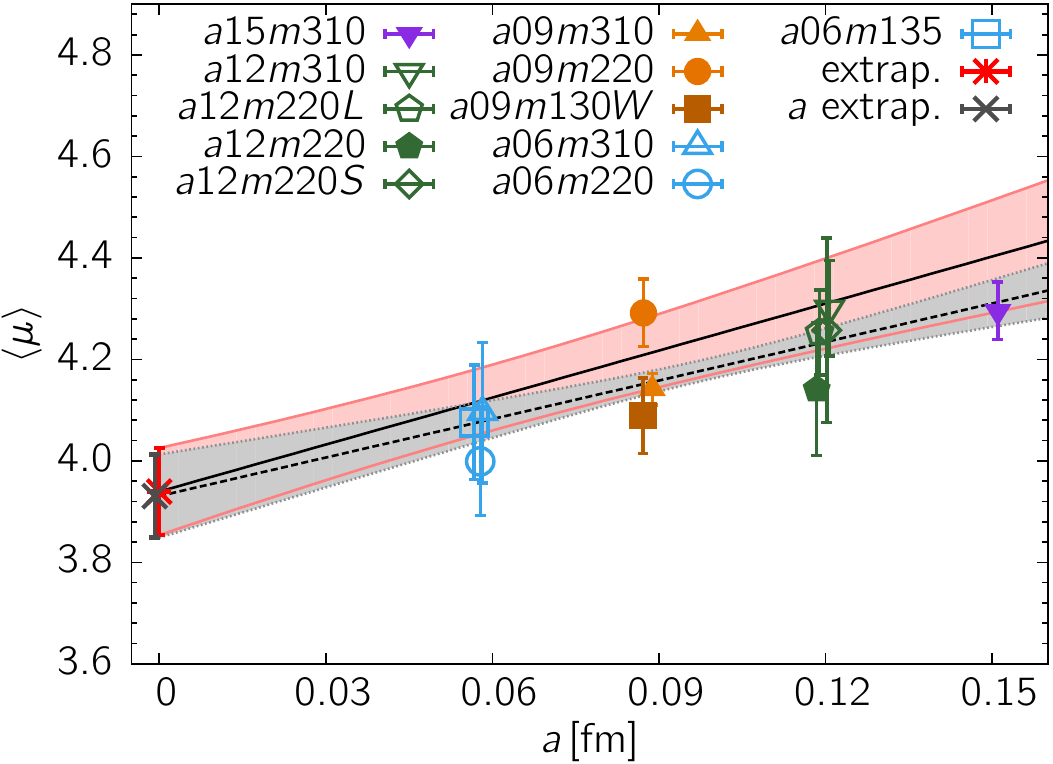}
    \includegraphics[width=0.32\linewidth]{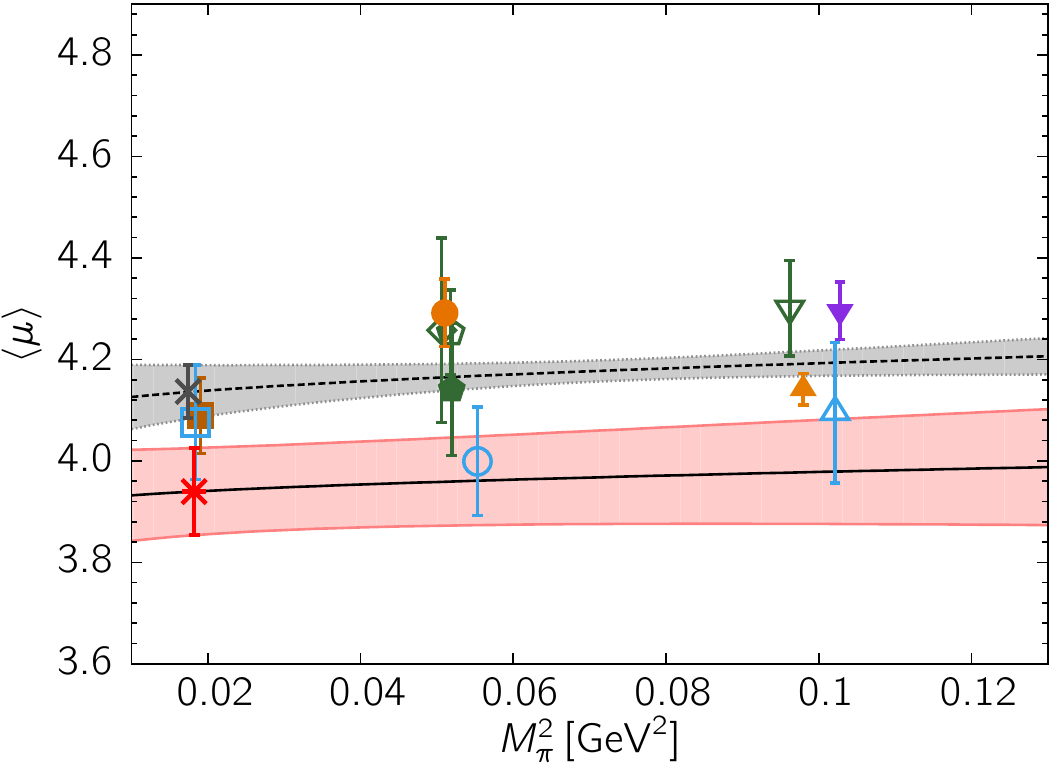}
    \includegraphics[width=0.32\linewidth]{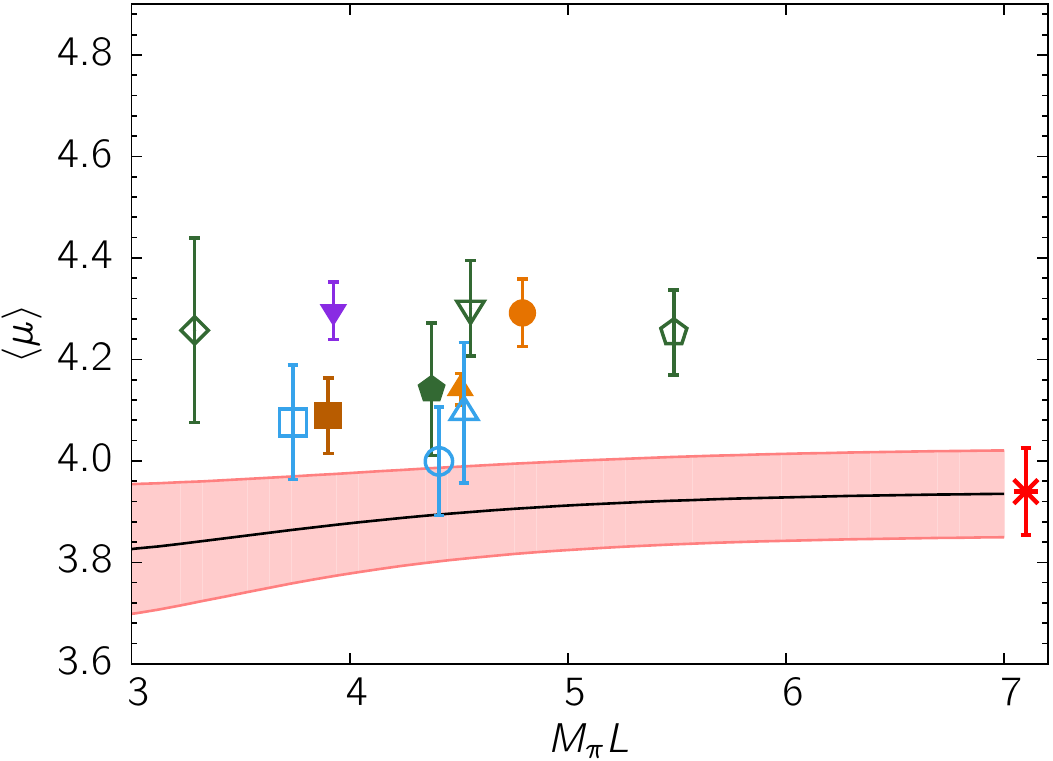}
}
\caption{\FIXME{fig:mu-extrap11} 
    The 11-point CCFV fits for $\mu^{p-n}$ to the dipole (top), $z^4$ (middle) and $z^{5+4}$ (bottom) data
    given in Table~\ref{tab:mu-results}.  Rest is the same as in
    Fig.~\protect\ref{fig:rE-extrap11}. 
}
\label{fig:mu-extrap11}
\end{figure*}

\begin{figure*}  
\centering
\subfigure{
\includegraphics[width=0.47\linewidth]{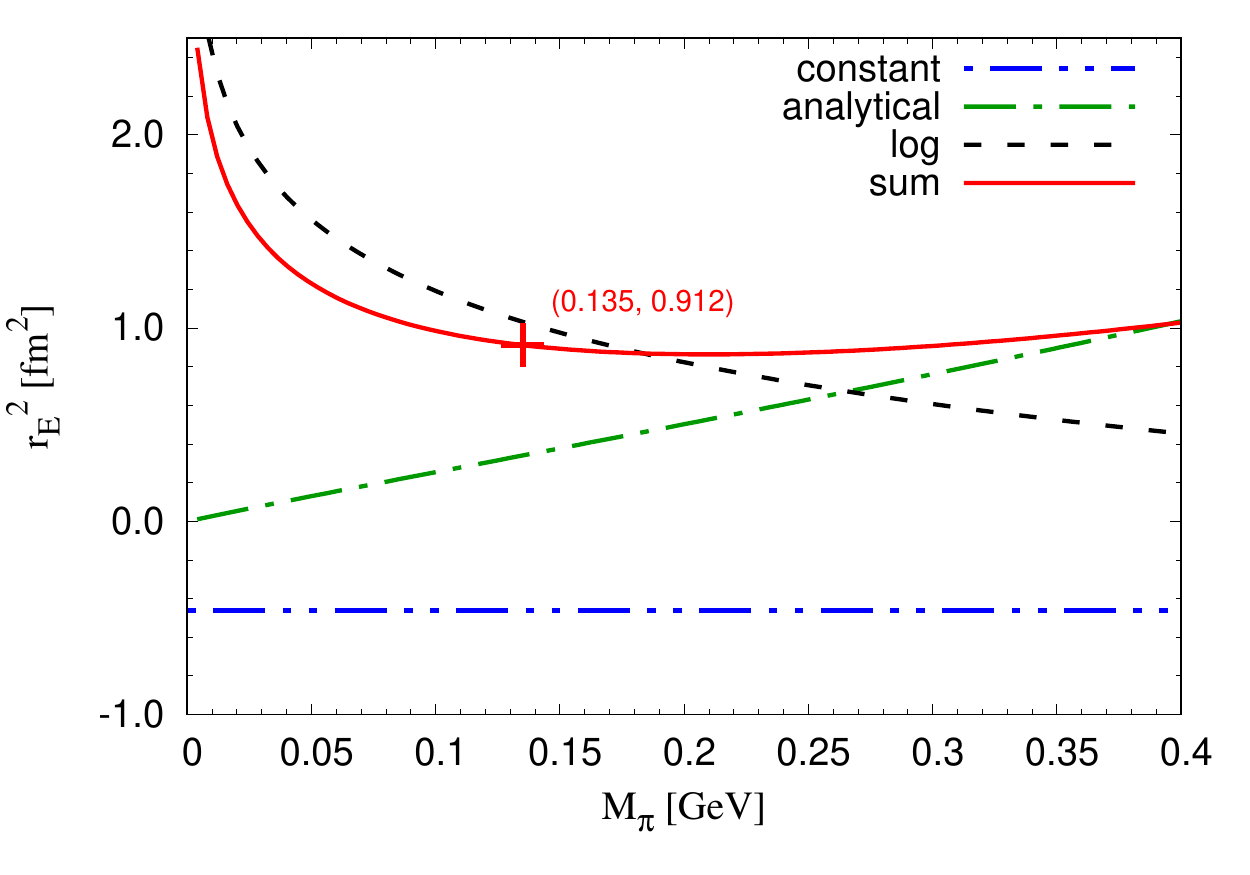}
\includegraphics[width=0.47\linewidth]{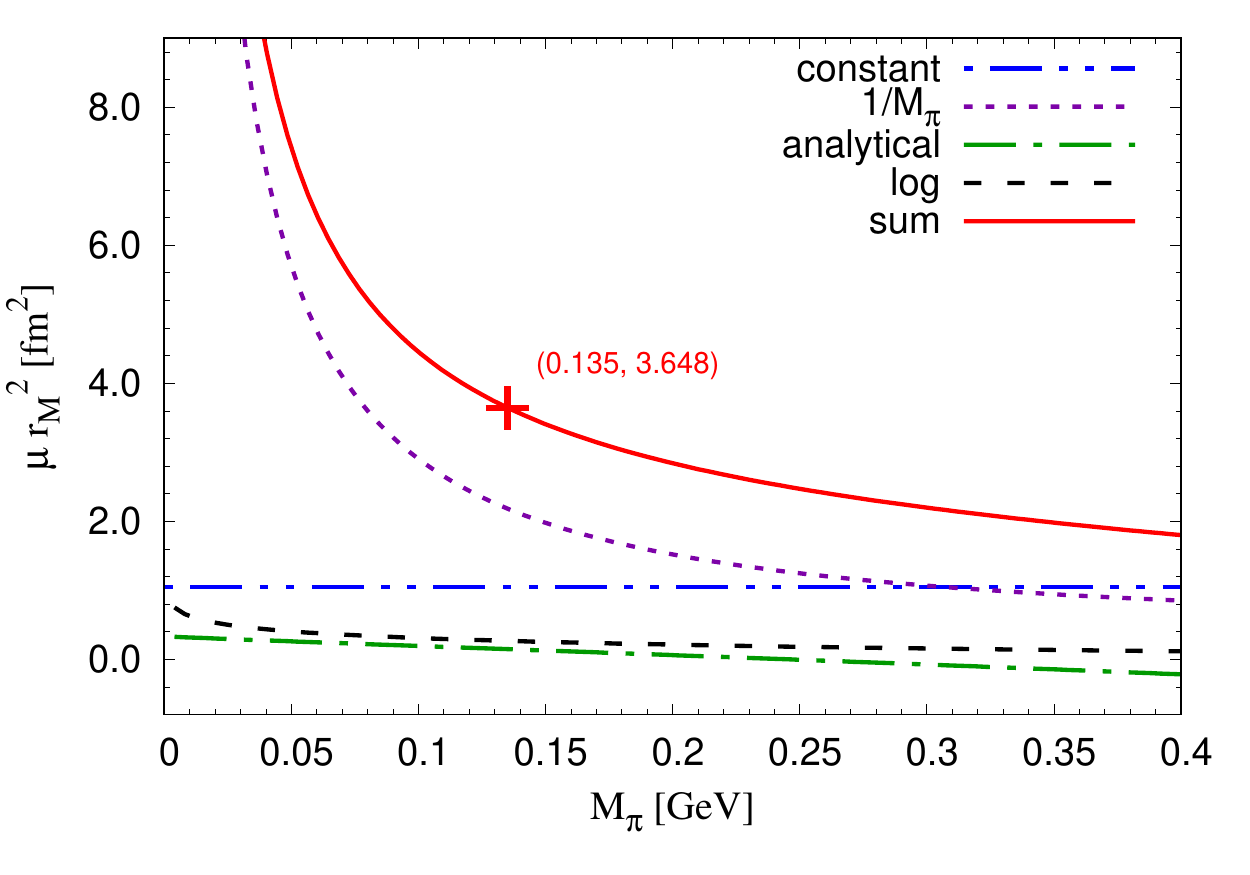}
}
\caption{\FIXME{fig:ChPT} The prediction of chiral perturbation theory
  for the isovector $\rEsq$ (left) and $\mu \rMsq$ (right) using the
  expressions given in Ref.~\protect\cite{Kubis:2000zd}.  The
  contribution of the sub-terms, ``constant'' (green dot-dash line)
  ``log'' (black dash line), ``analytical'' (blue dot-dot-dash line)
  and the ``$1/M_\pi$'', defined in the text, are shown
  separately. Their sum is shown by solid red line. The red plus sign
  marks the physical point $M_\pi=135$~MeV. The values of the LEC used
  in these fits are given in the text.}
\label{fig:ChPT}
\end{figure*}

To obtain results for $\rEsq$, $\rMsq$ and $\mu$ in the limits $a \to 0$,
$M_\pi \to 135$~MeV and $M_\pi L \to \infty$, we make a simultaneous
(CCFV) fit in these three variable to the data given in
Tables~\ref{tab:rE-results},~\ref{tab:rM-results}
and~\ref{tab:mu-results}. Given the spread in the lattice parameters of the 11
ensembles analyzed, we include the leading order correction term in
each of the three variables, i.e., fits with four free parameters,
$c_i^{E,M,\mu}$. The fit ansatz for the electric mean-square charge radius used is
\begin{align}
  \expv{r_E^2}(a,M_\pi,L) = &c_1^E + c_2^E a + c_3^E \ln(M_\pi^2/\lambda^2) + \nonumber \\
   &c_4^E \ln(M_\pi^2/\lambda^2) \exp(-M_\pi L) \,,
  \label{eq:rEsq-extrap}
\end{align}
where the mass scale $\lambda$ is chosen to be $M_\rho=775\,\MeV$ and
the form of the chiral and FV corrections are taken from Refs.~\cite{Bernard:1998gv,Kubis:2000zd,Gockeler:2003ay}. 
For the magnetic mean-square charge radius, we use 
\begin{equation}
  \expv{r_M^2}(a,M_\pi,L) = c_1^M + c_2^M a + \frac{c_3^M}{M_\pi} + \frac{c_4^M}{M_\pi} \exp(-M_\pi L)\,, 
  \label{eq:rMsq-extrap}
\end{equation}
where the leading dependence on $M_\pi$ is taken from Ref.~\cite{Bernard:1998gv,Kubis:2000zd}. Lastly, the 
ansatz used for the magnetic moment is 
\begin{align}
  \mu(a,M_\pi,L) = &c_1^\mu + c_2^\mu a + c_3^\mu M_\pi + \nonumber \\
   &c_4^\mu M_\pi\left(1-\frac{2}{M_\pi L}\right) \exp(-M_\pi L) \,. 
  \label{eq:mu-extrap}
\end{align}
where the forms of the chiral and finite volume correction terms are
taken from Ref.~\cite{Kubis:2000zd,Beane:2004tw}.  We express all
masses in units of GeV and the lattice spacing in fm.

In all three CCFV fit ansatz,
Eqs.~\eqref{eq:rEsq-extrap}--\eqref{eq:mu-extrap}, heavy baryon chiral
perturbation theory ($\chi$PT) has been used only to determine the
form of the leading order chiral correction. For example, for $\mu$,
$\chi$PT predicts the slope, $c_3^\mu$, of the linear dependence on
$M_\pi$ as $M_N g_A^2/(4 \pi F_\pi^2)$ with
$F_\pi=92.2$~MeV~\cite{Beane:2003xv}, however, we leave $c_3^\mu$ a free
parameter.  For $\rEsq$ and $\rMsq$, we do not have data at enough values of $M_\pi$ to test
the contribution of the different terms in the $\chi$PT
prediction~\cite{Kubis:2000zd} as discussed later in this section. To avoid over 
parameterization of the fit we, therefore, include only the nonanalytical term
in Eqs.~\eqref{eq:rEsq-extrap} and~\eqref{eq:rMsq-extrap}.  Our focus
is on obtaining estimates at $M_\pi=135$~MeV, and this is achieved by
relying on the data from the two physical mass ensembles to anchor the
chiral part of the fit.

In Tables~\ref{tab:rE-results},~\ref{tab:rM-results}
and~\ref{tab:mu-results}, we also give the results of the CCFV fits for the
following four combinations of the thirteen data points:
\begin{itemize}
\item
13-point fit. All the thirteen calculations as considered to be
independent, even though the a06m310 and a06m220 ensembles have been
analyzed twice with different smearing sizes.
\item
11-point fit. We use the average of the two values for $\rEsq$,
$\rMsq$ and $\mu$ on the $a06m310$ and $a06m220$ ensembles as these have 
been analyzed twice. In this averaging, we assume maximum
correlation between data
\item
10-point fit. We remove the coarsest ensemble, $a15m310$, from
the eleven data points defined above. 
\item
$10^\ast$-point fit. We remove the smallest volume ensemble,
  $a12m220S$, from the eleven data points defined above.
\end{itemize}
For each of these fits, we give results with (labeled extrap $c_4^X
\neq 0$) and without (labeled extrap $c_4^X = 0$) the finite volume
correction. The values of the coefficients are given in Table~\ref{tab:fitvalues}. 
In the limit $a \to 0$ and $M_\pi L \to \infty$, only the  terms proportional to $c_1^X$ and $c_3^X$ contribute. 
From these fits, we observe the following:
\begin{itemize}
\item
Of our estimate $\rEsq \approx 0.59$~fm${}^2$, roughly half 
comes from $c_1^E$ and the other half from $c_3^E$.  Compared to the
experimental value $\rEsq \approx 0.86$~fm${}^2$ (see
Eq.~\eqref{eq:isovectorradii}), about $0.27$ fm${}^2$ is missing.
\item
Of $\rMsq  \approx 0.46$~fm${}^2$, roughly 60\% comes from $c_1^M$ and the rest from $c_3^M$.
Compared to the experimental value $\rMsq \approx 0.85$~fm${}^2$ (see
Eq.~\eqref{eq:isovectorradii}), about $0.39$ fm${}^2$ is missing.
\item
There is a significant dependence of $\mu$ on the lattice spacing
$a$. As a result, we get a low value, $\mu \approx 4$ Bohr magneton,
in the continuum limit.
\item
The coefficient of the finite volume term is poorly
determined, which is reflected in the larger error estimates with
$c_4^X \neq 0$. In all cases, the two types of results overlap. To be
conservative, we quote all final results including the finite volume
term.
\end{itemize}

In Figs.~\ref{fig:rE-extrap11},~\ref{fig:rM-extrap11}
and~\ref{fig:mu-extrap11}, we show the CCFV fits versus $a$, $M_\pi$
and $M_\pi L$ for three analyses: dipole, $z^{4}$ and $z^{5+4}$. In
addition, we show fits versus a single variable $a$ or $M_\pi^2$ (gray
bands).  When the pink and gray bands are close or overlap, it means
that the dominant sensitivity of the CCFV fit is with respect to the 
single variable of the gray band.

For $\rEsq$, we also show the fit using the $\chi$PT expression given
in Ref.~\cite{Kubis:2000zd} as a solid red line in
Fig.~\ref{fig:rE-extrap11}.  The variation with $M_\pi^2$ in the
dipole, $z^4$ and the $z^{5+4}$ data is small over the range $135 <
M_\pi < 350$~MeV, and consistent with the prediction of $\chi$PT. The
singular behavior is expected to dominate for $M_\pi < 135$~MeV. 
As shown in Fig.~\ref{fig:ChPT}, 
over the range $135 < M_\pi < 350$~MeV, the decrease in the
``log'' part is partially compensated for by the increase in the
``analytical'' contribution as $M_\pi \to 0$.

\begin{figure}  
\centering
\subfigure{
\includegraphics[width=0.97\linewidth]{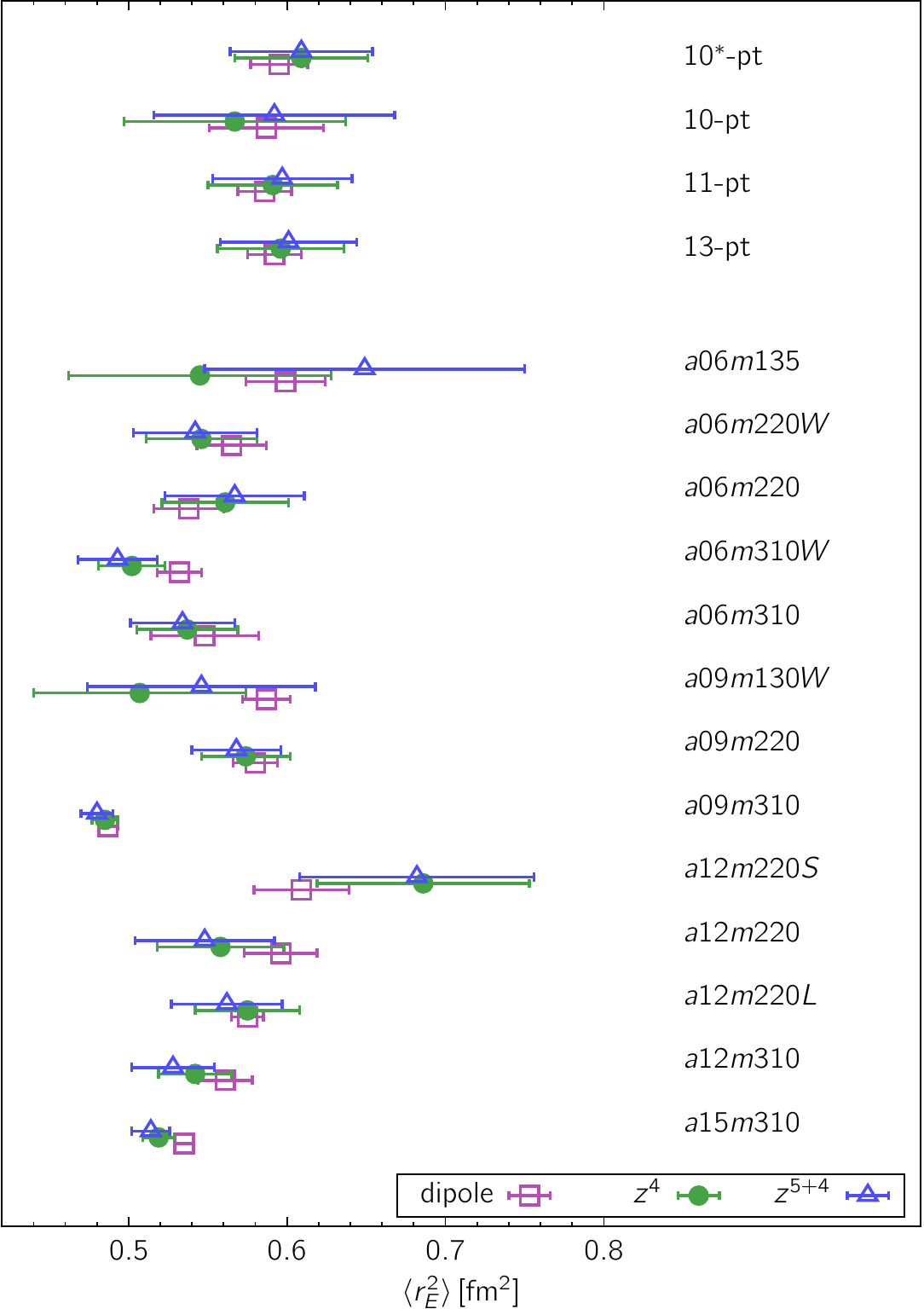}
}
\caption{\FIXME{fig:rEsummary} A summary of the results for $\rEsq$ in units of fm${}^2$ 
  presented in Table~\ref{tab:rE-results} from the 13 calculations and
  the four CCFV fits. In each case we show results for three $Q^2$ fits: the dipole, $z^{4}$ and $z^{5+4}$. }
\label{fig:rEsummary}
\end{figure}

The shape of the CCFV fit bands are similar for the dipole, $z^4$ and
the $z^{5+4}$, except that the $z$-expansion data and the fits have
larger errors. A visual overview of all 13 individual results for
$\rEsq$ and of the four CCFV fits is presented in
Fig.~\ref{fig:rEsummary}.  The variation with $a$ and $M_\pi$ in the
thirteen individual calculations is small and somewhat smaller in the
dipole than in the various $z$-expansion estimates.

\begin{table}[htbp]  
\caption{\FIXME{tab:fitvalues} Values of the parameters,
  $c_i^{E,M,\mu}$, defined in
  Eqs.~\protect\eqref{eq:rEsq-extrap},~\protect\eqref{eq:rMsq-extrap}
  and~\protect\eqref{eq:mu-extrap} for the 11-point fit used to obtain
  $\rEsq$, $\rMsq$ and $\mu$ in the continuum limit from the dipole
  and $z^4$ data. }
\label{tab:fitvalues}
\centering
\begin{ruledtabular}
\begin{tabular}{l|c|c|c|c}
$\rEsq$ & $c_1^E$    & $c_2^E$     & $c_3^E$         & $c_4^E$   \\
        &(fm${}^2$)  &(fm)         &(fm${}^2$)       &(fm${}^2$)  \\
\hline                                               
dipole  & 0.32(3)    & 0.62(12)    & -0.08(1)        & 0.33(26) \\
$z^4$   & 0.31(5)    & 0.50(23)    & -0.08(2)        & 0.11(69) \\
\hline                                               
\hline                                               
$\rMsq$ & $c_1^M$    & $c_2^M$     & $c_3^M$         & $c_4^M$ \\
        &(fm${}^2$)  &(fm)         &(fm${}^2$ GeV)   &(fm${}^2$ GeV)  \\
\hline                                               
dipole  & 0.31(3)    & 0.32(18)    & 0.024(6)        & -0.14(19) \\
$z^4$   & 0.28(6)    & 0.77(32)    & 0.023(15)       & -0.09(48) \\
\hline                                               
\hline                                               
$\mu$   & $c_1^\mu$  & $c_2^\mu$   & $c_3^\mu$       & $c_4^\mu$ \\
        &            &(fm${}^{-1}$)&(GeV${}^{-1}$)   &(GeV${}^{-1}$) \\
\hline
dipole  & 3.93(11)   & 2.28(89)    & 0.33(40)        & -44(39) \\
$z^4$   & 3.91(11)   & 3.10(98)    & 0.22(42)        & -51(45) \\
\end{tabular}
\end{ruledtabular}
\end{table}

\begin{figure}  
\centering
\subfigure{
\includegraphics[width=0.97\linewidth]{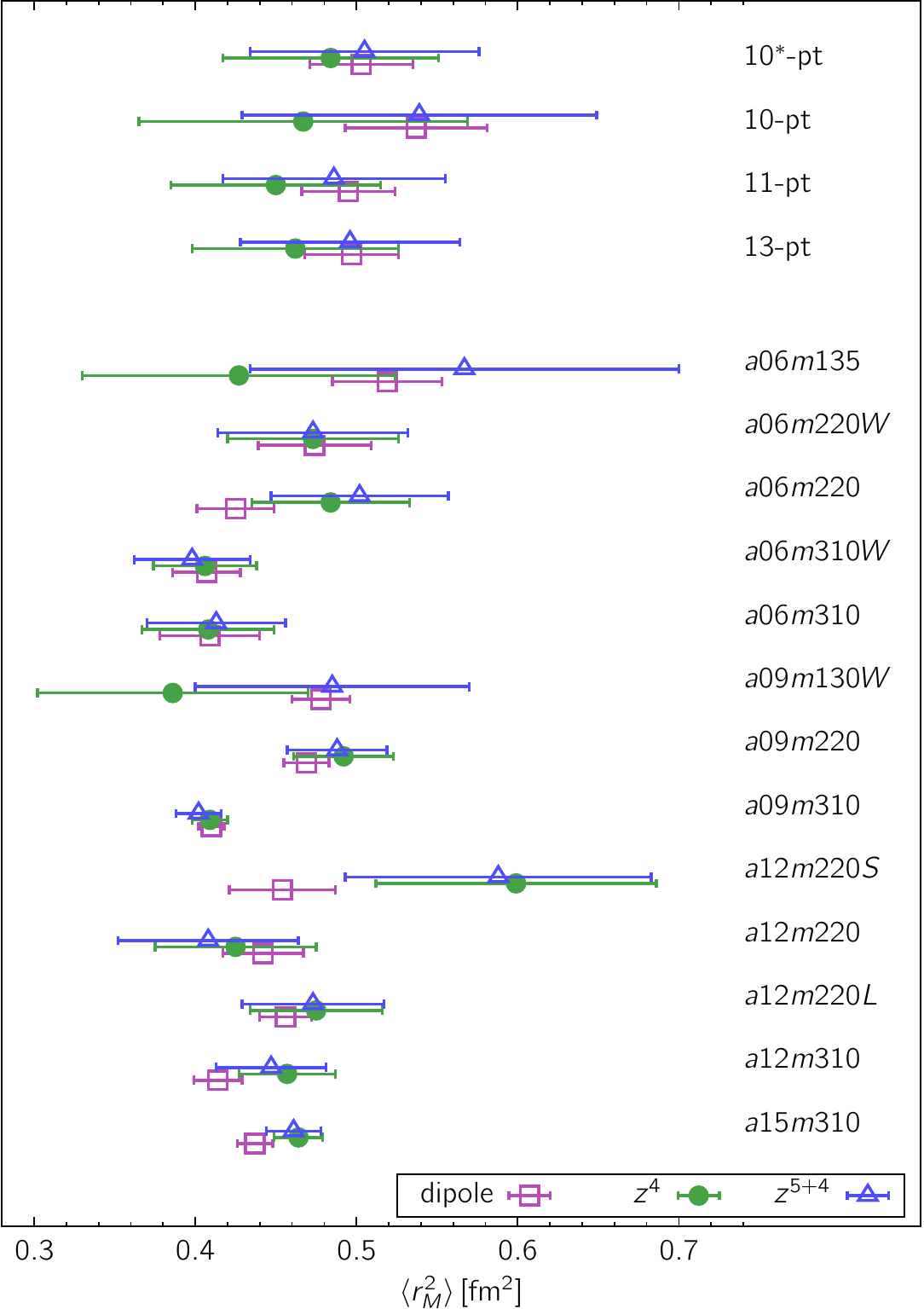}
}
\caption{\FIXME{fig:rMsummary} A summary of the results for $\rMsq$ in units of fm${}^2$
  presented in Table~\ref{tab:rM-results} from the 13 calculations and
  the four CCFV fits. In each case we show results for three $Q^2$
  fits: the dipole, $z^{4}$ and $z^{5+4}$. }
\label{fig:rMsummary}
\end{figure}

For $\rMsq$, the variation with $a$, $M_\pi$ and $M_\pi L$ for each of
the three cases, the dipole, $z^4$ and the $z^{5+4}$, is small as
shown in Figs.~\ref{fig:rM-extrap11} and~\ref{fig:rMsummary}.
Presumably, the expected $1/M_\pi$ chiral behavior (see
Eq.~\eqref{eq:rMsq-extrap}) sets in at $M_\pi < 135$~MeV.
Again, the CCFV fits for the three cases shown in
Fig.~\ref{fig:rM-extrap11} are similar.

\begin{figure}  
\centering
\subfigure{
\includegraphics[width=0.97\linewidth]{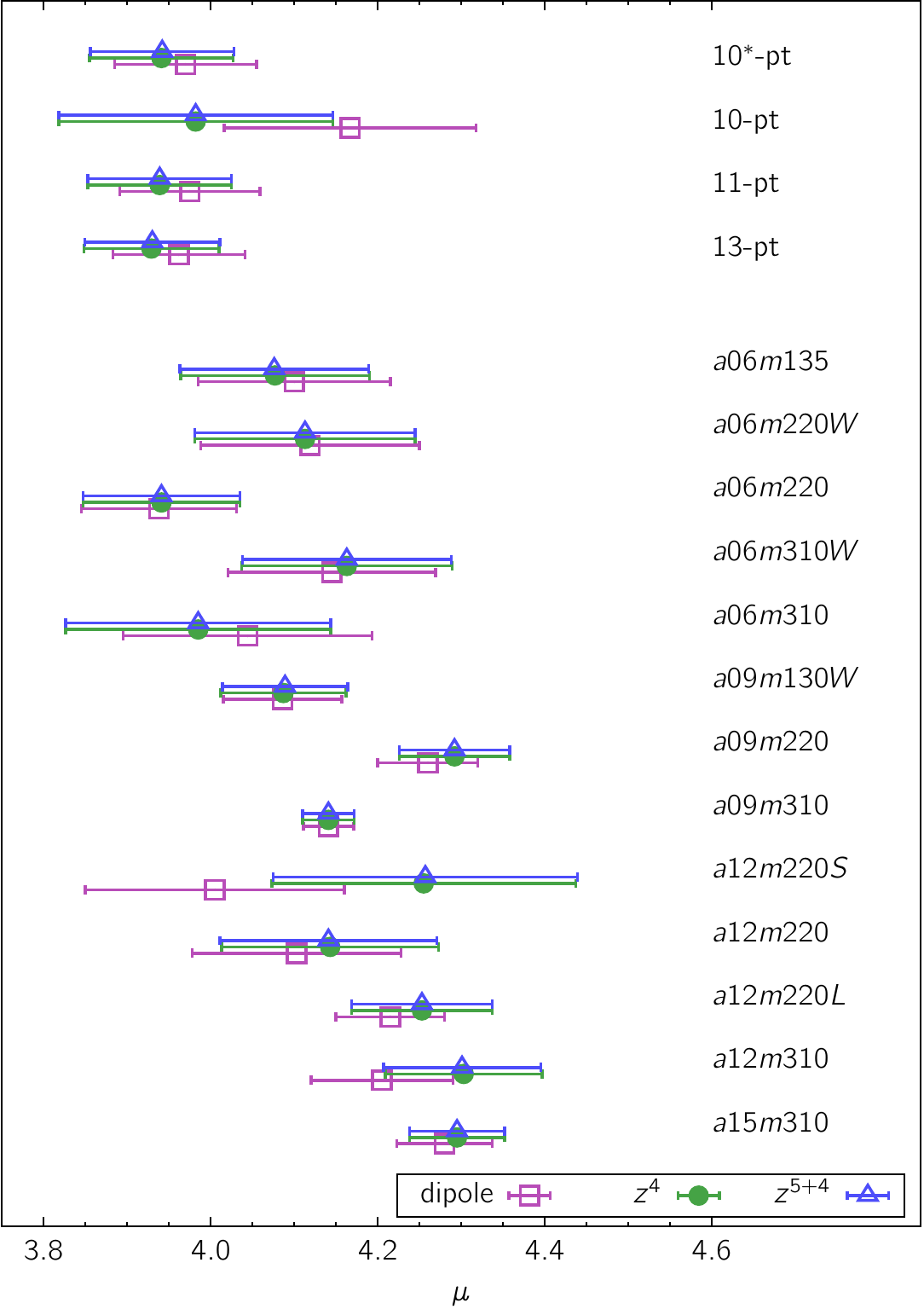}
}
\caption{\FIXME{fig:musummary} A summary of the results for $\mu$
  presented in Table~\ref{tab:mu-results} from the 13 calculations and
  the four CCFV fits. In each case we show results for three $Q^2$
  fits: the dipole, $z^{4}$ and $z^{5+4}$. }
\label{fig:musummary}
\end{figure}

The largest variation in $\mu$ is versus $a$ as shown in
Fig.~\ref{fig:mu-extrap11}. Again, the fit bands are similar for the
dipole, $z^4$ and the $z^{5+4}$ data. The positive slope versus $a$
lowers the continuum limit result with respect to the experimental value $\mu|_{\rm
  expt} = 4.7058$.  The size of the difference between lattice data
and experimental results suggests that discretization and other
systematic errors in $G_M(Q^2)$ are underestimated.  From the 
summary of the results presented in
Fig.~\ref{fig:musummary}, it is clear that the largest
uncertainty is in the smallest volume, $a12m220S$, and the two
physical mass, a09m130W and $a06m135$, points.

It is instructive to compare our data to the predictions of chiral
perturbation theory shown in detail Fig.~\ref{fig:ChPT}.  The chiral
expansion for the isovector $\rEsq$ and $\mu \rMsq$, given in
Ref.~\cite{Kubis:2000zd}, are shown as the sum of three terms: those
independent of $M_\pi$ (labeled constant), proportional to $\ln
M_\pi^2 / M_N^2$ (labeled log) and the remaining terms proportional to
powers of $M_\pi^2$ (labeled analytical). In making these plots, the
low energy constants (LEC) used are $c_4 = 3.4$~GeV${}^{-1}$, $c_6 = 4.77$, $d_6^r
= 0.74$, $e_{74}^r = 1.65$, $g_A = 1.276$ and $F_\pi =92$~MeV.

For $\rEsq$, the ``constant'' term is negative ($\approx -0.49$~fm${}^2$). In
the range $ M_\pi =$ 135-350~MeV, $\rEsq$ is approximately constant
at $0.9$fm${}^2$: in this interval, the growth in the $\ln
M_\pi^2 / M_N^2$ term compensates for the decrease in the
``analytical'' terms. Below $ M_\pi \approx 135$, the log term drives
the rise in the sum. As shown in Fig.~\protect\ref{fig:rE-extrap11},
lattice data are significantly smaller in magnitude but show a similar
small variation between $M_\pi = 135$--$350$~MeV.  Because of this
small variation, and having data at only three $M_\pi$ values, even 
including an additional analytical term proportional to just $M_\pi^2$
in our CCFV fits would over-parameterize the fit. Furthermore, over
this range, a simple $M_\pi^2$ term would equally well mimic the sum
of the log and the analytical terms. For this reason, we have included
only one of the possible $M_\pi$ dependent terms, the log, in our CCFV fits.

In the case of $\mu \rMsq$ shown in Fig.~\ref{fig:ChPT} (right), the
``log'' and ``analytical'' terms are small and the ``log'' shows
little variation.  The dominant contribution comes from the $1/M_\pi$
and ``constant'' terms. Since the $1/M_\pi$ term provides the largest
variation with $M_\pi$, we have only included it in the CCFV fit
defined in Eq.~\eqref{eq:rMsq-extrap}.

In short, even though the $\chi$PT based expressions used for both
$\rEsq$ and $\rMsq$, given in Eqs.~\eqref{eq:rEsq-extrap}
and~\eqref{eq:rMsq-extrap}, use only the leading chiral correction term 
from the expressions in Ref.~\cite{Kubis:2000zd}, the variation in our
data at three values of $M_\pi$ between 135--315~MeV is small, and 
including more terms would result in over-parameterization.  In this
situation, having lattice data at $M_\pi \approx 135$~MeV, is crucial for
controlling the uncertainty in the chiral fit to the lattice
data. Note that the errors we quote in the CCFV fit results are
comparable to those in the two physical mass points.

As is evident from the data in
Tables~\ref{tab:rE-results},~\ref{tab:rM-results}
and~\ref{tab:mu-results}, and shown in Fig.~\ref{fig:stabilityz}, the
$z$-expansion results without sum rules converge for $k \ge 4$.  Since
$\rEsq$, $\rMsq$ and $\mu$ should ideally be extracted from the small
$Q^2$ behavior, our final results are obtained as follows: We take the
$z^{4}$ result for the central value and the first error in it
represents the analysis uncertainty, i.e., including the ESC, $Q^2$ and
CCFV fits. We also quote a second systematic uncertainty to account
for having used just the leading order CCFV fits. This is taken to be the
largest of the following:
\begin{itemize}
\item
The difference between the two values on the physical mass ensembles, $a09m130W$ and 
$a06m135$. The second error estimate for $\rMsq$ is given by this difference. 
\item
The difference between the value at $a06m135$ and the continuum value given by the CCFV fit. 
This gives the second error estimate for $\rEsq$ and for $\mu^{p-n}$, which 
show the largest variation versus $a$.
\item
For the $z$-expansion, we also considered the difference between the
$z^3$ ($z^5$) and $z^4$ values.  These turn out to be smaller than the
estimates from the previous two cases.
\end{itemize}
The final results, obtained by applying this prescription to the
11-point CCFV fit values summarized in Tables~\ref{tab:rE-results},
\ref{tab:rM-results} and~\ref{tab:mu-results}, are given in
Table~\ref{tab:finalresults}.  For completeness, we also give the
results for the Dirac and Pauli radii derived from these $\rEsq$ and
$\rMsq$ using Eq.~\eqref{eq:rDP} in Table~\ref{tab:finalresults}.

\begin{table*}[htbp]  
\caption{\FIXME{tab:finalresults} Our final results for $\rEsq$,
  $\rMsq$ and $\mu$ from the 11-point CCFV fit to the dipole and the
  $z^4$ analysis data. The determination of the second error in these
  two estimates is explained in the text. The combined analysis, 
  defined by Eq.~\protect\eqref{eq:combined-fit} with the $z^4$ truncation 
  for the $Q^2$ behavior, has a single overall error. The bottom
  half of the table gives results for the Dirac, $\rDirac$, and Pauli,
  $\rPauli$, radii obtained using Eq.~\protect\eqref{eq:rDP}.}
\label{tab:finalresults}
\centering
\begin{ruledtabular}
\begin{tabular}{l|c|c|c|c|c}
                    & $\rEsq$        &  $\sqrt{\rEsq}$& $\rMsq$         & $\sqrt{\rMsq}$ &  $\mu$            \\
                    & (fm${}^2$)     &  (fm)          & (fm${}^2$)      & (fm)           & (Bohr Magneton)   \\
\hline
dipole fit          & 0.586(17)(13)  & 0.765(11)(8)   & 0.495(29)(41)   & 0.704(21)(29)  &  3.975(84)(125)   \\
\hline
$z^4$ fit           & 0.591(41)(46)  & 0.769(27)(30)  & 0.450(65)(102)  & 0.671(48)(76)  &  3.939(86)(138)   \\
\hline
Combined fit        & 0.564(114)     & 0.751(76)      & 0.459(189)      & 0.678(140)     &  3.922(83)       \\
\hline
\hline
                    & $\rDirac$      & $\sqrt{\rDirac}$ & $\rPauli$     &  $\sqrt{\rPauli}$ &               \\
                    & (fm${}^2$)     &  (fm)          & (fm${}^2$)      & (fm)           &                  \\
\hline
dipole fit          & 0.389(18)(15)  &  0.623(15)(12) & 0.531(44)(63)   & 0.729(30)(43)  &                  \\
\hline
$z^4$ fit           & 0.396(42)(49)  &  0.629(33)(37) & 0.469(90)(141)  & 0.685(66)(103) &                  \\
\hline
Combined fit        & 0.370(115)     &  0.609(94)     & 0.490(258)      & 0.700(184)     &                  \\
\end{tabular}
\end{ruledtabular}
\end{table*}

The central values for $\rE^{p-n}$, $\rM^{p-n}$ and $\mu^{p-n}$ are
about 17\%, 19\% and 16\% smaller than the phenomenological values
given in Eq.~\eqref{eq:isovectorradii} and the precise experimental value in Eq.~\eqref{eq:mu_expt}.
Estimates from the dipole and $z$-expansion fits, given in
Table~\ref{tab:finalresults}, are consistent, however, the errors in
$\rEsq$ and $\rMsq$ from the $z$-expansion fits are much larger, and
about half the difference from the experimental/phenomenological
values.  The errors in the dipole fits are small compared to the difference between the 
lattice and the phenomenological/experimental  estimates.  As
discussed in a number of places above, differences between the lattice
and phenomenological estimates can be accounted for if a linear combination
of the statistical and the various systematic errors is taken.

The encouraging results from our analysis are: (i) the data for both
$G_E(Q^2)$ and $G_M(Q^2)$ is seen to converge towards the Kelly
parameterization as $a$ and $M_\pi$ are decreased; (ii) the stability
of the $z$-expansion fits improves with statistical precision, however
constraints on the coefficients $a_k$ are still needed; (iii) while it
is hard to test the nonanalytical chiral behavior predicted in
Eqs.~\eqref{eq:rEsq-extrap} and~\eqref{eq:rMsq-extrap} with data at
only three values of $M_\pi^2$, having data at the two physical mass
ensembles anchors the CCFV fit and provides control over the
uncertainty in values obtained from the fits.

A weakness of the lattice analysis is that $G_M(0)$ cannot be
calculated directly due to kinematic constraints.  We have motivated
the use of a derived value of $G_M(0)$ to stabilize fits to
$G_M(Q^2)$.  Looking ahead, the most significant improvement needed
for extracting all three quantities with higher precision is
generating data at smaller values of $Q^2$. This, unfortunately,
requires ensembles with larger spatial volumes and/or new approaches
such as a lattice formulation of the Dirac action with twisted
boundary conditions~\cite{deDivitiis:2004kq,Frezzotti:2000nk}.  Both
options are beyond the scope of this work as they require new
simulations.

Two variants of the analysis presented above are described briefly next. 

\subsection{Analysis of $M_N^2 \rEsq$ and $M_N^2 \rMsq$ }
\label{sec:Dimless}
\FIXME{sec:Dimless} 

\begin{figure*} 
\centering
\subfigure{
\includegraphics[width=0.48\linewidth]{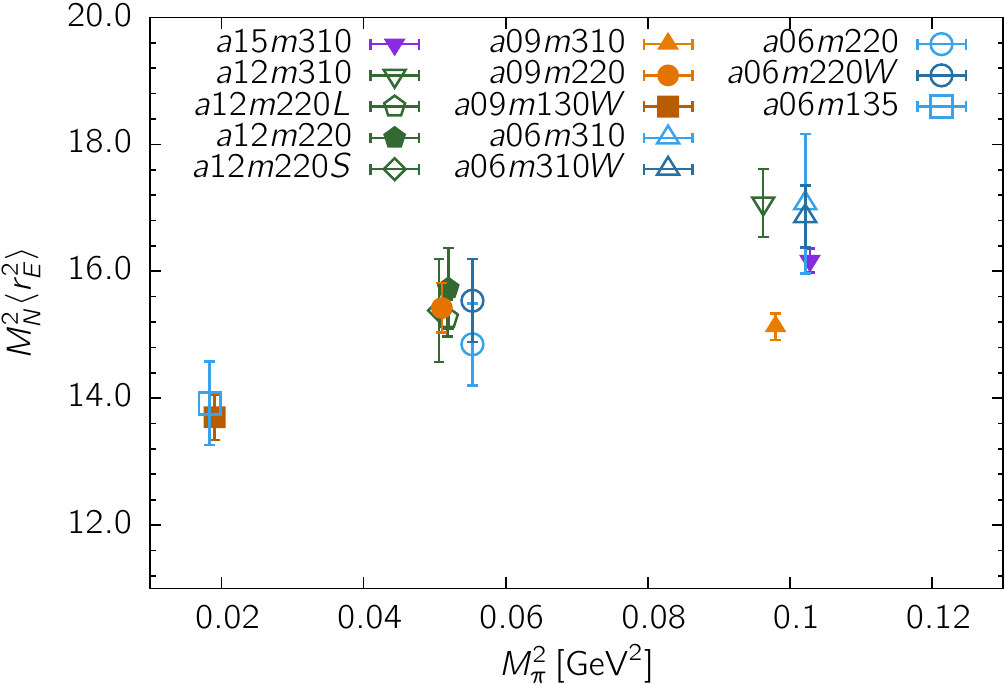}
\includegraphics[width=0.48\linewidth]{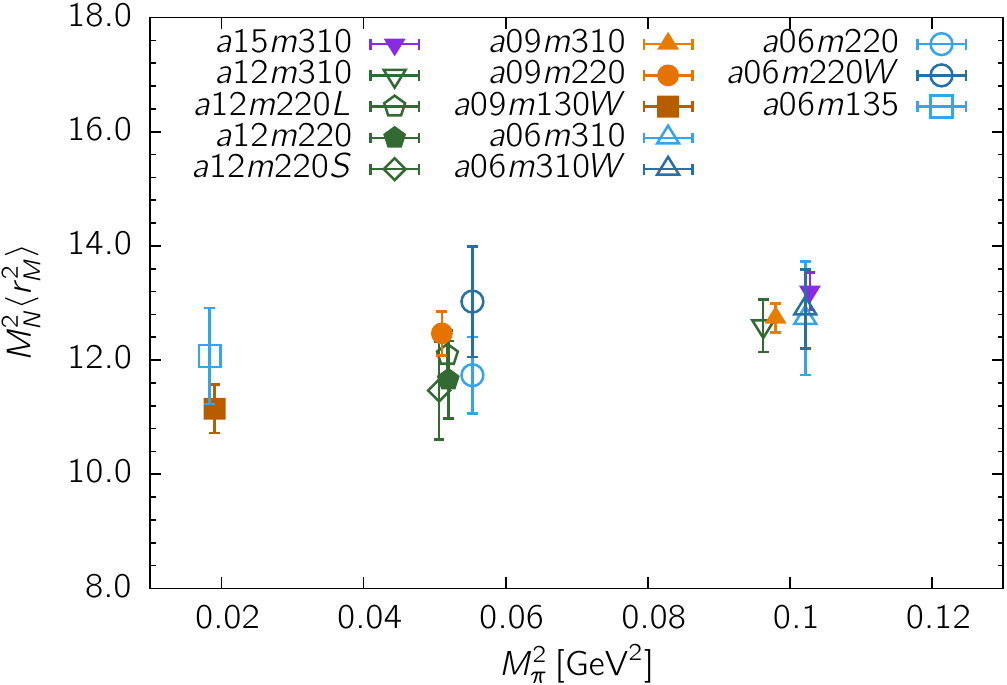}
}
\subfigure{
\includegraphics[width=0.48\linewidth]{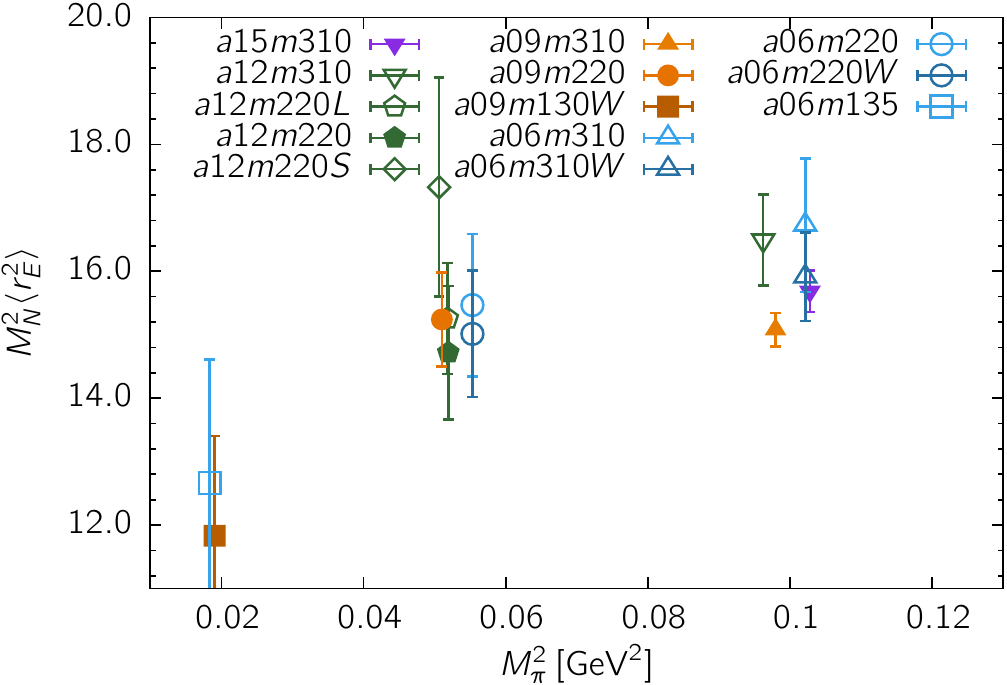}
\includegraphics[width=0.48\linewidth]{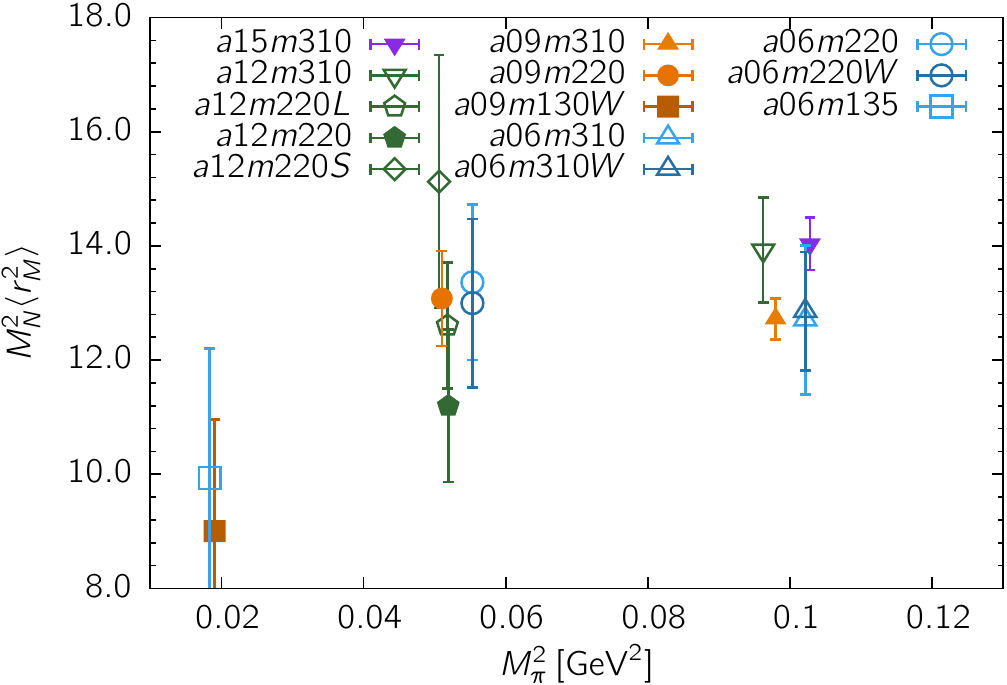}
}
\caption{\FIXME{fig:Dimless} The data for the 
dimensionless quantities $M_N^2 \rEsq$ and $M_N^2 \rMsq$ are
plotted versus $M_\pi^2$.  The top two panels show the data obtained using the 
dipole fit, and the lower two using the $z^4$ fit. 
\label{fig:Dimless}}
\end{figure*}

The dimensionless quantities $M_N^2 \rEsq$ and $M_N^2 \rMsq$ are
plotted in Fig.~\ref{fig:Dimless} versus $M_\pi^2$.  For comparison,
the phenomenological values for the isovector mean-square charge radii
given in Eq.~\eqref{eq:isovectorradii}, imply $M_N^2 \rEsq \approx
19.5$ and $M_N^2 \rMsq \approx 17.3$.  A priori, if some of the
systematics cancel in the product, then one would get smaller
variation with $a$ and $M_\pi$.  The data for $M_N^2 \rEsq$ and $M_N^2
\rMsq$ in Fig.~\ref{fig:Dimless} show that there is significant
variation with $M_\pi$.  Lacking a well motivated fit ansatz, a
reasonable option is to take the average of the values from the two
physical mass ensembles. These estimates are again low: $M_N^2
\rEsq|_{\rm dipole} = 13.75(32)$, $M_N^2 \rEsq|_{z^4} = 12.16(1.22)$,
$M_N^2 \rMsq|_{\rm dipole} = 11.34(38)$, and $M_N^2 \rMsq|_{z^4} =
9.40(1.50)$. Multiplying the results given in
Table~\ref{tab:finalresults} by $M_N^2=22.7$~fm${}^{-2}$ gives similar
values. Given that the errors are also similar and because these
estimates neglect possible $a$ dependence, we do not find this variant
of the analysis as providing an obvious improvement.

\subsection{Combined $Q^2-$CCFV Fit}
\label{sec:CombinedFit}
\FIXME{sec:CombinedFit} 

We also carried out a combined $Q^2-$CCFV fit to the $G_E(Q^2)$ and
$G_M(Q^2)$ data from the 13 calculations using a product of the
$z$-expansion for the $Q^2$ behavior and the functional forms 
for the CCFV ansatz given in
Eqs.~\eqref{eq:rEsq-extrap} and~\eqref{eq:rMsq-extrap}:
\begin{equation}
  G(z,\bm{\eta}) =  \sum_{k=0}^{M} d_k (\bm{\eta}) z^k \,.
  \label{eq:combined-fit}
\end{equation}
Here $\bm{\eta}$ represents the vector of variables in the CCFV fit,
and each coefficient $d_k$ of the $z$-expansion has a CCFV expansion
of the form given in Eq.~\eqref{eq:rEsq-extrap} or in
Eq.~\eqref{eq:rMsq-extrap}.  For example, for the four term CCFV
ansatz given in Eqs.~\eqref{eq:rEsq-extrap}, $\bm{\eta} = (1, a, \log
(M_\pi^2/\lambda^2), \log (M_\pi^2/\lambda^2) e^{-M_\pi L})$, the
combined fit has twenty parameters for the $z^4$ analysis. In
performing these fits, we used Gaussian priors with mean 0 and width
5, in their appropriate units, for all the parameters.  The resulting
central values of the parameters were within this range.

\begin{figure*} 
\centering
\subfigure{
\includegraphics[width=0.48\linewidth]{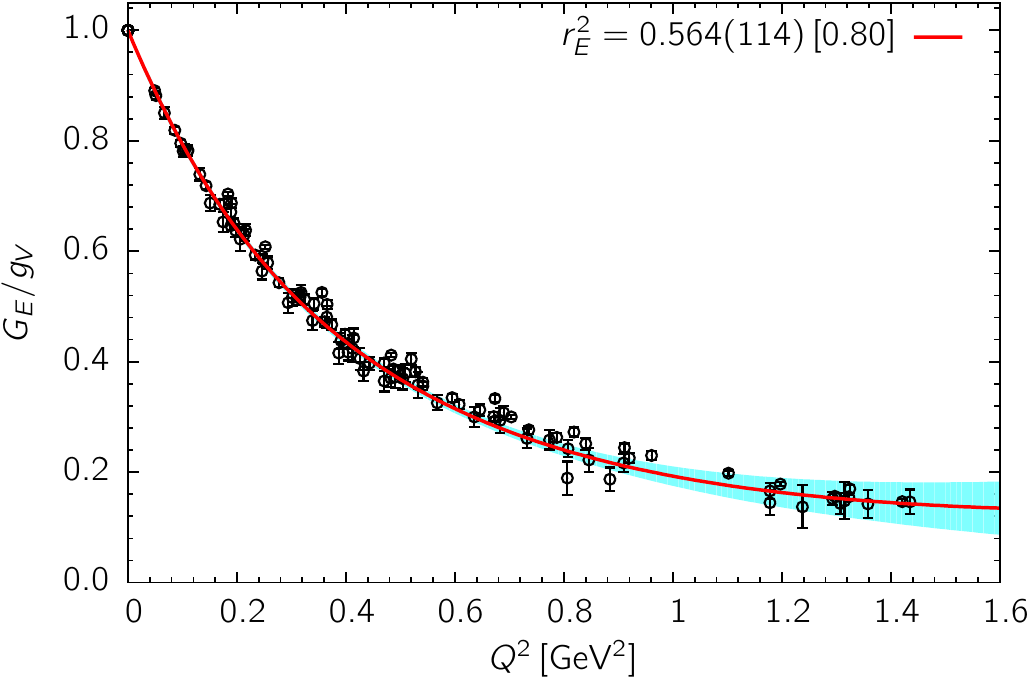}
\includegraphics[width=0.48\linewidth]{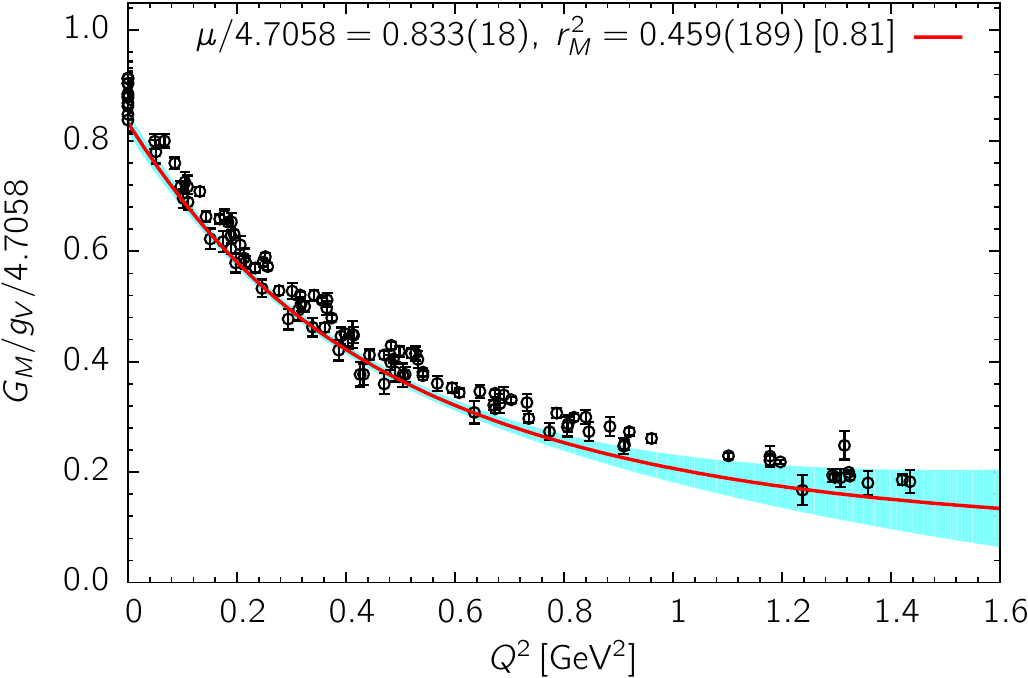}
}
\caption{\FIXME{fig:combined-fit} The data for $G_E(Q^2)/g_V$ (left)
  and $G_M(Q^2)/(g_V \times 4.7058)$ (right) from the thirteen
  calculations along with the combined $Q^2$-CCFV fit defined in
  Eq.~\protect\eqref{eq:combined-fit}. The solid red line and the
  turquoise error band show the $z^4$-CCFV fit neglecting the finite
  volume correction term.  The results for $\rEsq$ and $\rMsq$ (fm${}^2$) are
  given in the labels along with the [$\chi^2$/DOF] of the fit.
\label{fig:combined-fit}}
\end{figure*}

The central values of the results with and without the finite volume
term are consistent, however, the errors with the finite volume
correction term included are about a factor of two larger. In
Fig.~\ref{fig:combined-fit}, we show, for the $z^4$ case, the combined
fits neglecting the finite volume correction term. The results of these
combined fits are summarized in Table~\ref{tab:finalresults}, and found
to be consistent with those obtained by doing the $z^4$ and CCFV fits
separately (labeled $z^4$-fit).

\section{Comparison with Previous Work}
\label{sec:comparison}
\FIXME{sec:comparison} 

\begin{figure*} 
\centering
\subfigure{
\includegraphics[width=0.48\linewidth]{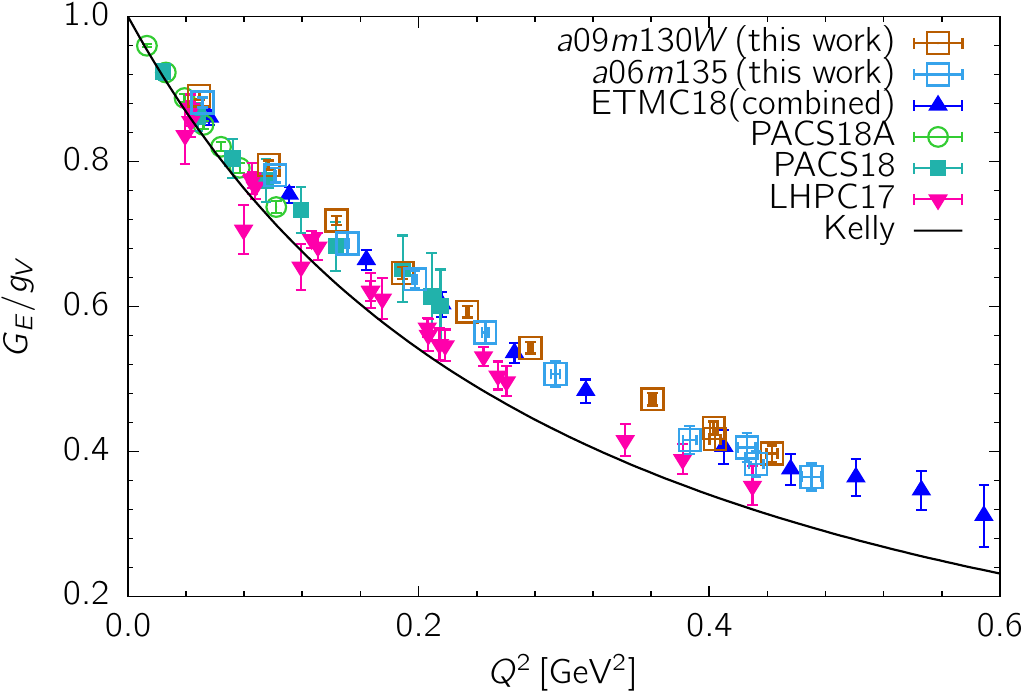}
\includegraphics[width=0.48\linewidth]{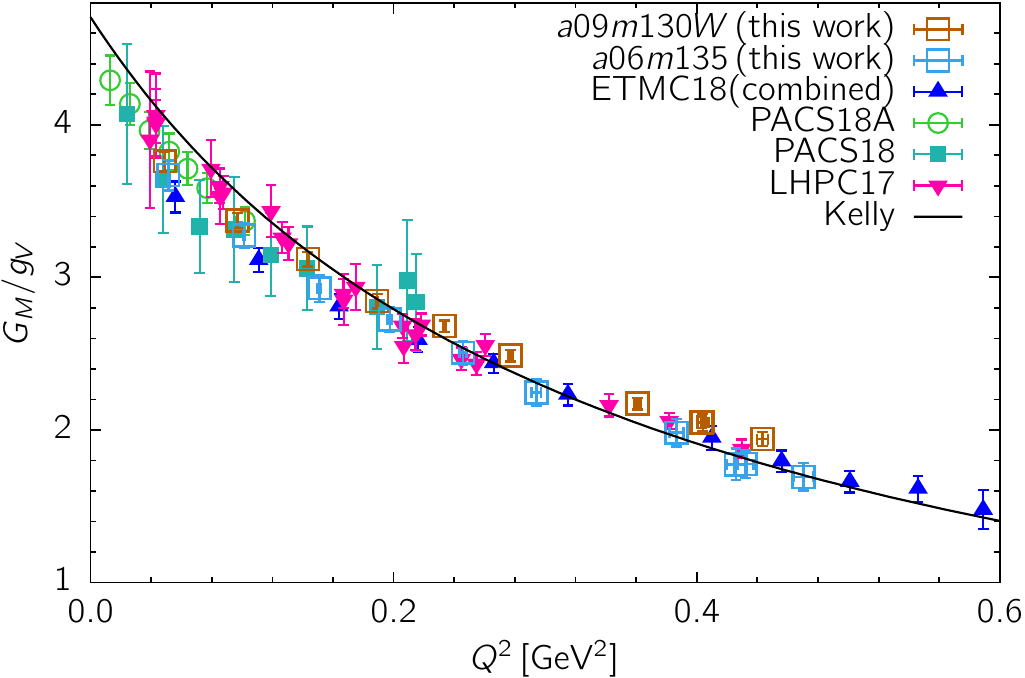}
}
\subfigure{
\includegraphics[width=0.48\linewidth]{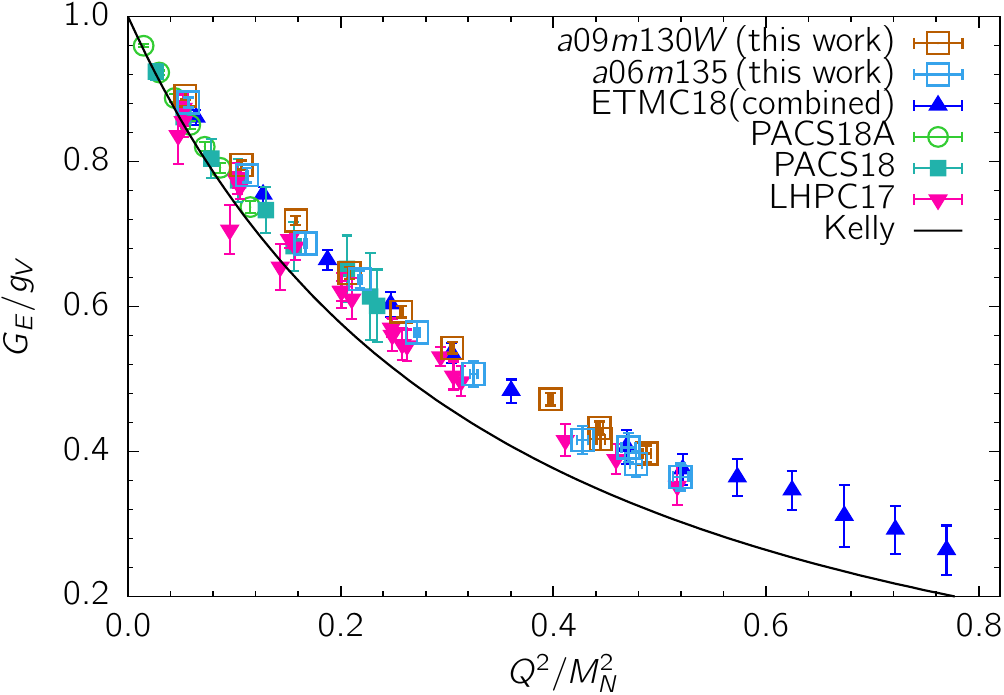}
\includegraphics[width=0.48\linewidth]{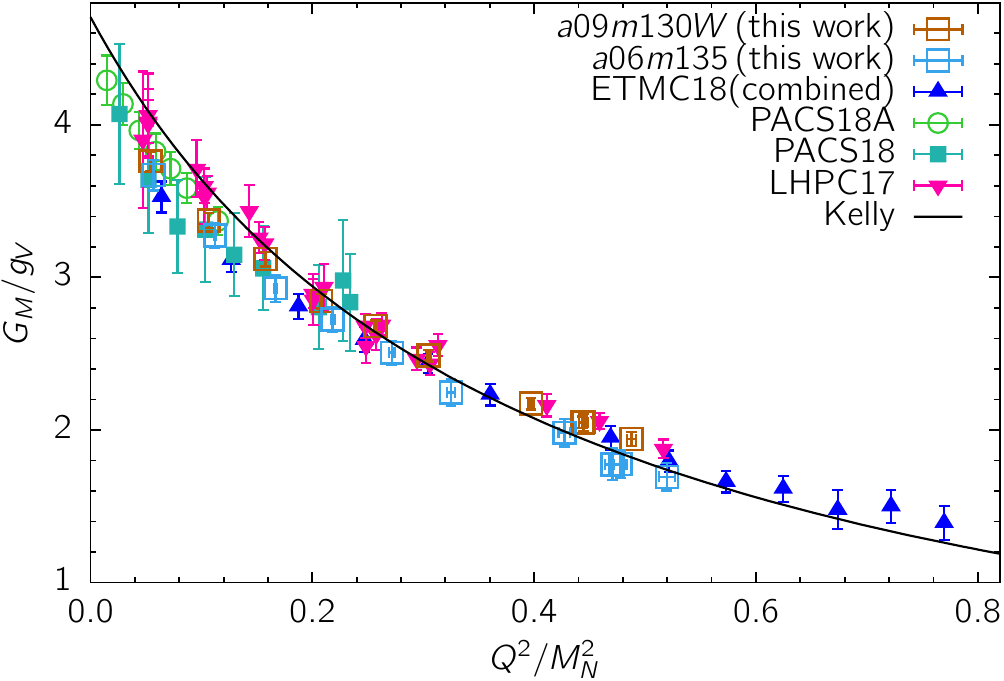}
}
\caption{\FIXME{fig:GEMcomp} Comparison of the data for the
  renormalized isovector $G_E(Q^2)$ and $G_M(Q^2)$ from collaborations
  that have published results at $M_\pi \approx 135$~MeV.  The lattice
  parameters of the various calculations are given in
  Table~\protect\ref{tab:comp}. The data are plotted as a function of
  $Q^2$, (top row) and $Q^2 / M_N^2$ (bottom row). The solid line is the Kelly fit to the
  experimental isovector data.
\label{fig:GEMcomp}}
\end{figure*}

There have been a number of lattice QCD calculations of electric and
magnetic isovector form factors of the nucleon. Recent ones include
LHPC'14~\cite{Green:2014xba}, Mainz'15~\cite{Capitani:2015sba},
LHPC'15~\cite{Green:2015wqa}, ETMC'17~\cite{Alexandrou:2017ypw},
LHPC'17~\cite{Hasan:2017wwt}, PACS'18~\cite{Ishikawa:2018rew},
PACS'18A~\cite{Shintani:2018ozy} and
ETMC'18~\cite{Alexandrou:2018sjm}.  In this work, we restrict the
comparison to calculations that have presented results at or near the
physical pion mass.  Their lattice parameters are given in
Table~\ref{tab:comp} and the data for $G_E(Q^2)$ and $G_M(Q^2)$ are
plotted in Fig.~\ref{fig:GEMcomp}.  We focus on comparing the data for
$G_E(Q^2)$ and $G_M(Q^2)$ as these are the primary quantities
calculated. Since the calculations have been done with different
lattice actions, the data, even at the physical pion mass, are only expected
to agree in the continuum limit.  We find that, in fact, they agree
remarkably well, much better than our analyses of various systematics
would indicate.

\begin{table*}[tbp]    
\caption{Lattice parameters of calculations that have presented
  results for $G_E(Q^2)$ and $G_M(Q^2)$ at or near the physical pion
  mass.  }
\label{tab:comp}
\begin{center}
\renewcommand{\arraystretch}{1.2} 
\begin{ruledtabular}
\begin{tabular}{l|cc|cc|ccc|c}
Ensemble ID                      & $a$ (fm)   & $M_\pi$ (MeV) & $L^3\times T$ & $M_\pi^{\rm val} L$ & $\tau/a$ & $N_\text{conf}$  & $N_{\rm meas}$   &   Action \\
\hline
$a09m130$  (this work)           & 0.0871(6)  & 138(1)   & $64^3\times 96$ & 3.90 & $\{8,10,12,14,16\}$  & 1290      & 165,120     & clover-on-2+1+1-HISQ  \\
$a06m135$  (this work)           & 0.0570(1)  & 136(2)   & $96^3\times 192$& 3.7  & $\{16,18,20,22\}$    & 675       &  43,200     & clover-on-2+1+1-HISQ  \\
\hline                                                                                                                                    
LHPC'17~\cite{Hasan:2017wwt}     & 0.093      & 135      & $64^3\times 64$ & 4.08 & $\{10, 13, 16\}$     & 442       &  56,576     & 2+1-clover          \\
\hline                                                                                                                                    
ETMC'18~\cite{Alexandrou:2018sjm}& 0.0809(4)  & 138(1)   & $64^3\times 128$& 3.62 & $\{12,14,16,18,20\}$ & 750       &  3K$-$48K   & 2+1+1-Twisted Mass  \\
ETMC'17~\cite{Alexandrou:2017ypw}& 0.0938(3)  & 130(1)   & $48^3\times 96$ & 2.98 & $\{10,12,14,16,18\}$ & 578$-$725 &  9K$-$64K   & 2-Twisted Mass  \\
ETMC'18~\cite{Alexandrou:2018sjm}& 0.0938(3)  & 130(2)   & $64^3\times 128$& 3.97 & $\{12, 14, 16\}$     & 333$-$1040&  5K$-$17K   & 2-Twisted Mass  \\
\hline                                                                                                                                    
PACS'18~\cite{Ishikawa:2018rew}  & 0.0846(7)  & 146      & $96^3\times 96$ & 6.01 & $\{15\}$             & 200       &  12,800     & 2+1-clover  \\
\hline                                                                                                                                    
PACS'18A~\cite{Shintani:2018ozy} & 0.0846(7)  & 135      &$128^3\times 128$& 7.41 & $\{10,12,14,16\}$    & 20        &  2.5K$-$10K & 2+1-clover         \\
\end{tabular}
\end{ruledtabular}
\end{center}
\end{table*}

\begin{table*}[tbp]    
\caption{Results for $\rE$, $\rM$, $\mu$ and the nucleon mass from
  published calculations at or near the physical pion mass.  The
  quantity used to set the lattice scale is given in the third column,
  with $r_0^2 F(r_0)$ and $r_1$ extracted from the heavy quark
  potential~\protect\cite{Sommer:2014mea}.
  ETMC'18~\cite{Alexandrou:2018sjm} results are derived from a single fit in $Q^2$
  to the combined 2- and 2+1+1-flavor data, i.e., neglecting the
  dependence on the number of flavors $N_f$ and the difference in the lattice spacing $a$. The LHPC'17~\cite{Hasan:2017wwt}
  results are from a single ensemble and taken from their analysis
  using the summation method to control ESC. The calculation of the
  scale used in LHPC'17 is given in Ref.~\protect\cite{Hasan:2019noy},
  and that by the PACS collaboration in
  Ref.~\protect\cite{Ishikawa:2015rho}.}
\label{tab:compR}
\begin{center}
\renewcommand{\arraystretch}{1.2} 
\begin{ruledtabular}
\begin{tabular}{l|c|c|c|c|cc}
Ensemble ID                      & $M_N$ (MeV)  & $a$ from           & $Q^2$ Fit  & $r_E$ (fm)   & $r_M$ (fm)   & $\mu$      \\
\hline                                                                                                            
$a09m130W$                       & 953(4)     & $r_1$              & $z^4$      & 0.769(27)(30)& 0.671(48)(76)& 3.94(9)(14)     \\
$a06m135$                        & 951(10)    & $r_1$              & $z^4$      & 0.765(11)(8) & 0.704(21)(29)& 3.98(8)(13)     \\
\hline                                                                                                            
LHPC'17~\cite{Hasan:2017wwt}     & 912(8)     & $M_\Omega$         & $z^5$      & 0.887(49)    &              & 4.75(15)          \\
\hline                                                                                                            
ETMC'18~\cite{Alexandrou:2018sjm}& 929(6)     & $r_0^2F(r_0)=1.65$ & dipole     & 0.802(19)(12)(1)  & 0.714(26)(88)(16)(${}^1_0$) & 3.96(14)(3)(7)(${}^1_0$)    \\
ETMC'17~\cite{Alexandrou:2017ypw}& 941(2)     & $r_0^2F(r_0)=1.65$ & dipole     & 0.808(30)(19)     & 0.732(36)(45)               & 4.02(21)(28)     \\
\hline                                                                                                            
PACS'18~\cite{Ishikawa:2018rew}  & 958(10)    & $M_\Omega$         & $z^8 | z^7$& 0.915(99)    & 1.437(409)   & 4.81(79)       \\
\hline                                                                                                            
PACS'18A~\cite{Shintani:2018ozy} & 942(11)    & $M_\Omega$         & dipole     & 0.875(15)(28) & 0.805(32)(274)  & 4.417(138)(317)        \\
\end{tabular}
\end{ruledtabular}
\end{center}
\end{table*}

All the data included in the comparison are shown in the upper two panels in
Fig.~\ref{fig:GEMcomp}. From the plot versus $Q^2$, we draw the following overall conclusions: 
\begin{itemize}
\item
The $G_E(Q^2)$ data approach the Kelly curve from above, while
$G_M(Q^2)$ from below for $Q^2 < 0.2$~GeV${}^2$.
\item
No significant dependence on the number of flavors or the lattice
spacing $a$ is manifest.
\item
The PACS'18A data at $Q^2 < 0.1$~GeV${}^2$, obtained using a large
volume, show a qualitatively different behavior and lie closer to the
Kelly curve. In this range of $Q^2$, the data are almost linear and
highly correlated.  They give a larger slope in both $G_M(Q^2)$ and
$G_E(Q^2)$ and thus larger $\rEsq$ and $\rMsq$.
\item
We can also compare data at $Q^2 \approx 0.05$ and 0.1~GeV${}^2$ from
our $a09m130W$ and $a06m135$ ensembles, from
ETMC'18~\cite{Alexandrou:2018sjm}, and the low error
LHPC'17~\cite{Hasan:2017wwt} points with that from PACS'18A. Given the
size of the statistical and systematic errors in individual data
points, it is not clear if the observed small differences are
significant at these two $Q^2$ values.
\end{itemize}

Two points become clear on plotting the data versus $Q^2/M_N^2$, as shown in the
bottom two panels of Fig.~\ref{fig:GEMcomp}. First, the collapse of
all data into a single curve over the whole range $Q^2/M_N^2 \lesssim
0.8$~GeV${}^2$ becomes even more pronounced. Second, the deviation of
this common curve from the Kelly curve is smaller.  Thus, not only do all
our data from the 13 calculations fall on a common curve when plotted
versus $Q^2/M_N^2$, as shown in Fig.~\ref{fig:Kelly}, but so do data
from four other collaborations using different lattice actions and
volumes.  A priori, a common curve would suggest that all the
systematics cancel, and the apparent differences between the various
calculations when the data are plotted versus $Q^2$ was largely a
consequence of how the lattice scale is set. 

Deviations from the Kelly curve are, however,  significant in $G_E(Q^2)$ for $Q^2
>0.1$~GeV${}^2$. Data for $G_M(Q^2)$ data undershoot for $Q^2 <
0.2$~GeV${}^2$ and are consistent with the Kelly curve above it.
These differences are a 2--3$\sigma$ effect, and comparable to the
size of the shift when the data are plotted versus $Q^2/M_N^2$ or
$Q^2$.  While an understanding of how the different systematics
contribute, and whether there is one that dominates requires future
more detailed calculations, we remind the reader that during the
course of our analyses, we have pointed out systematics, for example
due to ESC and the deteriorating signal in both the 2- and 3-point
correlation functions at large $\vec{q}^2$, could give rise to
uncertainties of this size. 

We have already shown that the pattern of ESC in our data is sensitive
to the value of $Q^2$. In particular, as discussed in
Sec.~\ref{sec:FFanalysis}, the ESC in correlators from which $G_E$ is
extracted increases with momentum and the convergence is from above. On the other hand, 
it is large at small $\vec{q}^2$ in correlators from which we
get $G_M$ and the convergence is from below. Thus, possible residual
ESC could account for the observed deviation from the Kelly
curve.

For $G_E$, there is a clear benefit to
performing calculations at small $Q^2$. As illustrated in
Figs.~\ref{fig:GE-ESC-a09m130W} and~\ref{fig:GE-ESC-a06m135} for the
physical mass ensembles, the ESC in $G_E(Q^2)$ is still small for
$\vec{n}^2=2$ corresponding to $Q^2 \approx 0.1$~GeV${}^2$. (Note that
for $\vec{q}^2 =0$, the ESC is essentially zero as the vector charge
is conserved and the local current has no $O(a)$ correction in forward
matrix elements.) There is, however, an increase in the ESC with
decreasing $a$ as shown in the bottom panels in
Fig.~\ref{fig:GE-ESC-p1}.  On the other hand, the ESC in $G_M(Q^2)$ is
large at small $Q^2$ as shown in Fig.~\ref{fig:GM-ESC-p1}, and the
resulting larger errors in $G_M(Q^2)$ reflect that uncertainty. In
contrast, the PACS'18A calculation indicates that the ESC is removed
in both form factors by using a tuned simple exponentially falling
smearing of sources for generating quark propagators compared to the
excited-state pattern that results from using a gauge invariant
Gaussian smearing used in our and the other four calculations summarized in Table~\ref{tab:comp}. Clearly,
the efficacy of the exponential source used by PACS'18A to remove
essentially all ESC needs to be validated.

The collapse of the data into a single curve indicates that finite
volume corrections are already small for $M_\pi L \ge 4$, and the main
advantage of the large volume used in the
PACS'18A~\cite{Shintani:2018ozy} study is it gives data at low
$Q^2$. These data for $Q^2 < 0.1$~GeV${}^2$ represent a qualitative
change in the behavior of both $G_M(Q^2)$ and $G_E(Q^2)$ which leads
to larger values for $\rEsq$ and $\rMsq$.  Note that since the
PACS'18A estimate $M_N=942(11)$~MeV is close to $M_N^{\rm
  phy}=939$~MeV, their data do not move with respect to the Kelly
curve when plotted versus $Q^2/M_N^2$ or $Q^2$.  The authors attribute the much
smaller errors, compared to the much higher statistics 
PACS'18~\cite{Ishikawa:2018rew} calculation, to the use of the all-mode-averaging
method and to a better tuned smearing ansatz (exponential) for the quark sources
used to calculate the quark propagators.  Since the advantage of
simulations on large volume lattices to get data at low $Q^2$, and
thus reliable estimates for $\rEsq$ and $\rMsq$, is obvious, it is
important to validate the relatively low statistics PACS'18A
calculation.

\section{Conclusions}
\label{sec:conclusions}
\FIXME{sec:conclusions} 

We have presented calculations of the isovector electric and magnetic
form factors, $G_E^{p-n}$ and $G_M^{p-n}$, using thirteen calculations on eleven ensembles of
2+1+1-flavors of HISQ~\cite{Follana:2006rc} fermions generated by the
MILC collaboration~\cite{Bazavov:2012xda}.  These ensembles are at
four lattice spacings, $a \approx 0.06$, $0.09$, $0.12$ and $0.15$~fm,
three values of pion masses, $M_\pi \approx 135$, $220$ and 310~MeV,
and the lattice size covers the range $3.3\, \lsim M_\pi L\,
\lsim5.5$. Each of these ensembles have been analyzed using $O(10^5)$
measurements using the truncated solver method with bias correction.
Using these high-statistics data we demonstrate control over
excited-state contamination and perform a simultaneous fit in lattice
spacing $a$, pion mass $M_\pi$ and lattice size $M_\pi L$ to get
results at the physical point that can be compared with experimental
values. 

Our work constitutes three improvements: 
\begin{itemize}
\item
The much higher statistics allowed us to understand and control ESC
better by keeping three states in the spectral decomposition of the
3-point correlation functions.
\item
Calculations at multiple values of $a$ and $M_\pi$ show that the
variations in the data versus these two parameters is small for $Q^2
\gtrsim 0.1$~GeV${}^2$ as illustrated in
Figs.~\ref{fig:GE-vsM},~\ref{fig:GE-vsa},~\ref{fig:GM-vsM},~\ref{fig:GM-vsa}
and in Figs.~\ref{fig:Kelly}. 
\item
We have presented first results with a CCFV fit to control the
lattice artifacts due to discretization, chiral and finite volume
effects. The data for $\rEsq$ and $\rMsq$ and the CCFV fits in
Figs.~\ref{fig:rE-extrap11} and~\ref{fig:rM-extrap11} show the
variation versus $M_\pi$ is small and consistent with the predictions
of chiral perturbation theory~\cite{Kubis:2000zd} as shown in
Fig.~\ref{fig:ChPT}.  In the $\chi$PT prediction, the nonanalytical
term in $M_\pi$, included in Eqs.~\eqref{eq:rEsq-extrap}
and~\eqref{eq:rMsq-extrap}, becomes significant only for $M_\pi <
135$~MeV, whereas over the range $350 > M_\pi > 135$~MeV, its growth
is compensated for by the decrease in the analytical corrections. With
such competing contributions in $M_\pi$, the data on the two physical
pion mass ensembles at $a\approx 0.09$ and $0.06$~fm play a
significant role in controlling the uncertainty. The CCFV fit in
Fig.~\ref{fig:mu-extrap11} shows a significant $a$ dependence in $\mu$
that leads to an underestimate by $\sim 16\%$.
\end{itemize}

Our final results for the mean-square charge radii, $\rEsq$ and
$\rMsq$ (or equivalently the Dirac, $\rDirac$, and Pauli, $\rPauli$,
radii derived from them), and the magnetic moment $\mu$ are given in
Table~\ref{tab:finalresults}. Using the dipole ansatz and the
$z$-expansion to fit the $Q^2$ dependence give consistent results,
however, the combined errors in the latter approach are about 2--3
times larger.  The central values for $\rEsq$, $\rMsq$ and $\mu$ are,
about 17\%, 19\% and 16\%, respectively, smaller than the
phenomenological values given in Eq.~\eqref{eq:isovectorradii} and the precise experimental value in 
Eq.~\eqref{eq:mu_expt}. The trend in the data for the form factors,
however, is towards the experimental values as the $Q^2$, lattice
spacing and the light quark mass are decreased.

With higher precision data, the major improvement observed has been in
the $z$-expansion estimates.  Including constraints
on the fit parameters, $|a_k| \lesssim 5$, the $z$-expansion fits for different
truncations became more consistent. Based on an analysis of the
experimental data with the same fit ans\"atze and evaluation of
various systematics in the lattice calculations, the extraction of
charge radii and magnetic moment could have $O(10\%)$ errors due to
the modeling of the $Q^2$ behavior.  Errors of similar size 
could also be due to statistics and ESC fits. 
Keeping in mind these estimates of the magnitude of
possible systematics, the total uncertainty in estimates given in
Tab.~\ref{tab:finalresults}, especially for the dipole fit, are likely
underestimated.  Consequently, we do not consider the current
deviations from the experimental values significant.

The magnitude of the systematic associated with what variable is used
to set the lattice scale is exposed by plotting the data versus
$Q^2/M_N^2$. As shown in Fig.~\ref{fig:Kelly} (bottom), our data from
the 13 calculations fall on a common curve when plotted versus
$Q^2/M_N^2$. In Fig.~\ref{fig:GEMcomp}, we further show that both
$G_E$ and $G_M$ from all lattice calculations done close to the
physical pion mass also collapse onto this curve. The shift in the
data when $G_E$ and $G_M$ are plotted versus $Q^2/M_N^2$ as compared
to $Q^2$ is a discretization effect, i.e., the scale obtained from
$M_N$ is different from that obtained by the various collaborations
using the quantities shown in Table~\ref{tab:compR}.  This is
remarkable considering that the number of quark flavors, lattice size
and lattice spacing are different in the various calculations.  Also,
the deviation of the combined lattice data from the Kelly curve is
significantly reduced.

Given our demonstration in Sec.~\ref{sec:comparison} that a large part
of the difference between data obtained by various collaborations is
an artifact of scale setting, the major advantage of the
PACS'18A~\cite{Shintani:2018ozy} calculation is that the large volume
provides data at $Q^2 < 0.1$~GeV${}^2$; the agreement between data
from different collaborations presented in Sec.~\ref{sec:comparison}
indicates that finite volume corrections are already small for $M_\pi
L \approx 4$. The PACS'18A data show no movement with respect to the
Kelly curve because the estimate of the nucleon mass is consistent
with the physical value. This may be because the lattice scale is set
using the Omega baryon mass, $M_\Omega$, which is likely correlated
with $M_N$, rather than indicating that discretization errors are
already small at $a \approx 0.09$~fm.  It is important to validate
their data at $Q^2 < 0.1$~GeV${}^2$ in future calculations and confirm
that the resulting estimates of $\rEsq$ and $\rMsq$ are consistent
with the experimental values.

To conclude, our analysis highlights three points. First, all lattice
data are remarkably consistent and the form factors show little
dependence on the number of flavors, lattice spacing, quark mass or
the lattice volume, at least for data with $M_\pi \lesssim 300$~MeV
and $M_\pi L \gtrsim 4$. Second, the size of the remaining deviations
in $G_E$ and $G_M$ between the lattice data and the Kelly curve are
consistent with the various quantifiable systematics such as
excited-state contamination and deteriorating signal at large
$\vec{q}$. Third, current results provide confidence that there are no
hidden systematics that afflict the calculations of form factors on
the lattice. 

With the lattice methodology in place, improved estimates for form
factors will be obtained in future high-statistics calculations that
provide data at $Q^2 < 0.1$~GeV${}^2$ and use nucleon interpolating
operators that have smaller excited-state contamination.


\appendix
\section{Lattice parameters}
\label{appendix:parameters}
\FIXME{appendix:parameters}

In this Appendix, we summarize, in Table~\ref{tab:ens}, the parameters
of the eleven ensembles used in the calculation.  Two ensembles,
$a06m310$ and $a06m220$ have been analyzed twice with different
smearing sizes as listed in Table~\ref{tab:cloverparams}, where we
give the parameters used in the generation of the clover
propagators. These two tables are essentially the same as in
Ref.~\cite{Gupta:2018qil}, and have been reproduced here to keep the
discussion self-contained.

\begin{table*}[tbp]    
\caption{Parameters, including the Goldstone pion mass $M_\pi^{\rm
    sea}$, of the eleven 2+1+1- flavor HISQ ensembles generated by the
  MILC collaboration and analyzed in this study are quoted from
  Ref.~\cite{Bazavov:2012xda}.  All fits are made versus $M_\pi^{\rm
    val}$ and finite-size effects are analyzed in terms of $M_\pi^{\rm
    val} L$.  Estimates of $M_\pi^{\rm val}$, the clover-on-HISQ pion
  mass, are the same as given in Ref.~\cite{Bhattacharya:2015wna} and
  the error is governed mainly by the uncertainty in the lattice
  scale. In the last four columns, we give, for each ensemble, the
  values of the source-sink separation $t_{\rm sep}$ used in the
  calculation of the three-point functions, the number of
  configurations analyzed, and the number of measurements made using
  the high precision (HP) and the low precision (LP) truncation of the
  inversion of the clover operator.  The smearing size used in the
  calculation of the quark propagator is given in
  Table~\protect\ref{tab:cloverparams}. }
\label{tab:ens}
\begin{center}
\renewcommand{\arraystretch}{1.2} 
\begin{ruledtabular}
\begin{tabular}{l|ccc|cc|cccc}
Ensemble ID & $a$ (fm) & $M_\pi^{\rm sea}$ (MeV) & $M_\pi^{\rm val}$ (MeV) & $L^3\times T$    & $M_\pi^{\rm val} L$ & $\tau/a$ & $N_\text{conf}$  & $N_{\rm meas}^{\rm HP}$  & $N_{\rm meas}^{\rm LP}$  \\
\hline
$a15m310 $      & 0.1510(20) & 306.9(5) & 320.6(4.3)     & $16^3\times 48$ & 3.93 &  $\{5,6,7,8,9\}$    & 1917 & 7668  & 122,688   \\
\hline
$a12m310 $      & 0.1207(11) & 305.3(4) & 310.2(2.8) & $24^3\times 64$ & 4.55 &  $\{8,10,12\}$      & 1013 & 8104  &  64,832   \\
$a12m220S$      & 0.1202(12) & 218.1(4) & 225.0(2.3) & $24^3\times 64$ & 3.29 & $\{8, 10, 12\}$     & 946  & 3784  &  60,544   \\
$a12m220 $      & 0.1184(10) & 216.9(2) & 227.9(1.9) & $32^3\times 64$ & 4.38 & $\{8, 10, 12\}$     & 744  & 2976  &  47,616   \\
$a12m220L$      & 0.1189(09) & 217.0(2) & 227.6(1.7) & $40^3\times 64$ & 5.49 & $\{8,10,12,14\}$    & 1000 & 4000  & 128,000   \\
\hline                                                                                                         
$a09m310 $      & 0.0888(08) & 312.7(6) & 313.0(2.8) & $32^3\times 96$ & 4.51 & $\{10,12,14,16\}$   & 2264 & 9056  & 114,896   \\
$a09m220 $      & 0.0872(07) & 220.3(2) & 225.9(1.8) & $48^3\times 96$ & 4.79 & $\{10,12,14,16\}$   & 964  & 7712  & 123,392   \\
$a09m130W$      & 0.0871(06) & 128.2(1) & 138.1(1.0) & $64^3\times 96$ & 3.90 & $\{8,10,12,14,16\}$ & 1290 & 5160  & 165,120   \\
\hline                                                                                                         
$a06m310 $      & 0.0582(04) & 319.3(5) & 319.6(2.2) & $48^3\times 144$& 4.52 & $\{16,20,22,24\}$   & 1000 & 8000  &  64,000   \\
$a06m31W      $ &            &          &            &                 &      & $\{18,20,22,24\}$   & 500  & 2000  &  64,000   \\
$a06m220 $      & 0.0578(04) & 229.2(4) & 235.2(1.7) & $64^3\times 144$& 4.41 & $\{16,20,22,24\}$   & 650  & 2600  &  41,600   \\
$a06m22W      $ &            &          &            &                 &      & $\{18,20,22,24\}$   & 649  & 2600  &  41,600   \\
$a06m135 $      & 0.0570(01) & 135.5(2) & 135.6(1.4) & $96^3\times 192$& 3.7  & $\{16,18,20,22\}$   & 675  & 2700  &  43,200   \\
\end{tabular}
\end{ruledtabular}
\end{center}
\end{table*}

\begin{table}[htbp]  
\caption{The parameters used in the
  calculation of the clover propagators.  The hopping parameter for
  the light quarks, $\kappa_l$, in the clover action is given by
  $2\kappa_{l} = 1/(m_{l}+4)$.  $m_l$ is tuned to achieve $M_\pi^{\rm
    val} \approx M_\pi^\text{sea}$. The parameters used to construct
  Gaussian smeared sources~\protect\cite{Gusken:1989ad}, $\{\sigma, N_{\text{KG}}\}$, are given in
  the fourth column where $N_{\text{KG}}$ is the number of
  applications of the Klein-Gordon operator and the width of the
  smearing is controlled by the coefficient $\sigma$, both in Chroma
  convention~\cite{Edwards:2004sx}.  The resulting root-mean-square
  radius of the smearing, defined as $\sqrt{\int r^2 \sqrt{S^\dag S}
    dr /\int \sqrt{S^\dag S} dr} $, is given in the last column.  }
  \label{tab:cloverparams}
\centering
\begin{ruledtabular}
\begin{tabular}{l|lc|c|c}
\multicolumn1c{ID}  & \multicolumn1c{$m_l$} &  $c_{\text{SW}}$ & Smearing    & RMS smearing \\
                    &                       &                  & Parameters  & radius       \\
\hline
$a15m310 $          & $-0.0893$  & 1.05094 & \{4.2, 36\}   & 4.69  \\
\hline
$a12m310 $          & $-0.0695$  & 1.05094 & \{5.5, 70\}   & 5.96  \\
$a12m220S$          & $-0.075$   & 1.05091 & \{5.5, 70\}   & 5.98  \\ 
$a12m220 $          & $-0.075$   & 1.05091 & \{5.5, 70\}   & 5.96  \\ 
$a12m220L$          & $-0.075$   & 1.05091 & \{5.5, 70\}   & 5.96  \\ 
\hline                                                        
$a09m310 $          & $-0.05138$ & 1.04243 & \{7.0,100\}   & 7.48  \\
$a09m220 $          & $-0.0554$  & 1.04239 & \{7.0,100\}   & 7.48  \\
$a09m130W$          & $-0.058$   & 1.04239 & \{7.0,100\}   & 7.50  \\
\hline                                                        
$a06m310 $          & $-0.0398$  & 1.03493 & \{6.5, 70\}   & 7.22  \\
$a06m310W     $     & $-0.0398$  & 1.03493 & \{12, 250\}   & 12.19  \\
$a06m220 $          & $-0.04222$ & 1.03493 & \{5.5, 70\}   & 6.22  \\
$a06m220W     $     & $-0.04222$ & 1.03493 & \{11, 230\}   & 11.24  \\
$a06m135 $          & $-0.044$   & 1.03493 & \{9.0,150\}   & 9.56  \\
\end{tabular}
\end{ruledtabular}
\end{table}


\section{Nucleon Mass}
\label{appendix:NucleonMass}
\FIXME{appendix:NucleonMass}

The masses of the nucleon ground and three excited states given by our
4-state fit are summarized in Table~\ref{tab:spectrum}. The ground
state masses are found to be stable under changes in the number of
states kept in the spectral decomposition of the two-point function
and the Euclidean time interval used in the fits since the data exhibit a
reasonable plateau in the effective mass plot for all the ensembles. On
the other hand, the excited-state energies are sensitive to the 
details of the fits. The main reason is the small number, 6--10, of points at
short times that are available to determine the six excited-state
parameters before the ground state dominates the two-point function.
This is particularly true of the $a15m310$, $a09m310$ and $a06m310W$
ensembles.  Overall, the estimates for the excited-state masses are
larger than values expected based on phenomenological arguments. For
example, the first excited-state mass for the ``Roper', and the $N\pi$
and the $N\pi\pi$ multiparticle states for our physical mass ensembles, should all
be between 1.3--1.7~GeV for our lattice parameters.  Having a reliable 
estimate of the first excited-state energy is the key variable in 
the 3${}^\ast$-fits to control ESC. 

In the fits to the nucleon two-point function, we find a strong
correlation between the excited-state energies and the
amplitudes. This poses a challenge: what priors to choose in the 3-
and 4-state fits, especially when there are near flat directions in
the parameter space. We chose priors with a large width and aimed for
stable first excited-state energy and amplitude. These are the most
important input for the analysis of the ESC as the $ 0 \leftrightarrow
1$ transition matrix elements are found to be the dominant artifact.
Note that the priors are used only to stabilize the fits and the errors are
given by the jackknife procedure.

\begin{figure*} 
\centering
\subfigure{
\includegraphics[width=0.46\linewidth]{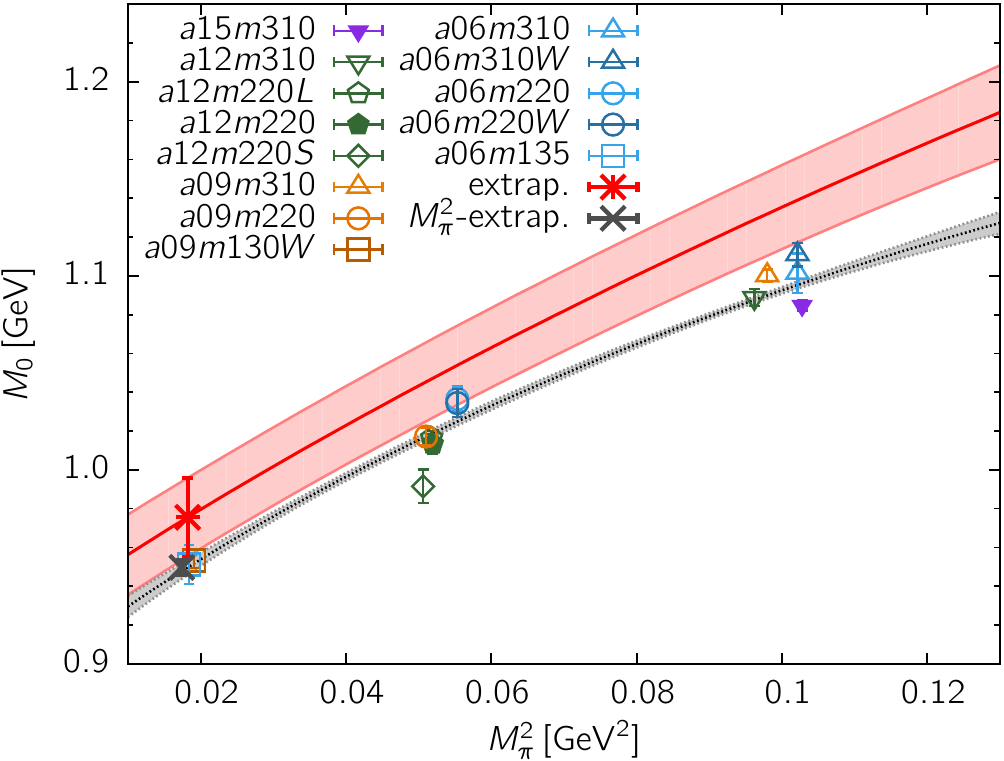}
\hspace{4mm}
\includegraphics[width=0.46\linewidth]{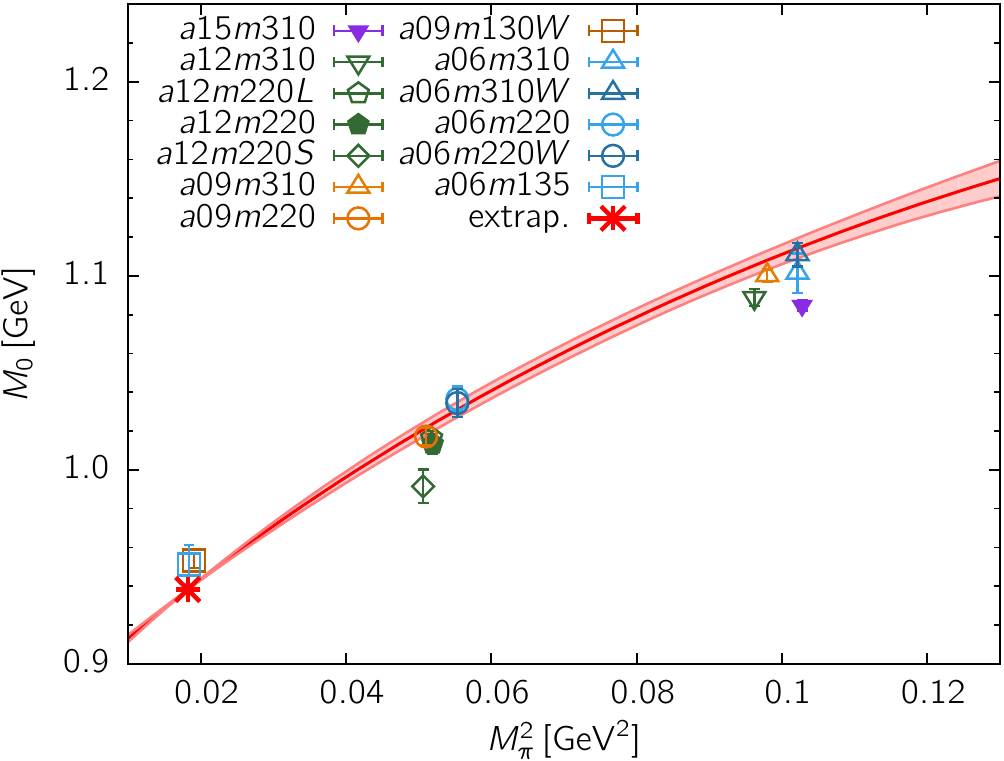}
}
\caption{\FIXME{fig:MN} CCFV fits to the ground state nucleon mass
  using Eq.~\eqref{eq:MNfitB}. In the right panel, the nucleon mass 
  is constrained to be $M_N^\text{phys}=939$~MeV at $M_\pi = 135$~MeV.
\label{fig:MN}}
\end{figure*}

The two physical mass ensembles give estimates for $M_N$ that are
about 13~MeV larger than the physical value $M_N^{\rm phy} =
939$~MeV. To investigate the dependence of $M_N$ on the lattice
spacing, pion mass and lattice size, we have carried out two fits:
\begin{align}
M_N &= c_0 + c_1 a + c_2 a^2 + c_3 M_\pi^2 + c_4 M_\pi^3 + c_5 M_\pi^2 e^{(-M_\pi L)}
\nonumber \\
M_N &= M_N^\text{phys} + c_1 a + c_2 a^2 + c_3 (M_\pi^2-(M_\pi^\text{phys})^2) 
\nonumber \\
    & \quad\quad     +  c_4 (M_\pi^3-(M_\pi^\text{phys})^3) + c_5 M_\pi^2 e^{(-M_\pi L)} \,,
\label{eq:MNfitB} 
\end{align}
where the second relation enforces $M_N^\text{phys}=939$~MeV at
$M_\pi=135$~MeV. The values of $a$ for the HISQ ensembles used to
convert the lattice data to GeV are taken from
Ref.~\cite{Bazavov:2012xda} and given in Table~\ref{tab:ens}.  The
fits to the ground-state nucleon mass $M_0$, given in
Table~\ref{tab:spectrum}, using Eqs.~\eqref{eq:MNfitB} are shown in
Fig.~\ref{fig:MN}, and the values of the fit parameters are given in
Table~\ref{tab:spec-ccfv}. Note that the $\chi$PT predicted value for the coefficient $c_4
= 3g_A^2/(32\pi F_\pi^2) = -5.716$ using $g_A=1.276$ and $F_\pi =
92.2\,\MeV$, whereas both fits give smaller values.

These fits imply that the first, unconstrained, CCFV fit to $M_N$ requires higher
order correction terms, but with just three values of the pion mass
and four values of $a$, most of the fits parameters $c_i$ are already
poorly determined as shown in Table~\ref{tab:spec-ccfv}.  While the
results for $M_0$ from the two physical mass ensembles are about
$13$~MeV larger than the physical value, the unconstrained fit gives
an even larger value $M_N=976(20)$~MeV. It is clear that better
control over systematics via calculations on a larger number of
ensembles is needed in future calculations.  The impact of the resulting 
mismatch between the scales set using $r_1$ calculated on the HISQ
ensembles and from $M_N$ calculated using the Wilson-clover fermions,
on the form factors is shown in Fig.~\ref{fig:Kelly} and discussed in
Sec.~\ref{sec:Q2lattice}.

\begin{table}[!htb]  
  \centering
  \caption{Nucleon ground and excited-state masses in $\GeV$ extracted from a 4-state fit.}
  \label{tab:spectrum}
  \begin{ruledtabular}
    \begin{tabular}{lllll}
ID       & $M_0$       & $M_1$      & $M_2$    & $M_3$    \\
\hline
a15m310  & 1.0848(28)  & 2.038(62)  & 2.40(8)  & 2.88(8)  \\
a12m310  & 1.0888(44)  & 1.576(89)  & 2.55(16) & 3.18(16) \\
a12m220L & 1.0165(35)  & 1.691(175) & 2.76(26) & 3.44(26) \\
a12m220  & 1.0133(52)  & 1.634(116) & 2.75(27) & 3.42(27) \\
a12m220S & 0.9915(86)  & 1.499(78)  & 3.00(22) & 3.66(22) \\
a09m310  & 1.1001(31)  & 2.065(128) & 3.61(32) & 4.78(33) \\
a09m220  & 1.0172(45)  & 1.718(91)  & 2.56(16) & 3.44(17) \\
a09m130W & 0.9532(39)  & 1.761(88)  & 2.98(15) & 3.81(15) \\
a06m310  & 1.1014(102) & 1.646(105) & 2.79(16) & 3.73(22) \\
a06m310W & 1.1109(61)  & 2.054(145) & 3.00(24) & 3.99(27) \\
a06m220  & 1.0365(65)  & 1.874(73)  & 3.05(11) & 3.96(19) \\
a06m220W & 1.0345(72)  & 1.816(144) & 2.70(24) & 3.69(31) \\
a06m135  & 0.9512(100) & 1.734(89)  & 3.01(13) & 4.01(17)
    \end{tabular}
  \end{ruledtabular}
\end{table}

\begin{table*}[!htb]  
  \centering
  \caption{Values for the parameters of the two CCFV fits to the nucleon mass defined in Eq.~\eqref{eq:MNfitB}.}
  \label{tab:spec-ccfv}
  \begin{ruledtabular}
    \begin{tabular}{lllllllll}
      Fit   & $c_0$          & $c_1$       & $c_2$         & $c_3$          & $c_4$         & $c_5$         & $M_N$         & $\chi^2$/dof \\
            & $[\GeV]$       & $[\GeV \fm^{-1}]$    & $[\GeV^{} \fm^{-2}]$  & $[\GeV^{-1}]$  & $[\GeV^{-2}]$ & $[\GeV^{-1}]$   & $[\GeV]$      & [$p$-value]\\
\hline                                               
      1     & 0.931(22)      & -0.273(391) & 0.34(1.96)    & 2.722(487)     & -2.2(1.3)     & -10.9(5.1)    & 0.9755(202)   & 0.45 [0.87] \\
\hline                                                     
      2     &                &  0.430(80)  & -3.14(53)     & 2.875(48)      & -2.6(1.3)     & -5.6(4.2)     & 0.939         &  0.81 [0.59]\\
    \end{tabular}
  \end{ruledtabular}
\end{table*}


\section{ESC in the extraction of the form factors}
\label{appendix:ESC}

\begin{figure*} 
\centering
\subfigure{
  \textcolor{white}{\rule{7.2cm}{2cm}}
\hspace{4mm}
\includegraphics[width=0.40\linewidth]{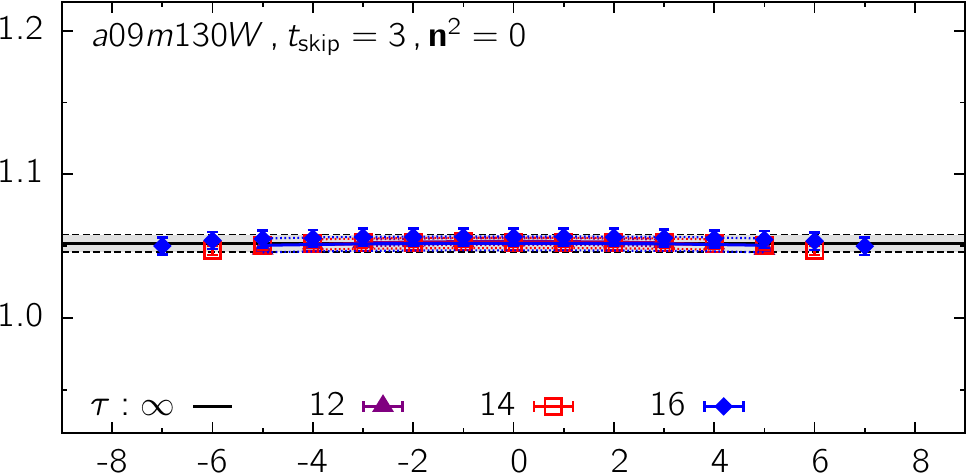}
}
\subfigure{
\includegraphics[width=0.40\linewidth]{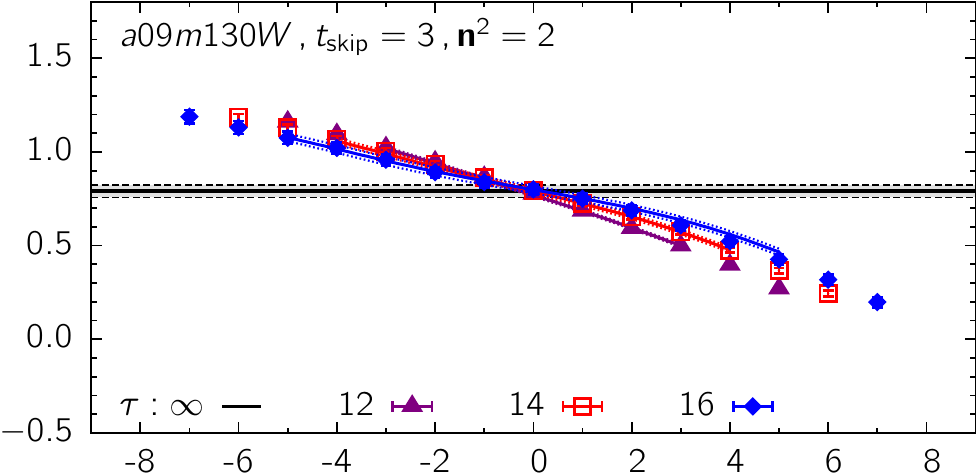}
\hspace{4mm}
\includegraphics[width=0.40\linewidth]{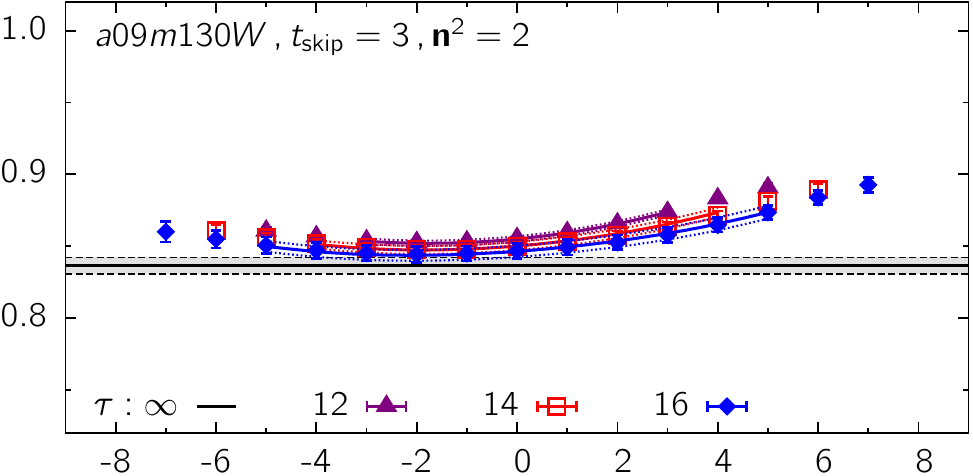}
}
\subfigure{
\includegraphics[width=0.40\linewidth]{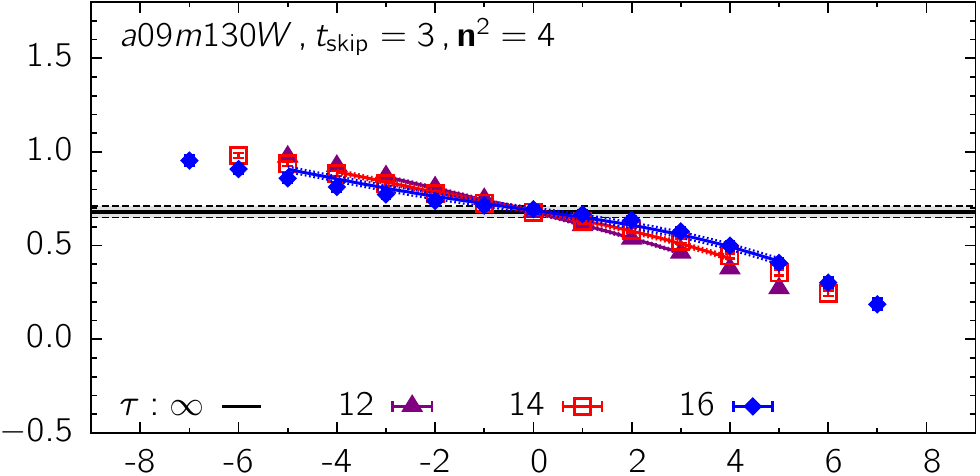}
\hspace{4mm}
\includegraphics[width=0.40\linewidth]{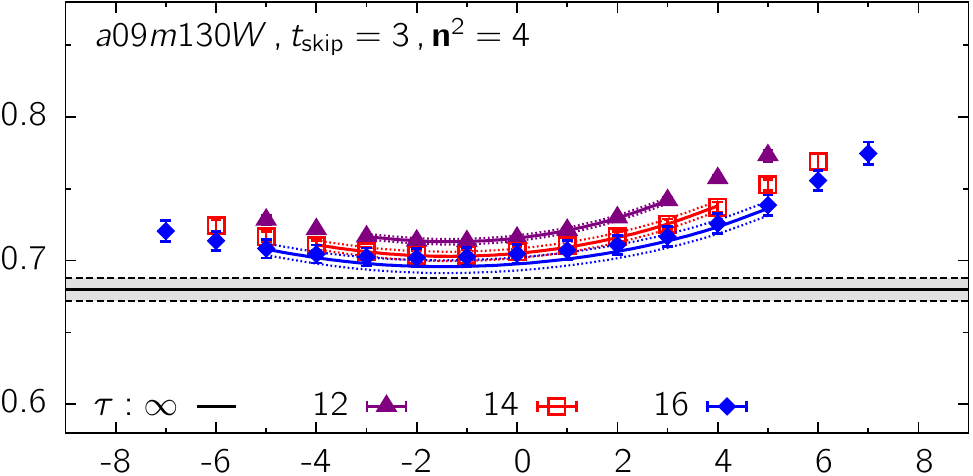}
}
\subfigure{
\includegraphics[width=0.40\linewidth]{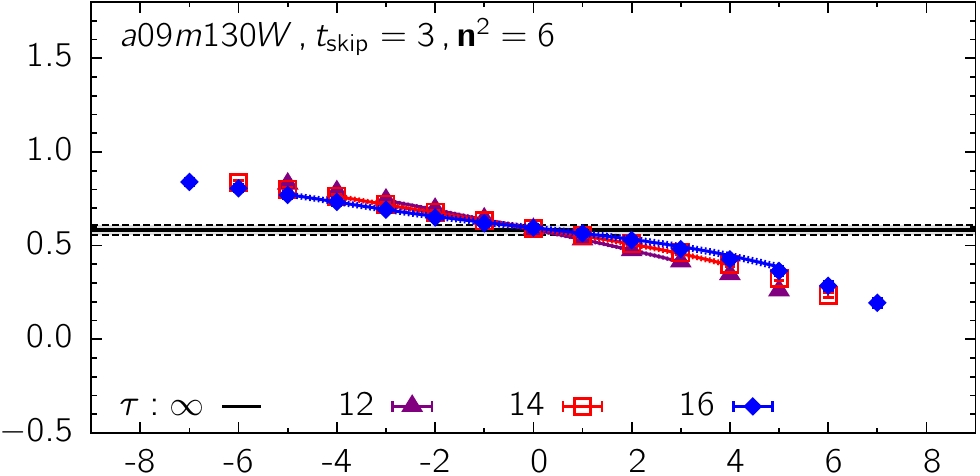}
\hspace{4mm}
\includegraphics[width=0.40\linewidth]{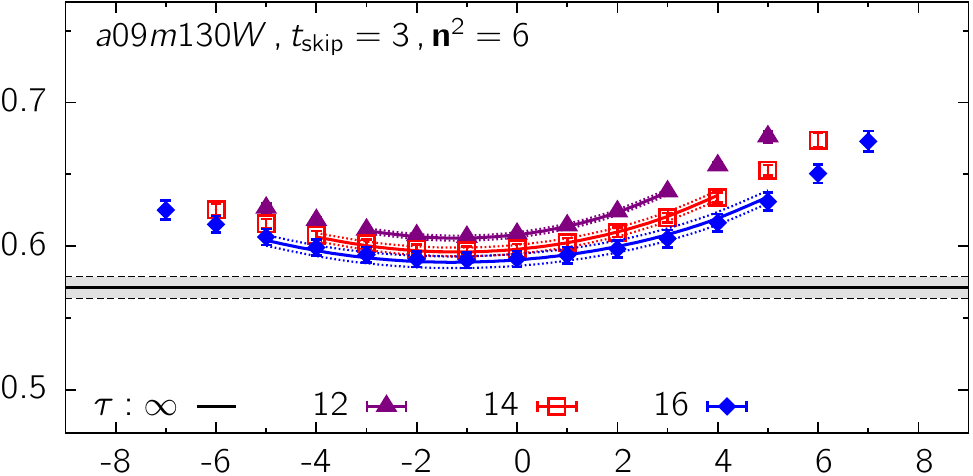}
}
\subfigure{
\includegraphics[width=0.40\linewidth]{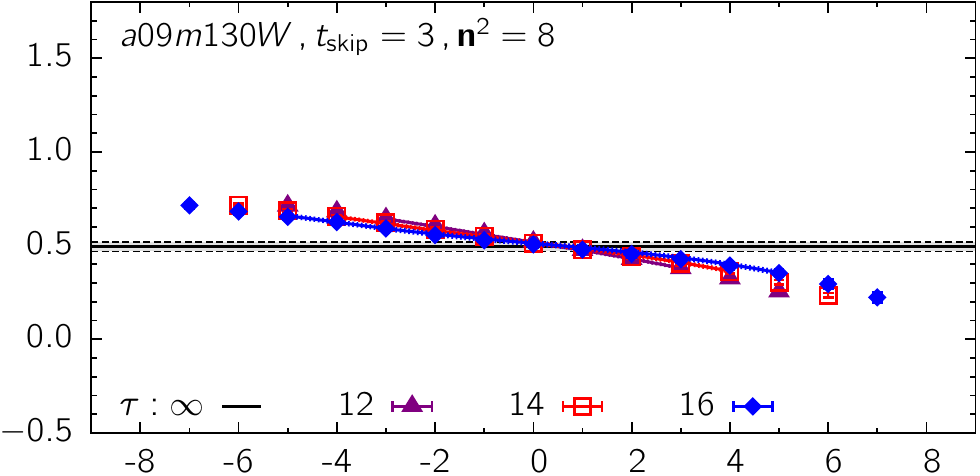}
\hspace{4mm}
\includegraphics[width=0.40\linewidth]{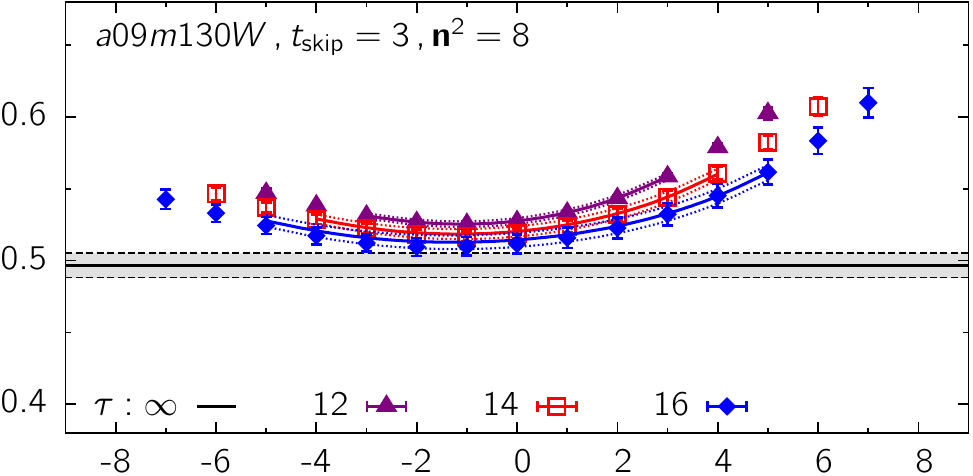}
}
\subfigure{
\includegraphics[width=0.40\linewidth]{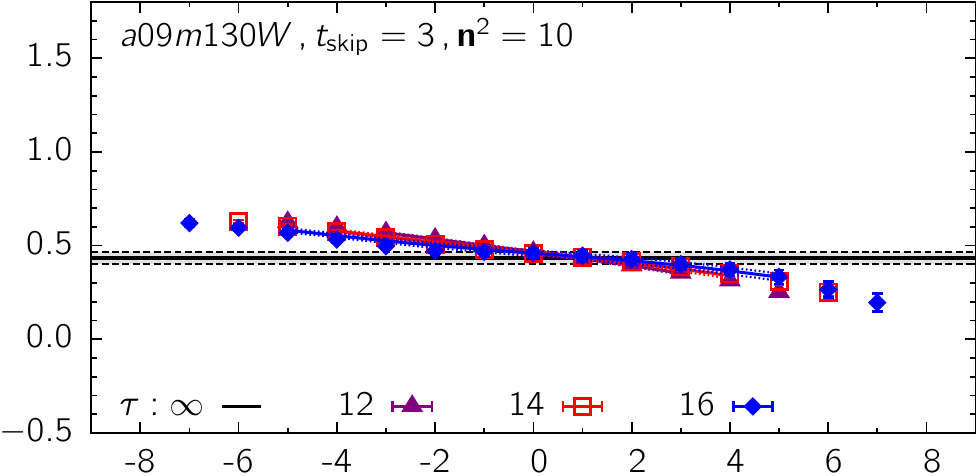}
\hspace{4mm}
\includegraphics[width=0.40\linewidth]{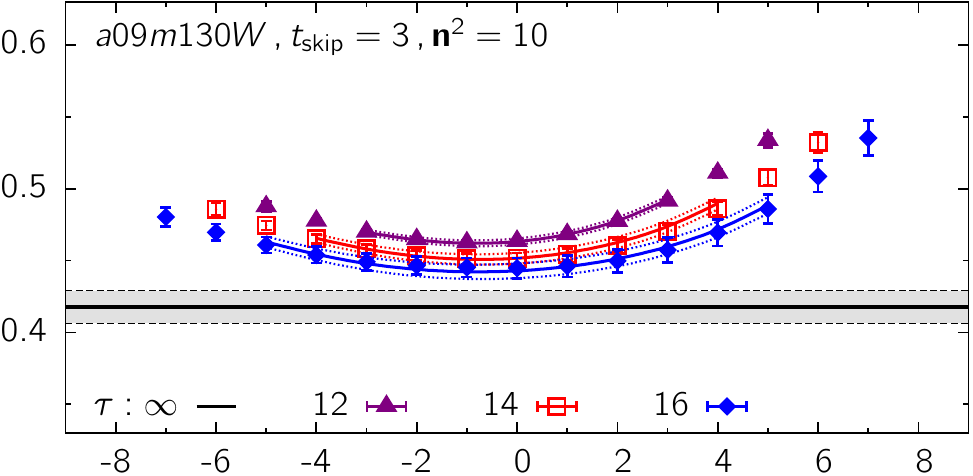}
}
\caption{\FIXME{fig:GE-ESC-a09m130W} Comparison of the ESC and the
  extraction of the unrenormalized isovector form factor $G_E$ from
  $\Im V_i$ as defined in Eq.~\protect\eqref{eq:GE1} (left panels),
  and from $\Re V_4$ defined in Eq.~\protect\eqref{eq:GE4} (right panels). 
  The  $a09m130W$ data are plotted versus $t-\tau/2$ for six values of the momenta, 
  ${\bf p}^2 = {\bf n}^2 (2\pi/La)^2$ with ${\bf n}^2 = 0,$ 2, 4,  6, 8, and 10.  
  The values of $\tskip$ and $\tau$ used in the 3${}^\ast$-state fits 
  are shown in the legends.  
  The horizontal gray band is the $\tsepi$ value, and the colored lines are the fit result 
  for $\tau=12, 14, 16$. The range of the y-axis is chosen to be the same
  for the left panels whereas the total interval $\Delta y=0.3$ is kept the
  same for the right panels.
\label{fig:GE-ESC-a09m130W}}
\end{figure*}

\begin{figure*} 
\centering
\subfigure{
  \textcolor{white}{\rule{7.2cm}{2cm}}
\hspace{4mm}
\includegraphics[width=0.40\linewidth]{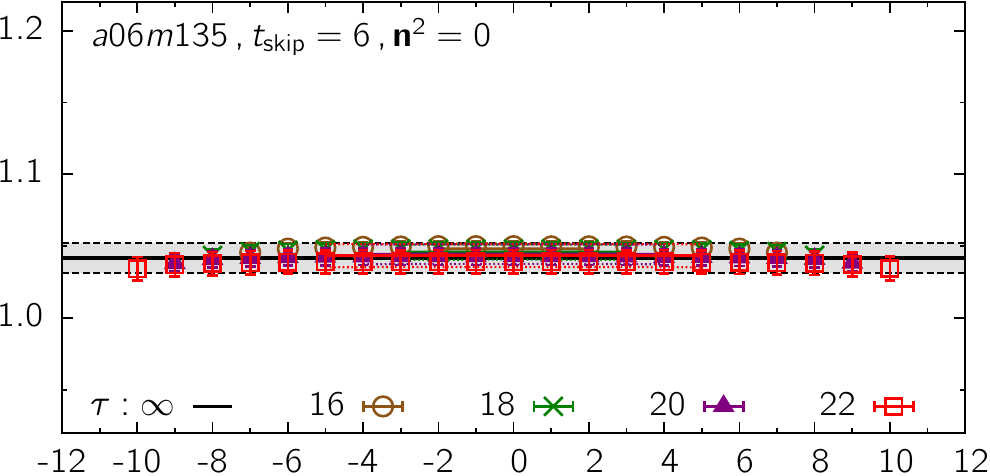}
}
\subfigure{
\includegraphics[width=0.40\linewidth]{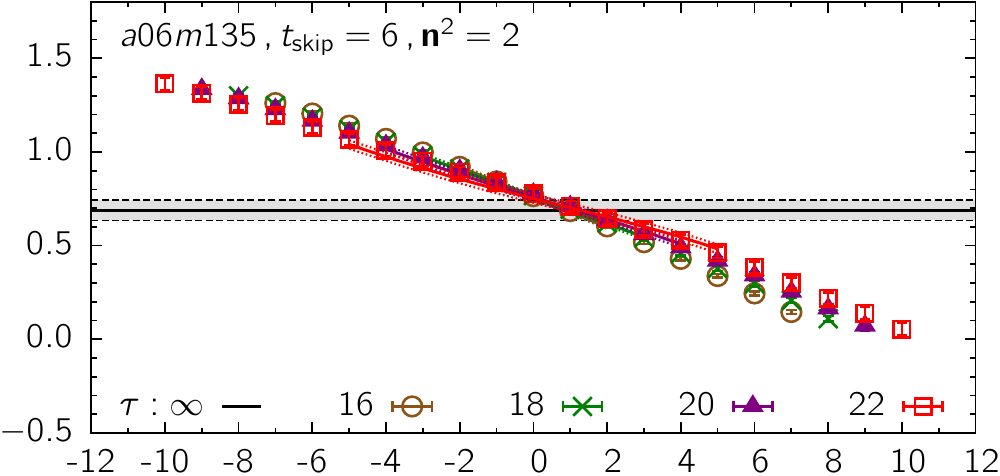}
\hspace{4mm}
\includegraphics[width=0.40\linewidth]{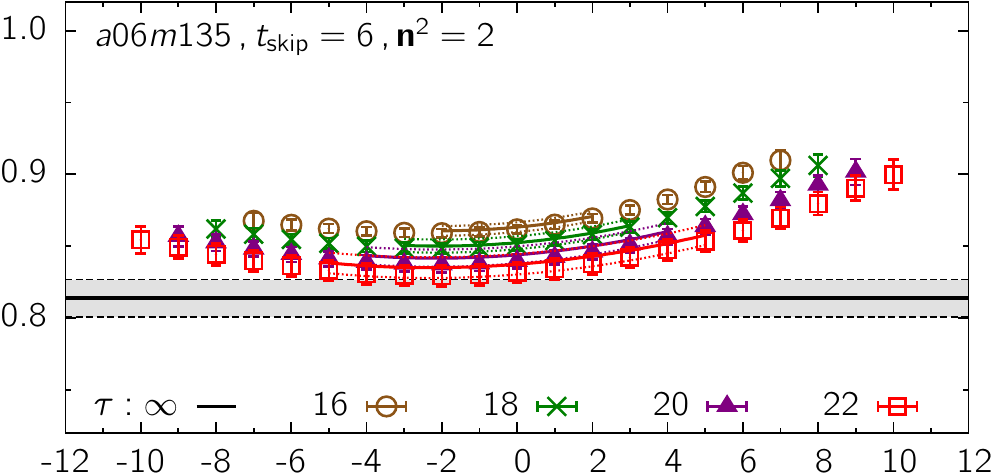}
}
\subfigure{
\includegraphics[width=0.40\linewidth]{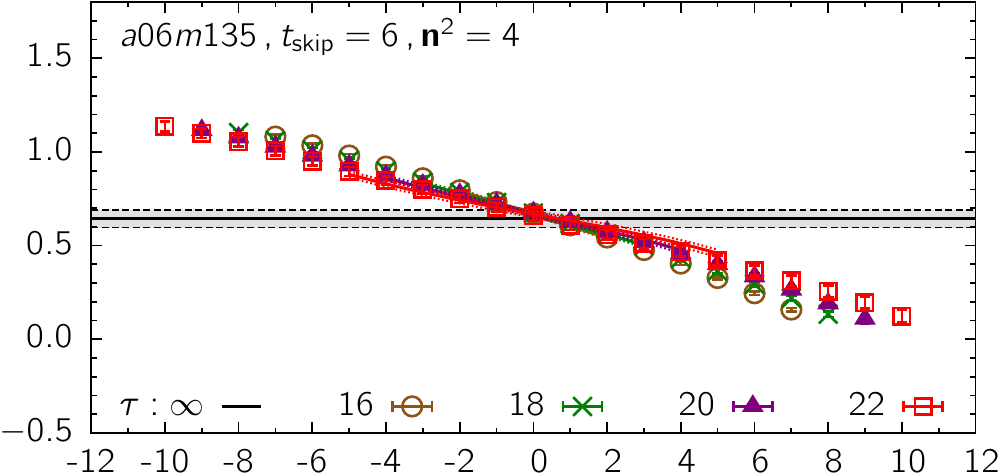}
\hspace{4mm}
\includegraphics[width=0.40\linewidth]{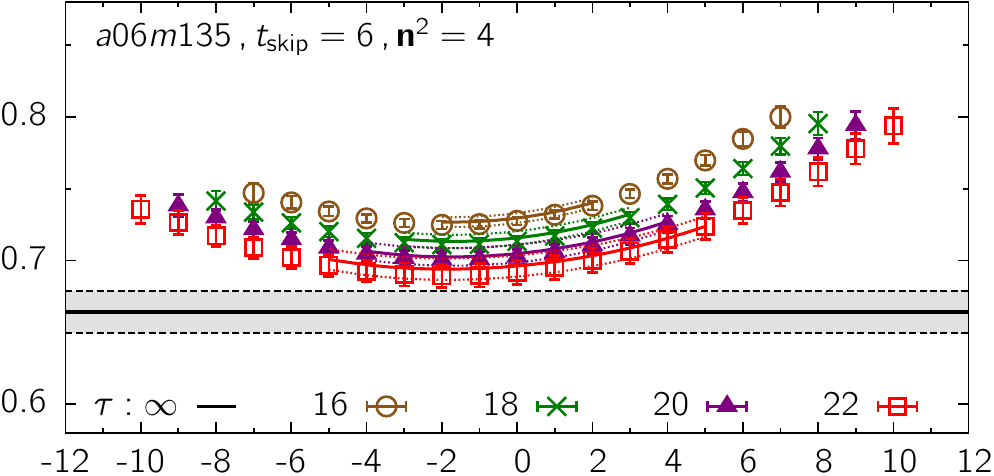}
}
\subfigure{
\includegraphics[width=0.40\linewidth]{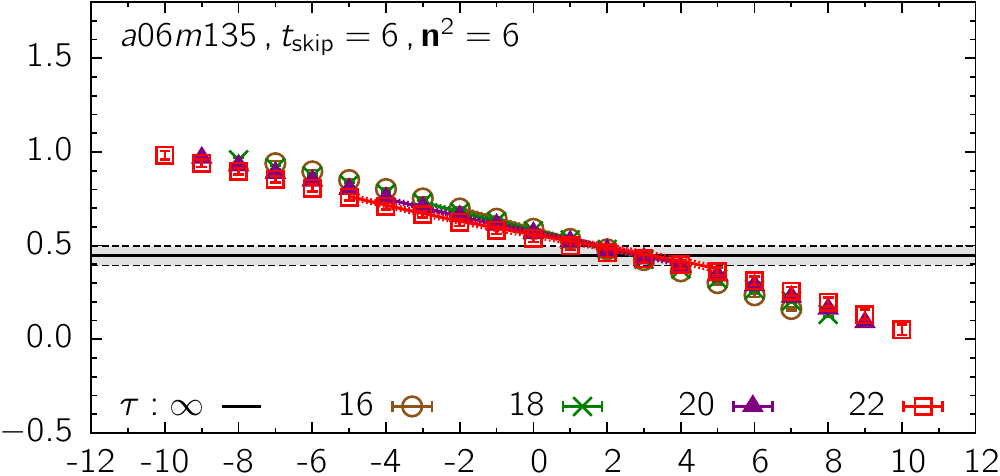}
\hspace{4mm}
\includegraphics[width=0.40\linewidth]{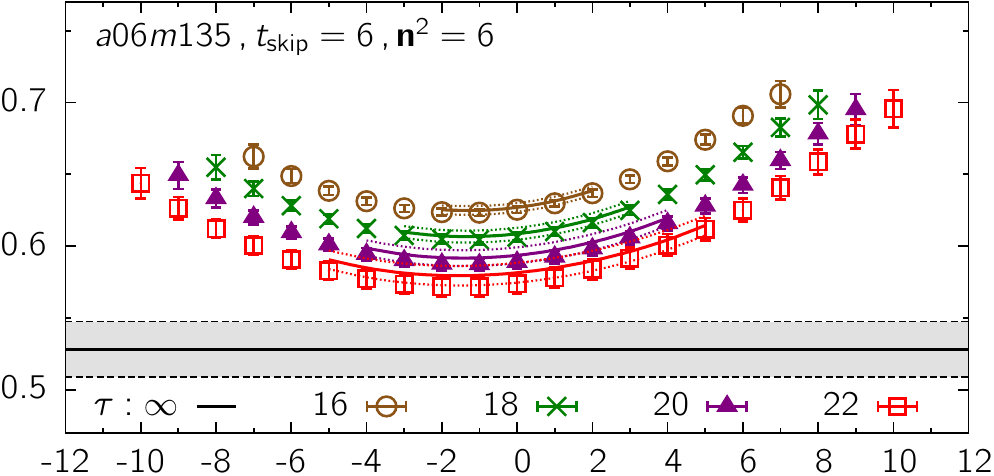}
}
\subfigure{
\includegraphics[width=0.40\linewidth]{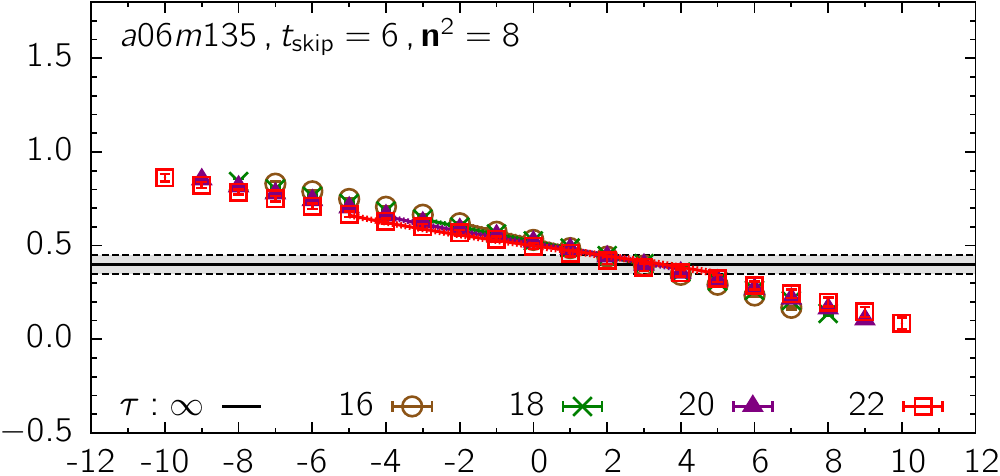}
\hspace{4mm}
\includegraphics[width=0.40\linewidth]{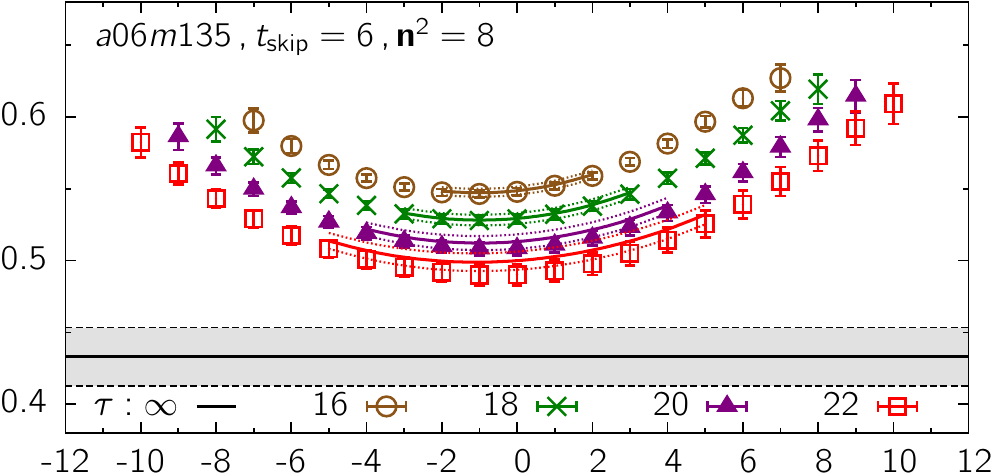}
}
\subfigure{
\includegraphics[width=0.40\linewidth]{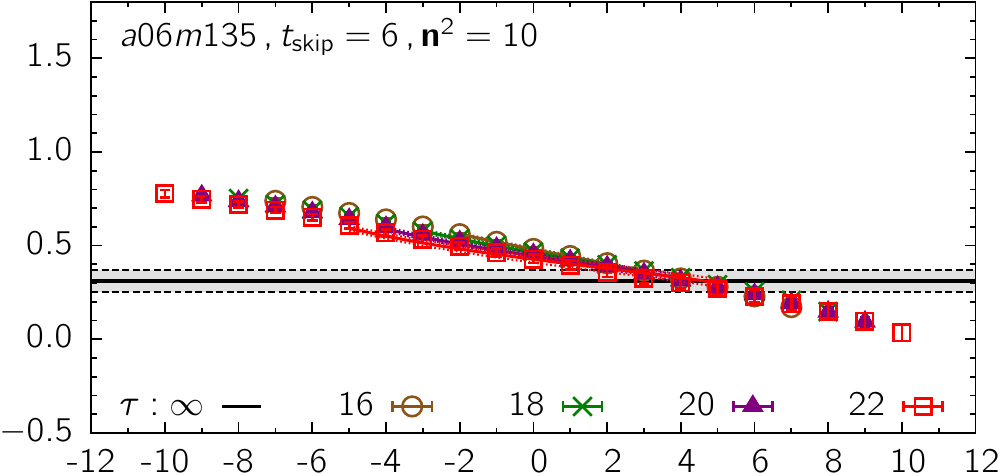}
\hspace{4mm}
\includegraphics[width=0.40\linewidth]{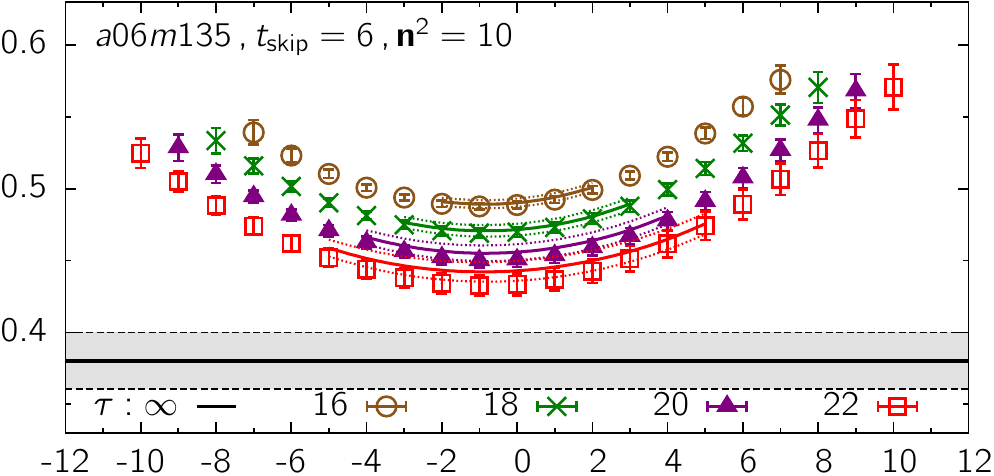}
}
\caption{\FIXME{fig:GE-ESC-a06m135} Comparison of the ESC and the
  extraction of the unrenormalized isovector form factor $G_E$ from
  $\Im V_i$ as defined in Eq.~\protect\eqref{eq:GE1} (left panels),
  and from $\Re V_4$ defined in Eq.~\protect\eqref{eq:GE4} (right
  panels).  The data from the $a06m135$ ensemble are plotted versus
  $t-\tau/2$ for six values of the momenta. The rest is the same as in
  Fig.~\protect\ref{fig:GE-ESC-a09m130W}.}
\label{fig:GE-ESC-a06m135}
\end{figure*}

\begin{figure*} 
\centering
\subfigure{
\includegraphics[width=0.47\linewidth]{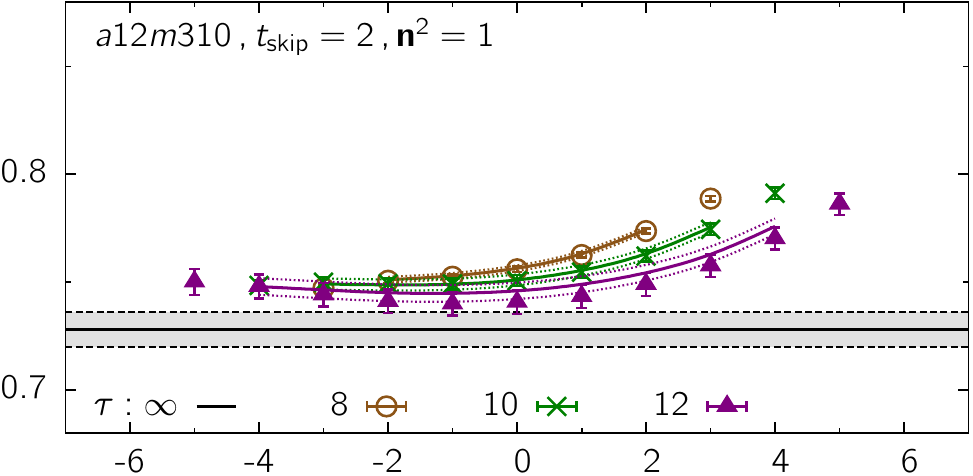}
\includegraphics[width=0.47\linewidth]{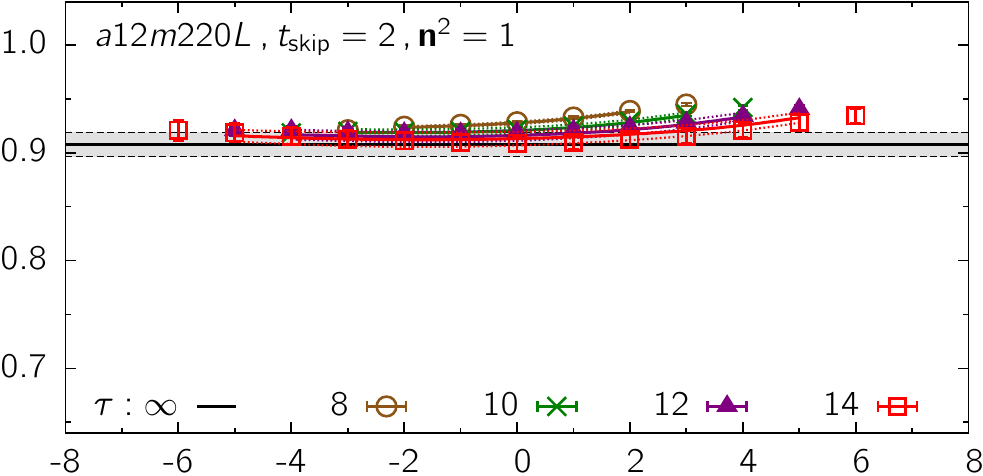}
}
\subfigure{
\includegraphics[width=0.47\linewidth]{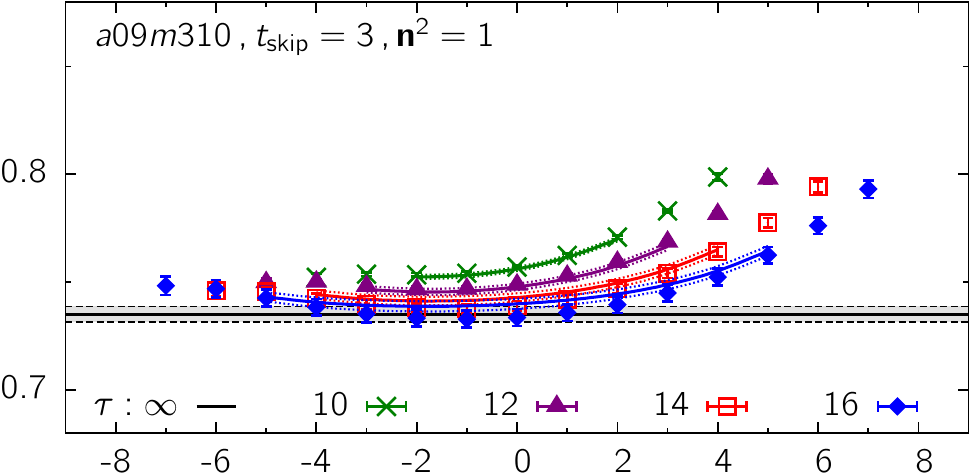}
\includegraphics[width=0.47\linewidth]{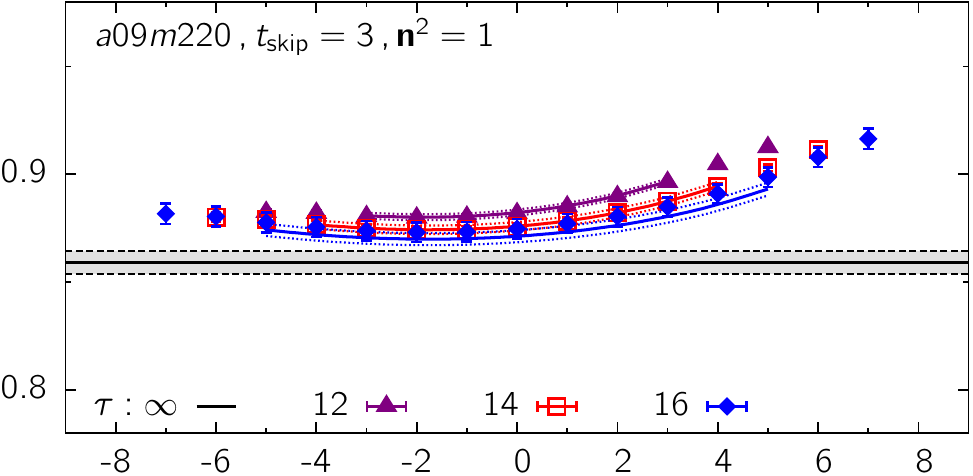}
}
\subfigure{
\includegraphics[width=0.47\linewidth]{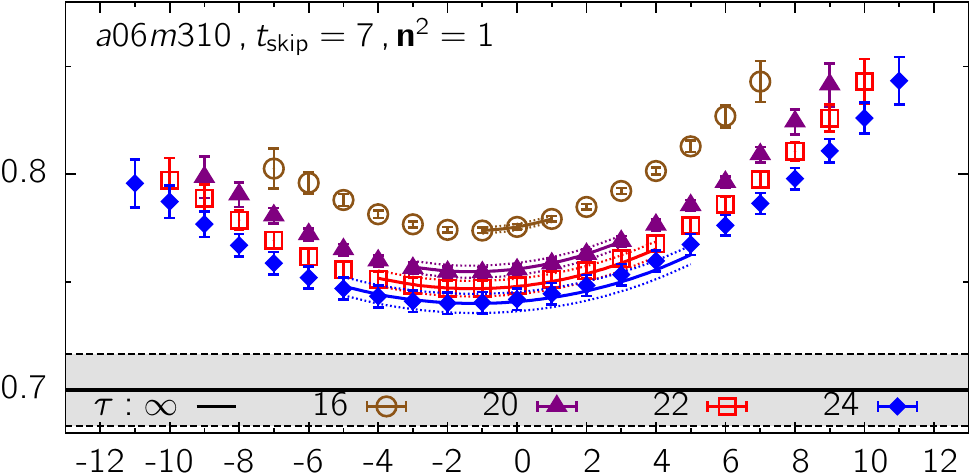}
\includegraphics[width=0.47\linewidth]{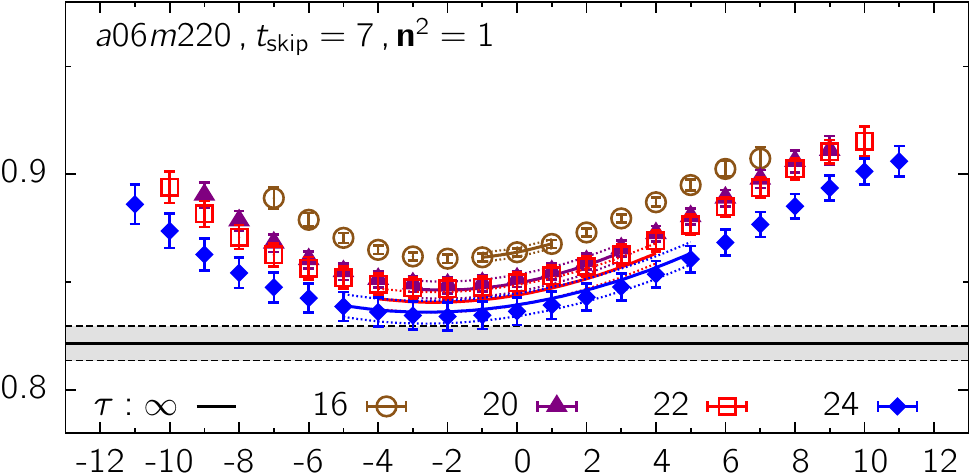}
}
\subfigure{
\includegraphics[width=0.47\linewidth]{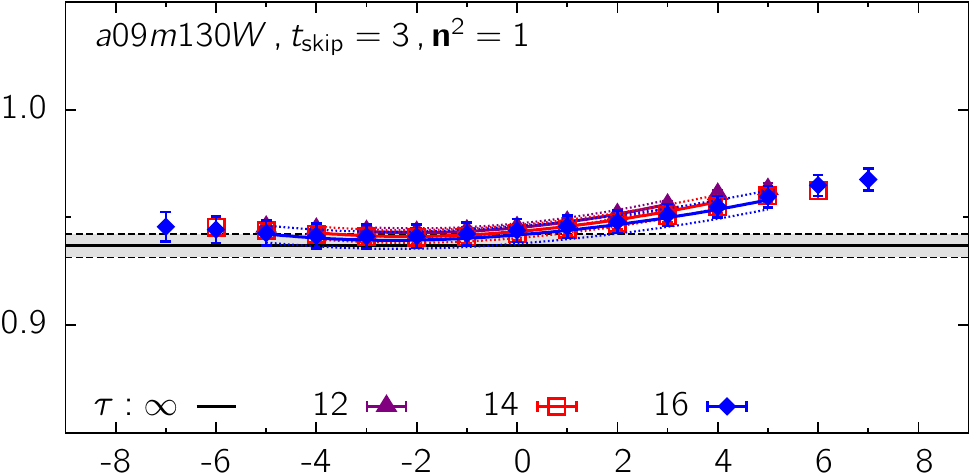}
\includegraphics[width=0.47\linewidth]{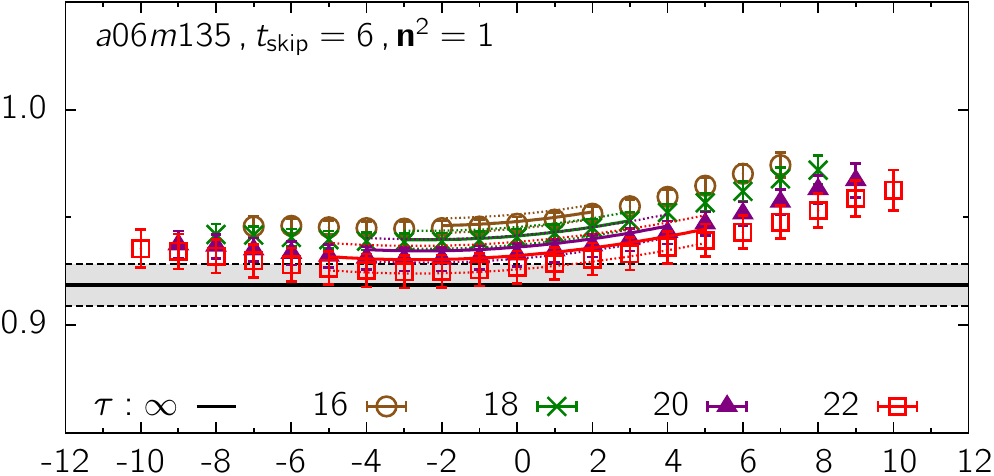}
}
\caption{\FIXME{fig:GE-ESC-p1} Data and the 3${}^\ast$-state fits to the unrenormalized
  electric form factor $G_E$ extracted from the $\Re V_4$ channel
  using Eq.~\protect\eqref{eq:GE4}.  The data with ${\bf p}^2 =
  (2\pi/La)^2$ for eight ensembles are shown as a function of $t-\tau/2$.  
  The y-axis total interval $\Delta y=0.2$ is the
  same in all the panels. The rest is the same as in
  Fig.~\protect\ref{fig:GE-ESC-a09m130W}.}
\label{fig:GE-ESC-p1}
\end{figure*}

\begin{figure*} 
\centering
\subfigure{
\includegraphics[width=0.47\linewidth]{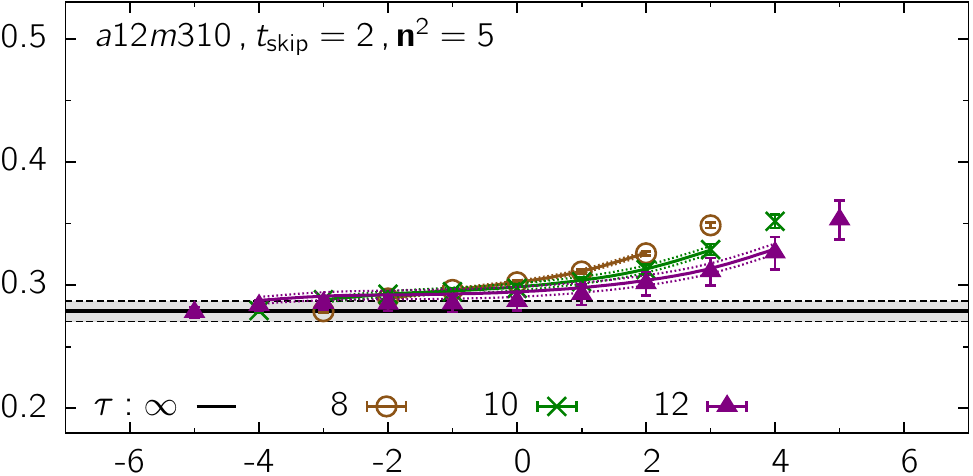}
\includegraphics[width=0.47\linewidth]{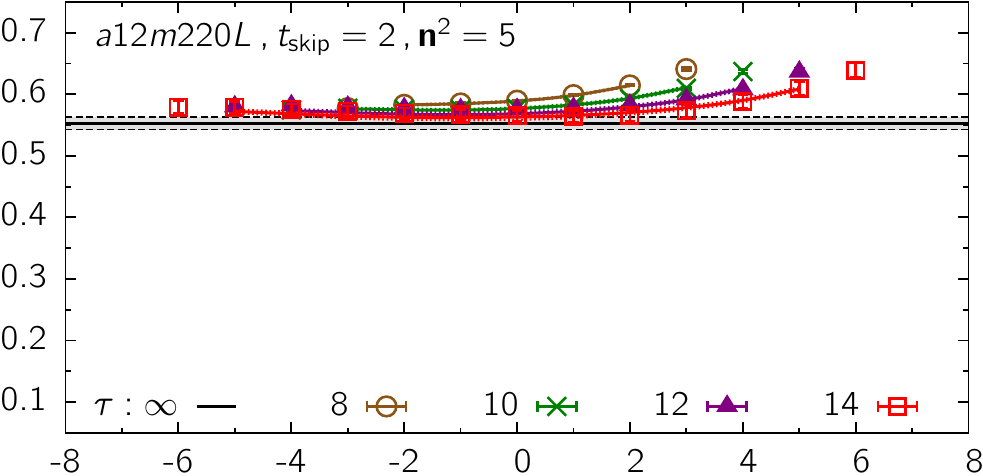}
}
\subfigure{
\includegraphics[width=0.47\linewidth]{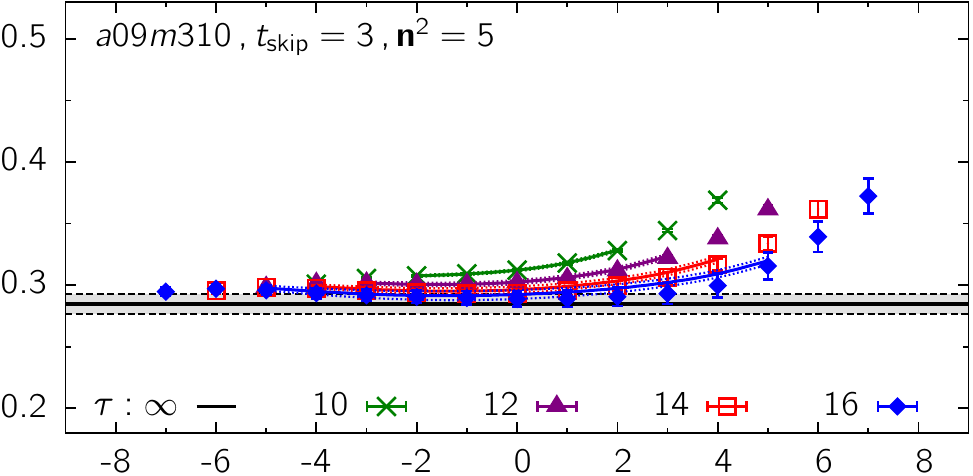}
\includegraphics[width=0.47\linewidth]{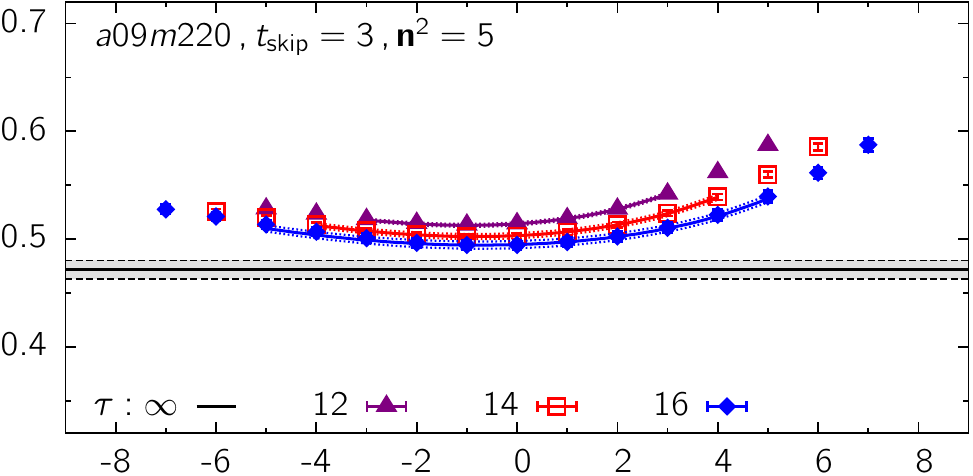}
}
\subfigure{
\includegraphics[width=0.47\linewidth]{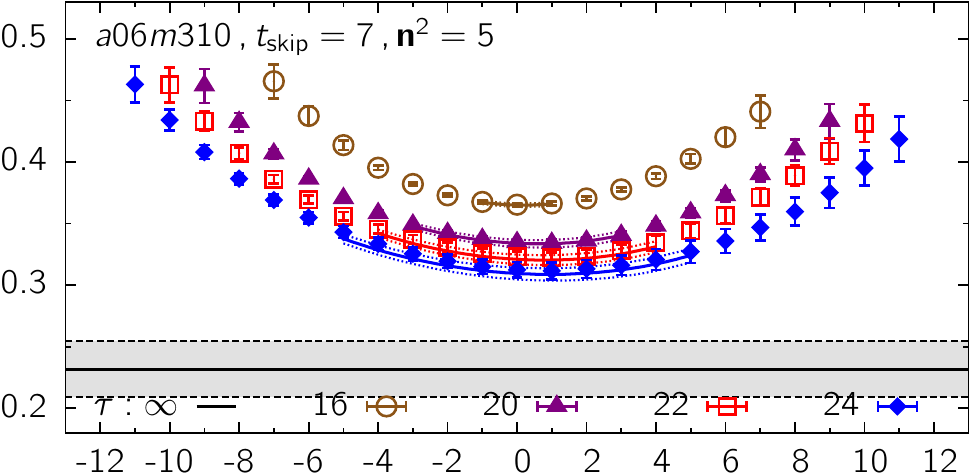}
\includegraphics[width=0.47\linewidth]{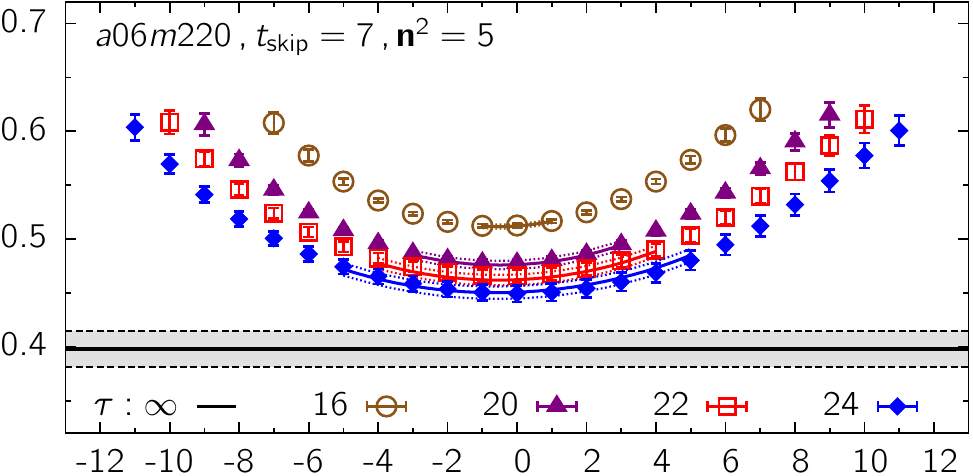}
}
\subfigure{
\includegraphics[width=0.47\linewidth]{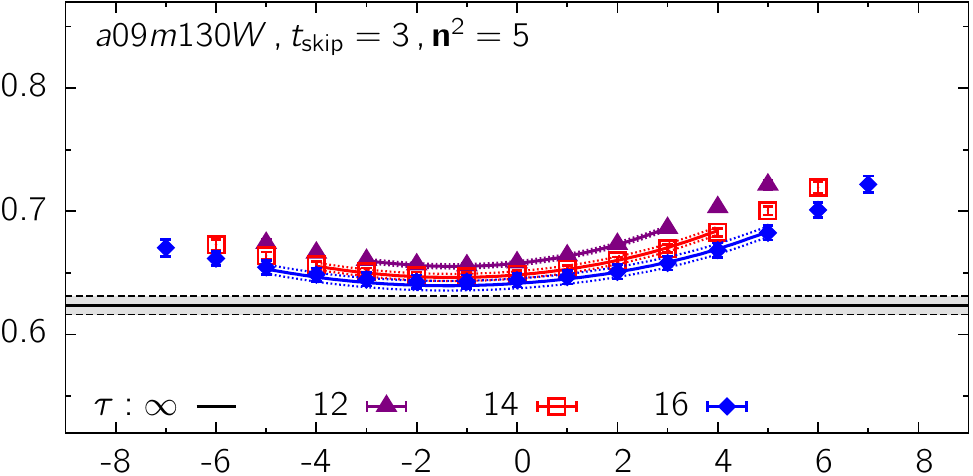}
\includegraphics[width=0.47\linewidth]{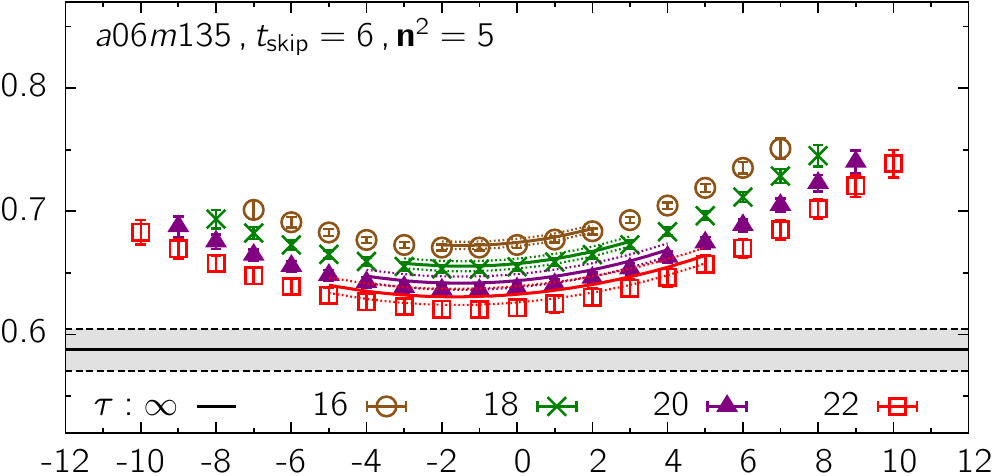}
}
\caption{\FIXME{fig:GE-ESC-p5} Data and the 3${}^\ast$-state fits to the unrenormalized isovector $G_E$ extracted
  from the $\Re V_4$ channel using Eq.~\protect\eqref{eq:GE4}.  The
  data with ${\bf p}^2 = 5 (2\pi/La)^2$ for eight ensembles are
  plotted versus $t-\tau/2$. The y-axis total interval $\Delta y=0.35$ is
  the same in all the plots.  The rest is the same as in
  Fig.~\protect\ref{fig:GE-ESC-a09m130W}.}
\label{fig:GE-ESC-p5}
\end{figure*}

\begin{figure*} 
\centering
\subfigure{
\includegraphics[width=0.47\linewidth]{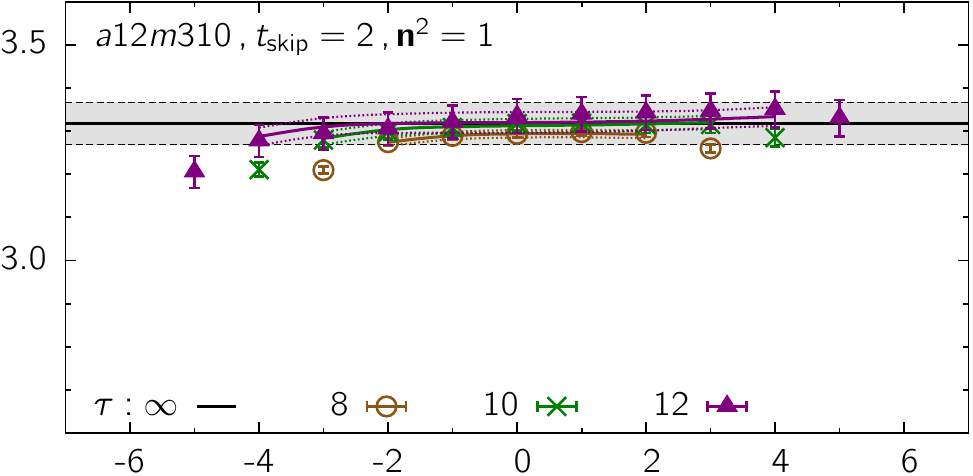}
\includegraphics[width=0.47\linewidth]{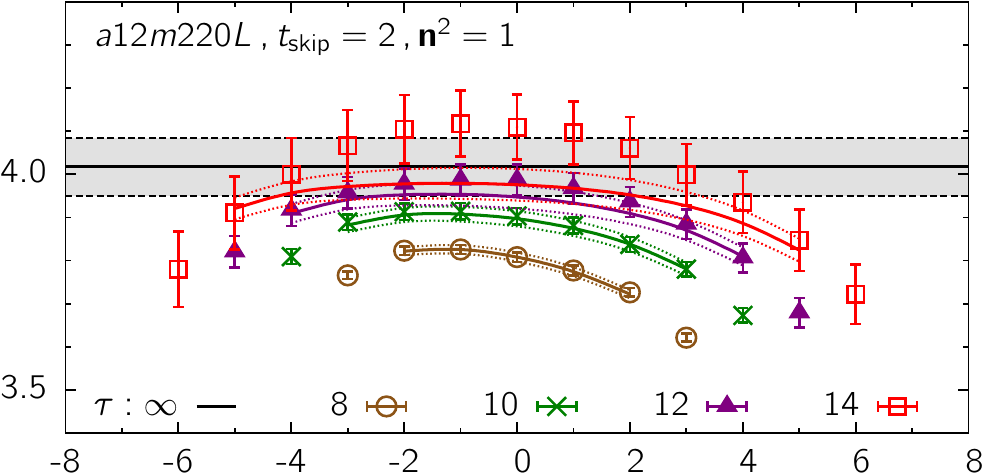}
}
\subfigure{
\includegraphics[width=0.47\linewidth]{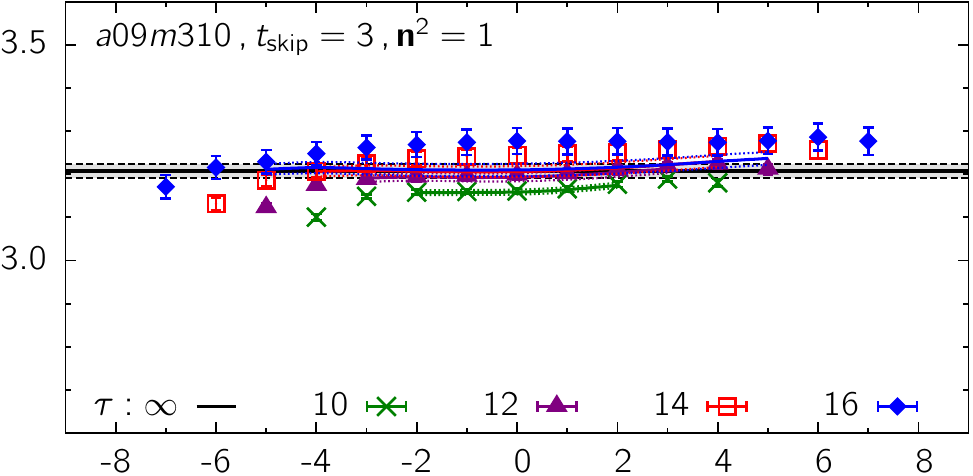}
\includegraphics[width=0.47\linewidth]{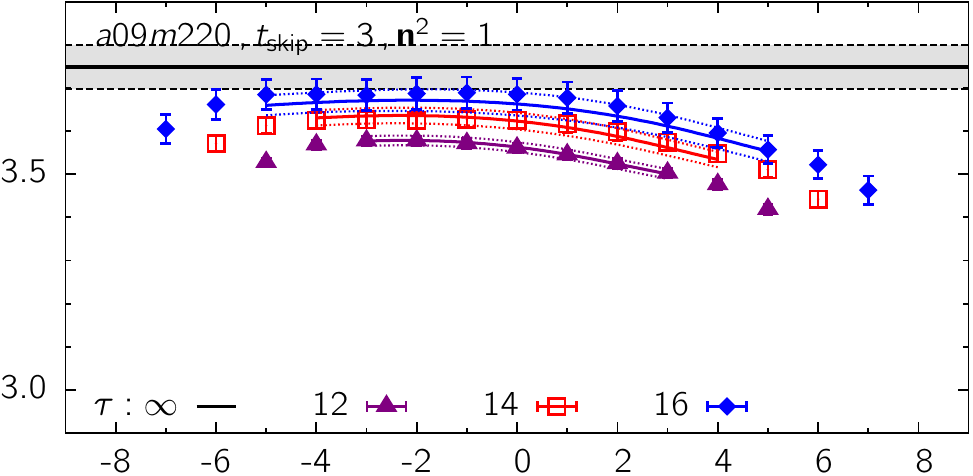}
}
\subfigure{
\includegraphics[width=0.47\linewidth]{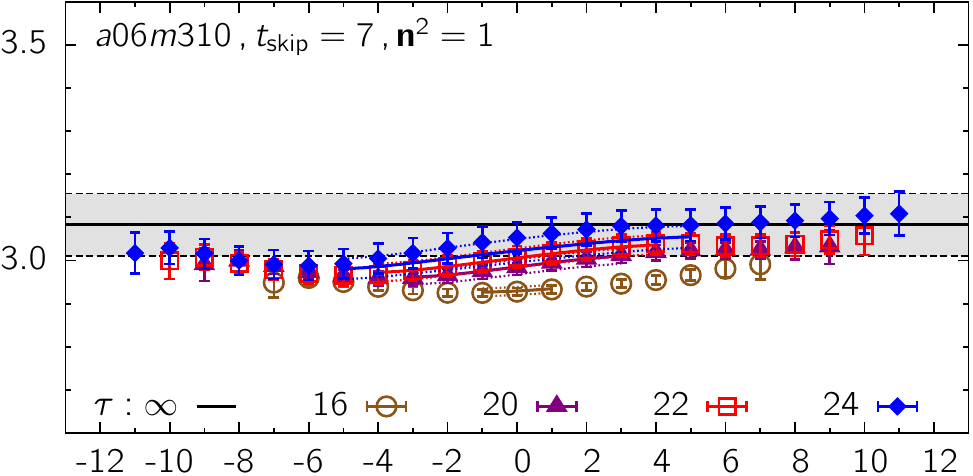}
\includegraphics[width=0.47\linewidth]{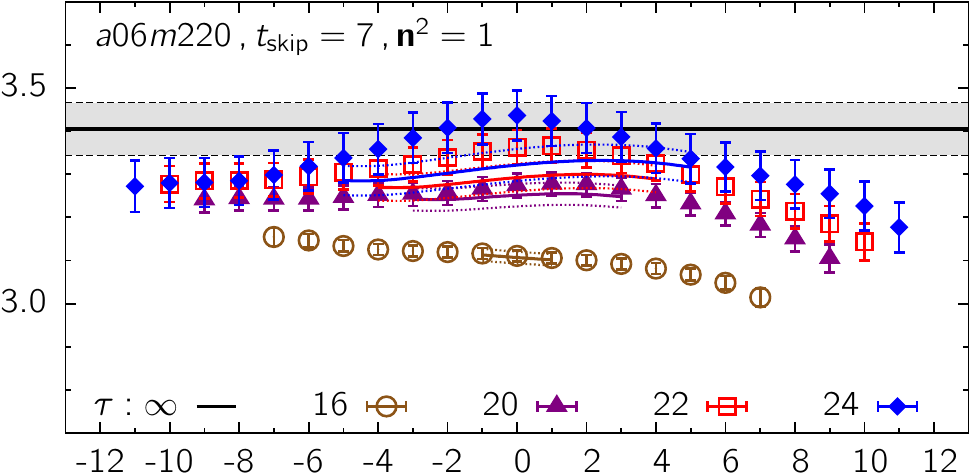}
}
\subfigure{
\includegraphics[width=0.47\linewidth]{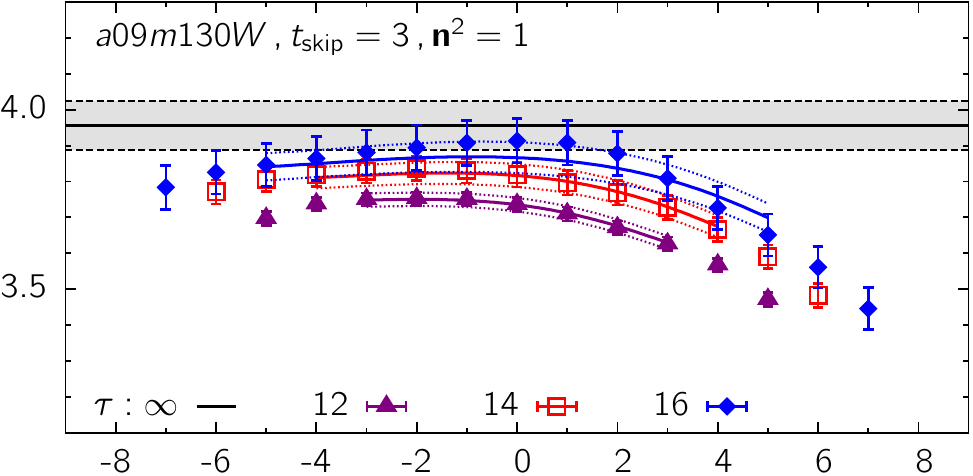}
\includegraphics[width=0.47\linewidth]{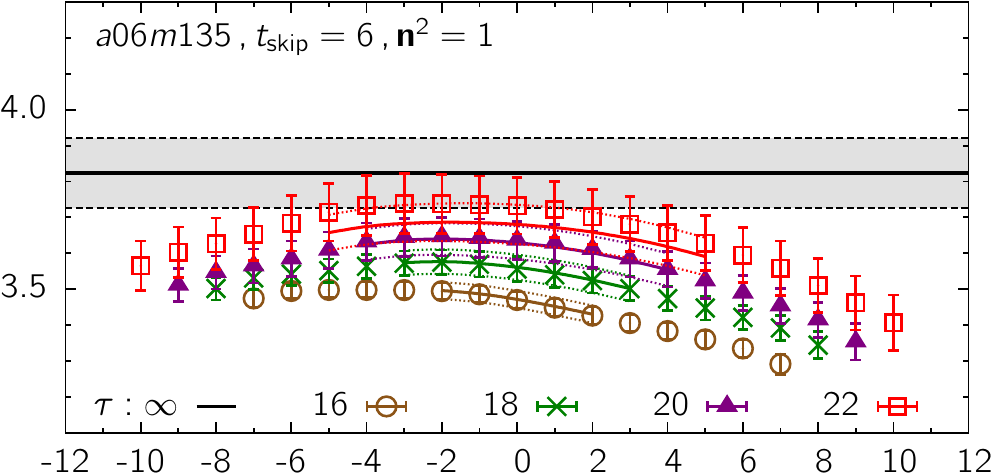}
}
\caption{\FIXME{fig:GM-ESC-p1} Data and the 3${}^\ast$-state fits to the unrenormalized
  isovector $G_M$ extracted from the $\Re V_i$ channels using
  Eq.~\protect\eqref{eq:GM1}.  The data for ${\bf p}^2 = (2\pi/La)^2$
  and for eight ensembles are plotted versus $t-\tau/2$.  The interval
  $\Delta y=1.0$ is the same for all the plots. The rest is the same
  as in Fig.~\protect\ref{fig:GE-ESC-a09m130W}. The size of the ESC is
  observed to increase as $M_\pi$ is decreased. }
\label{fig:GM-ESC-p1}
\end{figure*}

\begin{figure*} 
\centering
\subfigure{
\includegraphics[width=0.47\linewidth]{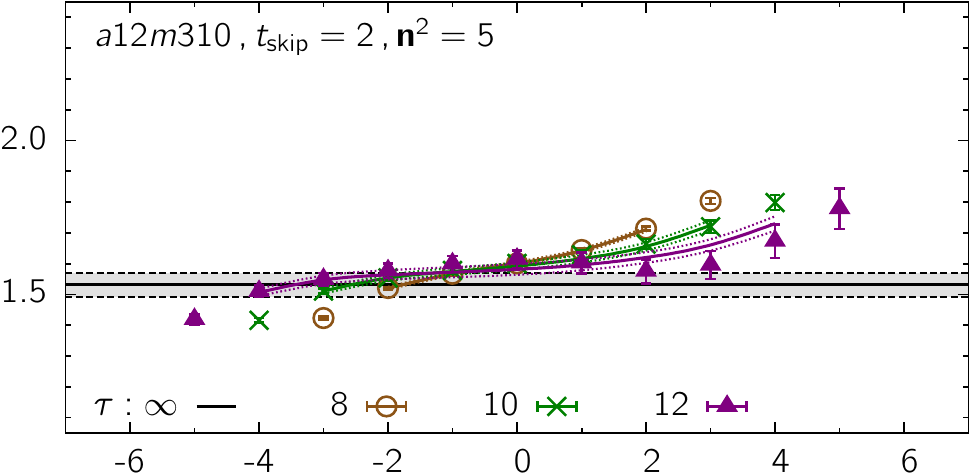}
\includegraphics[width=0.47\linewidth]{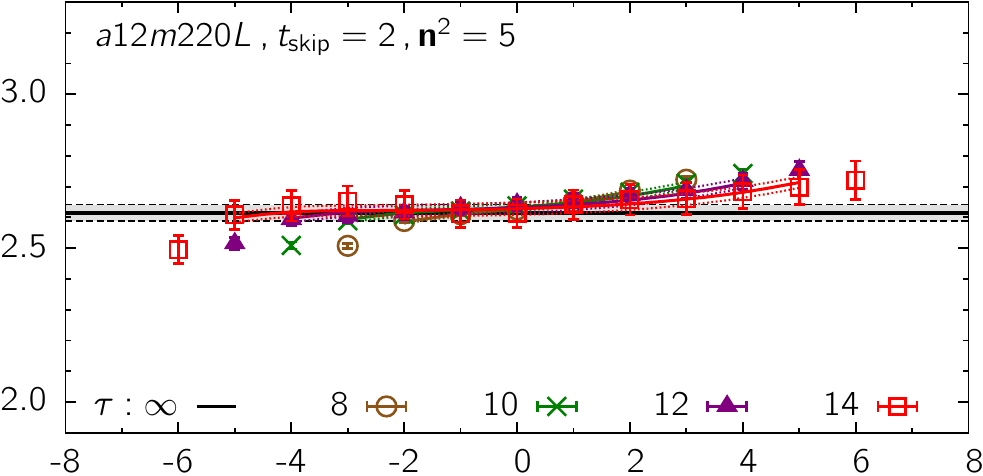}
}
\subfigure{
\includegraphics[width=0.47\linewidth]{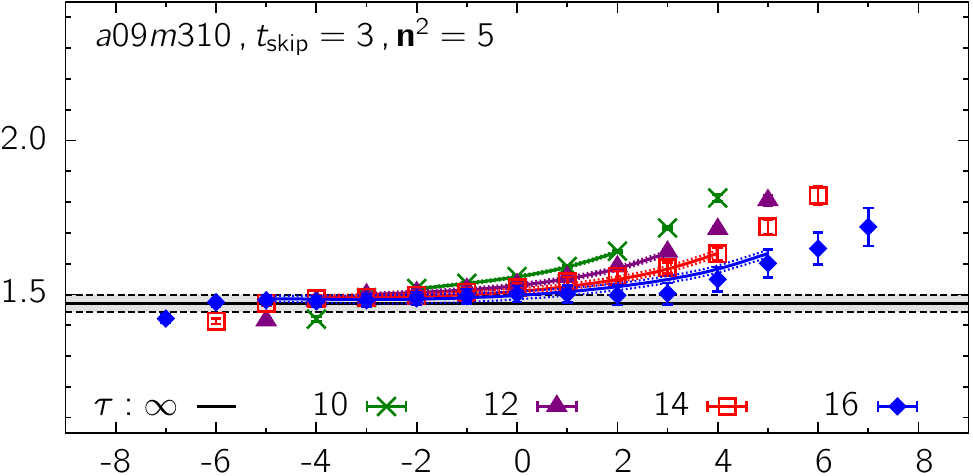}
\includegraphics[width=0.47\linewidth]{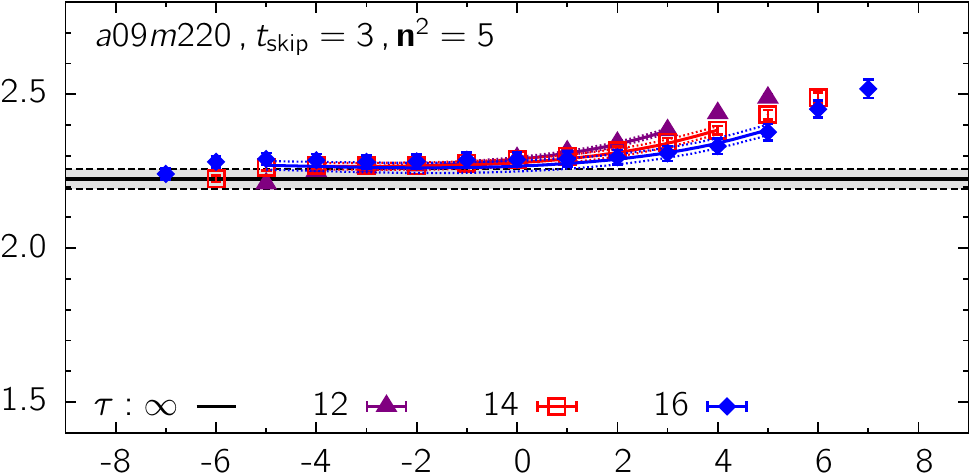}
}
\subfigure{
\includegraphics[width=0.47\linewidth]{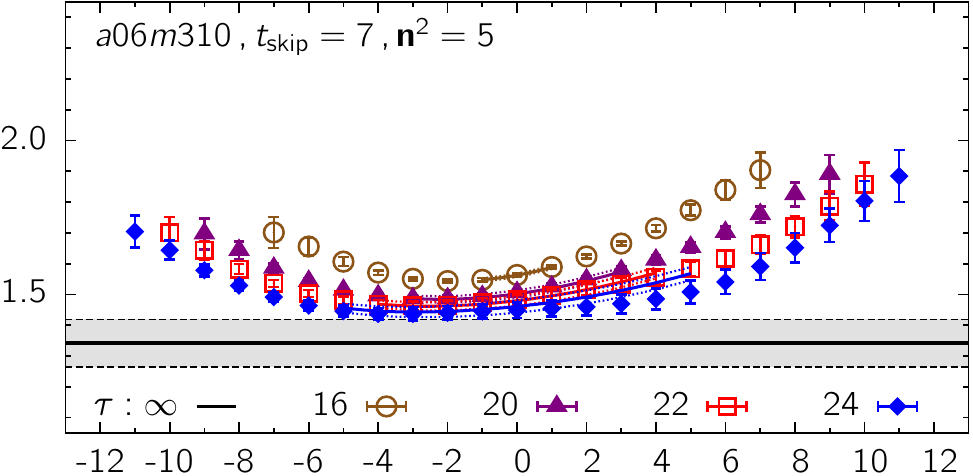}
\includegraphics[width=0.47\linewidth]{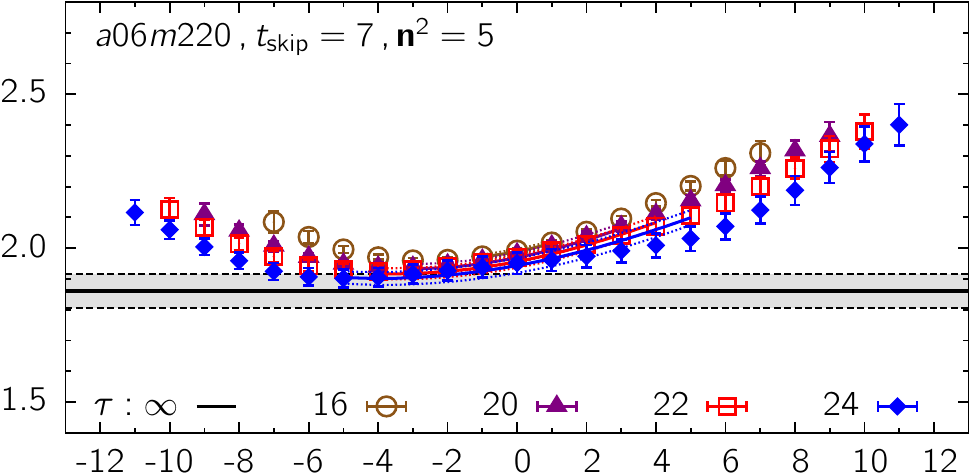}
}
\subfigure{
\includegraphics[width=0.47\linewidth]{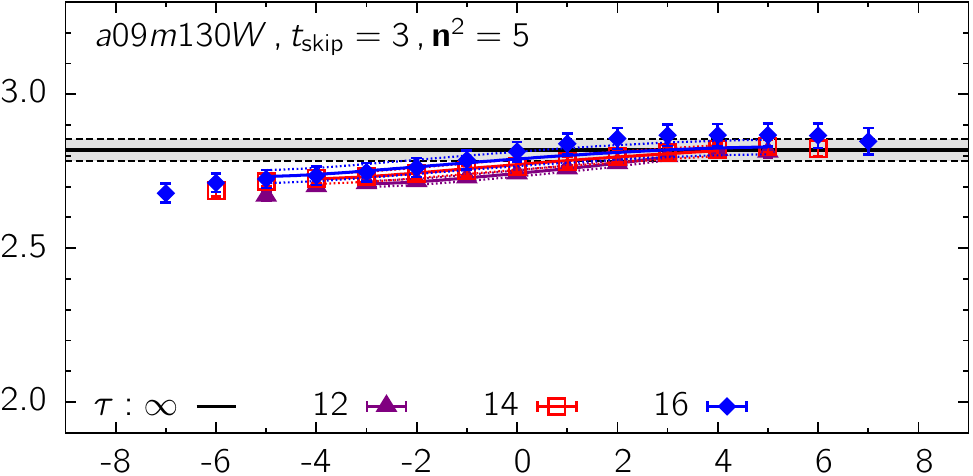}
\includegraphics[width=0.47\linewidth]{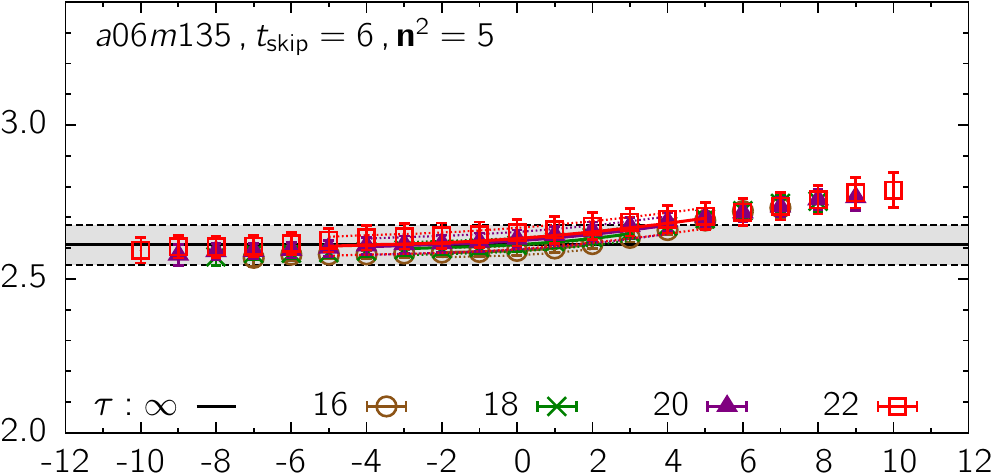}
}
\caption{\FIXME{fig:GM-ESC-p5} Data and the 3${}^\ast$-state fits to the unrenormalized
  isovector $G_M$ extracted from the $\Re V_i$ channels using
  Eq.~\protect\eqref{eq:GM1}.  The data for ${\bf p}^2 = 5
  (2\pi/La)^2$ and for eight ensembles are plotted versus $t-\tau/2$.
  The y-axis total interval $\Delta y=1.4$ is the same for all the
  plots. The rest is the same as in
  Fig.~\protect\ref{fig:GE-ESC-a09m130W}. The pattern of the ESC
  changes with momentum as can be seen by comparing with
  Fig.~\protect\ref{fig:GM-ESC-p1}.  }
\label{fig:GM-ESC-p5}
\end{figure*}

\begin{figure*} 
\centering
\subfigure{
\includegraphics[width=0.32\linewidth]{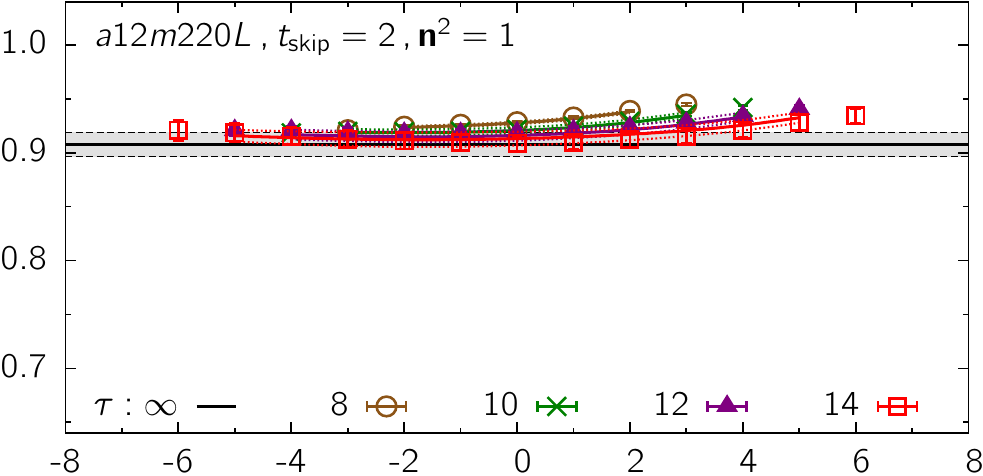}
\includegraphics[width=0.32\linewidth]{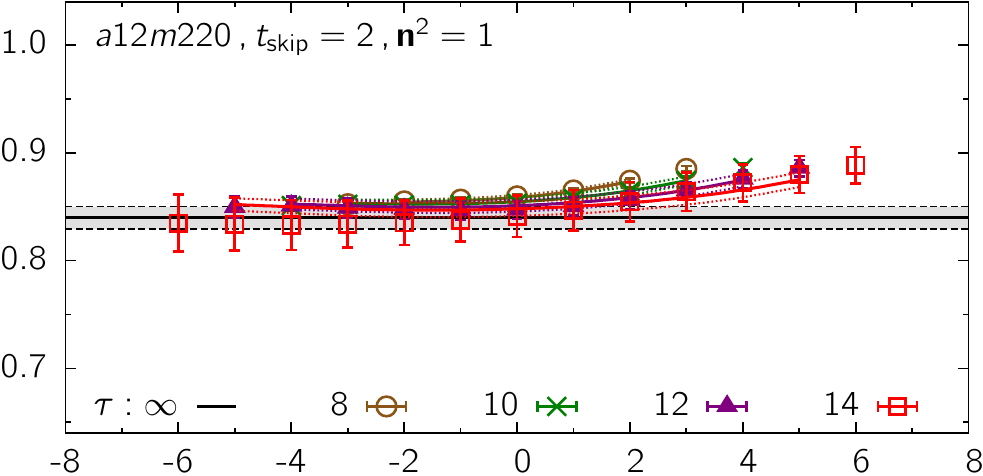}
\includegraphics[width=0.32\linewidth]{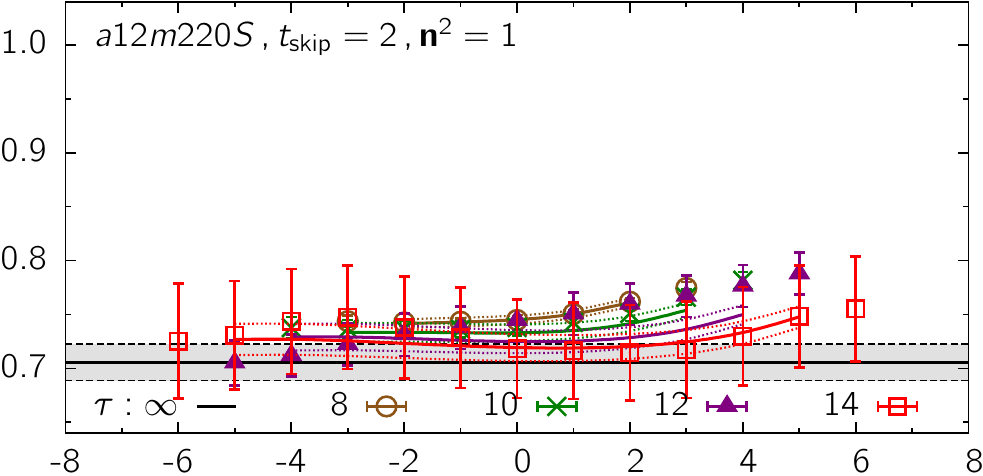}
}
\subfigure{
\includegraphics[width=0.32\linewidth]{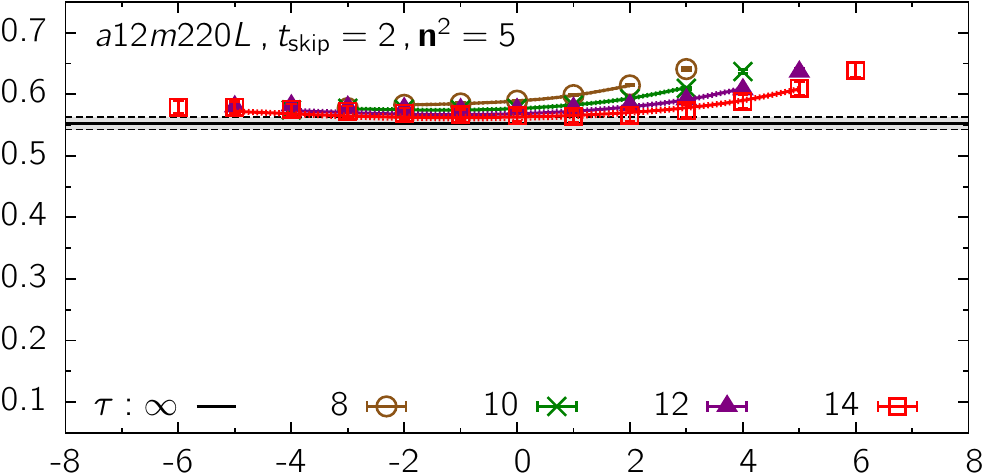}
\includegraphics[width=0.32\linewidth]{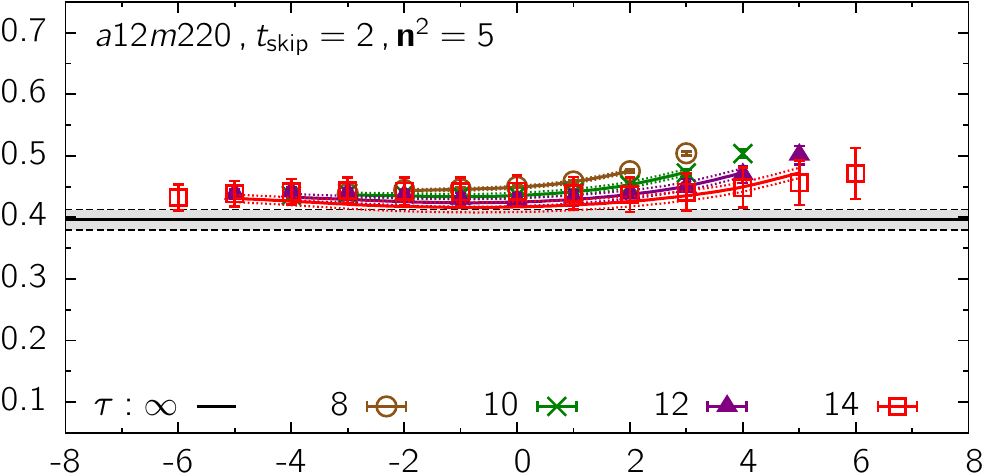}
\includegraphics[width=0.32\linewidth]{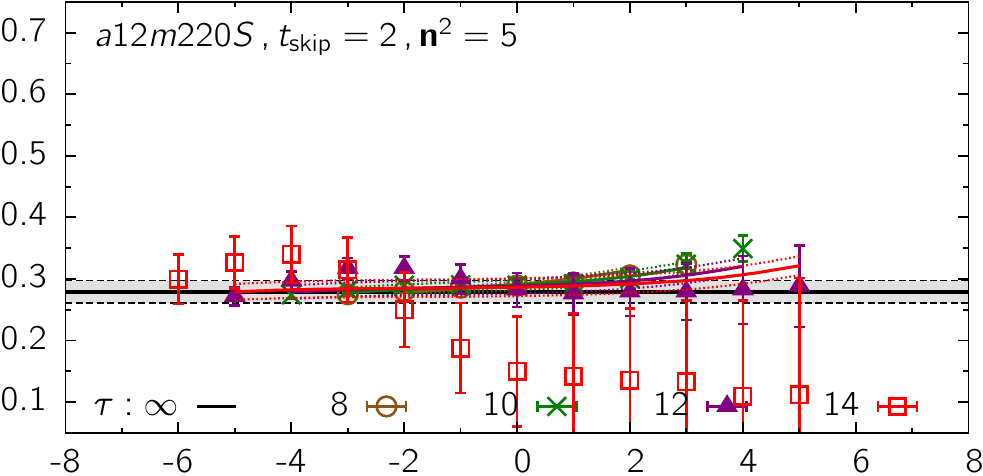}
}
\subfigure{
\includegraphics[width=0.32\linewidth]{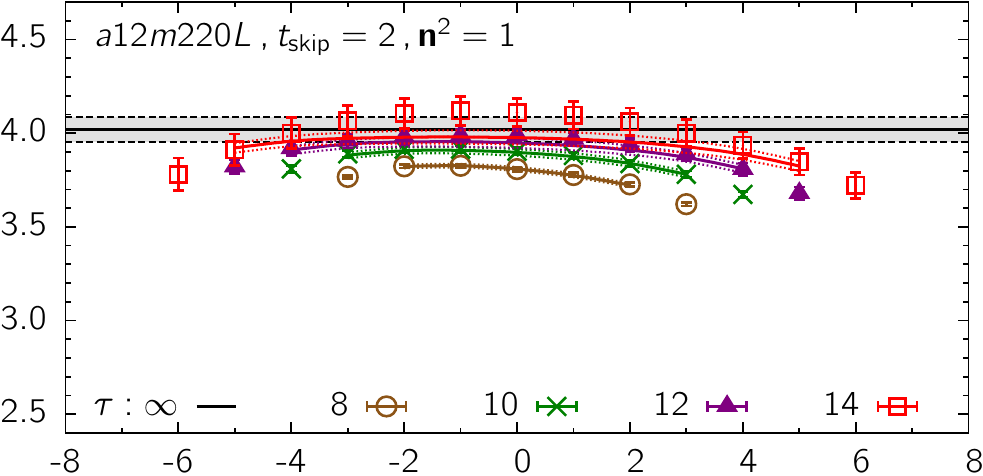}
\includegraphics[width=0.32\linewidth]{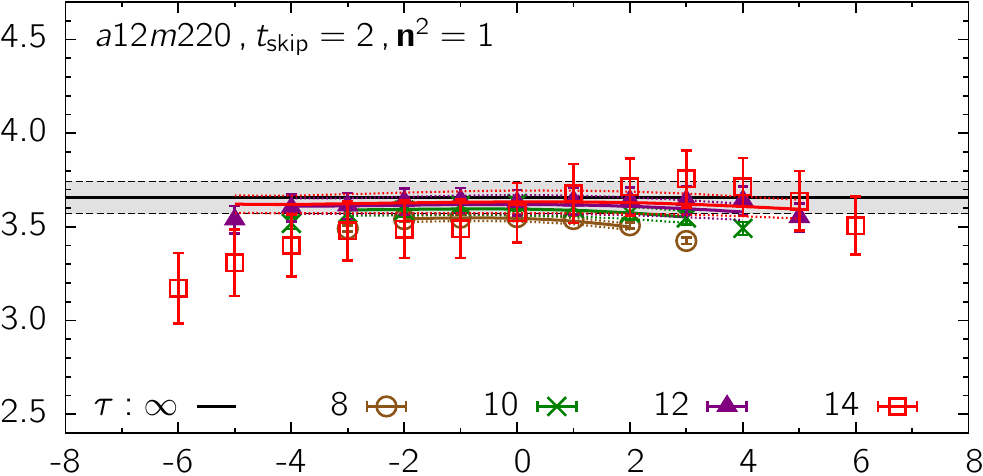}
\includegraphics[width=0.32\linewidth]{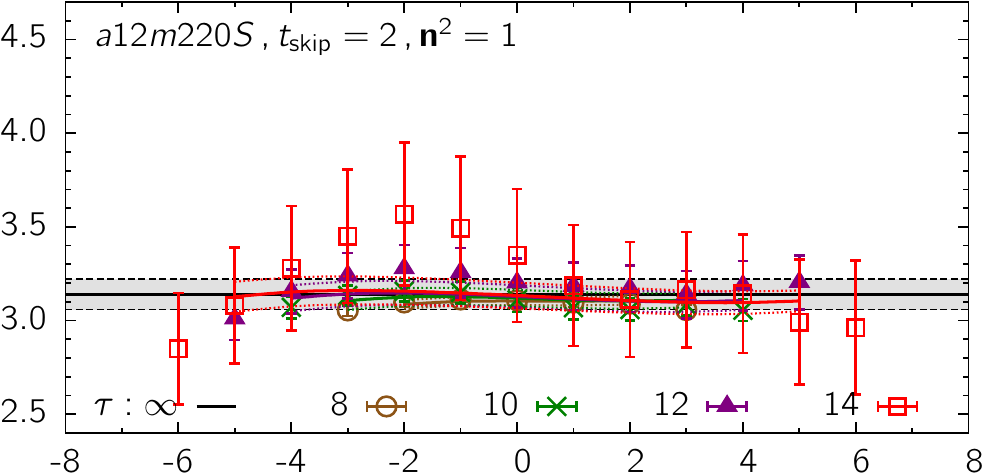}
}
\subfigure{
\includegraphics[width=0.32\linewidth]{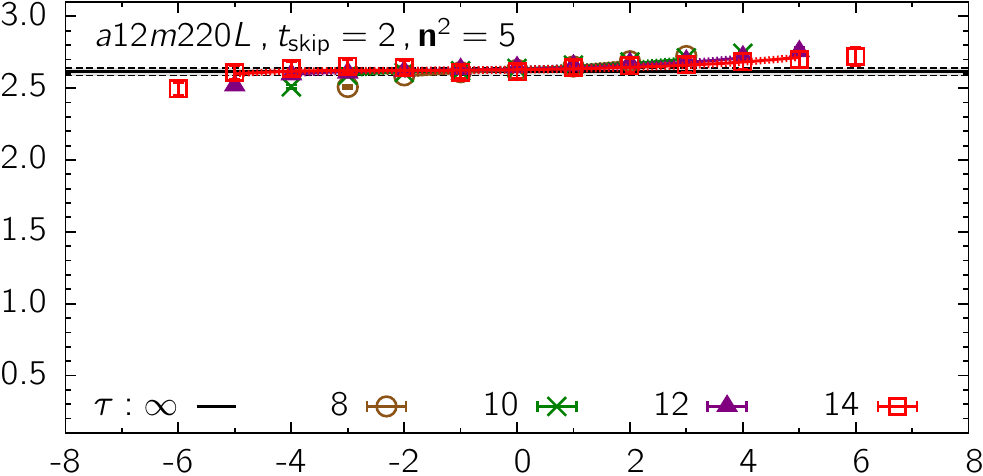}
\includegraphics[width=0.32\linewidth]{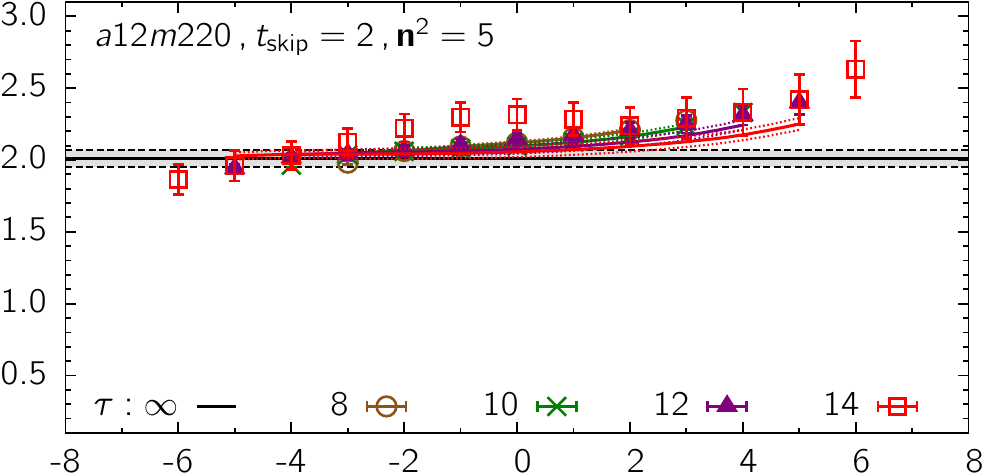}
\includegraphics[width=0.32\linewidth]{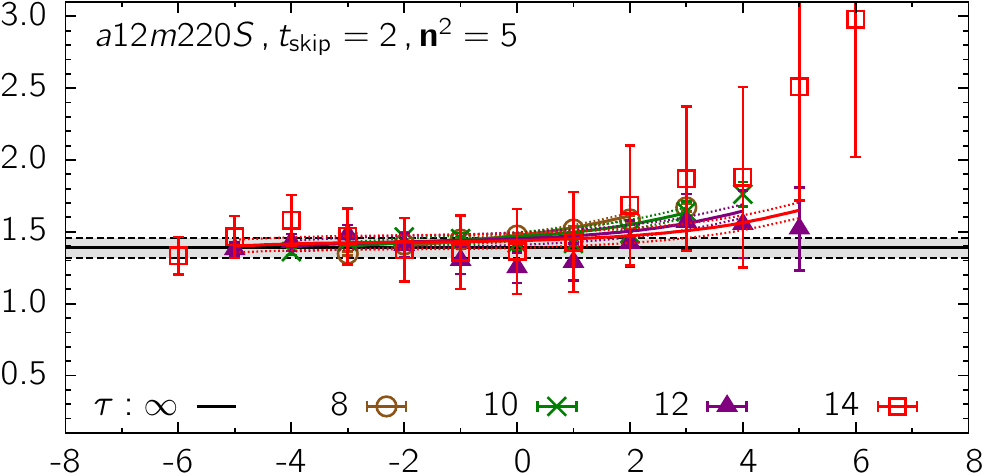}
}
\caption{\FIXME{fig:GM-ESC-vol} Data and the 3${}^\ast$-state fits to
  the unrenormalized isovector form factors $G_E^{V_4}$ and
  $G_M^{V_i}$ for the three ensembles $a12m220L$, $a12m220$ and
  $a12m220S$. The first two rows show $G_E$ for ${\bf n}^2=1$ and
  ${\bf n}^2=5$, while the last two show $G_M$. The plots for
  $a12m220L$ are the same as in
  Figs.~\ref{fig:GE-ESC-p1},~\ref{fig:GE-ESC-p5},~\ref{fig:GM-ESC-p1},
  and~\ref{fig:GM-ESC-p5}. Note that the data for fixed ${\bf n}$ but
  different $L$ cannot be compared since $Q^2$, and thus the value of
  the form factor, changes with the lattice size $L$. For fixed $a$
  and $M_\pi$, the data shifts to smaller values of $Q^2$ for a given
  $n^2$ as listed in Table~\protect\ref{tab:Q2-ALL}. Consequently, the
  quality of the signal improves with $L$. }
\label{fig:GM-ESC-vol}
\end{figure*}

\begin{figure*}[tpb] 
\centering
\subfigure{
\includegraphics[width=0.47\linewidth]{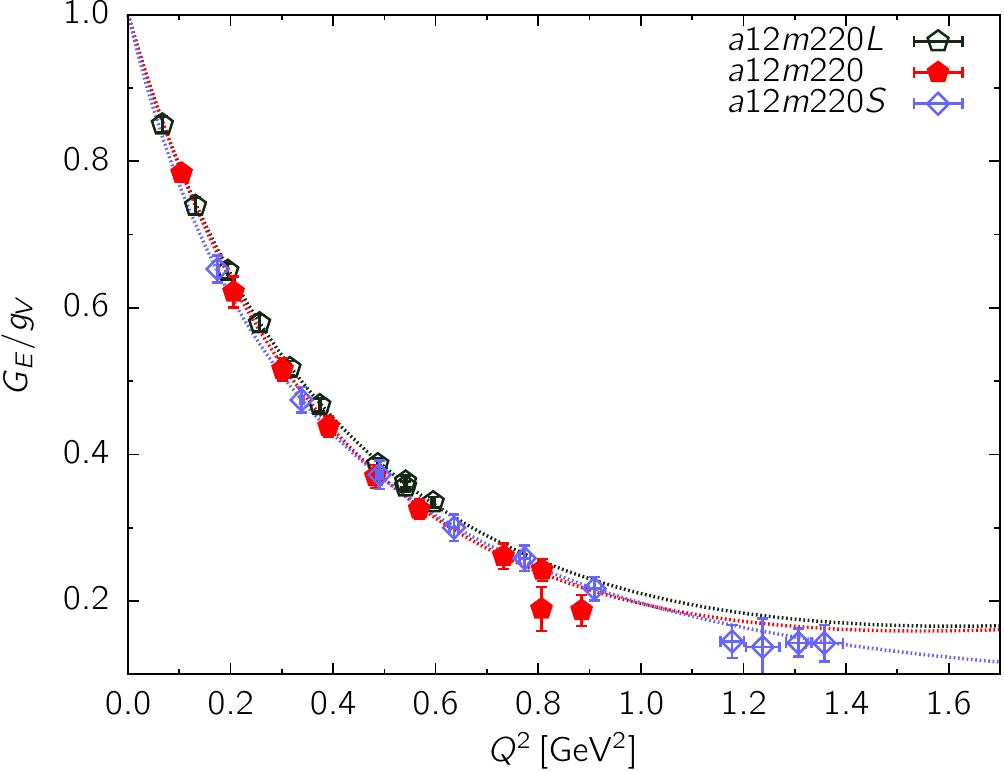} 
\includegraphics[width=0.47\linewidth]{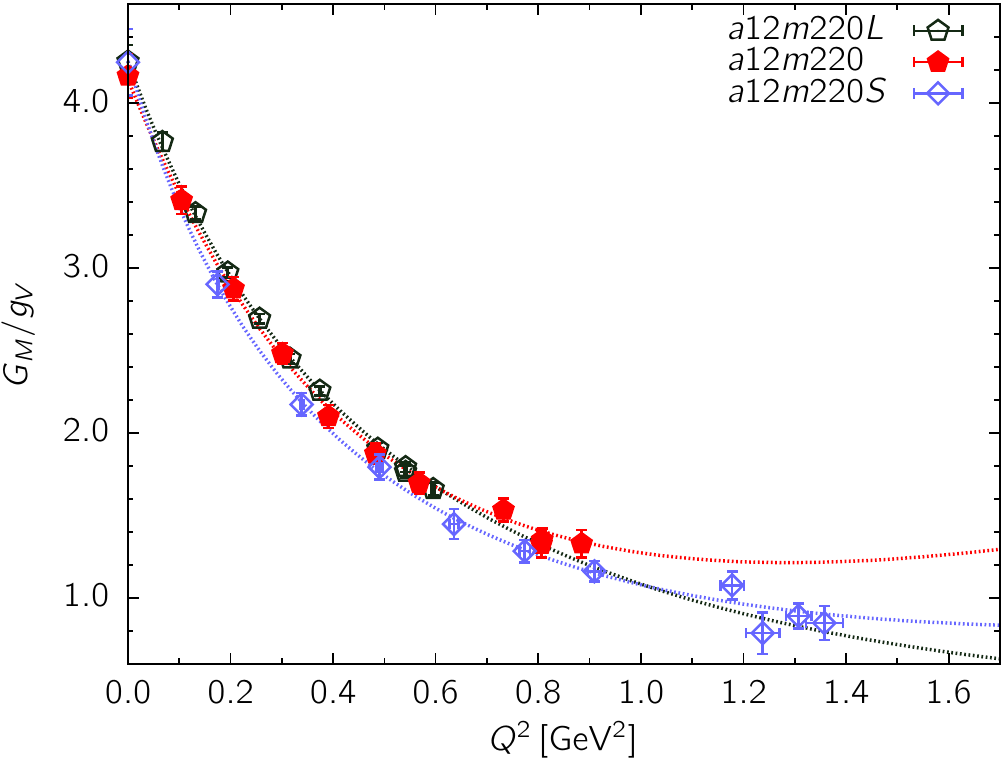}
}
\caption{\FIXME{fig:GE-GM-vsV} The data for the renormalized electric
  (left) and magnetic (right) form factors from the $a12m220S$,
  $a12m220$ and $a12m220L$ ensembles are plotted versus $Q^2$ to
  investigate possible dependence on the lattice volume. The dotted-dashed lines
  show the $z^4$ fits. For both form factors, the differences between
  the $a12m220$ ($32^3 \times 64$) and $a12m220L$ ($40^3 \times 64$)
  ensemble data are within the statistical uncertainty. The
  $G_M(Q^2)/g_V$ data from the smallest ($24^3 \times 64$) volume
  show a roughly $1\sigma$ difference. }
\label{fig:GE-GM-vsV}
\end{figure*}

\begin{figure*}   
\centering
\subfigure{
\includegraphics[height=1.39in,trim={0.0cm 0.03cm 0 0},clip]{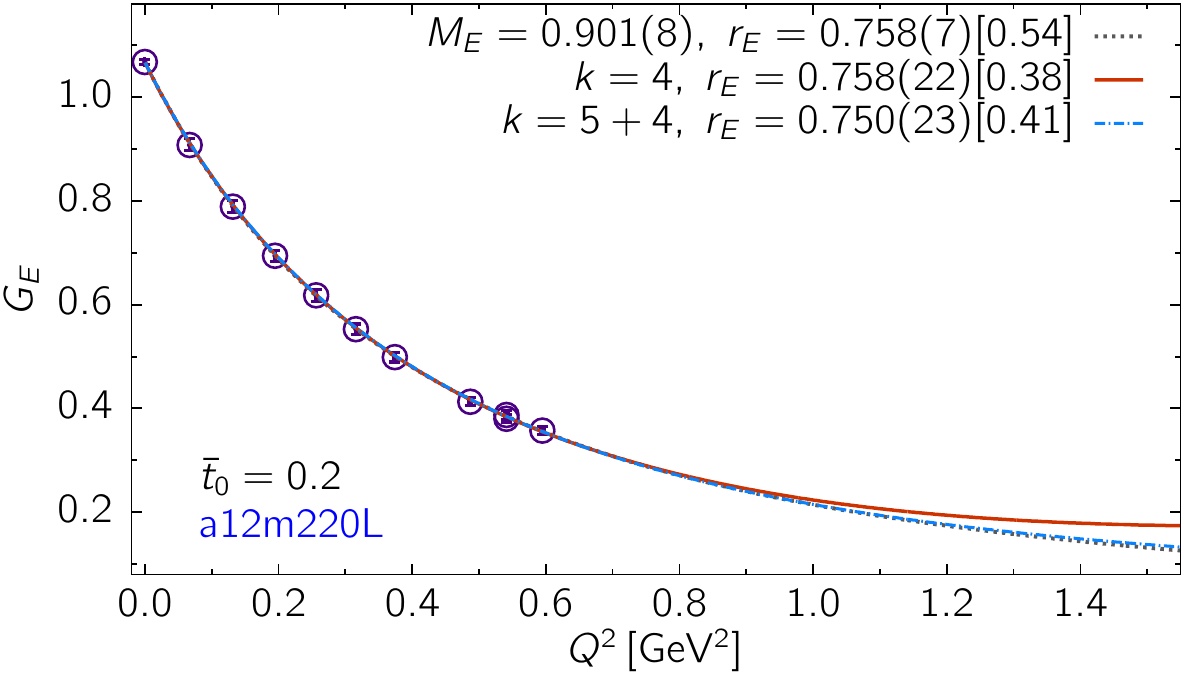}
\includegraphics[height=1.39in,trim={1.4cm 0.03cm 0 0},clip]{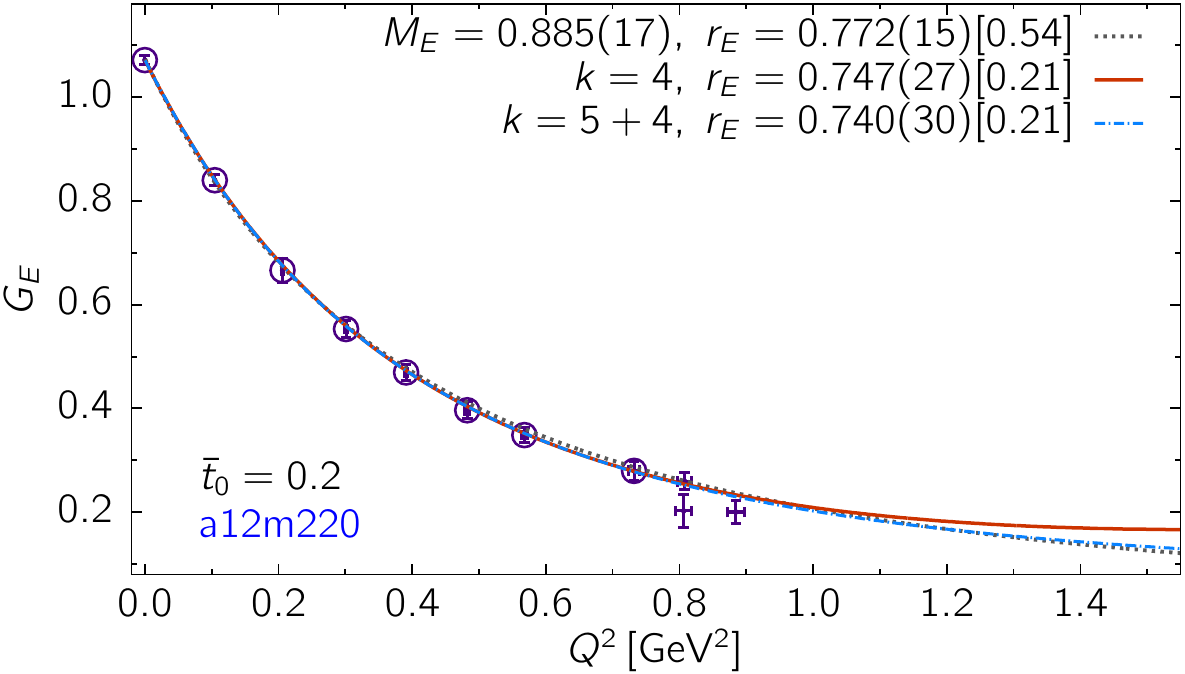}
\includegraphics[height=1.39in,trim={1.4cm 0.03cm 0 0},clip]{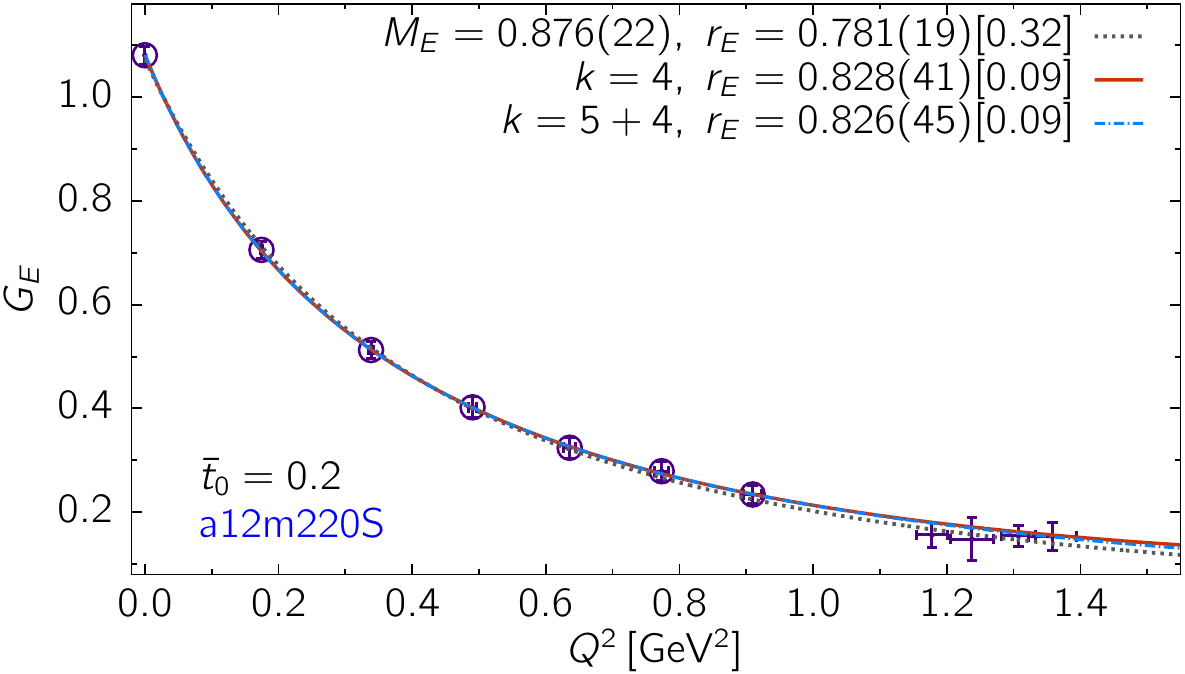}
}
\subfigure{
\includegraphics[height=1.39in,trim={0.0cm 0.03cm 0 0},clip]{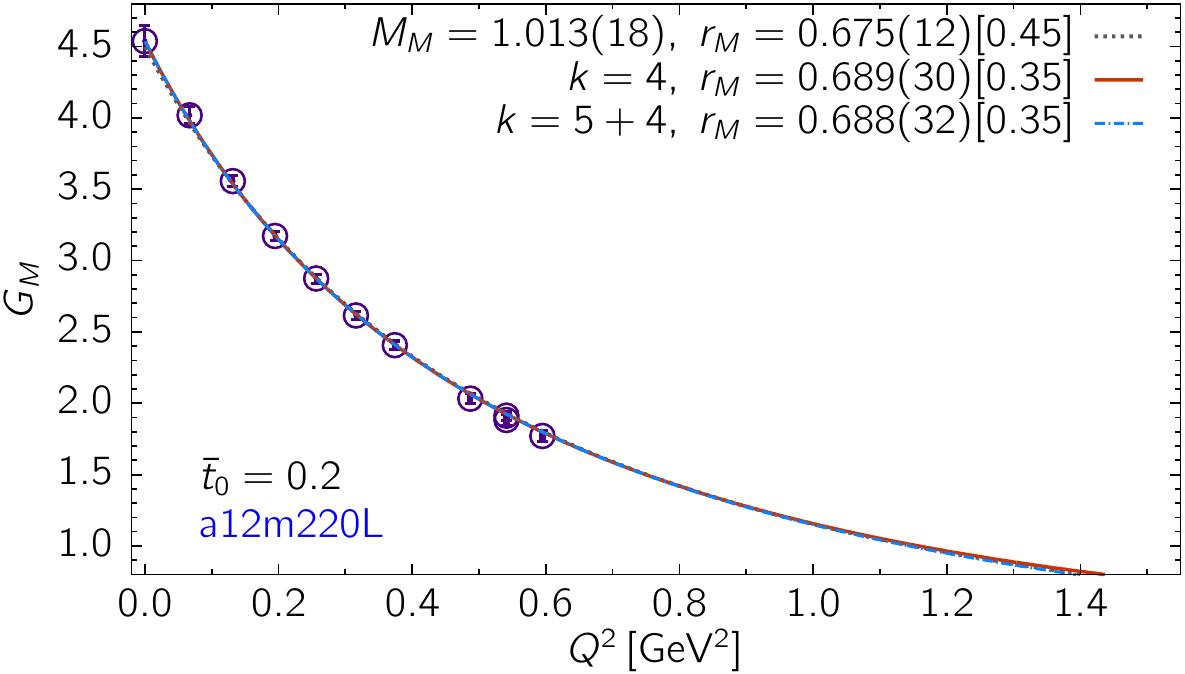}
\includegraphics[height=1.39in,trim={1.4cm 0.03cm 0 0},clip]{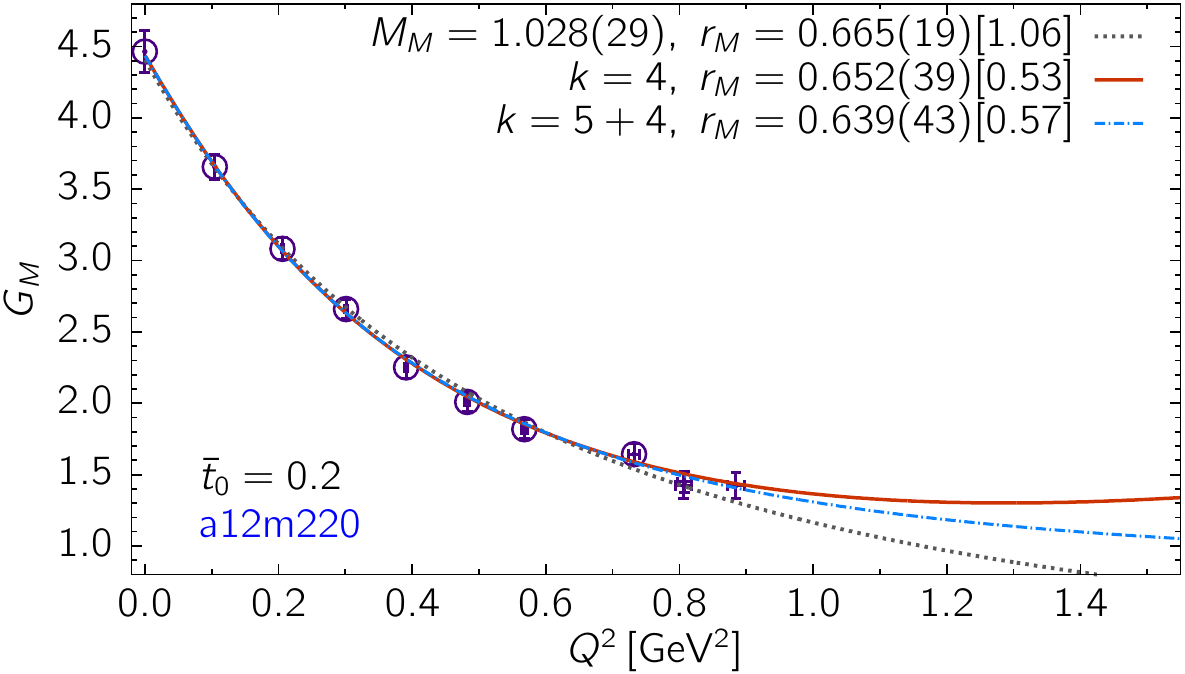}
\includegraphics[height=1.39in,trim={1.4cm 0.03cm 0 0},clip]{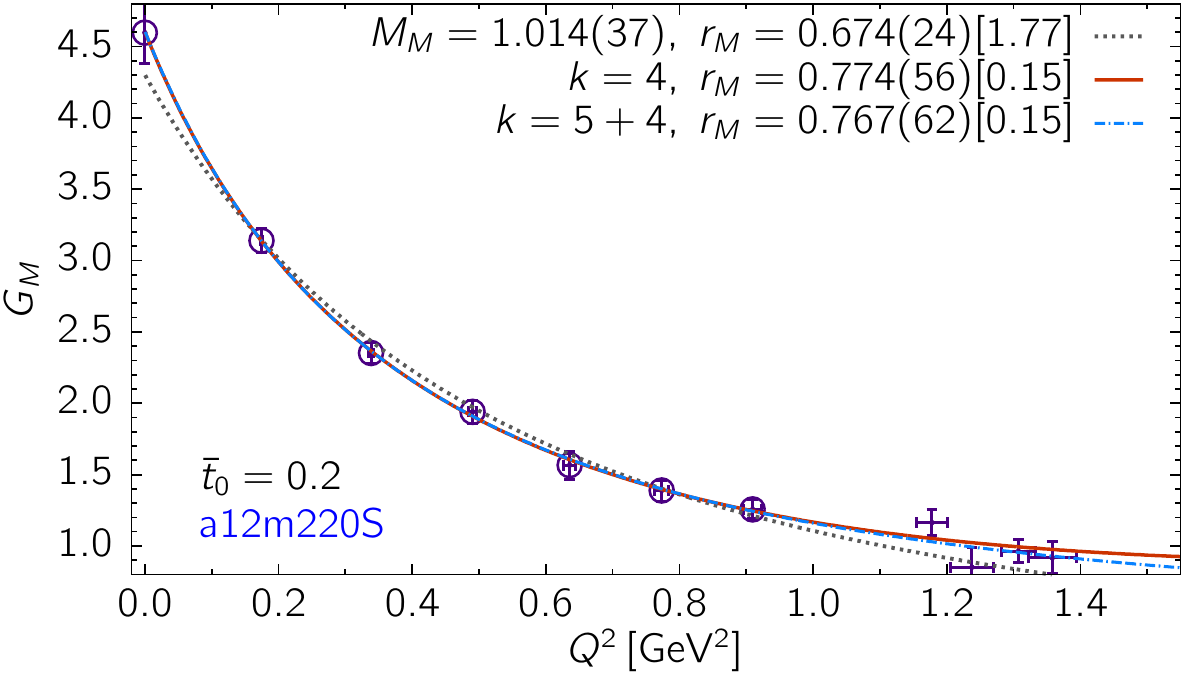}
}
\caption{\FIXME{fig:Vol-rErM} Comparison of results of the dipole,
  $z^4$ and $z^{5+4}$ fits to the unrenormalized isovector form
  factors $G_E(Q^2)$ (top) and $G_M(Q^2)$ (bottom) plotted versus $Q^2$
  (GeV${}^2$). The three ensembles, 
  $a12m220L$, $a12m220$ and $a12m220S$ have  different volumes, but
  the same lattice spacing $a\approx0.12\,\fm$ and pion
  mass $M_\pi \approx 220\,\MeV$. 
  The radii $\rE$ and $\rM$ are in units of fm and the masses $M_E$ and $M_N$ in GeV.}
\label{fig:Vol-rErM}
\end{figure*}

\begin{figure*} 
\centering
\subfigure{
\includegraphics[width=0.42\linewidth]{Figs_esc/ratio_V4_qsq1_a06m310}
\includegraphics[width=0.42\linewidth]{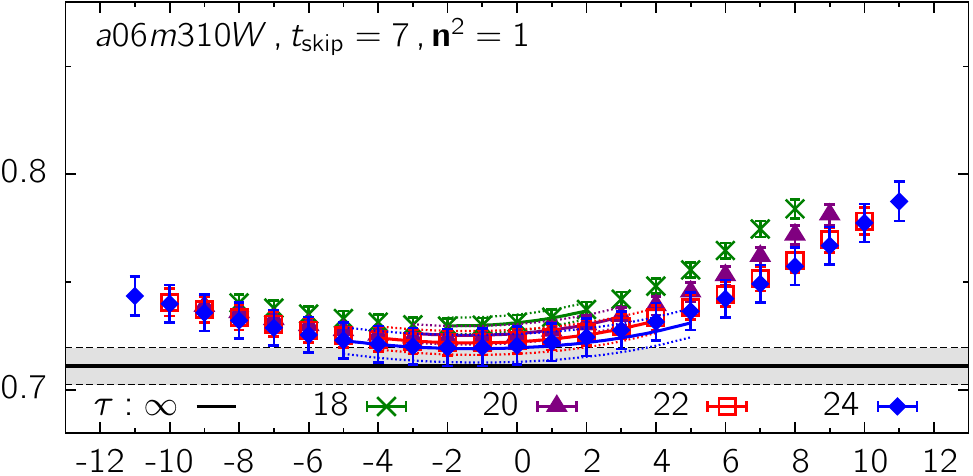}
}
\subfigure{
\includegraphics[width=0.42\linewidth]{Figs_esc/ratio_V4_qsq5_a06m310}
\includegraphics[width=0.42\linewidth]{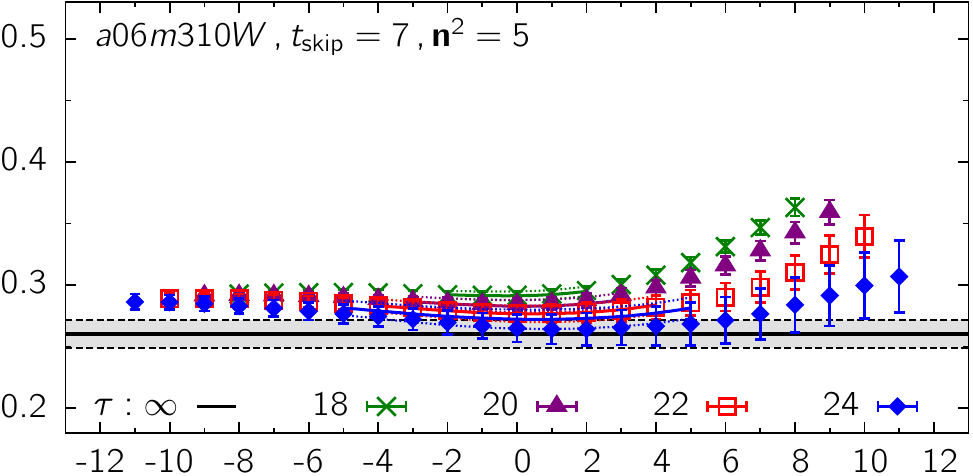}
}
\subfigure{
\includegraphics[width=0.42\linewidth]{Figs_esc/ratio_Vi_qsq1_a06m310}
\includegraphics[width=0.42\linewidth]{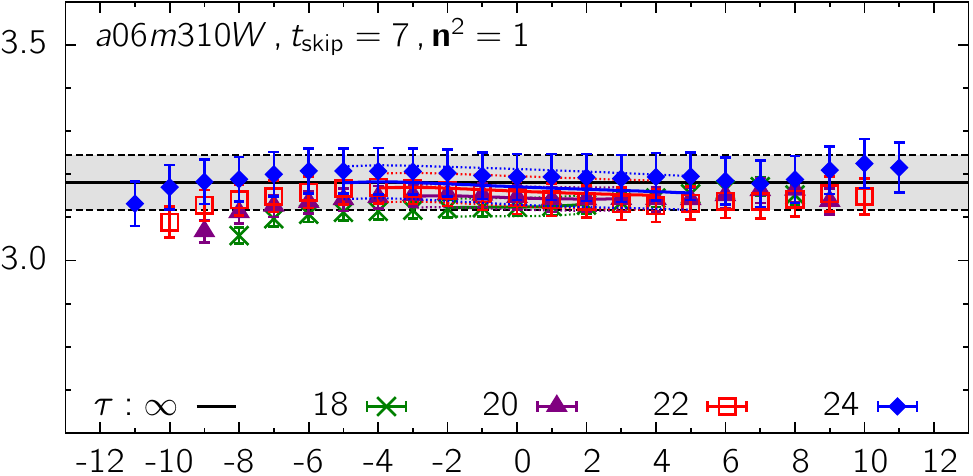}
}
\subfigure{
\includegraphics[width=0.42\linewidth]{Figs_esc/ratio_Vi_qsq5_a06m310}
\includegraphics[width=0.42\linewidth]{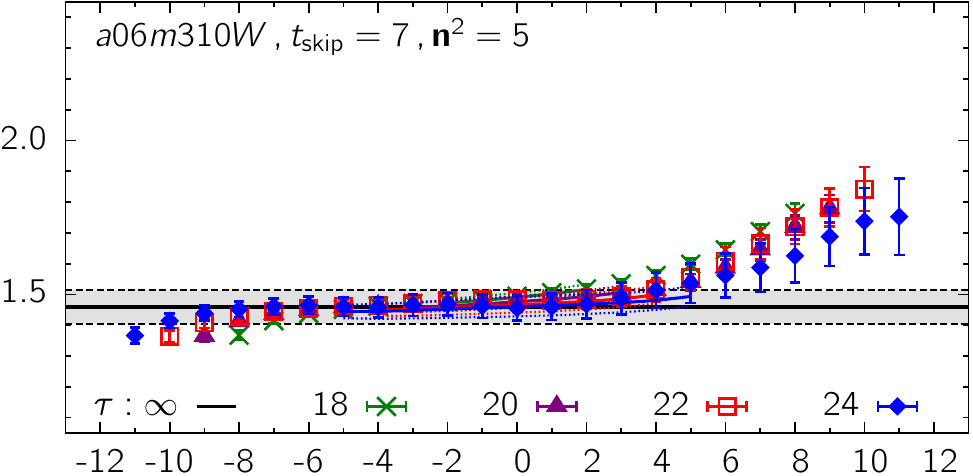}
}
\caption{\FIXME{fig:GM-ESC-smear310} Comparison of the signal and ESC
  versus the smearing size on the $a06m310$ ensemble. The top two rows
  show data for the unrenormalized isovector $G_E^{V_4}$ and the
  bottom two rows show $G_M^{V_i}$.  Plots on the left are with the
  smearing parameter $\sigma = 6.5$ and on the right with $\sigma=12$
  as defined in Table~\protect\ref{tab:cloverparams}. Plots in the
  first and third rows show data with ${\bf n}^2=1$ and in the second
  and fourth row with ${\bf n}^2=5$.}
\label{fig:GM-ESC-smear310}
\end{figure*}

\begin{figure*} 
\centering
\subfigure{
\includegraphics[width=0.42\linewidth]{Figs_esc/ratio_V4_qsq1_a06m220}
\includegraphics[width=0.42\linewidth]{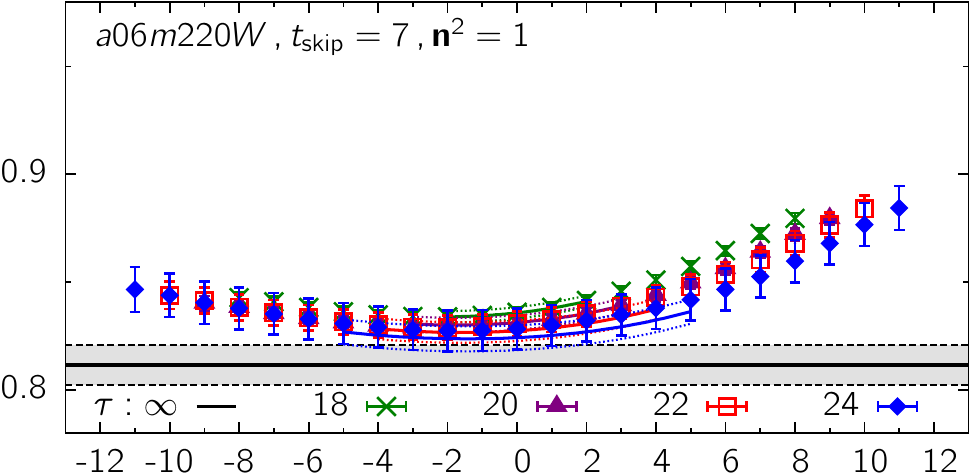}
}
\subfigure{
\includegraphics[width=0.42\linewidth]{Figs_esc/ratio_V4_qsq5_a06m220}
\includegraphics[width=0.42\linewidth]{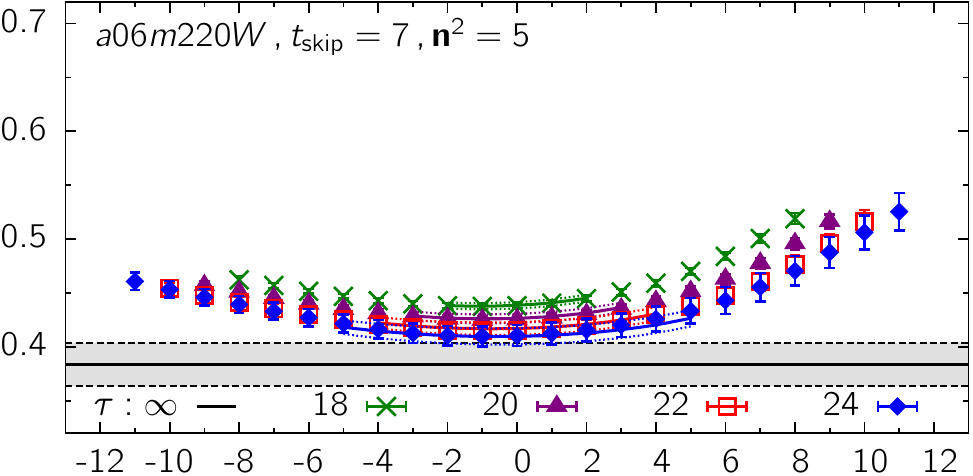}
}
\subfigure{
\includegraphics[width=0.42\linewidth]{Figs_esc/ratio_Vi_qsq1_a06m220}
\includegraphics[width=0.42\linewidth]{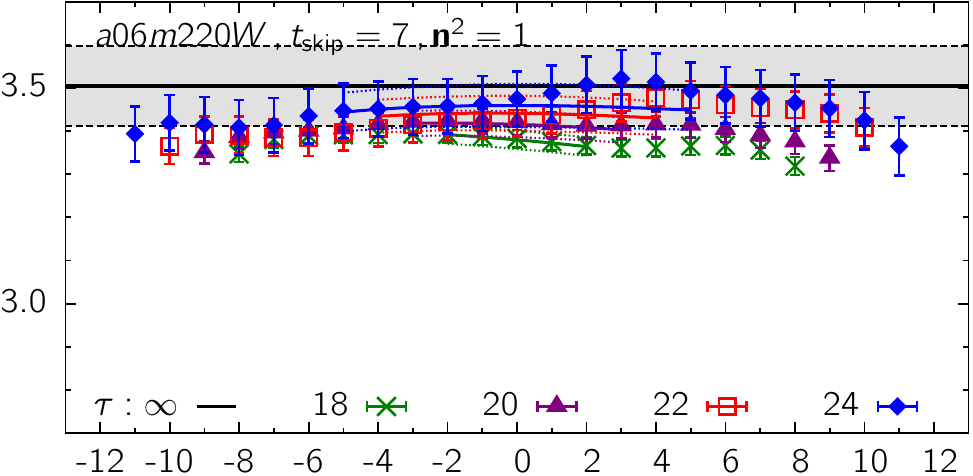}
}
\subfigure{
\includegraphics[width=0.42\linewidth]{Figs_esc/ratio_Vi_qsq5_a06m220}
\includegraphics[width=0.42\linewidth]{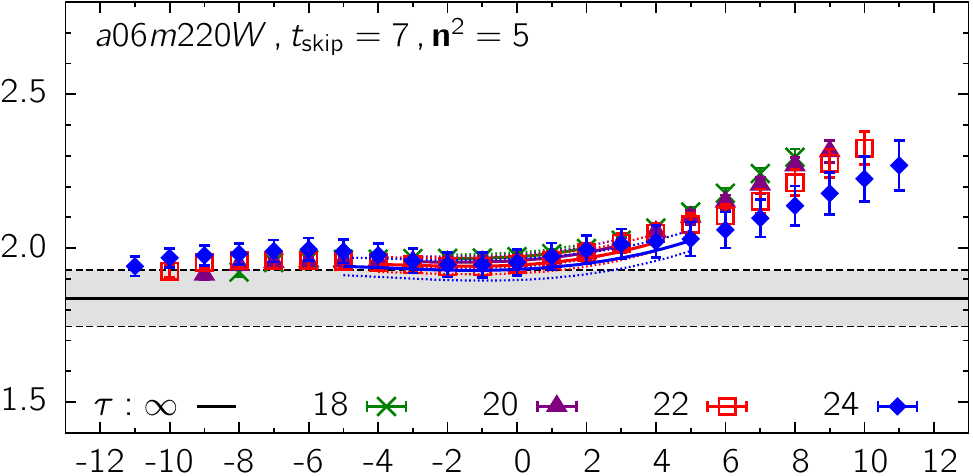}
}
\caption{\FIXME{fig:GM-ESC-smear220} Comparison of the signal and ESC
  versus the smearing size on the $a06m220$ ensemble. Plots on the
  left are with the smearing parameter $\sigma = 5.5$ and on the right
  with $\sigma=11$ as defined in
  Table~\protect\ref{tab:cloverparams}. The rest is the same as in
  Fig.~\protect\ref{fig:GM-ESC-smear310}.}
\label{fig:GM-ESC-smear220}
\end{figure*}

In this Appendix, we show the data and  
the $3^\ast$-state fits used to control the ESC in the
extraction of the electric and magnetic form factors. There are three sets of 
figures: 
\begin{itemize}
\item
The comparison of the ESC on the various ensembles and at different values of 
the momenta are shown in 
Figs.~\ref{fig:GE-ESC-a09m130W}--\ref{fig:GM-ESC-p5}.
\item
The improvement
in the quality of the signal with increase in the lattice size $L$ is shown in
Fig.~\ref{fig:GM-ESC-vol} using data from for the three ensembles
$a12m220L$, $a12m220$ and $a12m220S$. A study of finite size effects
in the form factors using these three ensembles are examined in
Fig.~\ref{fig:GE-GM-vsV}, and in the extraction of $\rEsq$ and $\rMsq$
using the dipole, $z^4$ and $z^{5+4}$ fits in Fig.~\ref{fig:Vol-rErM}.
\item
A comparison of the ESC with two different smearing sizes is shown in
Figs.~\ref{fig:GM-ESC-smear310} and~\ref{fig:GM-ESC-smear220} for the
$a06m310$ and $a06m220$ ensembles, respectively.
\end{itemize}

%


\begin{figure*}   
\centering
\subfigure{
\includegraphics[width=0.47\linewidth]{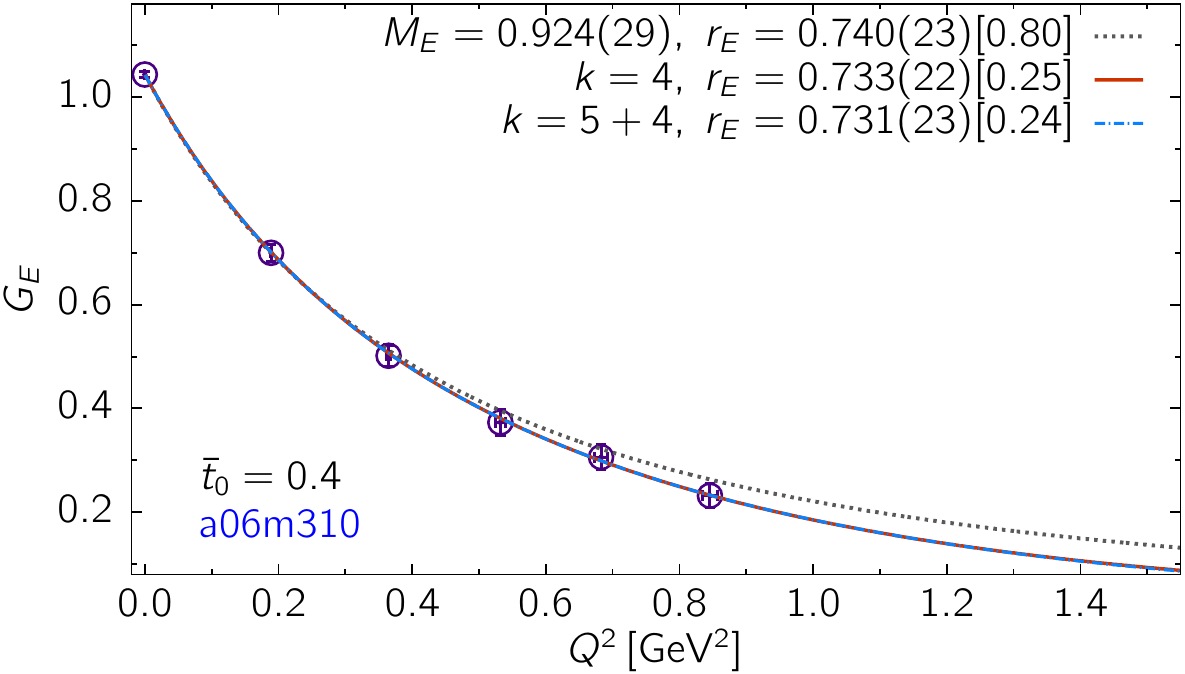}
\includegraphics[width=0.47\linewidth]{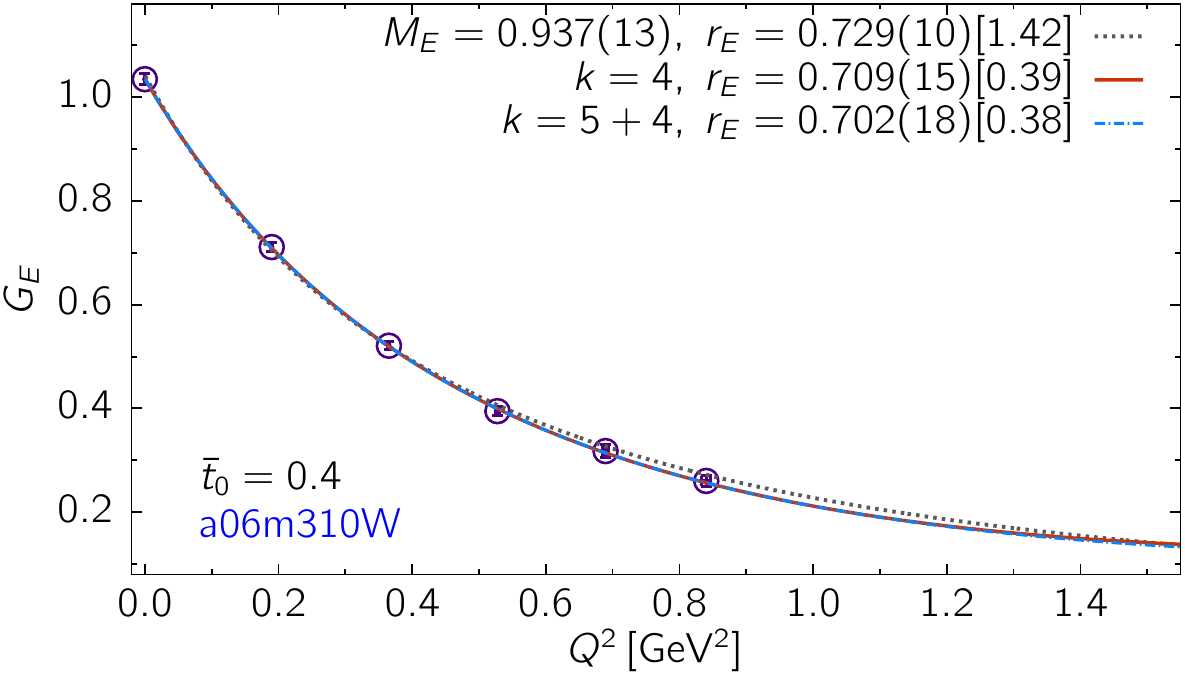}
}
\subfigure{
\includegraphics[width=0.47\linewidth]{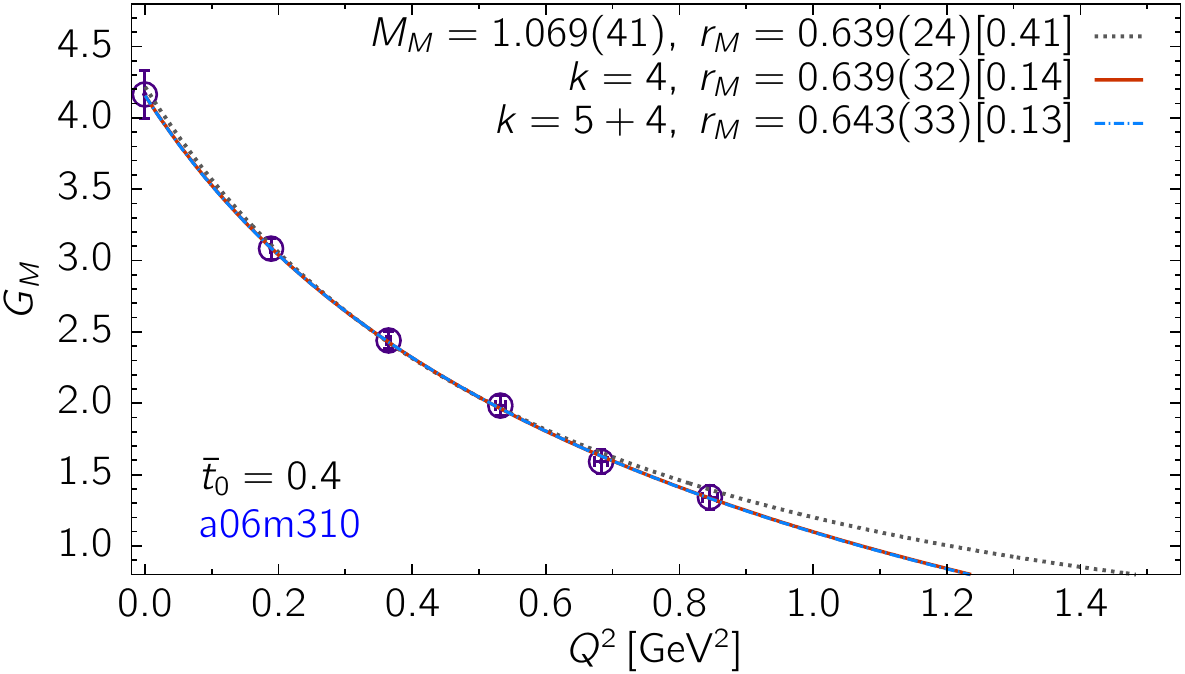}
\includegraphics[width=0.47\linewidth]{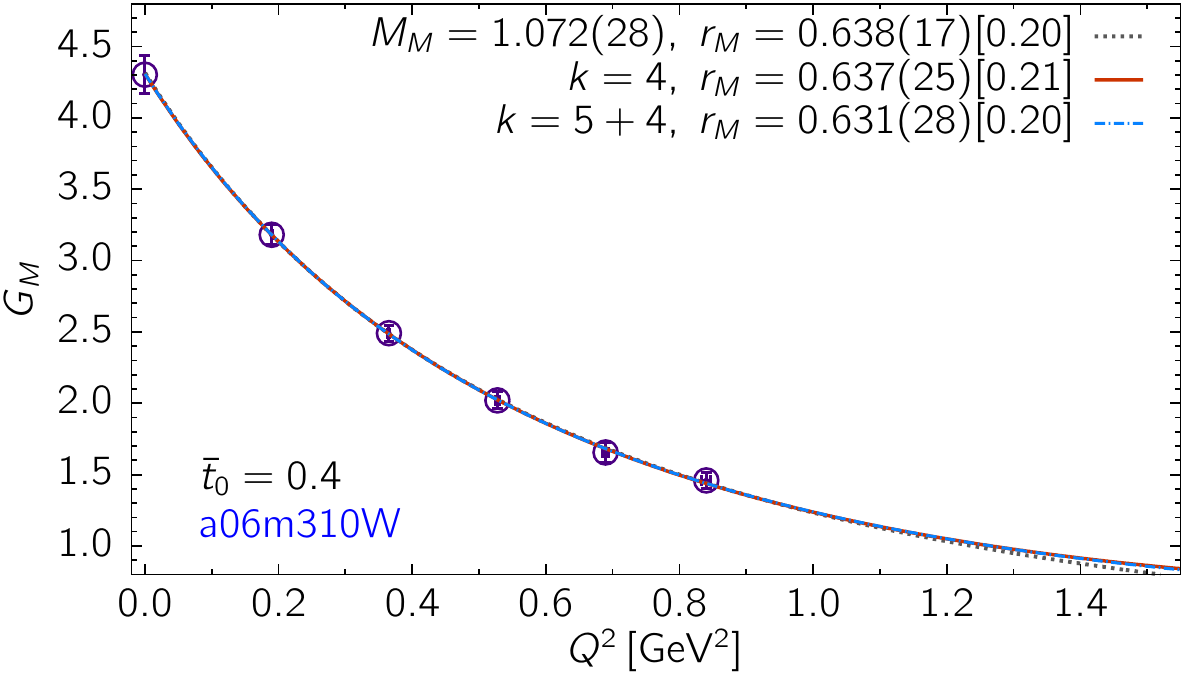}
}
\subfigure{
\includegraphics[width=0.47\linewidth]{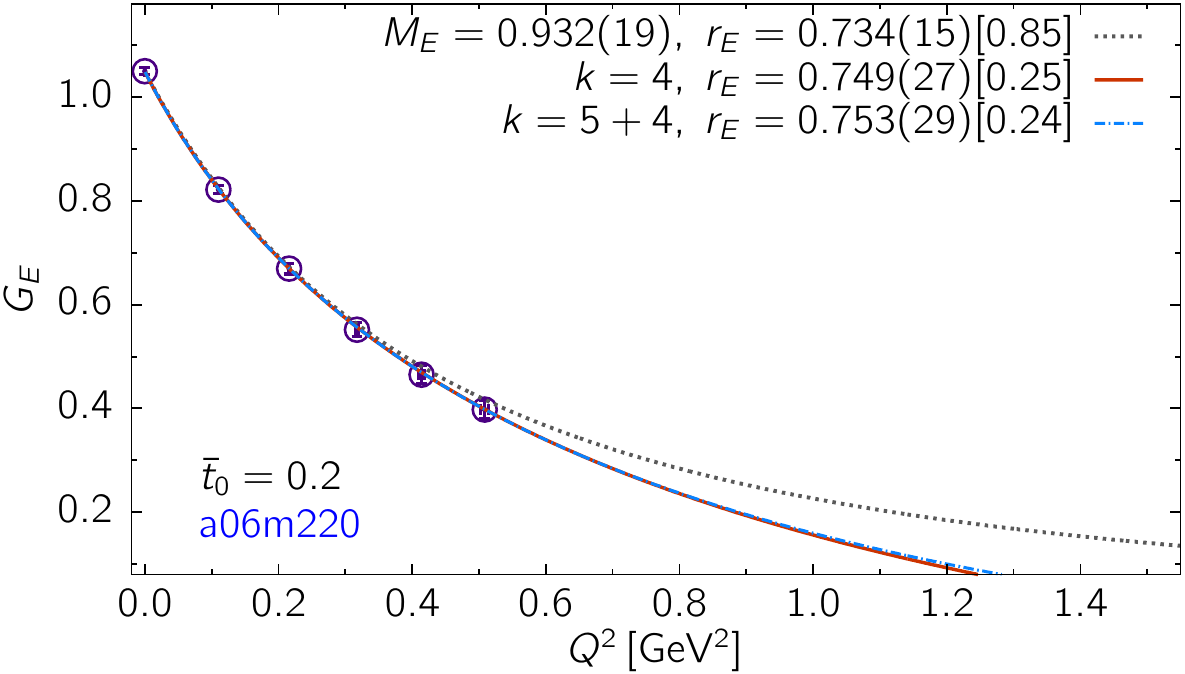}
\includegraphics[width=0.47\linewidth]{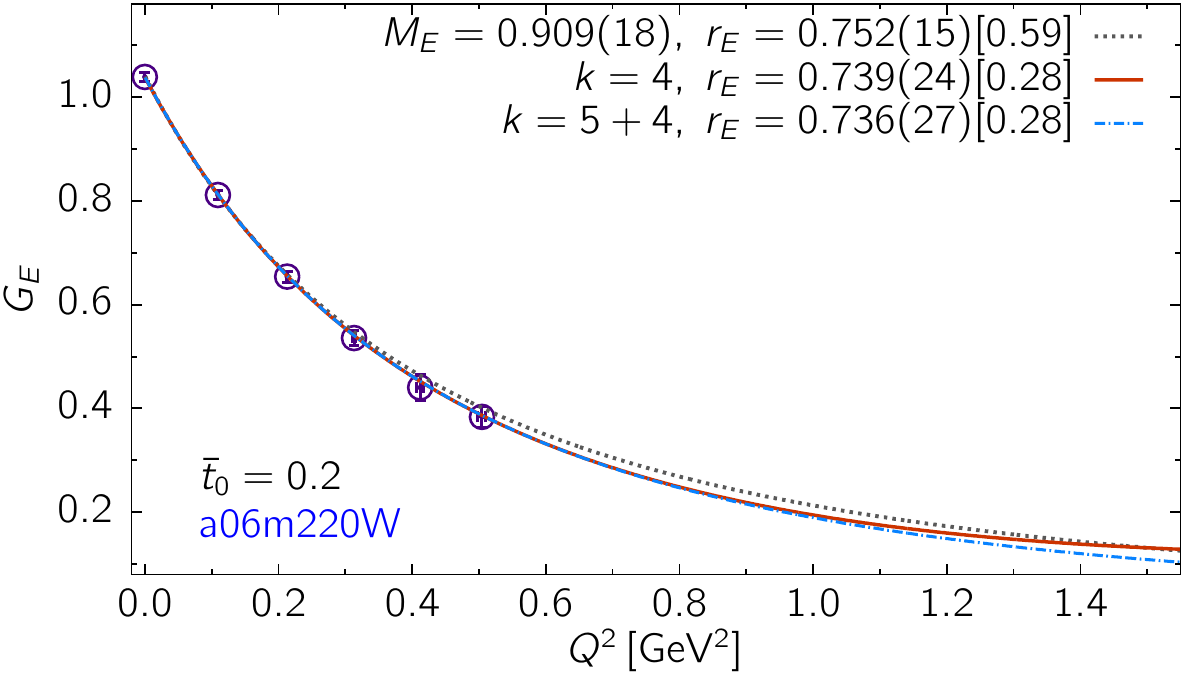}
}
\subfigure{
\includegraphics[width=0.47\linewidth]{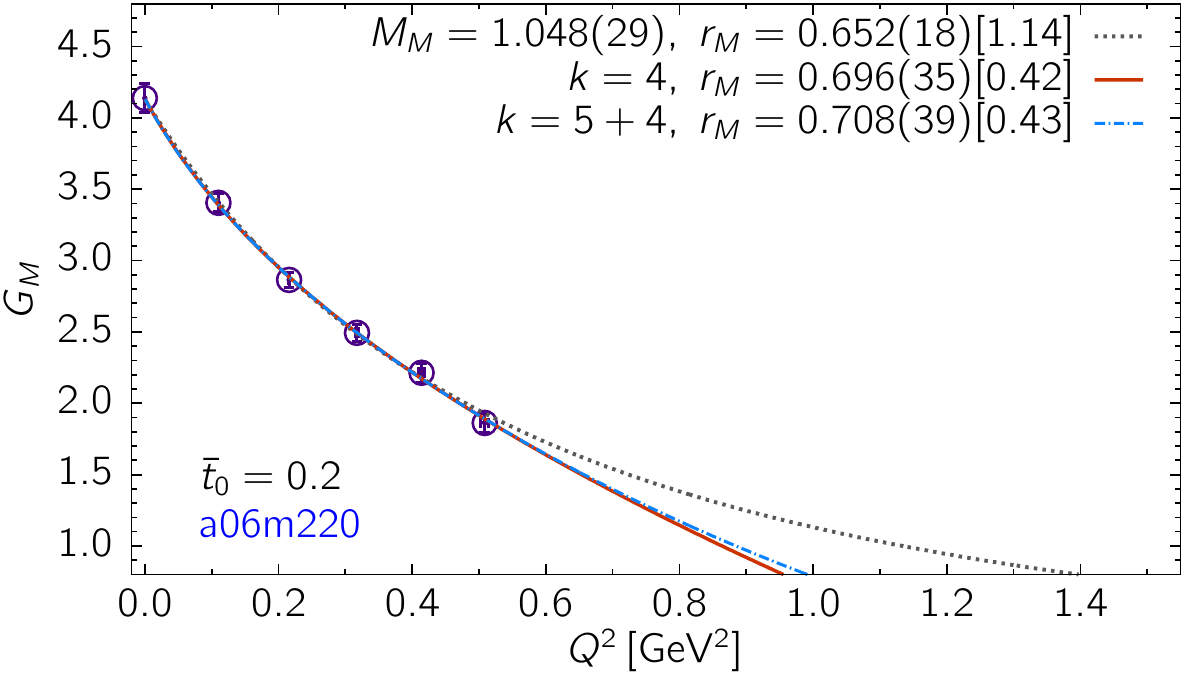}
\includegraphics[width=0.47\linewidth]{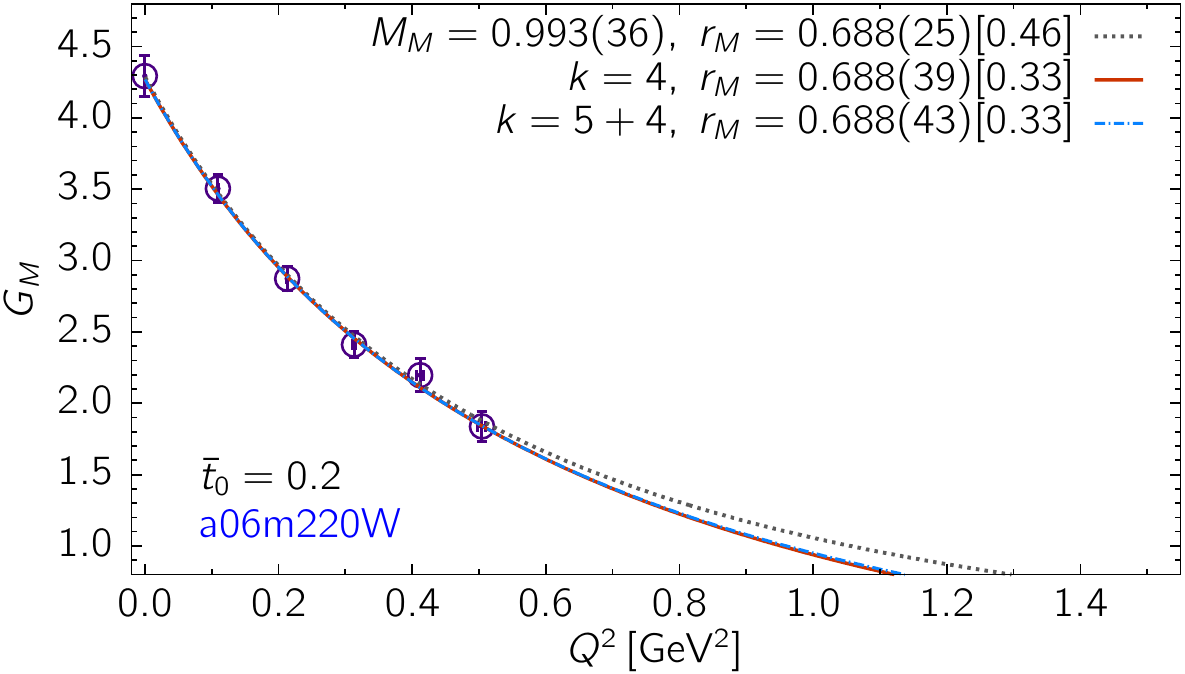}
}
\caption{\FIXME{fig:Smear-rErM} Comparison of results of the dipole,
  $z^4$ and $z^{5+4}$ fits to the unrenormalized isovector form
  factors $G_E(Q^2)$ and $G_M(Q^2)$ versus $Q^2$ in units of GeV${}^2$
  for the two different Gaussian smearing sizes given in
  Table~\ref{tab:cloverparams}. The top two rows show data from the
  $a06m310$ ensemble and the bottom two rows from the $a06m220$
  ensemble. In each row, the panels on the right show the data with
  the larger smearing size. The radii $\rE$ and $\rM$ are in units of fm and the masses $M_E$ and $M_N$ in GeV.}
\label{fig:Smear-rErM}
\end{figure*}
%


\section{Summary of Experimental form factors}
\label{appendix:expFF}

\begin{figure*}[tpb] 
\centering
\subfigure{
\includegraphics[width=0.47\linewidth]{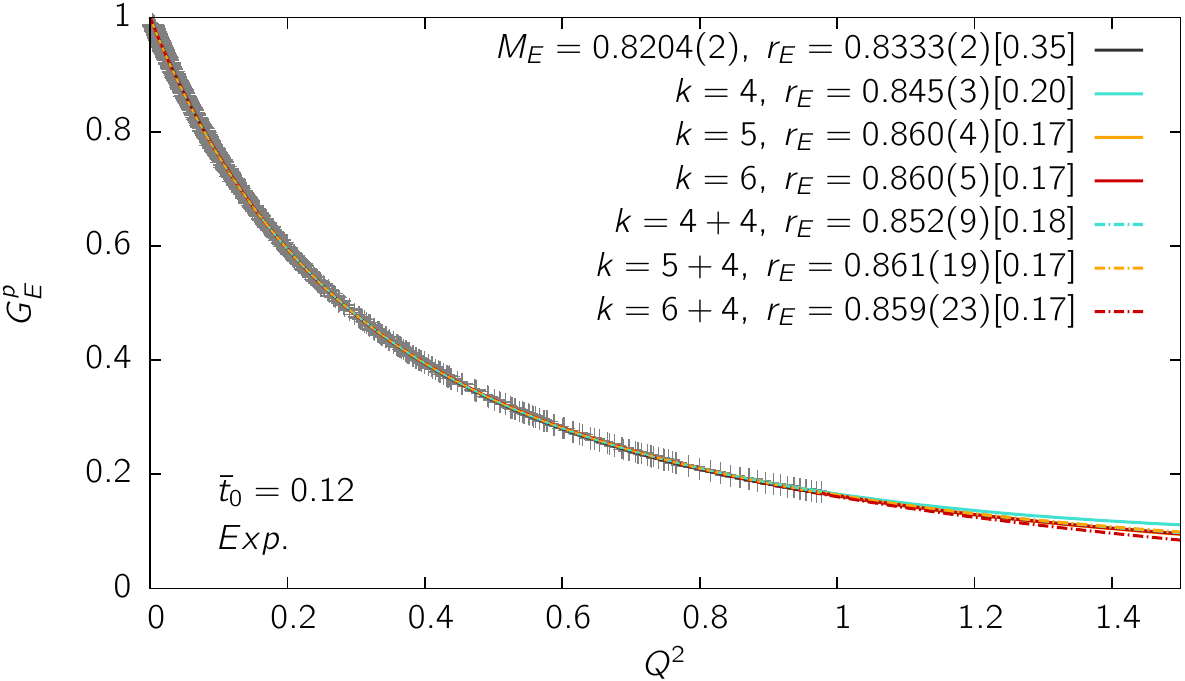}
\includegraphics[width=0.47\linewidth]{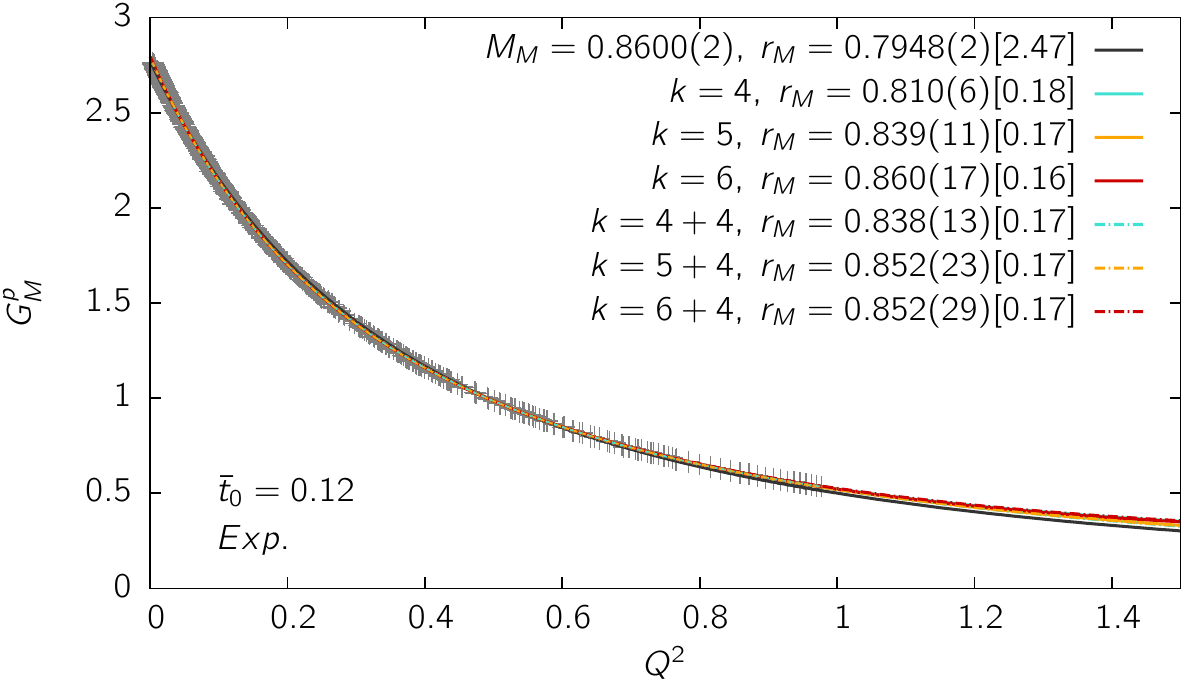}
}
\subfigure{
\includegraphics[width=0.47\linewidth]{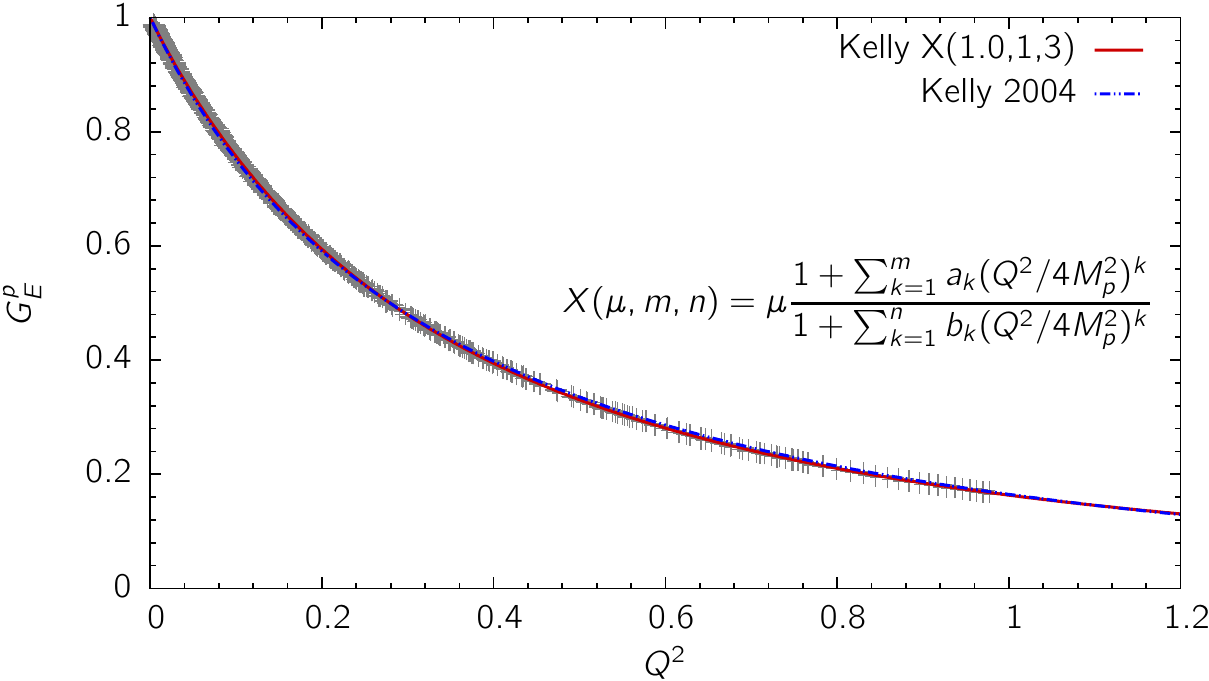}
\includegraphics[width=0.47\linewidth]{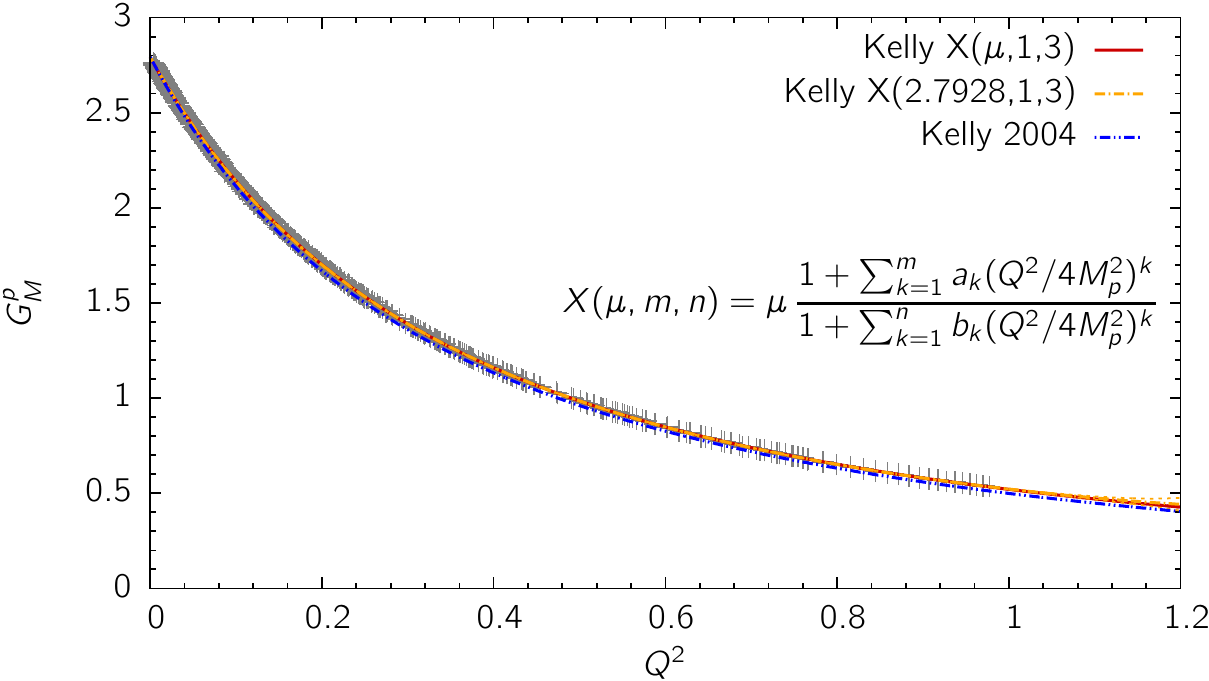}
}
\caption{\FIXME{fig:FFproton} The experimental data for the electric
  (left) and magnetic (right) form factors, $G_E^p(Q^2)$ and
  $G_M^p(Q^2)$, for the proton are plotted versus $Q^2$. These
  data~\cite{Yan:2018bez} are a rebinned version of the data from the
  A1 Collaboration at Mainz~\protect\cite{Bernauer:2013tpr} provided
  by Douglas Higinbotham~\protect\cite{Yan:2018bez}. The top row shows
  the results of the seven fits used by us to analyze the lattice
  data.  The bottom row shows the same data fit with the
  Kelly parameterization where  ``Kelly 2004'' refers to using the 
  parameters given in Ref.~\protect\cite{Kelly:2004hm}.  }
\label{fig:FFproton}
\end{figure*}

\begin{figure*}[tpb] 
\centering
\subfigure{
\includegraphics[width=0.47\linewidth]{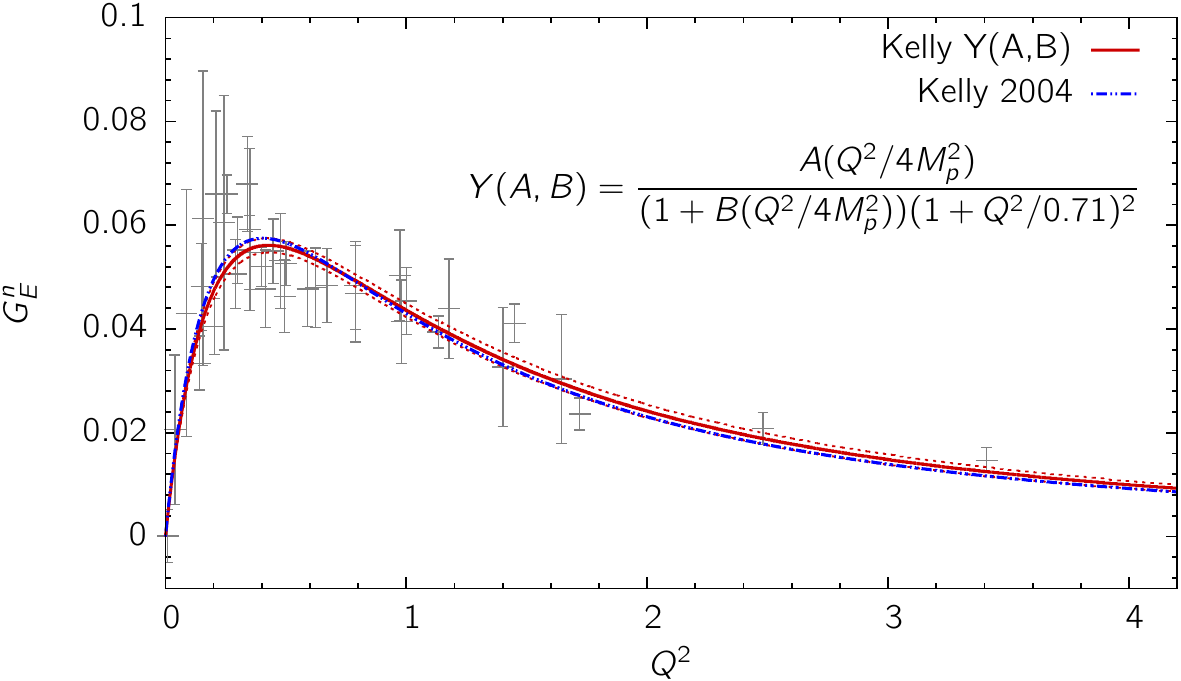}
\includegraphics[width=0.47\linewidth]{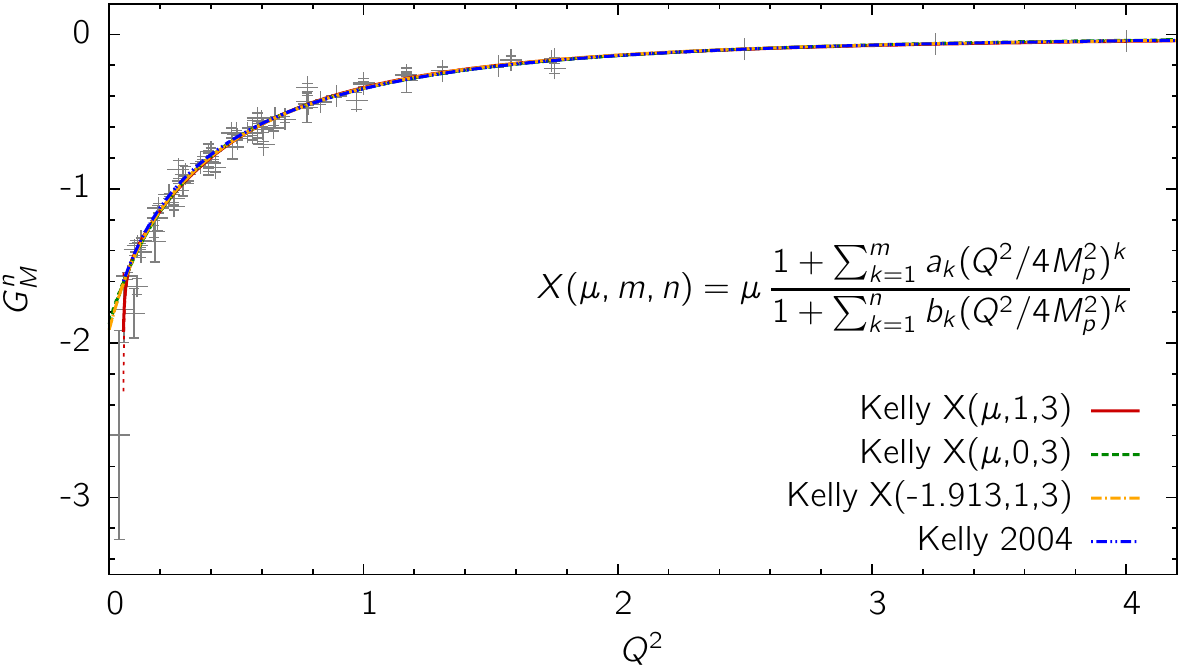}
}
\caption{\FIXME{fig:FFneutron} The data for the electric (left) and
  magnetic (right) form factors of the neutron, $G_E^n(Q^2)$ and
  $G_M^n(Q^2)$, plotted versus $Q^2$ (GeV${}^2$).  The $G_E^n(Q^2)$ data are
  compiled from
  Refs.~\protect\cite{Gentile:2011zz,Schiavilla:2001qe,Sulkosky:2017prr},
  and the $G_M^n(Q^2)$ data from
  Refs.~\protect\cite{Anderson:2006jp,Anklin:1994ae,Anklin:1998ae,Bartel:1973rf,Bartel:1972xq,Bartel:1972xq,Bermuth:2003qh,Bruins:1995ns,Budnitz:1969dt,Eden:1994ji,Esaulov:1987uc,Gao:1994ud,Golak:2000nt,Glazier:2004ny,Hanson:1973vf,Herberg:1999ud,Kubon:2001rj,Madey:2003av,Markowitz:1993hx,Meyerhoff:1994ev,Ostrick:1999xa,Passchier:1999cj,Plaster:2005cx,Rohe:1999sh,Stein:1966ke,Warren:2003ma,Zhu:2001md}.
Also shown are the fits with the Kelly parameterization. 
}
\label{fig:FFneutron}
\end{figure*}

In this Appendix, we collect in one place the experimental data for
the form factors for the proton and the neutron.  In
Fig.~\ref{fig:FFproton}, we show the data for $G_E^p(Q^2)$ and
$G_M^p(Q^2)$ compiled by Douglas Higinbotham~\cite{Yan:2018bez,Alarcon:2018zbz,Higinbotham:2019jzd}
from the cross sections provided in the Lee-Arlington-Hill supplemental material~\cite{Lee:2015jqa}, who
rebinned the original data obtained by the A1 Collaboration using the
MAMI beam at Mainz~\cite{Bernauer:2010wm,Bernauer:2013tpr}. The neutron data,
$G_E^n(Q^2)$, are collected from
Refs.~\cite{Gentile:2011zz,Schiavilla:2001qe,Sulkosky:2017prr}, and
$G_M^n(Q^2)$ from
Refs.~\cite{Anderson:2006jp,Anklin:1994ae,Anklin:1998ae,Bartel:1973rf,Bartel:1972xq,Bermuth:2003qh,Bruins:1995ns,Budnitz:1969dt,Eden:1994ji,Esaulov:1987uc,Gao:1994ud,Golak:2000nt,Glazier:2004ny,Hanson:1973vf,Herberg:1999ud,Kubon:2001rj,Madey:2003av,Markowitz:1993hx,Meyerhoff:1994ev,Ostrick:1999xa,Passchier:1999cj,Plaster:2005cx,Rohe:1999sh,Stein:1966ke,Warren:2003ma,Zhu:2001md}.
These are shown in Fig.~\ref{fig:FFneutron}. From these data, we 
evaluate the isovector form factors $G_E^{p-n}(Q^2)$ and
$G_M^{p-n}(Q^2)$ to which our lattice data are compared. 

The construction of the isovector form factors is done as follows: we
first fit the four sets of experimental data for $Q^2 \lesssim
1$~GeV${}^2$ using the Kelly parameterization as shown in
Figs.~\ref{fig:FFproton} and~\ref{fig:FFneutron}. Using the resulting
Kelly fits, we then construct the isovector combinations,
$G_E^p-G_E^n$ and $G_M^p - G_M^n$.  This parameterization is used throughout
the paper to compare the lattice data against.  From this procedure we
get
\begin{align}
r_E^{p-n}|_{\rm exp} &= 0.929(27) \,, \nonumber \\
r_M^{p-n}|_{\rm exp} &= 0.849(11) \,.
\label{eq:isovectorradii}
\end{align}
whereas, using the parameter values given in the original Kelly fit~\cite{Kelly:2004hm} gives
\begin{align}
r_E^{p-n}|_{\rm exp} &= 0.926(4) \,, \nonumber \\
r_M^{p-n}|_{\rm exp} &= 0.872(7) \,.
\label{eq:isovectorradiiKelly}
\end{align}
Lattice results for the isovector combination of the radii should be
compared to the values given in Eq.~\eqref{eq:isovectorradii}.  Note
that our less sophisticated analysis, which is also used to analyze the
lattice data, gives larger errors.

The results of the dipole fits to the proton data shown in
Fig.~\ref{fig:FFproton} give $r_E^p \sim 0.833$ and $r_M^p \sim
0.795$, which are roughly consistent with the careful analysis of
electron-experiment results~\cite{Yan:2018bez} given in
Eqs.~\eqref{eq:rE_expt} and~\eqref{eq:rM_expt} and the Kelly fits
shown in the bottom row of Fig.~\ref{fig:FFproton}.  Overall, the
dipole ansatz does a good job of fitting the experimental $G_E(Q^2)$
data, and the deviation is less than 1\% for $Q^2 < 1$~GeV${}^2$. The
dipole fit to $G_M(Q^2)$ is less good as shown by the large
$\chi^2/DOF$.

\begin{figure*}[tpb]         
\centering
\subfigure{
\includegraphics[width=0.32\linewidth]{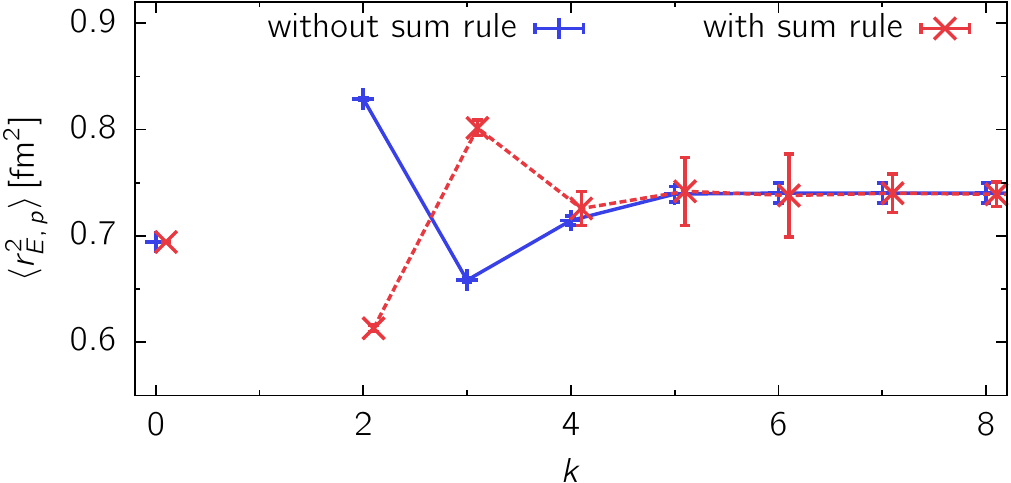}
\includegraphics[width=0.32\linewidth]{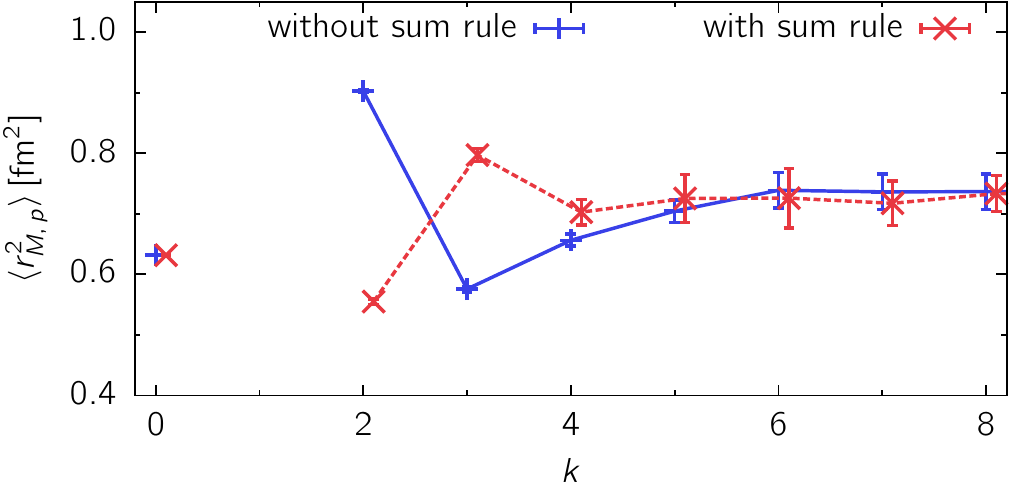}
\includegraphics[width=0.32\linewidth]{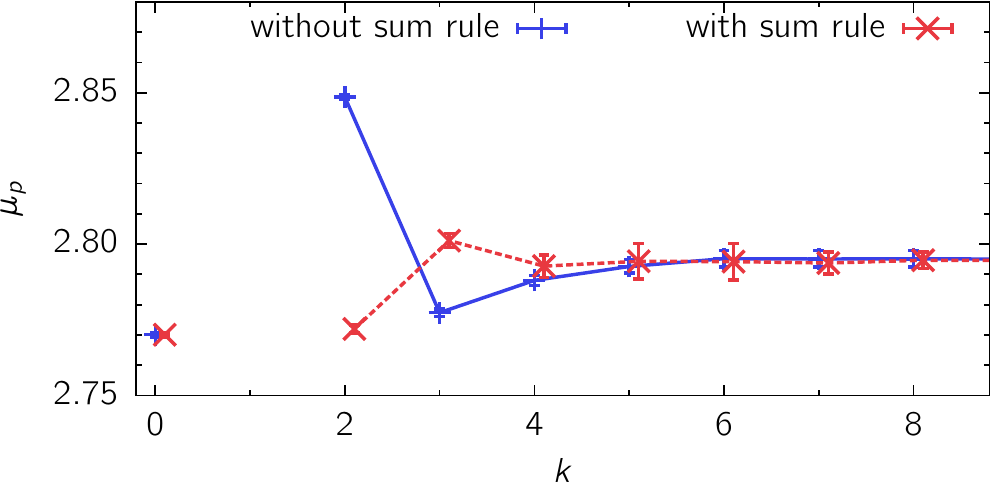}
}
\caption{\FIXME{fig:stabilityEXP} Estimates of $\rEsq$, $\rEsq$ and
  $\mu$, extracted from the experimental data for the proton, as a function of the order $z^k$ (blue) and
  $z^{k+4}$ (red) of the truncation of the $z$-expansion.  The dipole
  results is shown at $k=0$.  }
\label{fig:stabilityEXP}
\end{figure*}

The convergence of the $z$-expansion fits, with constraints on the
$a_k$, versus $k$ is shown in Fig.~\ref{fig:stabilityEXP}. Estimates
with $k \ge 5$ are stable for all three quantities.The results from
$z$-expansion fits, also shown in Fig.~\ref{fig:FFproton}, are
marginally larger than those from the dipole and differ by a few
percent from those in Eqs.~\eqref{eq:rE_expt}
and~\eqref{eq:rM_expt}. The errors in the $z$-expansion estimates are
larger, especially with the inclusion of the sum rules.  The overall
lesson from this exercise is that an uncertainty of $O(5\%)$ could be present 
in our analysis using either the dipole or the $z$-expansion
fits.

\begin{acknowledgments}
We thank the MILC Collaboration for providing the 2+1+1-flavor HISQ
lattices used in our calculations. The calculations used the Chroma
software suite~\cite{Edwards:2004sx}. R. Gupta thanks D. Higinbotham
for discussions and for providing the experimental data on the form
factors.  Simulations were carried out on computer facilities of (i)
the National Energy Research Scientific Computing Center, a DOE Office
of Science User Facility supported by the Office of Science of the
U.S. Department of Energy under Contract No. DE-AC02-05CH11231; and,
(ii) the Oak Ridge Leadership Computing Facility at the Oak Ridge
National Laboratory, which is supported by the Office of Science of
the U.S. Department of Energy under Contract No. DE-AC05-00OR22725;
(iii) the USQCD Collaboration, which are funded by the Office of
Science of the U.S. Department of Energy, and (iv) Institutional
Computing at Los Alamos National Laboratory.  T. Bhattacharya and
R. Gupta were partly supported by the U.S. Department of Energy,
Office of Science, Office of High Energy Physics under Contract
No.~DE-AC52-06NA25396.  T. Bhattacharya, R. Gupta, J.-C. Jang and
B.Yoon were partly supported by the LANL LDRD program.  The work of
H.-W. Lin is supported by the US National Science Foundation under
grant PHY 1653405 ``CAREER: Constraining Parton Distribution Functions
for New-Physics Searches''.
\end{acknowledgments}

\clearpage
%
\bibliography{ref} 

\end{document}